\documentclass[]{jkpaper}
\usepackage{tikz}
\usetikzlibrary{decorations.markings}
\usetikzlibrary{decorations.pathreplacing}
\usetikzlibrary{arrows.meta}
\usetikzlibrary{patterns}
\usetikzlibrary{decorations.pathmorphing}
\usepackage{tikz-cd}
\usepackage{bigints}
\usepackage{biblatex}
\usepackage{aas_macros}
\usepackage[outline]{contour}
\usepackage{hyperref}
\usepackage{url}
\usepackage{amssymb}
\usepackage{soul}

\theoremstyle{plain}

\theoremstyle{definition}

\newtheorem*{eg*}{Example}
\newtheorem{eg}{Example}

\newcommand{\OIST}{\raisebox{-0.08em}{\includegraphics[height=0.8em]{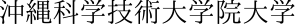}}}

\newcommand{\eps}{\varepsilon}

\DeclareMathOperator*{\SumInt}{
\mathchoice
{\ooalign{$\displaystyle\sum$\cr\hidewidth$\displaystyle\int$\hidewidth\cr}}
  {\ooalign{\raisebox{.14\height}{\scalebox{.7}{$\textstyle\sum$}}\cr\hidewidth$\textstyle\int$\hidewidth\cr}}
  {\ooalign{\raisebox{.2\height}{\scalebox{.6}{$\scriptstyle\sum$}}\cr$\scriptstyle\int$\cr}}
  {\ooalign{\raisebox{.2\height}{\scalebox{.6}{$\scriptstyle\sum$}}\cr$\scriptstyle\int$\cr}}
}

\newcommand{\stkout}[1]{\ifmmode\text{\sout{\ensuremath{#1}}}\else\sout{#1}\fi}

\definecolor{JulianBlue}{RGB}{30,100,200}
\definecolor{JulianRed}{RGB}{200,0,25}
\definecolor{JulianPurple}{RGB}{118,21,200}

\newcommand{\Observerdependence}{Observer\texorpdfstring{\textcolor{black!40!white}{-}}{-}dependence}
\newcommand{\observerdependent}{observer\texorpdfstring{\textcolor{black!40!white}{-}}{-}dependent}
\newcommand{\observerdependence}{observer\texorpdfstring{\textcolor{black!40!white}{-}}{-}dependence}

\title{Crossed products and quantum reference frames:\texorpdfstring{\\}{ }on the \observerdependence{} of gravitational entropy}
\author{Julian De Vuyst\texorpdfstring{\textsuperscript{1,a}}{}, Stefan Eccles\texorpdfstring{\textsuperscript{1,b}}{}, Philipp A.\ H\"ohn,\texorpdfstring{\textsuperscript{1,c}}{} and Josh Kirklin\texorpdfstring{\textsuperscript{1,2,d}}{}}
\institution{\texorpdfstring{\textsuperscript{1}}{}Qubits and Spacetime Unit,\texorpdfstring{\\}{ } Okinawa Institute of Science and Technology \emph{(}\OIST\emph{)},\texorpdfstring{\\}{ } 1919-1 Tancha, Onna-son, Kunigami-gun, Okinawa, Japan 904-0495
\texorpdfstring{\\\vspace*{1.5em}}{ } \texorpdfstring{\textsuperscript{2}}{}Perimeter Institute for Theoretical Physics,\\ 31 Caroline Street North, Waterloo, ON, N2L 2Y5, Canada}

\email{
\textsuperscript{a}\emaillink{julian.devuyst@oist.jp}\\
\textsuperscript{b}\emaillink{stefan.eccles@oist.jp}\\
\textsuperscript{c}\emaillink{philipp.hoehn@oist.jp}\\
\textsuperscript{d}\emaillink{jkirklin@pitp.ca}
}

\abstr{
A significant step towards a rigorous understanding of perturbative gravitational entropy was recently achieved by a series of works showing that a proper accounting of gauge invariance and observer degrees of freedom converts the Type III algebra of QFT observables in a gravitational subregion to a Type II crossed product, whose entropy reduces to the generalised entropy formula in a semiclassical limit. The observers thus used are also known as quantum reference frames (QRFs); as noted in our companion work~\cite{DeVuyst:2024pop}, using different QRFs result in different algebras, and hence different entropies -- so gravitational entropy is \observerdependent{}. Here, we provide an in-depth analysis of this phenomenon, with full derivations of many new results. Using the perspective-neutral QRF formalism, we extend previous constructions to allow for arbitrarily many observers, each carrying a clock with possibly degenerate energy spectra. We consider a semiclassical regime characterized by clocks whose energy fluctuations dominate over the fluctuations of the energy of the QFT. Unlike previous works, we allow the clocks and fields to be arbitrarily entangled. At leading order the von Neumann entropy still reduces to the generalised entropy, but linear corrections are typically non-vanishing and quantify the degree of entanglement between the clocks and fields. We also describe an `antisemiclassical' regime as the opposite of the semiclassical one, with suppressed fluctuations of the clock energy; in this regime, we show how the clock may simply be `partially traced' out when evaluating the entropy. Four explicit examples of \observerdependent{} entropy are then given, involving a gravitational interferometer, degenerate clock superselection, a semiclassical approximation applying to some clocks but not others, and differences between monotonic and periodic clocks.
}

\bibliography{refs}
\begin{document}
\maketitleandtoc
\section{Introduction}
\label{sect:intro}

Geometry and information are deeply intertwined. Nowhere is this more apparent than in the generalised entropy formula, which says that the entropy of a gravitational subregion is equal to $\frac{A}{4G_N}+S_{\text{inside}}$, where $A$ is the area of the boundary of the subregion, $G_N$ is Newton's constant, and $S_{\text{inside}}$ is the entropy of any degrees of freedom contained inside the subregion. Since its thermodynamical conceptions~\cite{Bekenstein:1972tm,Hawking:1975vcx}, this formula has been a guiding principle in quantum gravity (see~\cite{Ryu_2006,Lewkowycz_2013,Faulkner_2013,Engelhardt_2015,Strominger_1996,Meissner_2004} for a very small selection of results it has inspired). But it is misleadingly simple, for two reasons:
\begin{itemize}
    \item $S_{\text{inside}}$ is not well-defined, because the UV modes of the fields inside of the subregion contribute infinitely to it. More precisely, the von Neumann algebra of field operators with support in a subregion is Type III${}_1$, and thus has no well-defined von Neumann entropy functional~\cite{WittenRevModPhys.90.045003,Sorce:2023fdx,Takesaki1979}. One may impose a UV cutoff to convert this algebra to something of Type I, which has well-defined von Neumann entropies. But a simple UV cutoff would ultimately be incompatible with general covariance in a full theory of gravity, and the UV cutoff cannot be removed without $S_{\text{inside}}$ diverging. On the other hand, it was shown in~\cite{Susskind_1994,jacobson1994blackholeentropyinduced,FROLOV1997339,Frolov_1998} that a renormalisation of $G_N$ would allow the generalised entropy itself to be UV finite -- the divergent piece of the $S_{\text{inside}}$ term would be canceled by a similar divergence in the area term. This strongly indicates that the generalised entropy can be rigorously formulated without imposing a UV cutoff. But the underlying mechanism dictating how the appropriate renormalisation would be done has remained unclear.
    \item Perhaps more fundamentally, a proper definition of the physical subsystem associated with a gravitational subregion must take into account diffeomorphism invariance. Indeed, if one is not careful about this, diffeomorphisms could move degrees of freedom in and out of the subregion, which would be incompatible with the notion of a definite physical subsystem comprised of such degrees of freedom. The generalised entropy formula must ultimately be interpreted as giving the entropy of such a subsystem -- but much of the time it is employed without being very precise about this subtlety.
\end{itemize}
Resolving these issues is crucial not only for making sense of entropy in gravitational systems, but also for understanding the fundamental structure of spacetime in a quantum context. At first glance, the two problems seem to be more or less independent, but recent progress has shown that they are in fact tightly linked together by the notion of \emph{quantum reference frames} (QRFs), which simultaneously regularise the entropy and account for diffeomorphism invariance. Let us give a brief account of how this came about.

First, in~\cite{leutheusser2023causal,leutheusser2023emergent}, it was shown how an effective Type III${}_1$ algebra emerges from the large $N$ limit of the set of single trace operators acting on one of two copies of a holographic CFT in a thermofield double state; this was interpreted as the algebra of effective field theory operators with support on one side of the bulk dual black hole. Then,~\cite{Witten:2021unn} demonstrated that one may consistently also include the CFT Hamiltonian in the set of operators, and that doing so modifies the emergent algebra to be Type II${}_\infty$. The bulk interpretation of this construction is that one is taking the so-called \emph{crossed product} of the field theory algebra by the action of a Killing boost preserving the black hole horizon. This boost corresponds to the `modular flow' of the bulk vacuum state, which is what is responsible for the type conversion~\cite{Takesaki1979,Connes}. Unlike Type III algebras, Type II algebras \emph{do} have a well-defined entropy functional, at least up to a state-independent constant.

Next, it was explained in~\cite{Chandrasekaran:2022cip} how this construction extends beyond the holographic context; in particular, that paper considered a system consisting of an \emph{observer} carrying a clock inside of a static patch of de Sitter spacetime. The algebra of operators which act on the clock and the fields in the static patch, and which are invariant under the global de Sitter time evolution (which is a gravitational gauge symmetry) was shown to be given by the crossed product of the Type III${}_1$ static patch field algebra with respect to the action of this time evolution. Again, this is Type II${}_\infty$, which renders the entropy well-defined up to a state-independent constant (it was also shown that imposing a lower bound on the allowed energy of the clock further reduces the algebra to be Type II${}_1$, which can be understood as the condition that the algebra admits a maximally entangled state). Remarkably, it was then shown that (in a certain class of states consistent with a semiclassical approximation) the entropy functional for this algebra \emph{agrees} with the generalised entropy formula. Since~\cite{leutheusser2023causal,leutheusser2023emergent,Witten:2021unn,Chandrasekaran:2022cip}, there has been a flurry of follow-up works exploring extensions and applications of these results (see~\cite{Chandrasekaran_2023,Kudler-Flam:2024psh,Jensen:2023yxy,Ali:2024jkx,Klinger:2023tgi,Faulkner:2024gst,AliAhmad:2023etg,KirklinGSL,Kudler-Flam:2023qfl,Witten:2023xze, Kudler-Flam:2023hkl,Gesteau:2023hbq,Balasubramanian:2023dpj,Soni:2023fke,AliAhmad:2023etg,Aguilar-Gutierrez:2023odp,Gomez:2023wrq,Gomez:2023upk,Klinger:2023tgi,Klinger:2023auu} for a selection).

Thus, accounting for invariance under certain gravitational gauge symmetries kills two birds with one stone. One gets a rigorous and well-defined entropy functional that reproduces the generalised entropy (up to a state-independent constant), and one has a physical subsystem that is invariant under (a restricted set\footnote{In this paper, we will focus only on a single diffeomorphism, as was also done in the previous works we have described. The question of what should be done with the other diffeomorphisms is an important one. The situation simplifies somewhat when one considers perturbations around a spatially compact background with Killing symmetries which fix the subregion under consideration (such as the static patch in de Sitter spacetime), where, as we explained in~\cite{DeVuyst:2024grw}, linearisation stability conditions \cite{Deser:1973zza,Moncrief1,Moncrief:1976un,Moncrief:1978te,Moncrief:1979bg,Arms1,Arms:1979au,fischer1980structure,Arms:1982ea,Saraykar:1981qm,Higuchi:1991tk,Higuchi:1991tm,Losic:2006ht} play a key role. But, in the case of a general subregion in a general spacetime, this issue has not yet been fully addressed (though see \cite{Jensen:2023yxy,DeVuyst:2024grw} for some discussion).} of) diffeomorphisms.

In this paper, we will study how this construction depends on the properties of the clock used in the crossed product. As we first pointed out in~\cite{DeVuyst:2024pop} (see also~\cite{Fewster:2024pur,AliAhmad:2024wja}), such a clock is a QRF. These are dynamical objects constructed out of quantum degrees of freedom. They are essential for understanding subsystems in quantum theories with local symmetries, such as quantum gravity, where the archetypal QRFs are quantum clocks and rods~\cite{Goeller:2022rsx,Carrozza:2022xut,Carrozza:2021gju,Kabel:2024lzr,Kabel:2023jve,Susskind:2023rxm,Fewster:2024pur,Gomez:2023upk}. Various formalisms for working with QRFs have emerged~\cite{Aharonov:1967zza,Aharonov:1984,angeloPhysicsQuantumReference2011a,Giacomini:2017zju,delaHamette:2021oex,Hoehn:2023ehz,Hoehn:2019fsy,Hoehn:2020epv,Chataignier:2024eil,AliAhmad:2021adn,Hoehn:2021flk,delaHamette:2021piz,Vanrietvelde:2018pgb,Vanrietvelde:2018dit,Giacomini:2021gei,Castro-Ruiz:2019nnl,Suleymanov:2023wio,Carrozza:2024smc,Krumm:2020fws,delaHamette:2020dyi,Kabel:2024lzr,Kabel:2023jve,Carette:2023wpz,Loveridge:2019phw,loveridgeSymmetryReferenceFrames2018a,Bartlett:2006tzx,Castro-Ruiz:2021vnq,Bartlett:2006tzx}. There is not a one-fits-all formalism because QRFs may have different meanings and roles in different contexts, e.g.\ the role a QRF assumes in an operational quantum information scenario with agents is typically a distinct one from their role in gauge systems. These approaches thus differ in scope and this is reflected in differences in how they implement the relevant symmetries.  The particular context of a given problem should dictate which formalism is employed. Here, we are interested in computing entropies of quantum states satisfying the gravitational constraints; for this purpose the perspective-neutral formalism~\cite{DeVuyst:2024pop,delaHamette:2021oex,Hoehn:2023ehz,Hoehn:2019fsy,Hoehn:2020epv,Chataignier:2024eil,AliAhmad:2021adn,Hoehn:2021flk,Giacomini:2021gei,Castro-Ruiz:2019nnl,delaHamette:2021piz,Vanrietvelde:2018dit,Vanrietvelde:2018pgb,Hoehn:2021flk,Suleymanov:2023wio,Carrozza:2024smc} is the appropriate one, as it is distinguished by imposing constraints on states as well as operators, and it is the one we will use. This formalism encodes the physics of the full theory independently of any particular chosen reference frame, enabling a `perspective-neutral' description of subsystems. At the same time, it provides an explicit way to describe physics from `the perspective of' a QRF, via a so-called Page-Wootters reduction map, going back to~\cite{Page:1983uc,1984IJTP...23..701W}. The approach parallels the description of an abstract manifold in terms of local coordinate patches, where the manifold itself is perspective-neutral, and each set of local coordinates is a different frame perspective. There is a natural version of `quantum covariance', implemented by unitary maps switching between each of the frame perspectives. The formalism as we employ it here is thus a (simplified) version of quantum geometry, which is another reason why it is particularly suited to quantum gravity.

In gravity, diffeomorphism invariance prevents a na\"ive formulation of physical subsystems associated with fixed spacetime subregions. Gravitational QRFs allow us to evade this problem -- instead of focusing on fixed subregions, one instead defines \emph{dressed}, \emph{covariant}, \emph{dynamical} subregions, in terms of the quantum coordinates provided by the QRFs. The degrees of freedom inside the subregion are also defined relative to these quantum coordinates. An immediate consequence of this is that such a physical subsystem depends heavily on which QRF is used to define it -- and there are very many possible choices one can make for the QRF~\cite{Goeller:2022rsx,Carrozza:2022xut,Carrozza:2021gju}. For example, one could give an observer a clock, as in~\cite{Chandrasekaran:2022cip}. But one could use any other covariant dynamical degrees of freedom as a QRF: for example, the location at which a certain field takes on a particular value would give a quantum coordinate, and thus a QRF (for example as in~\cite{Kaplan:2024xyk}, or in~\cite{Chen:2024rpx,Kudler-Flam:2024psh} where a slow-rolling inflaton field was used). One can also think of edge modes~\cite{Donnelly_2016,Speranza:2017gxd,Freidel:2020xyx,Freidel:2020svx} as QRFs \cite{Carrozza:2021gju,Carrozza:2022xut,Kabel:2023jve,Araujo-Regado:2024dpr, AliAhmad:2024wja}. As we will explain, it is also possible to use multiple QRFs to define physical subsystems. Since using different QRFs (or different combinations of QRFs) results in different physical subsystems, it also leads to different values for entropies \cite{Hoehn:2023ehz}. It is our aim to explore various manifestations of these phenomena, which we refer to respectively as subsystem relativity \cite{Hoehn:2023ehz,Franzmann:2024rzj,AliAhmad:2021adn,delaHamette:2021oex,Castro-Ruiz:2021vnq,DeVuyst:2024pop}, and the \observerdependence{} of gravitational entropy.

Two aspects regarding entropy relativity are worthwhile to emphasise. First, the \observerdependence{} of gravitational entropy for a subregion does \emph{not} mean that the entropy for a black hole or a cosmological horizon depends on the QRF. Gravitational entropy relativity pertains to what different QRFs describe as the \emph{field} degrees of freedom and it turns out that these correspond to distinct gauge-invariant subalgebras. Horizon entropy, on the other hand, includes the entropy of \emph{all} degrees of freedom (in particular also the other clocks) in- or outside the horizon and this turns out not to depend on the QRF. It is the entropy associated with this gauge-invariant algebra that all observers agree on. 

Second, an \observerdependence{} of entropy can also arise for \emph{classical} reference frames, which are external in the sense that their dynamics is not included in the description. For instance, inertial and Rindler observers in Minkowski space assign different entropies to, say, the vacuum and this observation can be extended to general spacetimes \cite{Marolf:2003sq,Ju:2023bjl,Ju:2023dzo}. There is a key difference to the present case, however: for the classical, non-dynamical observers, the entropy relativity arises because they have access to \emph{distinct} spacetime regions, and thus different field degrees of freedom. By contrast, the entropy relativity we will present in Sec.~\ref{sect:observerDependence} is qualitatively very different: separate QRFs access the \emph{same} spacetime region and field degrees of freedom, and the entropy relativity rather traces back to the fact that the QRFs themselves are quantum subsystems contained in the region they describe.

This paper extends previous work by generalising the construction of subsystem algebras to systems with multiple quantum clocks, including those with degenerate energy spectra, and (briefly) periodic clocks. We analyze the \observerdependence{} of the entropy and its manifestation in various regimes, providing new insights into subsystem and entropy relativity in quantum gravity. We have already presented a selection of our findings in the companion paper~\cite{DeVuyst:2024pop}. However, one of the main goals of that paper was to also provide a useful pedagogical meeting point between the QRF and quantum gravity communities. In that spirit, we refer to it the reader who is taking their first steps in this burgeoning area. By contrast, the present paper will be much more technical and in-depth, with full derivations, and many novel results, examples and details. These are structured across the remaining sections, which we now outline to provide a roadmap for the reader.

In \textbf{Section~\ref{Section: quantum clocks as quantum reference frames}} we review the pertinent details of the perspective-neutral formalism as it applies to quantum clocks, and how they should be employed as temporal QRFs. First, we cover how to model clocks using covariant positive operator valued measures (POVMs) and coherent states, and describe the difference between ideal and non-ideal clocks (which are distinguished by whether or not they have the ability to record time with arbitrary precision). Then, we explain how to impose the gauge constraints on operators; in particular, the gauge-invariant operators of a clock and a subsystem decompose into so-called dressed observables and reorientations, and together these form a crossed product (or a subalgebra thereof). We describe how to also impose the gauge constraints on states, by means of a so-called group-averaging inner product. There are two important equivalent formulations of this procedure: the first, using refined algebraic quantisation (RAQ) \cite{Giulini:1998kf,Giulini:1998rk,Landsman:1993xe,AHiguchi_1991,Marolf:2000iq}, has historically been more often used by proponents of QRFs, while the latter, using `coinvariants', was employed in~\cite{Chandrasekaran:2022cip}; we describe both and explain how they are related. Then we explain how the Page-Wootters formalism may be used to `jump into the perspective' of a clock, and how one may transform between the perspectives of different clocks. We use these insights to then describe the notion of subsystem relativity that will cause the \observerdependence{} of gravitational entropy later. We also comment on how the algebra of gauge-invariant operators is represented at both the perspective-neutral and clock perspective levels, and describe subtleties related to the properties of this representation, in particular when the von Neumann nature of a gauge-invariant algebra survives the step from the kinematical to the gauge-invariant Hilbert space. Various technical details relevant to this Section may be found in Appendices~\ref{app_additionalnon-ideal} and~\ref{app_doublecheck}.

We then describe how the perspective-neutral formalism for quantum clocks applies to perturbative quantum gravity in \textbf{Section~\ref{sect:CLPW}}, describing how the Page-Wootters (PW) framework was implicitly used by Chandrasekaran-Longo-Penington-Witten (CLPW)~\cite{Chandrasekaran:2022cip} when describing states and observables on which constraints have been imposed, a fact which we summarise with the informal slogan
\begin{equation}
    \text{PW}=\text{CLPW}.
\end{equation}
This emphasises the role of QRFs in the definition of subsystems in quantum gravity. In particular, we reproduce the construction of a gauge-invariant gravitational subregion algebra as the crossed product of a type III QFT algebra with its modular flow.

Next, in \textbf{Section~\ref{Section: arbitrary number of observers}}, we obtain the density operator in this algebra corresponding to an arbitrary physical state of the full system. This proceeds by first constructing a trace on the algebra, and then deriving the appropriate density operator. In the interest of clarity, several details of the derivation are confined to Appendix~\ref{Appendix: density operator details}, and we focus at first on clocks with non-degenerate energy spectra. Unlike previous approaches (except for the companion work~\cite{DeVuyst:2024pop}), we allow for arbitrary entanglement between the clocks and the fields, and we also allow the use of multiple clocks in the definition of the subsystem. If one is using more than one clock, we explain how one may `partial trace' out a given clock to obtain the density operator of the subsystem that would be obtained by removing that clock. We also show how our formula reduces to that of previous work~\cite{Chandrasekaran:2022cip,Jensen:2023yxy}, in the case of a `classical-quantum' state as used in that work (although our formula applies much more generally). And we comment on a subtlety that arises when there are no `complementary clocks', i.e.\ clocks outside of the subsystem under consideration.

In \textbf{Section~\ref{sect:rhoDeg}}, we extend the analysis to degenerate clocks, i.e.\ those whose energy spectra are degenerate. In the interest of simplicity, we restrict attention to the case where the multiplicity of the energy is independent of the energy. We review how the perspective-neutral framework, including dressed observables and reorientations, extends to this case as described in~\cite{Hoehn:2020epv}. As we will explain, defining a physical subsystem in the presence of these degeneracies necessitates some extra operational input. In particular, in many physical scenarios, \emph{superselection} over the different degeneracy sectors of the clocks can occur. In different settings, this superselection either will or will not apply to each of the clocks. Depending on this, the physical subsystem one should consider is associated with different algebras of observables, and hence different density operators (and eventually different entropies). We show how to derive the density operators in a variety of different such cases.

Having found the density operators for the physical subsystems corresponding to gravitational subregions relative to quantum clocks, one can in principle compute the von Neumann entropies of these subsystems. However, for a general state this calculation is somewhat intractible, and to make progress it helps to restrict to states within certain regimes. The most powerful of these in the present context is the \emph{semiclassical} regime, which we describe in \textbf{Section~\ref{sec_entropy}}. This regime is characterized by a clock which behaves almost classically relative to the QFT, in the sense that the time it reads is very sharply peaked at a certain value, with negligible physical fluctuations. An alternative way to state this is that the fluctuations of the QFT Hamiltonian are suppressed relative to fluctuations of the clock Hamiltonian. We explain how the von Neumann entropy of the density operators found in the previous sections reduces to the generalised entropy of the gravitational subregion, at leading order in the semiclassical approximation. Unlike previous works such as~\cite{Chandrasekaran:2022cip,Jensen:2023yxy}, we do not require the clocks and fields to be in a product state; rather we allow arbitrary entanglement (so long as this is consistent with the semiclassical assumptions). We also explain how our result reduces to that of~\cite{Chandrasekaran:2022cip,Jensen:2023yxy} in the special case of the states investigated in those papers. We furthermore compute the linear corrections to the semiclassical approximation, finding that these do not vanish in general (but they do for states with no entanglement between the fields and the clocks, in agreement with~\cite{Faulkner:2024gst}). In the interest of clarity, many details of the derivations of these results are confined to Appendix~\ref{Appendix: semiclassical approximation details}.

Of course, semiclassical states form only a small subset of the full Hilbert space, and it is worth also considering states that fall under other regimes. One simple scenario, which we dub the `antisemiclassical' regime, is the opposite of the semiclassical regime in that it is characterised by clocks whose times are subject to such large fluctuations that they cannot be meaningfully used to aid in observations of the fields. In \textbf{Section~\ref{sect:antiSemiclassical}} we discuss this regime, and derive the density operators and von Neumann entropies of corresponding states. In particular, we show how these may be computed by  formulae that involve simply partially tracing out the antisemiclassical clock.

\textbf{Section~\ref{sect:observerDependence}} employs all of the results previously established in the paper to demonstrate several examples of the \observerdependence{} of gravitational entropy. First, we explore an explicit case study involving a superposition of clocks experiencing Shapiro time delay after moving past a massive body at different distances. This `gravitational interferometer' example was also discussed in the companion work~\cite{DeVuyst:2024pop}; here we give more details on it. Some mathematical technicalities may be found in Appendix~\ref{sect:BCH}. We then describe an example of how superselection over clock degeneracy sectors can lead to \observerdependence{} in the entropy, via a mechanism which is independent of those explored with the gravitational interferometer. Next, in a given system, some clocks may be subject to semiclassical approximation, while others are subject to an antisemiclassical approximation, while still others fall somewhere in between. This again leads to \observerdependent{} entropies: for the semiclassical clocks the generalised entropy formula applies, while for the others it does not. Up to this point in the paper, we focus on monotonic clocks, meaning those whose times are always strictly increasing. Our final example of \observerdependence{} in the entropy comes from introducing \emph{periodic} clocks, which were recently investigated in the perspective-neutral framework in~\cite{Chataignier:2024eil}. As we will discuss, periodic clocks dress field degrees of freedom in different ways to monotonic clocks, and this leads to different entropy functionals, with different statistical dependences on these degrees of freedom.

We end the paper in \textbf{Section~\ref{Section: Conclusion}} with a summary of its achievements, and a brief discussion of some open questions. In particular, we speculate on the consequences of accounting for a larger group of gravitational gauge symmetries, the inclusion of QRFs that transform under them, and how this will lead to more intricate manifestations of subsystem relativity and the \observerdependence{} of gravitational entropy.

\section{Quantum clocks as quantum reference frames}
\label{Section: quantum clocks as quantum reference frames}

In the remainder of this work, we will focus on monotonic quantum clocks as temporal QRFs associated with the (one-dimensional) translation group. This provides for a simple entry setting for the discussion of QRFs in quantum gravity and, as we shall see in Sec.~\ref{sect:CLPW}, encompasses recent studies of the type conversion of local quantum field algebras in perturbative quantum gravity, e.g.\ see \cite{Chandrasekaran:2022cip,Jensen:2023yxy,Kudler-Flam:2023qfl,Witten:2023xze, Kudler-Flam:2023hkl,Faulkner:2024gst,Kolchmeyer:2023gwa,Kolchmeyer:2024fly,Aguilar-Gutierrez:2023odp}.  In the absence of boundaries, these studies invoke local observers equipped with  quantum clocks to ``dress'' the QFT degrees of freedom with those of the clocks to deparametrise the QFT modular flow. When there is a boundary, an ADM Hamiltonian appears that is conjugate to a boundary time variable that similarly serves as a clock for dressing. The resulting dressed algebras are so-called crossed products and admit well-defined density matrices and entropies. These are well-defined, at least up to a state-independent constant -- and, as we will describe, in some cases a canonical choice exists to fix even this ambiguity.

We will see that such observers with clocks are nothing but QRFs and hence that the discussion of crossed products in gravity has a natural interpretation in terms of QRFs. This perspective will also permit us to generalise this construction to arbitrarily many observers and to explore the observer dependence of subregion density operators and entropies already summarised in \cite{DeVuyst:2024pop}. In our exposition of quantum clocks as temporal QRFs, we will largely follow \cite{Hoehn:2019fsy,Hoehn:2020epv}. Invoking the generalisation of that formalism to QRFs associated with general symmetry groups \cite{delaHamette:2021oex} to extend our results on crossed products, density operators and entropies to more general QRFs will be left for future work.\footnote{More general crossed product constructions have already been explored in \cite{Fewster:2024pur,AliAhmad:2024wja}, however using an inequivalent approach to QRFs that does not impose constraints on states unlike here and e.g.\ in \cite{Chandrasekaran:2022cip,Jensen:2023yxy,Kudler-Flam:2023qfl,Faulkner:2024gst}.}

\subsection{Modelling clocks with covariant POVMs}
\label{sect:POVMs}

Taken as a temporal QRF, a monotonic clock will be modelled by a Hilbert space $\mathcal{H}_C$ on which its Hamiltonian $H_C$ generates a unitary representation of the translation group $(\mathbb{R},+)$ via $U_C(t)=\exp(-it H_C)$. The clock's `frame orientation' is the time $\tau$ it reads, and our main concern will be to construct suitable orientation states and observables. For monotonic clocks we have $\tau\in\mathbb{R}$, so that orientations take values in the translation group. Ideally, we would like to construct an orientation observable conjugate to the clock Hamiltonian, $[T,H_C]=i$, so that we can think of the latter as generating monotonic ``reorientations'' of the clock. Of course, if $H_C$ is bounded, Pauli's argument tells us that such a $T$ cannot be self-adjoint \cite{Pauli1958}. Unruh and Wald \cite{Unruh:1989db} refined this observation (dropping canonical conjugacy), showing that bounded clock Hamiltonians do not admit self-adjoint monotonic time operators. The way out is to drop insisting on self-adjointness, and to invoke covariant positive operator-valued measures (POVMs) to build clock observables, and a consistent probability interpretation for the readings of the clock \cite{holevoProbabilisticStatisticalAspects1982,buschOperationalQuantumPhysics,Braunstein:1995jb,Smith2019quantizingtime,Hoehn:2019fsy,Hoehn:2020epv,Brunetti:2009eq,Loveridge:2019phw}.

For simplicity, we shall assume in the remainder that the spectrum $\sigma_C:=\rm{Spec}(H_C)$ is purely continuous and for clarity we begin with the case when the spectrum is further nondegenerate. Degenerate clock Hamiltonians, such as $H_C=p^2$ for some momentum $p$, will be discussed separately in Sec.~\ref{sect:rhoDeg}.\footnote{Periodic clocks ($\rm{U}(1)$ reference frames) can be treated similarly, however, the ensuing relational dynamics is somewhat more subtle \cite{Chataignier:2024eil}. We comment on this case in Subsec.~\ref{Subsection: periodic vs monotonic}.} 
Following \cite{Hoehn:2019fsy}, the clock states can then be constructed via\footnote{We are using the symmetric convention for $2\pi$ factors in Fourier transformations in contrast to \cite{Hoehn:2019fsy,Hoehn:2020epv}.}
\begin{eqnarray}\label{clockstates}
    \ket{t}= \frac{1}{\sqrt{2\pi}} \int_{\sigma_C}\dd{\epsilon}e^{ig(\epsilon)}e^{-it\epsilon}\ket{\epsilon}\,,
\end{eqnarray}
where $g(\epsilon)$ is an arbitrary phase, accounting for the fact that an operator conjugate to $H_C$ is unique only up to addition of arbitrary functions of $H_C$. They are covariant 
\begin{eqnarray}\label{clockcov}
    U_C(\tau)\ket{t}=\ket{t+\tau},
\end{eqnarray}
and furnish a resolution of the identity
\begin{eqnarray}\label{resid}
    \int_\mathbb{R}\dd{t}\ket{t}\!\bra{t}=\int_{\sigma_C}\dd{\epsilon}\ket{\epsilon}\!\bra{\epsilon}=\,\mathds{1}_C\,.
\end{eqnarray}
We may thus think of them as generalised (and distributional) coherent states associated with the translation group. 

In order to construct a probability measure, we define effect operators
\begin{eqnarray}
E_C(X)=\int_{X\subset\mathbb{R}}\dd{t}\ket{t}\!\bra{t}\geq0
\end{eqnarray}
for any (Borel) subset $X$ of the set of possible clock readings $\mathbb{R}$ (e.g.\ some time interval). Thanks to the normalisation in Eq.~\eqref{resid} and the fact that $E_C(\cup_iX_i)=\sum_iE_C(X_i)$ for any disjoint $X_i$ ($\sigma$-additivity), the effect operators form an operator-valued probability measure over the subsets of clock readings. For example, $\bra{\varphi}E_C(X)\ket{\varphi}$ is the probability that the clock reading lies in $X$ for the state $\ket{\varphi}\in\mathcal{H}_C$. Further thanks to Eq.~\eqref{clockcov}, we have for all $X$
\begin{eqnarray}
U_C(\tau)E_C(X)U_C^\dag(\tau)=E_C(X+\tau),\label{covPOVM}
\end{eqnarray}
and so the POVM is covariant with respect to the translation group generated by $H_C$.

The clock states need not be perfectly distinguishable
\begin{eqnarray}\label{eq:chi}
    \braket{t}{t'}= \frac{1}{2\pi} \int_{\sigma_C}\dd{\epsilon}e^{i\epsilon(t-t')}=:\chi(t-t')\,,
\end{eqnarray}
since, when $\sigma_C$ is just an interval, their overlap reads  
\begin{eqnarray}\label{fuzzy}
    \chi(t-t')=\begin{cases}
        \delta(t-t') &\sigma_C=\mathbb{R}\,,\\
        \frac{1}{2\pi}e^{i\epsilon_{\rm min}(t-t')}\left(\pi\delta(t-t')+i\rm{P}\frac{1}{t-t'}\right)&\sigma_C=[\epsilon_{\rm min},\infty),\\
        \frac{i}{2\pi}\frac{e^{i\epsilon_{\rm min}(t-t')}-e^{i\epsilon_{\rm max}(t-t')}}{t-t'}&\sigma_C=[\epsilon_{\rm min},\epsilon_{\rm max}]\,,
    \end{cases}
\end{eqnarray}
where $\rm{P}$ denotes the Cauchy principal value \cite{Hoehn:2019fsy}.\footnote{The term in parentheses in the middle case can also be understood as an $i\varepsilon$ prescription, since $\lim_{\varepsilon\to 0^+}\frac1{t-t'+i\varepsilon}=\rm{P}\frac1{t-t'}-i\pi\delta(t-t')$.} When $\sigma_C$ is given by a collection of disjoint intervals, the overlap will be given by a corresponding sum over these intervals of expressions as in the third line.  Clock readings are thus only perfectly distinguishable when the Hamiltonian is unbounded, $\sigma_C=\mathbb{R}$, a case sometimes referred to as an \emph{ideal} (or \emph{perfect}) \emph{clock}. Unless $\sigma_C$ is bounded in both directions, clock states are also distributional. For later purpose, we explain in App.~\ref{app_fourier} that a Fourier transform continues to exist even when $\sigma_C$ is bounded.

 We can define the moment operators associated with the covariant clock POVM as
\begin{eqnarray} \label{eq:clockObs}
    T^{(n)}:= \int_\mathbb{R}\dd{t}t^n\ket{t}\!\bra{t}\,.
\end{eqnarray}
Unless the clock is ideal, we have $T^{(n)}\neq (T^{(1)})^n$ and these operators are not self-adjoint but only symmetric \cite{Hoehn:2019fsy,holevoProbabilisticStatisticalAspects1982,buschOperationalQuantumPhysics}. In particular, the clock states $\ket{t}$ are only eigenstates of the moment operators when the clock is ideal. Nevertheless, they satisfy canonical commutation relations with the Hamiltonian \cite[Proof of Lemma 2]{Hoehn:2019fsy}
\begin{eqnarray}\label{cancon}
    [T^{(n)},H_C]=inT^{(n-1)}\, ,
\end{eqnarray}
and are thus conjugate to it. 

The sought-after clock orientation observable is given by the covariant POVM and all its moments. While these moments need not be self-adjoint, the effect operators $E_C(X)$ are, thereby yielding well-defined probabilities over the clock readings, even though the latter need not be perfectly distinguishable. Specifically, the non-monotonicity issue for clocks with bounded Hamiltonian reported by Unruh and Wald \cite{Unruh:1989db} is sidestepped thanks to the covariance of the POVM \cite{Hoehn:2019fsy}.

\subsection{Constraints, relational observable algebras and crossed products}\label{sseHphys}

Let us now add some system $S$ that may evolve relative to quantum clock $C$. The total (kinematical) Hilbert space is thus taken to be $\mathcal{H}_{\rm kin}=\mathcal{H}_S\otimes\mathcal{H}_C$. The system $S$ may itself be composite and its Hilbert space given by a tensor product of a collection of Hilbert spaces. In particular, later $S$ will contain additional clocks and it may also contain QFT Hilbert spaces. 

Evolution relative to $C$ means that we are considering  relational dynamics. This entails that we treat the total Hamiltonian, for simplicity assumed to be free of interactions between clock and system,\footnote{Relational dynamics in the presence of interactions is challenging and has been touched upon in \cite{Hohn:2011us,Smith2019quantizingtime,Hoehn:2023axh}.} as a gauge constraint
\begin{equation}\label{constraint}
    H = H_S+H_C.
\end{equation}
At this stage, we make no further assumption about the system Hamiltonian $H_S$. Thanks to our above assumptions on the clock Hamiltonian $H_C$, the constraint $H$ is a translation group generator on $\mathcal{H}_{\rm kin}$. We may interpret it as corresponding to time evolution of the combined $SC$ system in some (possibly fictitious) external time. By treating this as a gauge symmetry, we are taking this external time evolution to be physically inaccessible, instead aiming for an \emph{internal} evolution.

While this perspective can be applied to laboratory systems where external time is not fictitious, in gravitational systems without boundary (and more generally reparametrisation-invariant systems) as discussed later, this perspective is forced upon us by diffeomorphism invariance. The latter leads to Hamiltonian constraints and to what is sometimes known as the infamous problem of time \cite{Kuchar:1991qf,Isham:1992ms,andersonProblemTime2017}. It means nothing else than that in gravity degrees of freedom do not evolve relative to some background time, but relative to one another. 

The translation group $U_{SC}(t)=\exp(-itH)$ generated by the constraint can be viewed as corresponding to reorientations of an external clock frame, while Eqs.~\eqref{clockcov} and~\eqref{cancon} entail that the one generated by $H_C$ implements reorientations of the internal clock $C$. The former are gauge, the latter are physical since $[U_{SC}(t),\mathds{1}_S \otimes U_C(t')]=0$, hence $U_C(t')$ acts in a gauge-invariant way and changes the relation between clock and system.

Let us now consider the algebra of gauge-invariant observables. Let $\mathcal{A}_S\subseteq\mathcal{B}(\mathcal{H}_S)$ be some algebra acting on the system such that time evolution $U_S(t)=\exp(-itH_S)$ constitutes a group of automorphisms of it. In particular, $H_S$ need not necessarily be part of $\mathcal{A}_S$ in which case these autormorphisms are outer; this will be the case in perturbative quantum gravity below. In appendix~\ref{app_ginvalg}, we show that the gauge-invariant algebra of observables of the clock and system on $\mathcal{H}_{\rm kin}$ is given by relational observables and clock reorientations:
\begin{eqnarray}\label{gaugeinvalg}
    \mathcal{A}_{\rm inv}:=\left(\mathcal{A}_S \otimes \mathcal{B}(\mathcal{H}_C)\right)^{H}=\Big\langle O^\tau_C(a), \mathds{1}_S \otimes U_C(t)\,\Big|\,a\in\mathcal{A}_S,\,t\in\mathbb{R}\Big\rangle.
\end{eqnarray}
$\langle A,B\rangle$ denotes the algebra generated by operators of the form $A,B$. For the moment, we leave implicit what the topology is in which this algebra is closed. When $\mathcal{A}_S$ is a general $C^*$-algebra, closure will be with respect to the $C^*$-norm of the tensor product of Hilbert spaces $\mathcal{H}_R\otimes\mathcal{H}_S$, yielding the $C^*$-tensor product. When $\mathcal{A}_S$ is more specifically a von Neumann algebra,\footnote{Hence, $\mathcal{A}_S''=\mathcal{A}_S$, where $''$ denotes the bicommutant on $\mathcal{H}_S$.} the main focus later, closure will be defined in terms of the double commutant and we will highlight this by writing $\langle A,B\rangle''$ for the von Neumann algebra generated by $A,B$.\footnote{$\mathcal{A}_S$ could also be the algebra of polynomials in some basic set of generators and may thereby not be contained in the algebra of bounded operators, in which case the tensor product is an algebraic one.} $\{U_C(t)\}_{t\in\mathbb{R}}$ are the clock reorientations. Furthermore,
the relational observables describing the system property $a$ conditioned on the clock reading $\tau$ are given by a $G$-twirl (also known as `incoherent group averaging') over the gauge group:
\begin{eqnarray}\label{relobs}
O^\tau_C(a):=\int_\mathbb{R}\dd{t}U_{SC}(t)\left(a \otimes \ket{\tau}\!\bra{\tau}\right)U_{SC}^\dag(t)=\int_\mathbb{R}\dd{t}\ket{t}\!\bra{t}\otimes\sum_{n=0}\frac{i^n}{n!}(t-\tau)^n\left[a,H_S\right]_n,
\end{eqnarray}
where $\ket{\tau}$ are the clock states in Eq.~\eqref{clockstates} and $[\cdot,\cdot]_n$ denotes the $n$th nested commutator with $[a,H_S]_0=a$; note that the latter expression is an expansion in the moments of the clock POVM.
As shown in \cite{Hoehn:2019fsy}, these are quantisations of the classical relational observables for Hamiltonian constrained systems in \cite{Dittrich:2004cb}.\footnote{For the generalisation of relational observables to QRFs associated with general symmetry groups, see \cite{delaHamette:2021oex,Carette:2023wpz}, and for a related approach to relational observables associated with clocks see also \cite{Chataignier:2019kof,Chataignier:2020fys,Marolf:1994wh,Marolf:1994ss}.} For ideal clocks, the relational observables take a particularly simple form \cite{Hoehn:2019fsy}
\begin{equation}
    O_{C}^\tau(a) = e^{-i(T - \tau) H_S } \,a\, e^{i(T - \tau) H_S },
    \label{eq:idealDirac}
\end{equation}
since in this case the first moment operator $T=T^{(1)}$ is self-adjoint and the clock states are its eigenstates.

For later purpose we note in appendix~\ref{app_ginvalg} that if $\mathcal{A}_S$ is a von Neumann algebra, then so is the gauge-invariant algebra $\left(\mathcal{A}_S \otimes \mathcal{B}(\mathcal{H}_C)\right)^{H}$. In particular, if in addition the clock $C$ is ideal (and hence $\mathcal{H}_C=L^2(\mathbb{R})$), then this algebra equals the crossed product of $\mathcal{A}_S$ with the translation group generated by $H_S$ (e.g., see \cite{Chandrasekaran:2022cip,Jensen:2023yxy,Witten:2021unn}): 
\begin{equation}
    \mathcal{A}_S\rtimes_{U_S}\mathbb{R}\simeq\Big\langle e^{-iT H_S } \,a\, e^{iT  H_S },e^{-itH_C}\,\Big|\,a\in\mathcal{A}_S,\,t\in\mathbb{R}\Big\rangle''=\left(\mathcal{A}_S\otimes\mathcal{B}(L^2(\mathbb{R})_C)\right)^{H}.
\end{equation}
More generally, from the point of view of QRFs, we can give crossed products the following interpretation: 
\begin{quote}
    \emph{The crossed product of a von Neumann algebra $\mathcal{A}$ by a locally compact group $G$ is the von Neumann algebra of relational observables describing $\mathcal{A}$ relative to an \emph{ideal} QRF associated with $G$, and its reorientations.}
\end{quote} 
We can thus think of the gauge-invariant clock-system algebra in Eq.~\eqref{gaugeinvalg} as a generalisation of the crossed product to encompass non-ideal clocks, as well as the case when $\mathcal{A}_S$ is a $C^*$ algebra but not necessarily von Neumann. We emphasise that, at this stage, the von Neumann nature of these algebras (i.e.\ their closure) is defined in terms of the kinematical Hilbert space $\mathcal{H}_{\rm kin}$. We will be discussing the construction of the physical Hilbert space shortly and will later investigate whether the action of these algebras on this space remains von Neumann.  

Below, we shall be mostly concerned with such generalised crossed products of a type $\rm{III}_1$ von Neumann factor $\mathcal{A}_S$,\footnote{$\mathcal{A}_S$ being a factor means it has a trivial center, i.e.\ only $c$-numbers commute with all other elements in $\mathcal{A}_S$.} interpreted as an algebra containing regional QFT and graviton degrees of freedom (and possibly additional clocks), by a group associated with its modular flow.  That is, we will focus on the case that $H_S$ is the modular Hamiltonian of a cyclic separating vector $\ket{\Psi}\in\mathcal{H}_S$ for $\mathcal{A}_S$. This is the case in which a conversion to type II factors with well-defined density operators and entropies occurs (at least up to a state-independent constant)~\cite{Chandrasekaran:2022cip,Jensen:2023yxy,Witten:2023xze,Kudler-Flam:2023qfl,Kudler-Flam:2023hkl,Witten:2021unn,Faulkner:2024gst,Kudler-Flam:2024psh,Chen:2024rpx,Aguilar-Gutierrez:2023odp,Gomez:2023upk,Gomez:2023wrq}. In appendix~\ref{app_ginvalg}, we provide some intuition for this and explain that it is the addition of the clock reorientations to the dressed regional algebra that turns the modular flow inner and thereby is responsible for the type conversion (modular flow in type III algebras is always outer). But for now the discussion remains more general than von Neumann algebras.

\subsection{The physical Hilbert space}\label{ssec_Hphys}

Next, let us implement constraints also on states to construct the physical Hilbert space of gauge invariant states in line with our interpretation that the external evolution generated by $H$ is physically inaccessible. While this step is standard in gauge theories, gravity and string theory, it singles out the \emph{perspective-neutral approach} \cite{delaHamette:2021oex,Hoehn:2023ehz,Hoehn:2019fsy,Hoehn:2020epv,AliAhmad:2021adn,Vanrietvelde:2018dit,Vanrietvelde:2018pgb,Giacomini:2021gei,Castro-Ruiz:2019nnl,Krumm:2020fws,Hoehn:2021flk,Hohn:2018toe,delaHamette:2021piz,Hoehn:2023axh,Suleymanov:2023wio} we employ here from within the approaches to QRFs.\footnote{See \cite[Sec.~II]{Hoehn:2023ehz} for a discussion of how this step distinguishes the perspective-neutral approach from the quantum information one \cite{Bartlett:2006tzx,Castro-Ruiz:2021vnq} (going back to \cite{Aharonov:1967zza,Aharonov:1967zz}), the purely perspective-dependent one \cite{Giacomini:2017zju,delaHamette:2020dyi} and the operational one \cite{Loveridge:2019phw,Carette:2023wpz}, which do not implement constraints on states. The purely perspective-dependent one turns out to be equivalent to the perspective-neutral one for ideal QRFs (perfectly distinguishable orientation states), however, not for general non-ideal ones \cite{delaHamette:2021oex}. Within quantum information theory and the foundations of quantum theory there exist varying aims for the use of QRFs, which is why several approaches have developed, and there is not a one-fits-all approach. Here, we employ the one which applies to gauge theories and gravity.} We can view the kinematical Hilbert space $\mathcal{H}_{\rm kin}$ as the space of externally distinguishable states and the physical Hilbert space $\mathcal{H}_{\rm phys}$ as the one of internally distinguishable states from which all external frame information has been removed \cite[Sec.~II]{Hoehn:2023ehz}. We thus seek to implement a form of Dirac constraint quantisation \cite{diracLecturesQuantumMechanics1964,Henneaux:1992ig} and states formally satisfying a version of the Wheeler-DeWitt equation:
\begin{equation}\label{physstate}
    U_{SC}(t)|\Psi)=|\Psi)\,,\qquad\forall\,t\in\mathbb{R}\qquad\Rightarrow\qquad H|\Psi)=0.
\end{equation}
In the sequel, we shall be using round bras and kets, $|\cdot)$, to denote such physical states, while we reserve the standard bra-ket notation for kinematical or gauge-reduced states. 

A word of caution on the meaning of Eq.~\eqref{physstate} for the case of a noncompact gauge group as the translation group here: as zero will lie in the continuous spectrum of $H$, a state such as $|\Psi)$ corresponds to an improper eigenstate and thus cannot be normalisable in the original $\mathcal{H}_{\rm kin}$. To address this normalisation issue, we have to build a new inner product to turn the space of states formally obeying Eq.~\eqref{physstate} into a Hilbert space. Here, we will briefly recall two equivalent ways of constructing such a new Hilbert space $\mathcal{H}_{\rm phys}$ that will be convenient below: the procedure of co-invariants, also invoked by CLPW \cite{Chandrasekaran:2022cip}, and Refined Algebraic quantisation (RAQ) \cite{Giulini:1998kf,Giulini:1998rk,Landsman:1993xe,AHiguchi_1991,Marolf:2000iq}, usually employed in the perspective-neutral approach to QRFs \cite{delaHamette:2021oex,Hoehn:2019fsy}. Rather than viewing $\Ket{\Psi}$ as an element of $\mathcal{H}_{\rm kin}$, RAQ takes it as a distribution on it. 

The space of co-invariants is the space of gauge-equivalent kinematical states, i.e.\ $\ket{\psi}\sim\ket{\phi}$ iff $\ket{\psi}=U_{SC}(t)\ket{\phi}$ for some $t\in\mathbb{R}$. This space of equivalence classes does not possess a linear structure, unless we impose the \emph{additional} relation that the differences between gauge-equivalent states are equivalent to null states, i.e.\ $(\mathds{1}-U_{SC}(t))\ket{\psi}\hat{\sim} \,0$ for all $t\in\mathbb{R}$ and $\ket{\psi}\in\mathcal{H}_{\rm kin}$. This means in particular $H\ket{\psi}\hat{\sim}\,0$ for all $\ket{\psi}\in\mathcal{H}_{\rm kin}$. Quotienting the space of co-invariants by these null vectors yields a vector space, whose elements we denote by $|\Psi)$ and it is clear that gauge translations act trivially on this space, in that sense realising Eq.~\eqref{physstate}. An inner product on this space is given by group averaging
\begin{eqnarray}\label{physIP}
    \Braket{\Psi_1}{\Psi_2}:=\int_\mathbb{R}\dd{t}\bra{\psi_1}U_{SC}(t)\ket{\psi_2},
\end{eqnarray}
where $\ket{\psi_i}\in\mathcal{H}_{\rm kin}$ resides in the equivalence class $|\Psi_i)$ and the inner product under the integral is the standard one on $\mathcal{H}_{\rm kin}$.\footnote{Whether or not this physical inner product is positive semidefinite is in general unknown, as is convergence of the integral, though see \cite{Higuchi:1991tk,Higuchi:1991tm,Marolf:2000iq,Giulini:1998kf,Giulini:1998rk,Marolf:2008hg,} for some discussion on this. In what follows, we shall simply assume convergence and semidefiniteness.} The physical Hilbert space $\mathcal{H}_{\rm phys}$ is then given by completing in norm. This is essentially the route taken by CLPW \cite[App.~B.1]{Chandrasekaran:2022cip}.  Abstractly, this procedure defines a linear map $M:\mathcal{H}_{\rm kin}\to\mathcal{H}_{\rm phys}$ that implements the constraint in the sense that $MH=0$, and the core question is how to explicitly construct it. CLPW provide a method within their setup \cite[Sec.~4.2]{Chandrasekaran:2022cip} (see also Example~\ref{ex1} below, and Sec.~\ref{sect:PW=CLPW}).

To prepare the grounds for establishing its equivalence with a Page-Wootters reduction of gauge-invariant states constructed via RAQ, we note that we can think of the inner product as $(\Psi_1|\Psi_2)=\bra{\psi_1}M^\dag M\ket{\psi_2}=\bra{\psi_1}\Pi_{\rm phys}\ket{\psi_2}$, where (assuming we can take the integral inside the inner product) 
\begin{equation}\label{piphys}
    \Pi_{\rm phys}=\int_\mathbb{R}\dd{t}U_{SC}(t)
\end{equation}
is sometimes informally referred to as a coherent group averaging `projector', though it is not literally a projector (it is essentially $\delta(H)$). This formally defines an anti-linear `rigging map' \cite{Giulini:1998kf}, the core ingredient of RAQ,
\begin{equation}\label{eta}
\eta(\ket{\psi}):=\bra{\psi}\Pi_{\rm phys}
\end{equation}
from kinematical states into kinematical distributions. Their key property is that, thanks to the translation invariance of the integral measure, states in the image of $\eta$ are gauge-invariant, $U_{SC}(t)\eta(\ket{\psi}):=\eta(U_{SC}(t)\ket{\psi})=\eta(\ket{\psi})$.\footnote{Somewhat more accurately, we have $\eta:\mathcal{D}\to\mathcal{D}^*$, where $\mathcal{D}\subset\mathcal{H}_{\rm kin}$ is a linear dense subspace closed under gauge transformations and $\mathcal{D}^*$ is its algebraic dual (space of all linear functionals as opposed to only continuous ones). This leads to the triple $\mathcal{D}\hookrightarrow\mathcal{H}_{\rm kin}\hookrightarrow\mathcal{D}^*$, which inspires the name `rigging map'. $\mathcal{D}^*$ inherits an action of the gauge group via $U_{SC}(t)\xi[\phi]=\xi[U_{SC}^\dag(t)\phi]$, for any $\xi\in\mathcal{D}^*$ and $\phi\in\mathcal{D}$. Gauge invariance should be understood in the sense that $U_{SC}(t)\eta(\psi_1)[\psi_2]=\eta(\psi_1)[U_{SC}^\dag(t)\psi_2]=\eta(\psi_1)[\psi_2]$ for all $\psi_1,\psi_2\in\mathcal{D}$ and $t\in\mathbb{R}$. See \cite{Marolf:2000iq,Giulini:1998kf,Giulini:1998rk} for details.} It is for this reason that the states in the image also correspond to equivalence classes of kinematical states. We can thus set $(\Psi'|=\eta(\ket{\psi})$  for representatives of physical states, where for distinction from the co-invariant case we equip them with a $'$, and their inner product is given as before by group averaging, so\footnote{We could also define an inner product on the image of $\eta$ as $(\eta(\ket{\psi_2}),\,\eta(\ket{\psi_1})):=\eta(\ket{\psi_1})\left(\ket{\psi_2}\right)=\bra{\psi_1}\Pi_{\rm phys}\ket{\psi_2}$. Relative to the left inner product, we can again interpret $\eta^\dag\,\eta=\Pi_{\rm phys}$ as in the case of the co-invariant map $M$.} \begin{equation}
    (\Psi_1'|\Psi_2'):=\eta(\ket{\psi_1})\left(\ket{\psi_2}\right)=\bra{\psi_1}\Pi_{\rm phys}\ket{\psi_2}=(\Psi_1|\Psi_2)\,.
    \end{equation}
    Quotienting by null vectors and completing in norm then yields an equivalent representation of $\mathcal{H}'_{\rm phys}\simeq\mathcal{H}_{\rm phys}$ of the physical Hilbert space.

\begin{eg}\label{ex1}
    \emph{A simple example is given by the constraint $H=p_S+p_C$ for two ideal clocks on $\mathcal{H}_{\rm kin}=L^2(\mathbb{R}^2)$, with $p_S,p_C$ two momentum operators. 
    Defining $q_\pm:=q_C\pm q_S$, any kinematical state can be written as $\ket{\psi}=\int\dd{q}_+\dd{q}_-\psi(q_+,q_-)\ket{q_+,q_-}$, where $\ket{q_+,q_-}=\frac{1}{\sqrt{2}}\ket{q_C=1/2(q_++q_-),q_S=1/2(q_+-q_-)}$ are eigenstates of $q_\pm$.}
    
    \emph{Following the co-invariant way, we notice that $\ket{\psi}\sim U_{SC}(t)\ket{\psi}$ entails $\ket{q_+,q_-}\sim\ket{q_++t,q_-}$ and $(\mathds{1}-U_{SC}(t))\ket{\psi}\hat{\sim}\, 0$ implies $\ket{q_+,q_-}-\ket{q_++t,q_-}\hat{\sim}\,0$. As $U_{SC}(t)$ acts ergodically on the $q_+$ label and trivially on $q_-$, the conjunction of these relations means that for each fixed $q_-$, there is (up to normalisation) a unique surviving equivalence class of basis states; let us denote it by $|q_-)$. By Eq.~\eqref{physIP}, we have $(q_-|q_-')=\delta(q_--q_-')$. Physical states can thus be written as $|\Psi)=\int_\mathbb{R}\dd{q}_-\Psi(q_-)|q_-)$ with $\Psi$ square-integrable, so $\mathcal{H}_{\rm phys}\simeq L^2(\mathbb{R})$.}
    
    \emph{On the other hand, in momentum representation the `rigging map' Eq.~\eqref{eta} of RAQ reads 
    \begin{equation}
        (\Psi'|=\eta(\ket{\psi})=\int_\mathbb{R}\dd{p}_-\psi^*(0,p_-)\bra{p_+=0,p_-}.
    \end{equation}
    While such improper null eigenstates of $H=p_+$ are clearly not normalisable in $\mathcal{H}_{\rm kin}$, we have 
    \begin{equation}
        (\Psi_1'|\Psi_2')=\int_\mathbb{R}\dd{p}_-\psi_1^*(0,p_-)\psi_2(0,p_-),
    \end{equation}
whose absolute value is bounded if $\psi_1,\psi_2$ are bounded around $p_+=0$. Null states are those with $\psi(0,p_-)=0$ (i.e., by Fourier transformation, states such that $\int\dd{q}_+\psi(q_+,q_-)=0$). Quotienting by these, we again have $\mathcal{H}'_{\rm phys}\simeq L^2(\mathbb{R})$ in agreement with the construction via co-invariants.}  
\end{eg}

The observation that the co-invariant construction and RAQ lead to equivalent physical Hilbert spaces was also noted in the context of Jackiw-Teitelboim gravity in \cite{Held:2024rmg}.

In this section, we shall at times operate with the co-invariant and RAQ representations of physical states in parallel, emphasising how certain maps and expressions appear in the two. This is to facilitate exhibiting the relation between the subjects of gravitational crossed product algebras and QRFs, where the former so far invoked co-invariants and the latter RAQ. In later sections, for definiteness, we shall mostly invoke the co-invariant scheme and only sometimes comment on how expressions change in the equivalent RAQ formulation. 

Henceforth, when pursuing the RAQ formalism, we shall abuse notation somewhat, writing, as is standard in much of the physics literature on relational dynamics,
\begin{equation}\label{sloppy}
    |\Psi')=\Pi_{\rm phys}\ket{\psi},
\end{equation}
instead of Eq.~\eqref{eta}, keeping in mind however that physical states can be unnderstood as distributions on $\mathcal{H}_{\rm kin}$. As $\Pi_{\rm phys}$ is formally symmetric in the kinematical inner product, this does not cause problems. In particular, decomposing kinematical states in the clock and system energy bases, $\ket{\psi}=\int_{\sigma_C}\dd{\epsilon}\SumInt_{\sigma_S}\sum_{\lambda_E}\psi(\epsilon;E,\lambda_E)\ket{\epsilon}\otimes\ket{E,\lambda_E}$, where $\lambda_E$ is a possible degeneracy label for system energy $E$ and $\sigma_S$ denotes the spectrum of $H_S$, we can then write physical states formally as \cite{Hoehn:2019fsy}
\begin{equation}\label{physstateenergy}
    |\Psi')=\Pi_{\rm phys}\ket{\psi}=2\pi\SumInt_{\sigma_S\cap(-\sigma_C)}\sum_{\lambda_E}\psi(-E;E,\lambda_E)\ket{-E}\otimes\ket{E,\lambda_E},
\end{equation}
while the inner product becomes
\begin{equation}
    (\Psi_1'|\Psi_2')=2\pi\SumInt_{\sigma_S\cap(-\sigma_C)}\sum_{\lambda_E}\psi_1^*(-E,E,\lambda_E)\psi_2(-E;E,\lambda_E).
\end{equation}

\subsection{Physical representation of the algebra}

Although our gauge-invariant algebra $\mathcal{A}_{\rm inv}:=\left(\mathcal{A}_S \otimes \mathcal{B}(\mathcal{H}_C)\right)^{H}\subseteq\left(\mathcal{B}(\mathcal{H}_{\rm kin})\right)^{H}\subset\mathcal{B}(\mathcal{H}_{\rm kin})$ has been originally defined on the kinematical Hilbert space, we will be interested in its representation  on the space of physical states $\mathcal{H}_{\rm phys}$ (see App.~\ref{app_ginvalg} for detail). While the latter is a different Hilbert space, this action is defined because gauge-invariant observables ``commute'' with the co-invariant map $M$ above in the sense that 
\begin{equation}\label{coinvrep}
MO\ket{\psi}=r(O)M\ket{\psi}=r(O)|\Psi)\,,\qquad\forall\,O\in\mathcal{A}_{\rm kin}^H\text{ and }\ket{\psi}\in\mathcal{H}_{\rm kin}\,,
\end{equation}
where
\begin{equation}
   \mathcal{A}_{\rm kin}^H:= \left(\mathcal{B}(\mathcal{H}_{\rm kin})\right)^H\,.
\end{equation}
$r:\mathcal{A}_{\rm kin}^H\rightarrow \mathcal{B}(\mathcal{H}_{\rm phys})$ defines here the physical representation of the kinematical gauge-invariant algebra on the physical Hilbert space $\mathcal{H}_{\rm phys}$ defined the co-invariant way. Equivalently, kinematical gauge-invariant observables formally commute with the RAQ rigging map $\eta$ \cite{Giulini:1998kf,Giulini:1998rk,Marolf:2000iq}:
\begin{equation}\label{RAQrep}
   \Bra{\Psi'}r'(O^\dag):=\eta(O\ket{\psi})=\bra{\psi}O^\dag\Pi_{\rm phys}=\bra{\psi}\Pi_{\rm phys}O^\dag \,,
\end{equation}
where $r'$ now denotes the physical representation of kinematical gauge-invariant operators defined the RAQ way. Using the slightly sloppy notation from Eq.~\eqref{sloppy}, this becomes
\begin{equation}\label{sloppy2}
    r'(O)\Ket{\Psi'}=\Pi_{\rm phys}O\ket{\psi}=O\Pi_{\rm phys}\ket{\psi}
\end{equation}
and is formally consistent with the inner product: 
\begin{equation}
\Mel*{\Psi_1'}{r'(O)}{\Psi_2'}=\bra{\psi_1}O\Pi_{\rm phys}\ket{\psi_2}=\eta(O^\dag\ket{\psi_1})(\ket{\psi_2})=(r'(O^\dag)\Psi_1'|\Psi_2')\,.
\end{equation}
Formally we can thus write $r'(O)=O$ when it is clear that we mean that $O$ acts on physical states. One has to be careful with this, however, as $r'$ (equivalently $r$) is not generally faithful: there exist $O_1,O_2\in\mathcal{A}_{\rm kin}^H$, such that $O_1\neq O_2$ on $\mathcal{H}_{\rm kin}$, but $r'(O_1)=r'(O_2)$ \cite[Lemma 1]{Hoehn:2019fsy}. For instance, any observable in $\mathcal{A}_{\rm kin}^H$ that has exclusively support on energies inconsistent with solving the constraints will be mapped to zero under $r'$ (equivalently $r$).\footnote{Examples of such observables can be built diagonally in the energy bases of clock and system with exclusive support in the complement of the spectral intersection $\sigma_S\cap(-\sigma_C)$ in Eq.~\eqref{physstateenergy}.}  For the relational observables in Eq.~\eqref{relobs} $r'(O)=O$ on physical states formally trades the incoherent for coherent group averaging since
\begin{equation}\label{sloppy3}
    O^\tau_C(a)\Pi_{\rm phys}=\Pi_{\rm phys}(a\otimes\ket{\tau}\!\bra{\tau})\Pi_{\rm phys}\,.
\end{equation}
For notational simplicity, we shall henceforth write $r'(O)=O$ with the implicit understanding that this holds on physical states.
While the formulations in Eqs.~\eqref{coinvrep} and~\eqref{RAQrep} are cleaner, expressions such as Eqs.~\eqref{physstateenergy} and~\eqref{sloppy3} are somewhat more explicit.

The properties of this physical representation $r(\mathcal{A}_{\rm inv})$ (equivalently $r'$) may be quite different from those of $\mathcal{A}_{\rm inv}$ on $\mathcal{H}_{\rm kin}$. For example, as we shall see below in Sec.~\ref{ssec_vN}, while $\mathcal{A}_{\rm inv}$ is von Neumann provided $\mathcal{A}_S$ is, the same is no longer guaranteed for $r(\mathcal{A}_{\rm inv})$ because the notion of algebraic closure is now defined with respect to different structures (though in many cases of interest it will be von Neumann). Related to this, $r$ (equivalently $r'$) is not faithful as noted above. (It may be faithful on \emph{sub}algebras of $\mathcal{A}_{\rm kin}^H$, cf.~Sec.~\ref{ssec_vN}). This has repercussions for whether or not the frame reorientations $\mathds{1}_S \otimes U_C(\tau)$ in $\mathcal{A}_{\rm inv}$ (cf.~Eq.~\eqref{gaugeinvalg}) survive as independent observables on $\mathcal{H}_{\rm phys}$ (respectively $\mathcal{H}_{\rm phys}'$): when $H_S\in\mathcal{A}_S$ and $U_S(\tau)$ are inner automorphisms (standard QRF setup), then $H_C$ and $H_S$ are not independent operators in $r(\mathcal{A}_{\rm inv})$ and so the clock reorientations are contained in (the representation of) the relational observables on $\mathcal{H}_{\rm phys}$. By contrast, when $H_S\notin\mathcal{A}_S$ and $U_S(\tau)$ are outer automorphisms (case of perturbative quantum gravity and QFT below), then the clock reorientations remain independent within $r(\mathcal{A}_{\rm inv})$ and are not included in the relational observables.

For the physical representation of relational observables and reorientations, we shall henceforth write
\begin{equation}\label{physrep}
\mathcal{O}_{C}^\tau(a):=r(O^\tau_C(a))\,,\qquad\qquad V_C(\tau):=r(U_C(\tau))\,,
\end{equation}
in the language of co-invariants. By contrast, we continue to use $O_C^\tau(a),U_C(\tau)$ in the physical representation of the RAQ method, in line with our conventions above.

\subsection{``Jumping into a clock frame'': the Page-Wootters formalism}\label{ssec_PW}

Now that we have a QRF, we would like to ``jump into its perspective'' and describe the remaining degrees of freedom, hence the system $S$, relative to it. For clocks this means nothing else than describing the evolution of $S$ relative to $C$. There are two equivalent ways of doing so, both of which will be useful later on:
\begin{description}
    \item[gauge-invariant:] The first amounts to simply evaluating the relational observables $O_C^\tau(a)$ in Eq.~\eqref{relobs} in physical states $\Ket{\Psi}\in\mathcal{H}_{\rm phys}$ and letting the clock reading parameter $\tau$ run.
    \item[gauge-fixed:] The second conditions physical states on the clock reading, yielding a relational Schr\"odinger picture in which states of $S$ evolve in $\tau$. This is the Page-Wootters formalism \cite{Page:1983uc,1984IJTP...23..701W} (see \cite{Dolby:2004ak,Gambini:2008ke,giovannettiQuantumTime2015,moreva2014time,Smith2019quantizingtime,Smith:2019imm,Hoehn:2019fsy,Hoehn:2020epv,Chataignier:2024eil,Castro-Ruiz:2019nnl,Hoehn:2023axh} for some later developments).
\end{description}
We have essentially already seen the first path, so we introduce here briefly the Page-Wootters (PW) formalism. We will see in Sec.~\ref{sect:PWCLPW} that the procedure considered by CLPW \cite{Chandrasekaran:2022cip} (and adopted by JSS in \cite{Jensen:2023yxy}) to implement the constraint is equivalent to it. To this end, we will follow the  formulation of the PW-formalism in \cite{Hoehn:2019fsy,Hoehn:2020epv,Chataignier:2024eil},\footnote{For generalisations of the PW-formalism to QRFs associated with general symmetry groups and first steps in quantum field theory, see \cite{delaHamette:2021oex} and \cite{Hoehn:2023axh}, respectively.} which establishes its equivalence with the gauge-invariant formulation and thereby resolves three fundamental criticisms that Kucha\v{r} had raised against the formalism \cite{Kuchar:1991qf}, disrupting research on the topic for some time.\footnote{Kucha\v{r} argued that the PW-formalism would yield (a) conditional probabilities that are incompatible with the constraint, (b) incorrect transition probabilities, and (c) wrong localisation probabilities in relativistic settings \cite{Kuchar:1991qf}. (a) and (b) were resolved in \cite{Hoehn:2019fsy} through the equivalence with the gauge-invariant formulation, while (c) was resolved in \cite{Hoehn:2020epv} using the covariant clock POVMs for quadratic Hamiltonians exhibited in Sec.~\ref{sssec_degclock1}. Giovannetti et al.~\cite{giovannettiQuantumTime2015} proposed an earlier resolution of (b), which does lead to correct transition probabilities but modifies the constraint by including measurement interactions and ancilla systems and further relies on ideal clocks. The resolution in \cite{Hoehn:2019fsy} is not restricted to ideal clocks and solves the problem in a more general form than originally posed by Kucha\v{r}. The qualitative differences between the two resolutions of (b) have been further investigated in \cite{Hausmann:2023jpn}.} It also connects it with the QRF paradigm.

\subsubsection{Relational Schr\"odinger picture}\label{ssec_PWnondeg}

We will first exhibit the traditional formulation of the PW-formalism, which invokes solutions to the constraint, and thus is best expressed in RAQ. We will then translate it into the equivalent language of co-invariants.

The aim of PW was to construct a conditional probability interpretation of ``timeless'' physical states. This can be achieved by defining a reduction map $\mathcal{R}'_C(\tau):\mathcal{H}'_{\rm phys}\to\mathcal{H}_{|C}$, conditioning physical states on clock $C$ reading $\tau$ (the primes highlight that we are invoking the RAQ imposition of constraints): 
\begin{equation}\label{PWred}
    \mathcal{R}'_C(\tau):=\bra{\tau}\otimes \mathds{1}_S,
\end{equation}
where $\ket{\tau}$ is a clock state \eqref{clockstates} and $\mathcal{H}_{|C}\subseteq\mathcal{H}_S$ is the system subspace consistent with implementing the constraint (it is spanned by eigenstates of $H_S$ residing in the overlap $\sigma_S\cap(-\sigma_C)$, cf.~\eqref{physstateenergy}). It is the space of system states ``seen'' by clock $C$. For later, we note that the orthogonal projector onto this subspace is formally given by \cite{Hoehn:2019fsy}
\begin{equation}\label{projS}
    \Pi_{|C}=\bra{\tau}\Pi_{\rm phys}\ket{\tau}=\int_\mathbb{R}\dd{t}\chi^*(t)U_S(t)=\int_{\sigma_C}\dd{\epsilon}\delta(H_S+\epsilon),
\end{equation}
where $\chi(t)$ is the clock state overlap distribution \eqref{fuzzy} and $\Pi_{\rm phys}$ was defined in Eq.~\eqref{piphys}. Note that we have $\Pi_{|C}=\mathds{1}_S$ only for ideal clocks. In the case that the clock spectrum is an interval $\sigma_C=[\epsilon_{\rm min},\epsilon_{\rm max}]$, this takes the more explicit form
\begin{equation}
 \Pi_{|C}=\Theta(\epsilon_{\rm max}+H_S)\Theta(-\epsilon_{\rm min}-H_S).
\end{equation}

It is well-known that the conditional states 
\begin{equation}\label{PWstate}
\ket*{\psi_{|C}(\tau)}:=\mathcal{R}_C'(\tau)\Ket{\Psi'}
\end{equation}
solve the Schr\"odinger equation in the clock readings
\begin{equation}\label{Schrod}
    i \dv{}{\tau} \ket*{\psi_{|C}(\tau)} = i \dv{}{\tau} \bra{\tau'}U_C^\dagger(\tau - \tau') \otimes \mathds{1}_S\Ket{\Psi'}  = - \bra{\tau}H_C \otimes \mathds{1}_S\Ket{\Psi'} = H_S \ket*{\psi_{|C}(\tau)}.
\end{equation}
This remains true regardless of whether $H_S\in\mathcal{A}_S$. This time evolution is an internal, gauge-invariant one, as the derivation invokes the clock reorientations $ U_C(\tau)\otimes \mathds{1}_S$ and their equivalence with the system time evolution $\mathds{1}_C\otimes U_S(-\tau)$ on $\mathcal{H}_{\rm phys}'$.

The observation in \cite{Hoehn:2019fsy,Hoehn:2020epv,Chataignier:2024eil} is that the conditioning on clock readings in Eq.~\eqref{PWred} is nothing but a gauge fixing and its inverse, $\mathcal{R}'_{C}{}^\dag(\tau):\mathcal{H}_{|C}\to\mathcal{H}_{\rm phys}'$, is formally given by averaging over the gauge 
\begin{align}\label{PWinv}
    \mathcal{R}'_{C}{}^\dag(\tau)[\cdot] = \Pi_\text{phys}([\cdot] \otimes \ket{\tau}) =  \int_\mathbb{R} \dd{t}~U_S(t-\tau) [\cdot] \otimes \ket{t} .
\end{align}
The reduction map and its inverse are unitary and we have\footnote{Physical states are here taken in the form \eqref{sloppy}. If the more precise Eq.~\eqref{eta} was used instead, the reduction and its inverse would have to be taken in ``daggered'' form. The relation of the inner products would be unaffected.}
\begin{equation}\label{IPrel}
    (\Psi_1'|\Psi_2')=\Bra{\Psi_1'}\left(\mathds{1}_S \otimes \ket{\tau}\!\bra{\tau}\right)\Ket{\Psi_2'}_{\rm kin}=\braket*{\psi^1_{|C}(\tau)}{\psi^2_{|C}(\tau)},
\end{equation}
where the right expression is the standard inner product on $\mathcal{H}_{|C}$ (inherited from $\mathcal{H}_S$). By the intermediate expression, we mean the matrix element of $\mathds{1}_S \otimes \ket{\tau}\!\bra{\tau}$ in the physical states $\Ket{\Psi_i'}$ viewed as kinematical distributions and thus evaluated in the kinematical inner product. This expression follows from the left hand side, using the definitions in Eqs.~\eqref{physIP} and~\eqref{piphys} and the formal identity $\Pi_{\rm phys}\left(\mathds{1}_S \otimes \ket{\tau}\!\bra{\tau}\right)\Pi_{\rm phys}=\Pi_{\rm phys}$ \cite{Hoehn:2019fsy,delaHamette:2021oex}, which follows from the resolution of the identity in Eq.~\eqref{resid}. This intermediate expression highlights that the right side is just a gauge-fixing of the physical inner product. We can interpret the reduction map $\mathcal{R}'_C(\tau)$ as a ``quantum coordinate map'' into the perspective of the temporal QRF $C$. The subscript $|C$ for objects in its image thus stands for ``relative to clock $C$''.

\begin{figure}
    \centering
    \includegraphics[width=.55\linewidth]{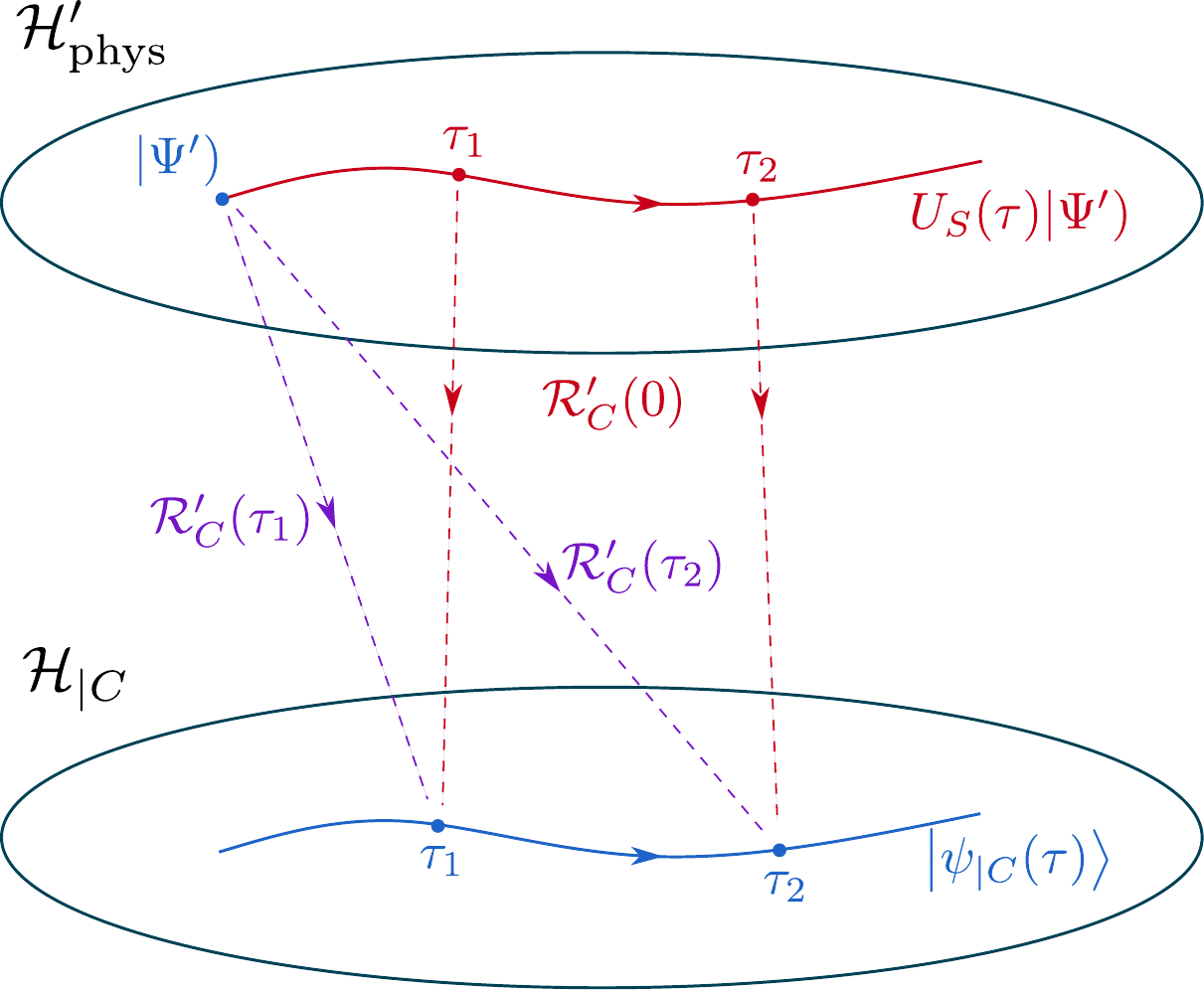}
    \caption{The \textcolor{JulianBlue}{Schr\"odinger evolution} of $\ket*{\psi_{\vert C}(\tau)}$ stemming from $\Ket{\Psi'}$ can be obtained in two ways. The \textcolor{JulianPurple}{tangential decoding} is achieved by acting with $\mathcal{R}'_{\vert C}(\tau)$ on $\Ket{\Psi}$ for different values of $\tau$. Alternatively, the \textcolor{JulianRed}{transversal decoding} first requires the trajectory $U_S(\tau) \Ket{\Psi'}$ for different values of $\tau$ on which we then act with the same $\mathcal{R}'_{\vert C}(0)$.}
    \label{fig: Schrodinger evolution}
\end{figure}

 Given that the conditioning on the clock reading is a gauge fixing, one might now be worried that the Schr\"odinger evolution in $\tau$ described by Eq.~\eqref{Schrod} is one tangential to a gauge orbit and hence unphysical. At the same time, we argued above that this same evolution should be physical, and so transversal to a gauge orbit, because it invokes the clock reorientations. While at first these statements seem to contradict one another, both are in fact correct. To see this, we note that Eq.~\eqref{PWinv} implies that physical states can also be written as
\begin{equation}
    \Ket{\Psi'}=\int_\mathbb{R}\dd{t}~\ket*{\psi_{|C}(t)}\otimes \ket{t}.
\end{equation}
This is behind the slogan ``time from entanglement'', though we emphasise that this entanglement is defined with respect to a tensor product structure on $\mathcal{H}_{\rm kin}$ as opposed to one on $\mathcal{H}_{\rm phys}'$ \cite{Hoehn:2019fsy,AliAhmad:2021adn,Hoehn:2023axh}. This way of writing physical states makes transparent that the full Schr\"odinger trajectory $\{\ket*{\psi_{|C}(\tau)}\,|\,\tau\in\mathbb{R}\}$ is encoded both tangentially along a gauge orbit, as well as transversally to it, see Fig.~\ref{fig: Schrodinger evolution}. It is encoded tangentially because we can obtain the full trajectory from the same physical state $\Ket{\Psi'}$ (which corresponds to one gauge orbit) just by running through all gauges $\tau$ in $\mathcal{R}'_C(\tau)\Ket{\Psi'}$. Conversely, it is encoded transversally also because the same trajectory can be obtained by gauge fixing every state in the \emph{physical} Hilbert space trajectory $\{(U_S(\tau) \otimes \mathds{1}_C)\Ket{\Psi'}\,|\,\tau\in\mathbb{R}\}$ (which satisfies the Schr\"odinger equation on $\mathcal{H}_{\rm phys}'$) in the same way, e.g.\ with $\mathcal{R}'_C(0)$. When we say that the Schr\"odinger evolution is physical, we have this transversal evolution in mind.

Let us now translate the Page-Wootters formalism into the language of co-invariants. To this end, we recall that $\Ket{\Psi}=M\ket{\psi}$, $\ket{\psi}\in\mathcal{H}_{\rm kin}$, and $M^\dag M=\Pi_{\rm phys}$. Hence, to obtain the same relational Schr\"odinger picture on $\mathcal{H}_{|C}$ via a reduction $\mathcal{R}_C(\tau):\mathcal{H}_{\rm phys}\rightarrow\mathcal{H}_{|C}$ with
\begin{equation}\label{PWredcoinv}
    \ket*{\psi_{|C}(\tau)}=\mathcal{R}_C(\tau)\Ket{\Psi}\,,
\end{equation}
where the unprimed notation now clarifies that we are working with co-invariants, we must have
\begin{equation}\label{MR1}
    \mathcal{R}_C(\tau)M=\mathcal{R}_C'(\tau)\,\Pi_{\rm phys}\,.
\end{equation}
The ``dagger'' of this relation has to be defined with respect to the appropriate inner product. Since we have $(\Psi_1'|\Psi_2')=\bra{\psi_1}\Pi_{\rm phys}\ket{\psi_2}=\bra{\psi_1}\mathcal{R}_C'{}^\dag(\tau)\mathcal{R}'_C(\tau)\Pi_{\rm phys}\ket{\psi_2}=\braket*{\psi^1_{|C}(\tau)}{\psi^2_{|C}(\tau)}$ and this must be equal to $\bra{\psi_1}M^\dag\mathcal{R}_C^\dag(\tau)\mathcal{R}_C(\tau) M\ket{\psi_2}=\bra{\psi_1}M^\dag M\ket{\psi_2}=(\Psi_1|\Psi_2)$ and $\Pi_{\rm phys}$ is not a projector, this leads to the inverse relation for $\mathcal{R}_C^\dag(\tau):\mathcal{H}_{|C}\rightarrow\mathcal{H}_{\rm phys}$ with
\begin{equation}\label{MR2}
    M^\dag\mathcal{R}_C^\dag(\tau)=\mathcal{R}_C'{}^\dag\,.
\end{equation}
The inner product is then preserved and it is clear that the two formulations of the PW-formalism are equivalent.

\subsubsection{Observable reduction}\label{ssec_obsred}

Let us now clarify how the physical Hilbert space representation of gauge-invariant observables reduces into $C$'s perspective. 
It was shown in \cite{Hoehn:2019fsy} that relational observables gauge fix as follows for $a\in\mathcal{A}_S$:
\begin{equation}\label{obsredthm}
\mathcal{R}'_C(\tau){O}_C^\tau(a)\mathcal{R}'_C{}^\dag(\tau)=\Pi_{|C}\,a\,\Pi_{|C}\,.
\end{equation}
Using Eqs.~\eqref{MR1},~\eqref{MR2} and~\eqref{physrep}, as well as $r(O)M=MO$ for $O\in\mathcal{A}_{\rm kin}^H$, the left hand side becomes
\begin{equation}
\begin{split}\label{equivred}
\mathcal{R}'_C(\tau){O}_C^\tau(a)\mathcal{R}'_C{}^\dag(\tau) &=\mathcal{R}'_C(\tau)O_C^\tau(a)M^\dag\mathcal{R}_C^\dag(\tau)=\mathcal{R}'_C(\tau)M^\dag\mathcal{O}_C^\tau(a)\mathcal{R}_C^\dag(\tau)\\
&=\mathcal{R}_C(\tau)\mathcal{O}_C^\tau(a)\mathcal{R}_C^\dag(\tau)\,,
\end{split}
\end{equation}
so that we have an equivalent reduction result in co-invariant and RAQ language.
The gauge-invariant algebra therefore reduces equivalently in both languages as 
\begin{equation}\label{redalg}
    \mathcal{R}_C(\tau)\,r(\mathcal{A}_{\rm inv})\mathcal{R}_C^{\dag}(\tau) = \Pi_{|C}\mathcal{A}_{S|C}\Pi_{|C}=\mathcal{R}_C'(\tau)\mathcal{A}_{\rm inv}\mathcal{R}'_C{}^\dag(\tau),
\end{equation}
where $\mathcal{A}_{S|C}$ is generated by $a\in\mathcal{A}_S$ and $U_S(t)$, $t\in\RR$. Note that $\Pi_{|C}\in\mathcal{A}_{S|C}$ because it is a bounded function of $H_S$. 

Returning to the original motivation by Page and Wootters to define a conditional probability interpretation for physical states, it is now clear from Eqs.~\eqref{IPrel},~\eqref{obsredthm} and \eqref{equivred} that the evolution of a system observable $a\in\mathcal{A}_S$ in relational Schr\"odinger states $\ket*{\psi_{|C}(\tau)}$ is equivalent to that of the relational observable $\mathcal{O}_C^\tau(a)$ (respectively $O_C^\tau(a)$) in the ``timeless'' physical state $\Ket{\Psi}$ {(respectively $\Ket{\Psi'}$)} corresponding to $\ket*{\psi_{|C}(\tau)}$ via Eqs.~\eqref{PWstate} and~\eqref{PWredcoinv},
\begin{equation}
    (\Psi|\mathcal{O}_C^\tau(a)|\Psi)=\langle\psi_{|C}(\tau)|a|\psi_{|C}(\tau)\rangle=\Bra{\Psi'}O_C^\tau(a)\Ket{\Psi'}\,.
\end{equation} 
This can in particular be used to formulate consistent conditional probabilities that have an equivalent gauge-invariant and gauge-fixed form \cite{Hoehn:2019fsy,delaHamette:2021oex}.

For later, we note that even when $C$ is ideal (and so $\Pi_{|C}=\mathds{1}_S$) and $\mathcal{A}_S$ von Neumann, so that $\mathcal{A}_{\rm inv}$ is isomorphic to the crossed product $\mathcal{A}_S\rtimes_{U_S}\RR$, the intermediate expression $\Pi_{|C}\mathcal{A}_{S|C}\Pi_{|C}$ in Eq.~\eqref{redalg} is not unitarily equivalent to the crossed product algebra.  Since $\mathcal{R}_C$ is unitary, this means that $r$ is not faithful for $\mathcal{A}_{\rm inv}$. Indeed, a standard `gauge-fixed' form of the crossed product is generated by $a$ and bounded functions of $H'-H_S$ for some unbounded Hamiltonian $H'$ (e.g., see \cite{Chandrasekaran:2022cip,Jensen:2023yxy}), i.e.\ one of an ideal clock. Such an $H'$ is missing from  the right hand side in Eq.~\eqref{redalg}, which already tells us that an additional complementary ideal clock will be needed to render $r$ faithful. This is related to a puzzle discussed by CLPW in \cite{Chandrasekaran:2022cip}, and we will come back to this subtlety below.

Since the reduction map is unitary, the relations in Eqs.~\eqref{obsredthm} and~\eqref{equivred} tell us that the representation of relational observables on physical states, $\mathcal{O}_C^\tau(a)$, always constitutes a unital $*$-homomorphism from $\Pi_{|C}\mathcal{A}_S\Pi_{|C}$ into $r(\mathcal{A}_{\rm inv})$, \emph{regardless} of whether $C$ is ideal (see \cite{Hoehn:2019fsy,delaHamette:2021oex} for further discussion). By contrast, $O_C^\tau(a)$ only constitutes a homomorphism from $\mathcal{A}_S$ (and even $\Pi_{|C}\mathcal{A}_S\Pi_{|C}$) into the kinematical gauge-invariant algebra $\mathcal{A}_{\rm inv}$ when the clock $C$ is ideal because otherwise the $G$-twirl does not preserve products \cite{delaHamette:2021oex,Hoehn:2019fsy}. In other words, relational observables only preserve algebraic structure on physical states.

In appendix~\ref{app_bounded}, we use the reduction result in Eqs.~\eqref{obsredthm} and~\eqref{equivred} to further show that relational observables are complete on the physical Hilbert space. Specifically, we show that every bounded operator on $\mathcal{H}_{\rm phys}$ can be obtained from a bounded kinematical or system operator via the relational observable construction in Eq.~\eqref{relobs} and that this relational observable is also kinematically bounded. 

\subsection{QRF transformations: changing to a different clock}\label{ssec_QRFchanges}

A core output of the recent wave of QRF research is the formulation of a quantum frame covariance of physical properties. This relies on transformations between different QRFs and their ``internal perspectives'' on the physics. The key difference to classical frame transformations is that one is now able to transform between reference frames that may be in relative superposition or entangled with other systems and this leads to interesting physical consequences.
Such transformations have appeared in somewhat different contexts and formulations \cite{Giacomini:2017zju,Vanrietvelde:2018dit,Vanrietvelde:2018pgb,Hohn:2011us,Hoehn:2019fsy,Hoehn:2020epv,Hoehn:2023ehz,delaHamette:2020dyi,AliAhmad:2021adn,delaHamette:2021oex,Castro-Ruiz:2019nnl,Castro-Ruiz:2021vnq,Giacomini:2021gei,Carette:2023wpz,Krumm:2020fws,Kabel:2024lzr}. Here, we briefly review the QRF transformations of the perspective-neutral approach \cite{Vanrietvelde:2018dit,Vanrietvelde:2018pgb,Hoehn:2019fsy,Hoehn:2020epv,Hoehn:2023ehz,AliAhmad:2021adn,delaHamette:2021oex,Castro-Ruiz:2019nnl,Giacomini:2021gei}, as it is the one applicable to the present case of gauge systems with constraints. For our purposes, we restrict to the special case of clock transformations \cite{Hoehn:2019fsy,Hoehn:2020epv,Chataignier:2024eil} (with precursors \cite{Hohn:2011us,Hohn:2018iwn,Hohn:2018toe,Castro-Ruiz:2019nnl}) and refer to \cite{delaHamette:2021oex} for general symmetry groups.

The sought-after clock transformations are not related to the clock reorientations $\mathds{1}_S \otimes U_C(\tau)$. The latter change the orientation of the given temporal QRF $C$, i.e.\ the reading of the clock, and thus the relation between the clock and evolving system. This is a physical transformation and changes the state in the physical Hilbert space. Instead, we would now like to change to a \emph{different} clock subsystem and ``jump into its new internal perspective''. This transformation will amount to a change of \emph{description} of any state in $\mathcal{H}_{\rm phys}$ (respectively $\mathcal{H}_{\rm phys}'$), but not a change of state itself. Indeed, it will take the form of a change of gauge.

 To this end, we will assume additional structure in the system $S$, namely that it contains at least an additional clock subsystem. Thus, we replace $C\to C_1$ and $S\to SC_2$ and consider now 
\begin{eqnarray}\label{2clockconstraint}
    \mathcal{H}_{\rm kin}=\mathcal{H}_{S}\otimes\mathcal{H}_{1}\otimes\mathcal{H}_2,\qquad\qquad \text{with}\qquad\qquad H=H_{S}+H_{1}+H_2,
\end{eqnarray}
henceforth simplifying the notation by writing just $i$ instead of $C_i$ in subscripts. The argument in App.~\ref{app_ginvalg} entails that we can write the gauge-invariant algebra
\begin{equation}
    \mathcal{A}_{\rm inv}:=\left(\mathcal{A}_S \otimes \mathcal{B}(\mathcal{H}_{1})\otimes\mathcal{B}(\mathcal{H}_{2})\right)^{H}
\end{equation} 
on $\mathcal{H}_{\rm kin}$ equivalently as being generated by relational observables and reorientations relative to either $C_1$ or $C_2$ (cf.~Eq.~\eqref{gaugeinvalg})
\begin{eqnarray}
 \mathcal{A}_{\rm inv}&=&\Big\langle O^{\tau_1}_{C_1}(a),\mathds{1}_S \otimes U_{1}(t)\otimes \mathds{1}_{2}\,|\,a\in\mathcal{A}_S \otimes \mathcal{B}(\mathcal{H}_{2}),\,t\in\mathbb{R}\Big\rangle\nonumber\\
 &=&\Big\langle O^{\tau_2}_{C_2}(a),\mathds{1}_S \otimes \mathds{1}_{1}\otimes U_{2}(t)\,|\,a\in \mathcal{A}_S \otimes \mathcal{B}(\mathcal{H}_{1}),\,t\in\mathbb{R}\Big\rangle,
\end{eqnarray}
where relational observables are defined as before in Eq.~\eqref{relobs} but now also include the ones describing the two clocks relative to one another. Again, when $\mathcal{A}_S$ is von Neumann, then so is $\mathcal{A}_{\rm inv}$, and when $\mathcal{A}_S$ is a Type $\rm{III}_1$ factor, then so is $\mathcal{B}(\mathcal{H}_i)\otimes\mathcal{A}_S$, $i=1,2$, in which case $\mathcal{A}_{\rm inv}$ is also a factor (cf.~appendix~\ref{app_ginvalg}). In the special case of an ideal clock $C_1$, $\mathcal{A}_{\rm inv}$ is the crossed product of $\mathcal{A}_S \otimes \mathcal{B}(\mathcal{H}_{2})$ by the translation group.

We begin again in RAQ language, before translating the observation into the language of co-invariants. The ``quantum coordinate map'' into the perspective of clock $C_1$ when it reads $\tau_1$ becomes 
\begin{equation}\label{R1}
    \mathcal{R}'_{1}(\tau_1)=\bra{\tau_1}_1\otimes \mathds{1}_2\otimes \mathds{1}_S.
\end{equation} 
It is a unitary $\mathcal{R}'_1(\tau_1):\mathcal{H}'_{\rm phys}\to\mathcal{H}_{|1}$, where $\mathcal{H}_{|1}:=\Pi_{|1}\left(\mathcal{H}_{S}\otimes\mathcal{H}_2\right)\subseteq\mathcal{H}_{S}\otimes\mathcal{H}_2$ is the Hilbert space of $SC_2$ relative to clock $C_1$. It is the subspace of the kinematical Hilbert space factors $\mathcal{H}_{2}\otimes\mathcal{H}_S$ consistent with the constraint $H$ and the corresponding projector is given via Eq.~\eqref{projS} 
\begin{equation}\label{pi1}
    \Pi_{|1}:=\Pi_{|C_1}=\int_\mathbb{R}\dd{t}\chi^*_1(t)U_S(t)\otimes U_2(t),
\end{equation}
where $\chi_1(t)=\bra{t}0\rangle_1$. When $\sigma_1$ is just an interval $\sigma_1=[\epsilon_{\rm min}^1,\epsilon_{\rm max}^1]$, we have explicitly
\begin{equation}
    \Pi_{|1}=\Theta(\epsilon^1_{\rm max}+H_S+H_2)\Theta(-\epsilon^1_{\rm min}-H_S-H_2).
\label{eq:reducedProjector}
\end{equation}
The reduction into $C_2$-perspective is defined analogously.

The transformation from the perspective of $C_1$ into that of $C_2$  is thus given by \cite{Hoehn:2019fsy}
\begin{eqnarray}\label{clockchange}
    V'_{1\to2}(\tau_1,\tau_2):=\mathcal{R}'_2(\tau_2)\circ\mathcal{R}_1'{}^\dag(\tau_1)=\int_\mathbb{R}\dd{t}\ket{t+\tau_1}_1\otimes\bra{\tau_2-t}_2\otimes U_S(t).
\end{eqnarray}
It is clear that $\ket*{\psi_{|2}(\tau_2)}=V'_{1\to2}(\tau_1,\tau_2)\ket*{\psi_{|1}(\tau_1)}$, where $\ket*{\psi_{|i}(\tau_i)}=\mathcal{R}_i'(\tau_i)\Ket{\Psi'}$ is the state of $C_jS$ relative to $C_i$, $i\neq j$, which evolves in $\tau_i$ according to the Schr\"odinger equation with Hamiltonian $H_S+H_j$. The compositional form of the temporal QRF change is reminiscent of how one changes between different coordinate systems on a manifold -- they are ``quantum coordinate changes'':
\begin{equation}\label{clockchangefig}
    \begin{tikzcd}[column sep=8em, row sep=3em]
        & \mathcal{H}'_\text{phys} \arrow[swap]{dl}{\mathcal{R}_1'(\tau_1)} \arrow{dr}{\mathcal{R}_2'(\tau_2)} & \\
        \mathcal{H}_{|1}=\Pi_{|1}\left(\mathcal{H}_{S}\otimes\mathcal{H}_2\right) \arrow[swap]{rr}{V_{1 \to 2}'(\tau_1,\tau_2) \,=\, \mathcal{R}'_2(\tau_2) \circ \mathcal{R}_{1}'{}^\dag(\tau_1)} & & \mathcal{H}_{|2}=\Pi_{|2}\left(\mathcal{H}_{S}\otimes\mathcal{H}_1\right)
    \end{tikzcd}
\end{equation}
In analogy to how a manifold is an intrinsic object independent of what coordinates we choose, we can likewise view $\mathcal{H}_\text{phys}'$ as an intrinsic state space encoding the internal physics independently of the choice of QRF. 
$\mathcal{H}_\text{phys}'$ (and similarly the algebra $\mathcal{B}(\mathcal{H}_{\rm phys}')$) is thus a QRF perspective-neutral structure, encoding and linking all internal QRF perspectives. This provides this approach to QRF covariance with its name. It can be argued that this covariance structure constitutes a quantum extension of classical covariance structures akin to those underlying special covariance \cite[Sec.~II]{Hoehn:2023ehz} and \cite{delaHamette:2021oex}.

From the preceding discussion in Sec.~\ref{ssec_PW} it is clear that the same structure appears in co-invariant language, where the QRF transformations now read
\begin{equation}
    V_{1\to2}(\tau_1,\tau_2)=\mathcal{R}_2(\tau_2)\circ\mathcal{R}_1^\dag(\tau_1)\underset{\eqref{MR1}}{=}\mathcal{R}'_2(\tau_2)M^\dag\,\mathcal{R}_1^\dag(\tau_1)\underset{\eqref{MR2}}{=}V'_{1\to2}(\tau_1,\tau_2)
\end{equation}
and yield the same diagram as Eq.~\eqref{clockchangefig} with all primes dropped.

$V'_{1\to2}$  is a controlled, hence nonlocal unitary. This can also be seen by its action on observables, e.g.\ for $a\in\mathcal{A}_S$, we have \cite{Hoehn:2019fsy}
\begin{eqnarray}
    V'_{1\to2}(\tau_1,\tau_2)\Pi_{|1}\left(a\otimes \mathds{1}_2\right)\Pi_{|1}\left(V'_{1\to2}(\tau_1,\tau_2)\right)^\dag&=&\mathcal{R}'_2(\tau_2)\,{O}_{C_1}^{\tau_1}(a \otimes \mathds{1}_2)\mathcal{R}'_2{}^\dag(\tau_2)\label{wrongpersp}\\
    &=&\Pi_{|2}O_{C_1}^{\tau_1}( a)=\Pi_{|2}\int_\mathbb{R}\dd{t}U_{S1}(t)\left(a \otimes \ket{\tau_1}\!\bra{\tau_1}_1 \right)U_{S1}^\dag(t).\nonumber
\end{eqnarray}
Note that $O_{C_1}^{\tau_1}(a)$ commutes with $\Pi_{|2}$ as the latter is a function of $H_1+H_2$. Thus, when $C_1$ is ideal and so $\Pi_{|1}=\mathds{1}_{S2}$, a local system operator relative to $C_1$ transforms into a composite $C_1S$ operator relative to $C_2$. This is a reflection of subsystem relativity, which we will discuss in the next section, and which is at the heart of all QRF relative physical properties.
Thus, more generally
\begin{equation}\label{projalgeb}
    \mathcal{R}'_2(\tau_2)\mathcal{A}^H_{SC_1}\mathcal{R}_2'{}^\dag(\tau_2)=\Pi_{|2}\mathcal{A}^H_{SC_1}=\mathcal{R}_2(\tau_2)r(\mathcal{A}^H_{SC_1})\mathcal{R}_2^\dag(\tau_2)\,,
\end{equation}
where $\mathcal{A}^H_{SC_1}=\left(\mathcal{A}_S \otimes \mathcal{B}(\mathcal{H}_1)\otimes \mathds{1}_2\right)^H$ and an identity factor $\mathds{1}_2$ has been implicitly dropped in $\Pi_{|2}\mathcal{A}^H_{SC_1}$.

We noted above that the PW reduction map constitutes a gauge fixing. The clock frame transformation therefore takes the form of a change of gauge. This does not mean, however, that a choice of clock frame is equivalent to a choice of gauge or that the QRF dependence of physical properties is a gauge artifact. Indeed, we saw above that there also exists an equivalent manifestly gauge-invariant manner to ``jump'' into a QRF perspective, namely to dress kinematical operators with the chosen clock frame. In Sec.~\ref{ssec_subsystem relativity}, we will see that these lead to distinct gauge-invariant algebras\footnote{There also exists an equivalent gauge-invariant version of QRF transformations that acts on these algebras of observables and transforms them into one another \cite{Hoehn:2023ehz,delaHamette:2021oex} (see also \cite{Carrozza:2021gju,Goeller:2022rsx} for their formulation in classical gauge theory and gravity). This is a manifestation of a certain equivalence between gauge-fixings and dressings. However, we will not be using this version of QRF transformations in this work and hence do not review them here.} and that a QRF transformation amounts to a change of tensor product structure on $\mathcal{H}_{\rm phys}$. Rather, the choice of QRF is a choice of convention to split the kinematical degrees of freedom into redundant and non-redundant ones. For a gauge fixing procedure this means a convention for {which} kinematical degrees of freedom one may fix, however, not yet \emph{how} to fix them; a choice of QRF thus corresponds to an entire family of possible gauge fixings but it is not a gauge itself. This is linked to our discussion above that the Schr\"odinger evolution is physical. 

\subsection{Subsystem relativity}\label{ssec_subsystem relativity}

We started with an assumed tensor product structure (TPS) on the kinematical Hilbert space, $\mathcal{H}_{\rm kin}=\mathcal{H}_S \otimes \mathcal{H}_{1}\otimes\mathcal{H}_{2}$. Given that this is the space of states that are distinguishable relative to an external (possibly fictitious) frame, the TPS on it corresponds to an externally distinguishable subsystem structure. But our internal physics happens on the space $\mathcal{H}_{\rm phys}$ of internally distinguishable states, and as we have seen this is a distinct Hilbert space that in the present case is also not a subspace of $\mathcal{H}_{\rm kin}$; it thus does not inherit the kinematical TPS. We are now interested in an internally distinguishable subsystem structure and TPS on $\mathcal{H}_{\rm phys}$. Such a structure is crucial for defining a gauge-invariant notion of entanglement and entropies, but, as we will see, it depends on the choice of QRF and corresponds to its internal perspective \cite{Hoehn:2023ehz,AliAhmad:2021adn,delaHamette:2021oex,Castro-Ruiz:2021vnq}.

This can be seen in two equivalent ways \cite[Sec.~III]{Hoehn:2023ehz}. For later purpose, we will formulate everything here in the co-invariant language; from the preceding sections it is clear that all following statements take an equivalent form in RAQ language.  Let us first consider the case that $C_1,C_2$ are both ideal, so that $\Pi_{|1}$ and $\Pi_{|2}$ are each equal to the identity. Since the ``quantum coordinate maps'' are unitary, the diagram~\eqref{clockchangefig} tells us that each reduction map $\mathcal{R}_i(\tau_i)$, $i=1,2$, is nothing but a TPS on $\mathcal{H}_{\rm phys}$ and, because $V_{1\to2}$ is a nonlocal unitary, the change of clock amounts to a change of TPS on the physical Hilbert space.\footnote{More precisely, a TPS on an abstract Hilbert space $\mathcal{H}$ is an equivalence class of unitaries, $U:\mathcal{H}\to\bigotimes_\alpha\mathcal{H}_{\alpha}$, such that $U_1\sim U_2$ if $U_2U_1^{-1}$ is a product of local unitaries $\otimes_\alpha U_\alpha$ (and possibly permutations of subsystem factors). Since $\mathcal{R}_i(\tau_i+t)=\left(U_j(t)\otimes U_S(t)\right)\mathcal{R}_i(\tau_i)$ via the equivalence of reorientations and subsystem evolution on $\mathcal{H}_{\rm phys}$, we have that $\mathcal{R}_i(\tau_i)$ defines the same equivalence class (and so TPS on $\mathcal{H}_{\rm phys}$) for all $\tau_i\in\mathbb{R}$. Hence, the TPS depends only on the choice of QRF, but not on its orientation to which we might gauge fix.} In other words, an \emph{ideal} internal QRF perspective is a TPS on $\mathcal{H}_{\rm phys}$. 

Suppose now that $C_1,C_2$ are non-ideal with $\sigma_i:=\sigma_{C_i}=[\epsilon_{\rm min}^i,\epsilon_{\rm max}^i]$ so $\Pi_i\neq \mathds{1}_{Sj}$. We show in appendix~\ref{app_non-idealclockTPS} that the reduced Hilbert space in $C_1$-perspective of diagram \eqref{clockchangefig} admits the following decomposition
\begin{equation}\label{non-idealdecompeq}
    \mathcal{H}_{|1}=\mathcal{H}_{S2}^-\oplus\left(\Pi_S(\mathcal{H}_S) \otimes \mathcal{H}_{2}\right)\oplus\mathcal{H}_{S2}^+,
\end{equation}
where $\Pi_S$ is the projector onto the subspace of $\mathcal{H}_S$ spanned by eigenstates with $-\epsilon_{\rm max}^1-\epsilon_{\rm min}^2\leq E_S\leq-\epsilon_{\rm min}^1-\epsilon^2_{\rm max}$ and $\mathcal{H}^\pm_{2S}$ are direct sums/integrals of tensor products between $C_2$ and $S$ subspaces when $\sigma_S:=\rm{Spec}(H_S)$ is discrete/continuous. (Later, $H_S$ will be a QFT modular Hamiltonian and thus feature a continuous spectrum.) That is, for non-ideal clocks, $\mathcal{R}_i(\tau_i)$ defines instead a direct sum/integral of TPSs on $\mathcal{H}_{\rm phys}$ and the clock change amounts to a change of such a direct sum/integral of TPSs.

Second, the QRF relativity of the internal subsystem structure can also be seen at the gauge-invariant, algebraic level.  To this end, we can ask which system observables can be measured relative to both clock frames if each clock has no operational access to the other clock degrees of freedom. These reside in the intersection of the two gauge-invariant kinematical subalgebras, describing $S$ and the clock $C_1$ or $C_2$, respectively,
\begin{eqnarray}\label{subalgebras}
    \mathcal{A}_{SC_1}^{H}:=\left(\mathcal{A}_S\otimes\mathcal{B}(\mathcal{H}_{1})\otimes \mathds{1}_{2}\right)^{H},\qquad\qquad\mathcal{A}_{SC_2}^{H}:=\left(\mathcal{A}_S \otimes \mathds{1}_{1}\otimes\mathcal{B}(\mathcal{H}_{2})\right)^{H}.
\end{eqnarray}
We have, using that $\mathcal{A}^{H}=\mathcal{A}\cap\big\langle H\big\rangle'$, where $\big\langle H\big\rangle'\subset\mathcal{B}(\mathcal{H}_{\rm kin})$ denotes the commutant of (bounded functions of) the constraint,
\begin{eqnarray}
   \mathcal{A}_{SC_1}^{H} \cap\mathcal{A}_{SC_2}^{H}=\left(\mathcal{A}_{SC_1}\cap\mathcal{A}_{SC_2}\right)^{H}=\left(\mathcal{A}_S \otimes \mathds{1}_{1}\otimes \mathds{1}_{2}\right)^{H}.
\end{eqnarray}
Thus, the algebra of observables accessible to both clock frames is equal to the algebra of gauge-invariant observables that are \emph{internal} to the system. The intuition behind this is simple \cite{Hoehn:2023ehz}: $\mathcal{A}_{SC_1}^{H}$ does not contain reorientations of $C_2$ but commutes with them and vice versa with $C_1$ and $C_2$ interchanged. Hence, observables in the intersection must commute with reorientations of both clocks but cannot include reorientations of either. This means that the intersection is given by relational observables such that $a$ in Eq.~\eqref{relobs} already commutes with $H_S$ and so $O_{C_i}^{\tau_i}(a\otimes \mathds{1}_j)= a\otimes \mathds{1}_{1}\otimes \mathds{1}_{2}$, $i=1,2$, $i\neq j$. These are also the only observables that are purely $S$-local relative to both frames \cite{Hoehn:2023ehz}.

The two subalgebras in Eq.~\eqref{subalgebras} are thus distinct. An extreme example is provided when $\mathcal{A}_S$ is a type $\rm{III}_1$ von Neumann algebra and $H_S$ is a modular Hamiltonian for it with an ergodic action, a situation that will appear in perturbative quantum gravity (e.g.\ in de Sitter space \cite{Chandrasekaran:2022cip,Chen:2024rpx}) below. Then the only $S$ observables that commute with the constraint are $c$-numbers. In other words, in that case, the two frames do not share \emph{any} common QFT observables and we have
\begin{equation}\label{trivial}
    \mathcal{A}_{SC_1}^{H} \cap\mathcal{A}_{SC_2}^{H}=\CC \mathds{1}_{S12}.
\end{equation}

Nevertheless, the two subalgebras \eqref{subalgebras} are isomorphic when $C_1$ and $C_2$ have identical spectra. For example, when both clock are ideal, both algebras are isomorphic to the crossed product $\mathcal{A}_S\rtimes_{U_S}\mathbb{R}$ and hence also isomorphic to one another. However, they are not isomorphic more generally. As noted in Sec.~\ref{ssec_obsred}, when the clock is ideal, relational observables constitute a homomorphism from the system algebra $\mathcal{A}_S$ into $\mathcal{A}_{\rm inv}$, while this is not true when the clock is non-ideal. Thus, for instance, when $C_1$ is ideal but $C_2$ is not, the two subalgebras are not isomorphic.

Moreover, note that $[\mathcal{A}_{SC_1}^H,\mathcal{A}_{SC_2}^H]\neq0$ because the dressed $S$ operators in one algebra do not commute with the dressed $S$ operators in the other. Hence, even in the extreme case that the algebras overlap trivially -- as they sometimes do in QFT, Eq.~\eqref{trivial} -- the two gauge-invariant algebras describing $S$ relative to the two clocks do not define independent subsystems and acting with operators from one algebra will affect the outcomes of measurements from the other. 
 
 It is clear that these observations carry over to the  representation of these algebras on the physical Hilbert space, i.e.\
\begin{equation}\label{physoverlap}
r(\mathcal{A}_{SC_1}^{H})\cap r(\mathcal{A}_{SC_2}^{H})=r((\mathcal{A}_S \otimes \mathds{1}_{1}\otimes \mathds{1}_{2})^{H}).
\end{equation}
It is in particular somewhat simpler to see that $r(\mathcal{A}_{SC_1}^{H})$ and $r(\mathcal{A}_{SC_2}^{H})$ are isomorphic only when $C_1$ and $C_2$ have identical spectra. Owing to the unitarity of $\mathcal{R}_i$, the relation in Eq.~\eqref{redalg} applied to this case tells us that
\begin{equation}\label{eq263}
r(\mathcal{A}_{SC_i}^{H} )\simeq\Pi_{|i}\mathcal{A}_{S|i}\Pi_{|i}, \quad i=1,2,
\end{equation}
where $\mathcal{A}_{S|i}$ is generated by $a\in\mathcal{A}_S$ and $U_{Sj}(t)$ with  $j\neq i$. Since $\mathcal{A}_{S|1}$ and $\mathcal{A}_{S|2}$ are isomorphic and $\Pi_{|i}\in\mathcal{A}_{S|i}$, we have that the physical representations $r(\mathcal{A}_{SC_1}^{H} )$ and $r(\mathcal{A}_{SC_2}^{H})$ are isomorphic only provided the projectors $\Pi_{|1}$ and $\Pi_{|2}$ are identical functions of $H_S+H_2$ and $H_S+H_1$, respectively.\footnote{Otherwise, suppose $\Pi_{|1}=f(H_S+H_2)$ and $\Pi_{|2}=g(H_S+H_1)$ for $f,g$ distinct. Then the projector $I-f\in\mathcal{A}_{S|i}$ maps to zero in one representation but not the other, i.e.\ $\Pi_{|1}(\mathds{1}_{S2}-f(H_S+H_2))\Pi_{|1}=0$, but $\Pi_{|2}(\mathds{1}_{S1}-f(H_S+H_1))\Pi_{|2}\neq0$.} By Eq.~\eqref{pi1} this means that the spectra of $C_1,C_2$ coincide. Thus, the system (later the QFT degrees of freedom) will appear differently relative to a fuzzy clock than it does to a sharp one.

In the special case when the clock $C_j$ is ideal and $\mathcal{A}_S=\mathcal{B}(\mathcal{H}_S)$ is the full system algebra (hence a Type $\rm{I}$ factor), we have $r(\mathcal{A}_{SC_1C_2}^H)=\mathcal{O}_{i|j}\vee\mathcal{O}_{S|j}$\footnote{$\mathcal{A}_1\vee\mathcal{A}_2$ denotes the smallest algebra containing both $\mathcal{A}_1$ and $\mathcal{A}_2$.} with $[\mathcal{O}_{i|j},\mathcal{O}_{S|j}]=0$, for either $j=1$ or $j=2$ and $i\neq j$, where $\mathcal{O}_{i|j}=\big\langle\mathcal{O}_{C_j}^{\tau_j}(a)\,|\,a\in\mathcal{B}(\mathcal{H}_i)\otimes \mathds{1}_{Sj}\big\rangle$ is the algebra of relational observables on the physical Hilbert space describing $C_i$ relative to $C_j$, and similarly $\mathcal{O}_{S|j}=\big\langle\mathcal{O}_{C_j}^{\tau_j}(a)\,|\,a\in \mathcal{A}_S \otimes \mathds{1}_{12}\big\rangle$ is the algebra of relational observables on $\mathcal{H}_{\rm phys}$ describing $S$ relative to $C_j$. In this case, $\mathcal{O}_{i|j}$ and $\mathcal{O}_{S|j}$ induce the TPS $\mathcal{R}_j(\tau)$ associated with $C_j$ on $\mathcal{H}_{\rm phys}$ (e.g., see \cite{Hoehn:2023ehz} for the finite-dimensional case). When $C_j$ is non-ideal, then $\mathcal{O}_{i|j}$ and $\mathcal{O}_{S|j}$ do not close as algebras\footnote{For $\mathcal{O}_{S|j}$ this can be seen from Eq.~\eqref{eq263}, by noting that $\Pi_{|j}\mathcal{A}_S\Pi_{|j}$ is not closed under products because $\Pi_{|j}\notin\mathcal{A}_S$.} and typically do not commute \cite{AliAhmad:2021adn}.

The fact that different QRFs associate different gauge-invariant subalgebras and Hilbert space tensor factors with the same (kinematical) system $S$ means that the notions of subsystem and subsystem locality depend on the choice of QRF and with it essentially all physical properties that derive from a subsystem partition. For example, in quantum mechanics it was shown that not only entanglement and superpositions (originally observed in \cite{Giacomini:2017zju}), but also entropies, interactions, and thermodynamical properties depend on the choice of QRF \cite{Hoehn:2023ehz}; in particular, subsystem  thermality and temperature are affected by QRF changes in a way that is independent of classical frame changes in quantum field theory, which lead to phenomena such as the Unruh effect. Subsystem relativity can further be understood as a generalisation of the relativity of simultaneity in special relativity \cite{Hoehn:2023ehz}; much like the latter explains the \emph{a priori} counter-intuitive phenomena of special relativity, so too is subsystem relativity the source of the novel QRF-dependent physics. Our observation in Sec.~\ref{sec_entropy} (expanding on \cite{DeVuyst:2024pop}) that entanglement entropy associated with gravitational subregions is \observerdependent{} is a manifestation and (perturbative) quantum gravity extension of this.

\subsection{Kinematical vs.\ physical von Neumanness}\label{ssec_vN}

Let us assume again that $\mathcal{A}_S$ is a von Neumann algebra. We noted before that then so are both $\mathcal{A}_{SC_i}^H$, $i=1,2$, and $\mathcal{A}_{SC_1C_2}^H$. This statement refers to the gauge-invariant kinematical level, i.e.\ the closure\footnote{Thanks to the bicommutant theorem, this can be any of the bicommutant, weak or strong closure.} of the algebra is defined in terms of the structure of $\mathcal{H}_{\rm kin}$. What about the representations of these algebras on the physical Hilbert space $\mathcal{H}_{\rm phys}$, which is not a subspace of $\mathcal{H}_{\rm kin}$ and so closure will be defined differently? To answer this question, we will here again invoke the setting of co-invariants for later purpose, but it is clear that the conclusions are equivalent in RAQ language. We will now argue that, while $r(\mathcal{A}_{SC_i}^H)$ remains a von Neumann algebra, this is not {necessarily} true for $r(\mathcal{A}_{SC_1C_2}^H)$ when $\mathcal{A}_S$ is a Type $\rm{III}_1$ factor. In other words, for the physical von Neumanness of the observable algebras under consideration it is crucial that there is a (kinematically) complementary clock on which this algebra has trivial support. Indeed, later the system $S$ will contain many more clocks.

To see that $r(\mathcal{A}_{SC_1}^H)$ (and similarly $r(\mathcal{A}_{SC_2}^H)$) remains von Neumann, we can invoke Eq.~\eqref{projalgeb} which entails that reduction into the complementary clock $C_2$'s perspective gives
\begin{equation}\label{wrong2}
    \mathcal{R}_2(\tau_2)\,r(\mathcal{A}_{SC_1}^H)\mathcal{R}^\dag_2(\tau_2)=\Pi_{|2}\mathcal{A}_{SC_1}^H
\end{equation}
with $\Pi_{|2}$ and $\mathcal{A}_{SC_1}^H$ commuting. Note that $\Pi_{|2}$ is only contained in $\mathcal{A}_{SC_1}^H$ when $H_S$ generates inner automorphisms of it; otherwise $\Pi_{|2}\mathcal{A}_{SC_1}^H$ is not a subalgebra of $\mathcal{A}_{SC_1}^H$. However, the right-hand side is clearly a subalgebra of $\mathcal{B}(\mathcal{H}_{|2})$ and since this constitutes a $*$-homomorphism $\pi:\mathcal{A}_{SC_1}^H\to\mathcal{B}(\mathcal{H}_{|2})$ it is a representation of $\mathcal{A}_{SC_1}^H$ on $\mathcal{H}_{|2}$. $\mathcal{R}_2$ is unitary and so our physical representation is isomorphic to it, $r(\mathcal{A}_{SC_1}^H)\simeq\Pi_{|2}\mathcal{A}_{SC_1}^H$. But we also have that $\Pi_{|2}\mathcal{A}_{SC_1}^H$ is (ultra-)weakly closed in $\mathcal{B}(\mathcal{H}_{|2})$.\footnote{Let $a_i\in\mathcal{A}\subset\mathcal{B}(\mathcal{H})$ be a net of operators. It converges weakly to the limit operator $a\in\mathcal{B}(\mathcal{H})$ if the matrix elements $\bra{\phi}a_i\ket{\psi}\to\bra{\phi}a\ket{\psi}$ converge for all $\phi\in\mathcal{H}^*$ and $\psi\in\mathcal{H}$. $\mathcal{A}$ is weakly closed in $\mathcal{B}(\mathcal{H})$ if it contains the limit operator of every net of operators in $\mathcal{A}$ that converges in $\mathcal{B}(\mathcal{H})$. It is ultraweakly closed if tensoring in an identity factor $\mathcal{A}\to\mathcal{A}\otimes\mathds{1}$ on an additional infinite dimensional (separable) Hilbert space does not change these convergence and closure properties.} Indeed, since $\mathcal{A}_{SC_1}^H$ is a von Neumann algebra, it is (ultra-)weakly closed on $\mathcal{H}_S\otimes\mathcal{H}_1$, as well as on any subspace that it leaves invariant, like $\mathcal{H}_{|2}=\Pi_{|2}\left(\mathcal{H}_S\otimes\mathcal{H}_1\right) $. Hence, $\Pi_{|2}\mathcal{A}_{SC_1}^H$ is a von Neumann algebra and therefore also $r(\mathcal{A}_{SC_1}^H)$.

We conclude that $r$ is an ultraweakly continuous representation of $\mathcal{A}_{SC_1}^H$ (and similarly $\mathcal{A}_{SC_2}^H$) on the physical Hilbert space. The projector $\Pi_{|2}$ on the right hand side of Eq.~\eqref{wrong2} might at first suggest that this representation may not be faithful\footnote{The representation is faithful when its kernel is $\{0\}$.} when $C_2$ is non-ideal, so that the projector is non-trivial. However, it turns out that every non-trivial ultraweakly continuous representation of a von Neumann \emph{factor} is faithful
\cite[Lecture 22, Proposition 3]{LurievNalgebras}. 

In the remainder of this work, we shall be interested in the case when $\mathcal{A}_S$ is a Type $\rm{III}_1$ von Neumann factor, associated with the quantum field and graviton observables inside a gravitational subregion, and when $H_S$ generates modular flow. We noted before that in this case both $\mathcal{A}_{SC_i}^H$, $i=1,2$, are Type $\rm{II}$ factors. Thus, the physical representation $r(\mathcal{A}_{SC_1}^H)$ (and similarly $r(\mathcal{A}_{SC_2}^H)$) is faithful regardless  of whether the complementary clock is ideal or non-ideal, and in particular it remains a factor.

Let us now explain why the physical representation $r$ of $\mathcal{A}_{SC_1C_2}^H$ fails to be von Neumann when $H_S$ acts ergodically on $\mathcal{A}_S$ and its commutant $\mathcal{A}_S'$ on $\mathcal{H}_S$ (as for the Type $\rm{III}_1$ static patch algebras in de Sitter space \cite{Chandrasekaran:2022cip,Chen:2024rpx}).  The full commutant $(\mathcal{A}_{SC_1C_2}^H)'$ in $\mathcal{B}(\mathcal{H}_{\rm kin})$ is generated by $a'\in\mathcal{A}_S'$ and bounded functions of the constraint $H$. Only the latter are gauge invariant because the ergodic action of $H_S$ means that only $c$-numbers from $\mathcal{A}_S'$ commute with the constraint. The physical representation $r$ of the gauge-invariant component\footnote{By this we mean $(\mathcal{A}_{SC_1C_2}^H)'\wedge\mathcal{B}(\mathcal{H}_{\rm kin})^H$, the largest gauge invariant algebra whose elements are in both $(\mathcal{A}_{SC_1C_2}^H)'$ and $\mathcal{B}(\mathcal{H}_{\rm kin})^H$} of $(\mathcal{A}_{SC_1C_2}^H)'$ thus coincides with the identity factor $\CC\mathds{1}$.  The double commutant on the physical Hilbert space then returns the full Type $\rm{I}$ factor $r(\mathcal{A}_{SC_1C_2}^H)''=\mathcal{B}(\mathcal{H}_{\rm phys})$. That is, $r(\mathcal{A}_{SC_1C_2}^H)\subsetneq r(\mathcal{A}_{SC_1C_2}^H)''$ is not closed under the bicommutant and therefore not a von Neumann algebra.

This observation links with challenges exhibited by CLPW in \cite[Sec.~4.3]{Chandrasekaran:2022cip} when discussing a single observer with a clock in de Sitter space. We will return to this topic in Sec.~\ref{subsec: need for complementary frame}.

\section{Realisation in perturbative quantum gravity: PW = CLPW}
\label{sect:CLPW}

\emph{The quick reader mainly interested in our derivations of density operators and entropies for arbitrarily many observers/clock QRFs can skip to Sec.~\ref{Section: arbitrary number of observers}.}

Thus far we have not been specific about the system $S$. It could have been another mechanical system or it could in fact have been a quantum field subsystem.  We will now become more concrete and inquire under which conditions we can realise Hamiltonian constraints of the type in Eq.~\eqref{2clockconstraint} 
\begin{equation}\label{2clockconstraint2}
    H=H_S+H_1+H_2
\end{equation}
in quantum gravity. (Below we will generalise this to an arbitrary number of clocks.) While such constraints and clocks arise naturally in minisuperspace, e.g.\ see \cite{kiefer2007quantum,Ashtekar:1993wb,Bojowald:2010qpa},\footnote{For the use of clock QRFs in that case, see for example \cite{Hohn:2018toe,Hoehn:2020epv,Hohn:2011us}.} we are here after the more interesting case when $S$ is a QFT encoding gravitons, as well as possibly several species of matter degrees of freedom. Specifically, we are interested in scenarios in which $H_S$ is a modular Hamiltonian and a type transition of the $S$ algebra occurs along with an intrinsic UV-regularisation of entanglement entropies. In this regard, we have to clarify suitable conditions under which \begin{itemize} 
\item[(a)] it is legitimate to consider only a single constraint, where the gauge diffeomorphism group of gravity is infinite-dimensional, and 
\item[(b)] modular flow aligns with a temporal spacetime diffeomorphism. 
\end{itemize}
This will thus be a hybrid scenario of quantum mechanical clock QRFs and a field-theoretic system. For a full account of the bulk diffeomorphism group, field-theoretic frames are needed \cite{Goeller:2022rsx,Carrozza:2022xut,Hoehn:2023axh}.

\subsection{Subregions and constraints in perturbative quantum gravity}
\label{sect:linearisationConditions}

The fact that there are no interactions between the clocks and $S$ in Eq.~\eqref{2clockconstraint2} already suggests that, in a gravitational setting, we must be ignoring backreaction and thus considering a perturbative regime. The background spacetime will also permit us to associate the system $S$ with the field degrees of freedom inside some subregion defined relative to that background structure in the spirit of the algebraic approach \cite{haag2012local,WittenRevModPhys.90.045003} (though see \cite{Witten:2023xze} for a background-independent proposal). 

In the remainder of this work, we thus consider quantum gravity around a background in the $\kappa\to0$ limit, where $\kappa=\sqrt{32\pi G_N}$. Accordingly, we shall be mostly interested in the degrees of freedom of the linearised theory, though, as we explain below, the crossed product generating constraints are of second-order in $\kappa$ \cite{Jensen:2023yxy,Kudler-Flam:2023qfl,DeVuyst:2024grw,Kaplan:2024xyk}. This background can be a vacuum spacetime, or one sourced by matter, and the gravity theory may be general relativity or some more general gravity theory \cite{DeVuyst:2024grw}. For the moment, we assume that this background possesses a Killing horizon as this is the case when it is known that the modular flow of a KMS state of the QFT has a geometric interpretation as a spacetime diffeomorphism, addressing (b) above, namely the one corresponding to the boost Killing vector field generating the horizon \cite{haag2012local,Bisognano:1975ih,WittenRevModPhys.90.045003,Sewell:1982zz,sorce2024analyticity}. We comment on more general proposals below.

In this background, we consider two subregions $\mathcal{U}$ and $\mathcal{U}'$ that are causal complements of one another. For example, in \cite{Chandrasekaran:2022cip}, $\mathcal{U},\mathcal{U}'$ are two complementary static patches in de Sitter space or the left and right exterior region in a maximally extended asymptotically flat Schwarzschild spacetime. In \cite{Kudler-Flam:2023qfl} this was extended to several other types of black hole spacetimes and in \cite{Jensen:2023yxy} it was proposed that $\mathcal{U}$ could be a general subregion in a generic spacetime. 

We denote the QFT Hilbert space of the global spacetime encoding gravitons and matter degrees of freedom with $\mathcal{H}_S$ and identify the algebra $\mathcal{A}_\mathcal{U}\subset\mathcal{B}(\mathcal{H}_S)$ generated by all the bounded QFT operators with exclusive support in $\mathcal{U}$ and invariant under \emph{linearised} bulk diffeomorphisms, as well as any matter gauge symmetry, with the system algebra, i.e.
\begin{equation}
    \mathcal{A}_S=\mathcal{A}_\mathcal{U}\,.
\end{equation}
At leading order in $\kappa$, only the graviton field will transform non-trivially under spacetime diffeomorphisms if the background is a vacuum one \cite{Jensen:2023yxy,DeVuyst:2024grw}. In that case, one can extract the gauge-invariant graviton data via standard methods, such as transverse-traceless modes. When the background is sourced by a non-trivial matter content, typically both matter and gravitons will transform non-trivially under linearised diffeomorphisms \cite{DeVuyst:2024grw}, and gauge-invariant observables can in this case be formulated using dressing methods such as in \cite{Donnelly:2015hta,Donnelly:2016rvo,Frob:2022ciq,Frob:2023gng}.
In line with the general expectation for asymptotically scale invariant quantum field theories, we shall henceforth assume that $\mathcal{A}_S$ is a Type $\rm{III}_1$ von Neumann factor \cite{Fredenhagen:1984dc,Buchholz:1986bg,Buchholz:1995gr,haag2012local,Yngvason_2005}. Hence, its commutant $\mathcal{A}_S'$ is also a Type $\rm{III}_1$ factor and Haag duality \cite{haag2012local,WittenRevModPhys.90.045003} tells us that we can identify it with the corresponding algebra $\mathcal{A}_{\mathcal{U'}}$ associated with the causal complement $\mathcal U'$.

So far we have described the linearised theory in $\kappa$. In order to obtain the modular crossed product algebra, we in fact have to go one order higher because on-shell of the background equations of motion, the first-order constraint associated with any Killing field  is identically a boundary term
\begin{equation}
    C^{(1)}[\kappa\xi]=\int_\Sigma C^{(1)}_{\kappa\xi}=\int_{\partial\Sigma}B^{(1)}_{\kappa\xi}\,,
\end{equation}
where $C_{\kappa\xi}^{(1)}$ is the first-order constraint current associated with Killing field $\xi$, $B^{(1)}_{\kappa\xi}$ is some boundary term (the first-order ADM Hamiltonian in general relativity) and $\Sigma$ is a global Cauchy slice \cite{DeVuyst:2024grw} (see also \cite{fischer1980structure,Abbott:1981ff} for the case of pure vacuum general relativity). In particular, for spatially closed spacetimes, $C^{(1)}[\kappa\xi]$ vanishes \emph{identically}.

\begin{figure}
    \centering
    \begin{tikzpicture}[scale=1.16]
        \begin{scope}[red!40!white]
            \foreach \i in {0.3,1,1.85,2.6} {
                \draw[postaction={decorate,decoration={markings,mark=at position .57 with {\arrow[scale=1.1]{stealth}}}}] (2,2) .. controls ({2-0.8*\i},{2-0.8*\i}) and ({2-0.8*\i},{0.8*\i-2}) .. (2,-2);
                \draw[postaction={decorate,decoration={markings,mark=at position .57 with {\arrow[scale=1.1]{stealth}}}}] (2,2) .. controls ({2+0.8*\i},{2-0.8*\i}) and ({2+0.8*\i},{0.8*\i-2}) .. (2,-2);

                \draw[postaction={decorate,decoration={markings,mark=at position .57 with {\arrow[scale=1.1]{stealth}}}}] (6,-2) .. controls ({6-0.8*\i},{0.8*\i-2}) and ({6-0.8*\i},{2-0.8*\i}) .. (6,2);
                \draw[postaction={decorate,decoration={markings,mark=at position .57 with {\arrow[scale=1.1]{stealth}}}}] (6,-2) .. controls ({6+0.8*\i},{0.8*\i-2}) and ({6+0.8*\i},{2-0.8*\i}) .. (6,2);
            }
        \end{scope}
        \draw (0,0) -- (2,2) -- (6,-2) -- (8,0) -- (6,2) -- (2,-2) -- (0,0);
        \draw[decorate,decoration={snake,amplitude=0.7pt,segment length=3pt},gray] (2,2) .. controls (3,1.9) and (5,1.9) .. (6,2);
        \draw[decorate,decoration={snake,amplitude=0.7pt,segment length=3pt},gray] (2,-2) .. controls (3,-1.9) and (5,-1.9) .. (6,-2);
        \draw[very thick, blue] (4,0) .. controls (3,0.3) and (1,0.3) .. (0,0);
        \draw[very thick, blue] (4,0) .. controls (5,-0.3) and (7,-0.3) .. (8,0) node[midway,below,blue!60!black] {\large$\Sigma$};
        \fill (8,0) circle (0.07) node[right] {$i^0$};
        \node[above right] at (7,1) {$\mathscr{I}^+$};
        \node[below right] at (7,-1) {$\mathscr{I}^-$};
    \end{tikzpicture}
    \caption{Depicted is the maximally extended Schwarzschild spacetime together with its time Killing flow lines. The Killing boost Hamiltonian $H_\xi$ simultaneously maps the Cauchy slice $\Sigma$ forwards in time in the right region and backwards in time in the left region. The role of the clocks can here be played by the ADM Hamiltonian defined the right and left asymptotic boundary $H_{\rm ADM}^{(2)}=H_R-H_L$ with the relative minus sign accounting for the opposite directions of the Killing vector $\xi$ in both regions.}
    \label{Figure: Schwarzschild both side boost}
\end{figure}
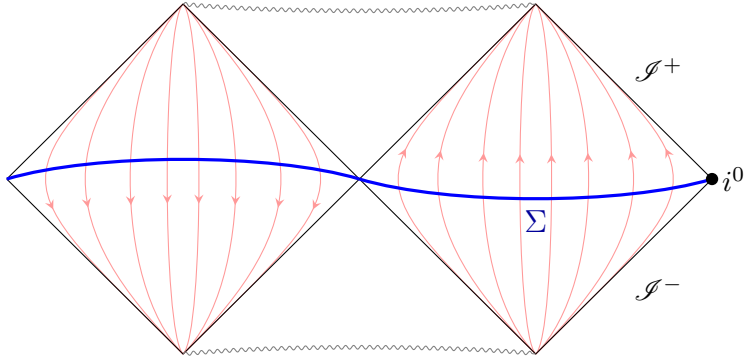

The constraint associated with $\xi$ only has a non-trivial bulk contribution at second order in $\kappa$ 
\begin{equation}\label{2ndordercon}
    C^{(2)}[\kappa\xi]=H_\xi-\int_{\partial\Sigma} B^{(2)}_{\kappa\xi}\,,
\end{equation}
where (on-shell of the first-order equations of motion and boundary conditions) $H_\xi$ is given by the Killing boost Hamiltonian 
\begin{equation}\label{boostgen}
H_\xi=\int_\Sigma (\epsilon^0)^\mu T_{\mu\nu}\xi^\nu\,,
\end{equation}
with $\epsilon^0$ the volume form of the background metric $g_0$ and $T_{\mu\nu}$ a stress-energy tensor of all fields, including gravitons \cite{DeVuyst:2024grw}. The crux is that $\beta H_\xi$ for some $\beta\geq0$  is the modular Hamiltonian associated with some global QFT state $\ket{\psi}_S\in\mathcal{H}_S$ that is cyclic and separating for $\mathcal{A}_S=\mathcal{A}_\mathcal{U}$ \cite{haag2012local,Bisognano:1975ih,WittenRevModPhys.90.045003,Sewell:1982zz,sorce2024analyticity}, and we identify it as the system Hamiltonian:
\begin{equation}
    H_S=H_\xi\,.
\end{equation}
The boundary piece can be interpreted as the second-order ADM Hamiltonian \cite{DeVuyst:2024grw,Kudler-Flam:2023qfl,Faulkner:2024gst}, 
\begin{equation}
    H_{\rm ADM}^{(2)}=\int_{\partial\Sigma}B^{(2)}_{\kappa\xi}\,,
\end{equation}
and it is given by a sum of Hamiltonians, one for each disconnected piece of the asymptotic boundary. For example, in the maximally extended asymptotically flat Schwarzschild spacetime, there is a left and a right asymptotic boundary and $H_{\rm ADM}^{(2)}=H_R-H_L$ \cite{Chandrasekaran:2022cip}, so that the constraint reads
\begin{equation}
    C^{(2)}[\kappa\xi]=H_S-H_R+H_L
\end{equation}
and reproduces the form Eq.~\eqref{2clockconstraint2} with the left and right ADM Hamiltonians assuming the role of the clock Hamiltonians. The relative minus sign between $H_R$ and $H_L$ traces back to the fact that the Killing flow generated by the boost field $\xi$ is future-pointing in the right region and past-directed in the left one (see Fig.~\ref{Figure: Schwarzschild both side boost}). It is important that $H_{R,L}$ is \emph{linear} in the second-order metric perturbations \cite{DeVuyst:2024grw,Kudler-Flam:2023qfl}, which means that its spectrum is unbounded in both directions and so $H_L,H_R$ correspond to \emph{ideal} clock QRFs on the boundary.\footnote{More precisely, this requires a regularisation of the ADM Hamiltonians, $H_{R,L}\to H_{R,L}-M_0$ to account for the diverging black hole mass $M_0$ as $G_N\to0$. $H_{R,L}$ are not necessarily non-degenerate on the Hilbert space of second-order fluctuation variables. However, one can refactorise the Hilbert space of the second-order fluctuations such that $H_{R,L}$ act only on a $L^2(\mathbb{R})$ factor of it, thereby turning it into an ideal clock Hamiltonian. See \cite[Sec.~5]{Chandrasekaran:2022cip} for discussion. } 

\begin{figure}
    \centering
    \begin{tikzpicture}[scale=1.16]
        \begin{scope}[red!40!white]
            \foreach \i in {0.3,1,1.85,2.6} {
                \draw[postaction={decorate,decoration={markings,mark=at position .57 with {\arrow[scale=1.1]{stealth}}}}] (2,2) .. controls ({2+0.8*\i},{2-0.8*\i}) and ({2+0.8*\i},{0.8*\i-2}) .. (2,-2);

                \draw[postaction={decorate,decoration={markings,mark=at position .57 with {\arrow[scale=1.1]{stealth}}}}] (6,-2) .. controls ({6-0.8*\i},{0.8*\i-2}) and ({6-0.8*\i},{2-0.8*\i}) .. (6,2);
            }
        \end{scope}
        \draw (2,2) -- (6,-2) -- (6,2) -- (2,-2) -- (2,2);
        \draw (2,2) -- (6,2);
        \draw (2,-2) -- (6,-2);
        \draw[very thick, blue] (4,0) .. controls (5,-0.3) and (5.5,-0.3) .. (6,-0.3) node[midway,below,blue!60!black] {\large$\Sigma$};
        \draw[very thick, blue] (4,0) .. controls (3,0.3) and (2.5,0.3) .. (2,0.3);
        \node[right] at (6,0) {$r=0$};
    \end{tikzpicture}
    \caption{The static patch of de Sitter space depicted with its Killing flow lines. The killing flow generated by $H_\xi$ moves the cauchy slice $\Sigma$ forwards in time in the right patch but backwards in time in its complement, the left patch. The clocks measuring the boost time would be $H_1$ in the right patch and $H_2$ in the left patch respectively. One can think of them being localised along the $r=0$ worldlines. The backwards evolution in the complement leads to a relative minus sign between the two clock Hamiltonians.}
    \label{Figure: de Sitter boost}
\end{figure}
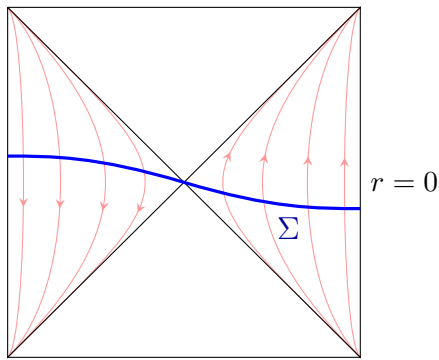

When the spacetime is spatially closed, such as de Sitter space, on the other hand, no intrinsic gravitational ADM boundary clocks are present and the second-order constraint would just be given by $H=H_S$. It turns out that the modular flow generated by $H_S$ acts ergodically on $\mathcal{A}_\mathcal{U}=\mathcal{A}_S$ \cite{Chandrasekaran:2022cip,Chen:2024rpx} and this means that the content of $\mathcal U$ that is invariant under this constraint is trivial
\begin{equation}
    \mathcal{A}_S^H=\mathbb{C}\,.
\end{equation}
To remedy this issue, CLPW introduced an observer equipped with a quantum clock in each of the two complementary static patches $\mathcal U,\mathcal U'$ such that each timelike worldline connects the past with the future tip of the respective static patch \cite{Chandrasekaran:2022cip}. This shifts the stress-energy tensor in Eq.~\eqref{boostgen} by the clock contributions, $H_\xi\to H_\xi+H_1-H_2$, where $i=1,2$ labels the clock Hamiltonians in the two patches. In this case, there is no reason for $H_i$ to be unbounded or non-degenerate and so the clocks need not be ideal. When the worldlines coincide with Killing flow lines generated by the cosmological horizon boost vector field, these clocks measure that boost time and the relative minus sign between the clock Hamiltonians again accounts for the relative orientation between the direction of time evolution in the two patches \cite{DeVuyst:2024grw,Jensen:2023yxy,Chandrasekaran:2022cip,Witten:2023xze} (see Figure~\ref{Figure: de Sitter boost}). This once more reproduces a Hamiltonian constraint of the form Eq.~\eqref{2clockconstraint2} and in this case one does obtain non-trivial boost-invariant algebras for both $\mathcal U,\mathcal U'$ from the crossed product construction. Clearly, the observers' clocks are QRFs of exactly the type we described in Sec.~\ref{Section: quantum clocks as quantum reference frames}.

Operationally, we can assume that each observer can probe the observables in the vicinity of their worldline. The timelike tube theorem \cite{Borchers1961,Araki:1976zv,Strohmaier:2023hhy,Strohmaier:2023opz,Witten:2023qsv} asserts that they can then also access the entire algebra associated with the `timelike envelope' of their wordline. This is the causal region defined by all the events that can be reached by deforming the original worldline, while keeping its endpoints fixed and preserving the timelike nature of the worldline. For an observer in de Sitter space, this envelope will agree with its entire static patch. The timelike tube theorem thereby supports an operational interpretation of the entropies that will be associated with such regions. 

Let us now come to two related questions, namely (a) above and:
\begin{itemize}
    \item[(c)] why should one impose second-order constraints, Eqs.~\eqref{2clockconstraint2} and~\eqref{2ndordercon}, when we are taking the $\kappa\to0$ limit and are thus interested in the linear-order theory?
\end{itemize}
We have discussed these questions and answer options at length in \cite{DeVuyst:2024grw} and we refer to that work for details. In a nutshell, there is an essentially unambiguous answer for spatially closed spacetimes in general relativity with isometries: a linearised solution is a valid approximation to an exact solution  \emph{if and only if} all second-order Killing symmetry generators, called Taub charges \cite{Taub1,taub2011variational}, vanish \cite{fischer1980structure,Arms:1982ea,Marsden_lectures}. The boost constraint happens to be one of those and crucially there is just a \emph{finite number} of them, one for each independent Killing field. When the first-order in $\kappa$ fields do not satisfy these constraints, the linearised solution cannot be integrated to an exact one. However, we can safely ignore all other infinitely many second-order diffeomorphism constraints as far as a consistent embedding of the linearised theory into the full non-linear theory is concerned.  The constraints Eqs.~\eqref{2clockconstraint2} and~\eqref{boostgen} are thus to be understood as one of a finite number of necessary and sufficient stability conditions for perturbation theory to avoid spurious solutions. This is the output of the subject of linearisation instabilities \cite{Deser:1973zza,Moncrief1,Moncrief:1976un,Arms1,Arms:1979au,Arms:1986vk,Saraykar:1981qm}. While these are classical results, there are strong arguments for also imposing these constraints as stability conditions in the quantum theory \cite{Moncrief:1978te,Moncrief:1979bg,Losic:2006ht,Higuchi:1991tk,Higuchi:1991tm,Kaplan:2024xyk}. Furthermore, the original instability theorems pertain to general relativity, however, it appears likely that similarly strong results  hold in arbitrary generally covariant theories in spatially closed spacetimes \cite{DeVuyst:2024grw}. Finally, the isometries  are now typically broken down by the presence of the observers to a form $\mathbb{R}\times G$, for some compact group $G$ \cite{Chandrasekaran:2022cip}, where $\mathbb{R}$ corresponds to the boost translations. Given the compactness of $G$ and the direct product structure, expanding the QRFs to account for $G$ \cite{delaHamette:2021oex} will however not affect the crossed product type transition or the discussion qualitatively otherwise. In that sense, it is legitimate to ignore them here.\footnote{In the operational approach to QRFs, which does not impose constraints on states, these isometries have been encompassed in the crossed products of \cite{Fewster:2024pur}; such crossed products were also considered in \cite{AliAhmad:2024eun}. Encompassing them in an approach that does impose constraints on states also, as likely appropriate in gravity, would build upon \cite{delaHamette:2021oex}. }

In the absence of isometries or the presence of boundaries (with interesting boundary conditions), no instabilities arise and no second-order constraint is necessary in order to ensure consistency of the first-order theory \cite{DeVuyst:2024grw,Deser:1973zzb,choquet1979maximal,fischer1980structure,Arms:1982ea,Marsden_lectures}.
One must thus argue differently why one would still want to consider second-order constraints of the form Eq.~\eqref{2clockconstraint2} and why considering only a single constraint or finite number of them, as done in practice, is legitimate. As we argue in \cite{DeVuyst:2024grw}, this will depend on whether one is also interested in certain second-order observables, such as in \cite{Kudler-Flam:2023qfl,Faulkner:2024gst,KirklinGSL}, though strictly speaking this case still stands on arguably less robust grounds than the spatially compact one with isometries. 

It is also not known in general under which conditions modular flow of a QFT algebra admits a geometric interpretation. However, there is a `geometric modular flow conjecture' asserting that the diffeomorphism generated by a vector field that appears boost-like near the boundary of a general subregion in a spacetime with or without isometries will agree infinitesimally with the modular flow of \emph{some} KMS state and so can be used for building crossed products \cite{Jensen:2023yxy,sorce2024analyticity}.

In the sequel, we shall simply work with constraints of the form Eq.~\eqref{2clockconstraint2}, without further specifying whether we work in a spacetime with or without boundary and whether or not an exact boost Killing field exists.  We note the various levels of justification for considering such constraints mentioned above (with an essentially unambiguous justification for spatially closed spacetimes with boost isometries). Henceforth, we  also assume that $H_S$ is a modular Hamiltonian, while $H_i$ could thus be an internal or ADM boundary clock Hamiltonian, depending on the situation.

\subsection{PW = CLPW}
\label{sect:PWCLPW}

Let us now clarify more precisely that what CLPW \cite{Chandrasekaran:2022cip} and follow-up works such as JSS \cite{Jensen:2023yxy} call an observer is precisely a temporal QRF according to the perspective-neutral approach reviewed in the previous section. We will briefly demonstrate that the description of constrained states and observables used in these works is equivalent to the (expanded form of the) Page-Wootters (PW) formalism underlying that approach, as reviewed in Sec.~\ref{ssec_PW}. This is the basis for the informal slogan in the title of this section. We will build upon this observation in subsequent sections to generalise the construction to arbitrarily many observers/QRFs. 

\subsubsection{Map to coinvariant representation is Page-Wootters reduction}\label{sect:PW=CLPW}

\begin{figure}
\centering
    \begin{tikzcd}[column sep=3em, row sep=3em]
    & \mathcal{H}_\text{ext}
    \ar[blue, dr,"\prod_{i}\Pi_i"]
    \ar[orange, dl,"H^{\text{ext}}=0", swap]
    &
    \\
    \mathcal{H}^{\text{ext}}_{\text{phys}} 
    \ar[blue, dr, "\prod_{i}\Pi_i"']
    && 
    \mathcal{H}_\text{kin}
    \ar[orange, dl,"H=0"]
    \\
    & \mathcal{H}_\text{phys}
    &
    \\
    &\phantom{\mathcal{H}}&
    \end{tikzcd}
    \qquad 
    \begin{tikzcd}[column sep=3em, row sep=3em]
    & \mathcal{A}^\text{ext}_{SC}
    \ar[blue, dr,"\prod_{i}\Pi_i"]
    \ar[orange, dl,"H^{\text{ext}}=0", swap]
    &
    \\
    \mathcal{A}^{H^\text{ext}}_{SC} 
    \ar[blue, dr, "\prod_{i}\Pi_i"']
    && 
    \mathcal{A}_{SC}
    \ar[orange, dl,"H=0"]
    \\
    & \mathcal{A}^H_{SC}
    \ar[black, d, "r"]
    &
    \\
    &
    r(\mathcal{A}^H_{SC})
    &
    \end{tikzcd}
    \caption{Commutative diagram showing two orders in which constraints may be imposed.  CLPW \cite{Chandrasekaran:2022cip} and JSS \cite{Jensen:2023yxy} move counterclockwise from the top, imposing non-idealness (bounded energy) of the clocks only after the constraint is implemented.  Our approach starts with the kinematical Hilbert space $\mathcal{H}_{\rm kin}$, where clocks already have the final restriction on their energy spectra in place, and imposes the constraint on this.  This diagram shows a comparison at the gauge invariant level, culminating on the physical Hilbert space. An analogous diagram  instead implements the gauge constraint via gauge fixing, for instance culminating in the reduced perspective of frame $C_2$, as we do in the main text below. The diagram on the right shows the similar two orders of imposing the constraints on the algebras.}
    \label{fig: commutative constraints}
\end{figure}
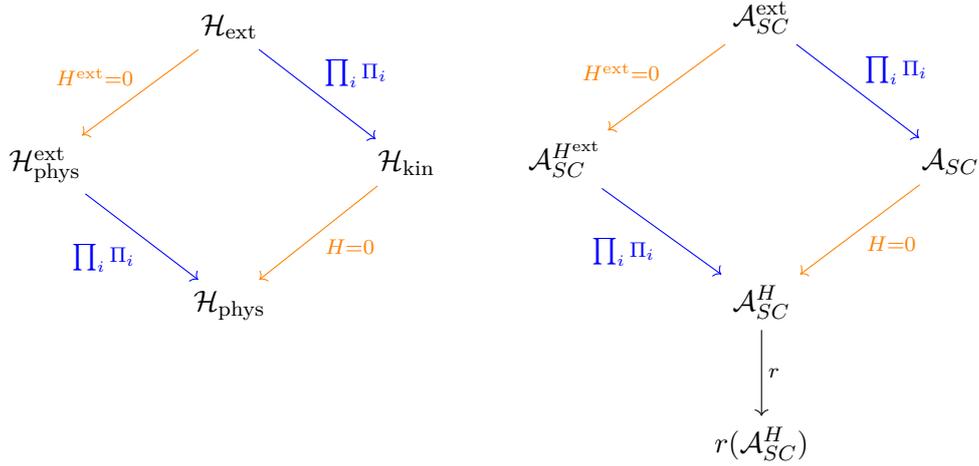

In both \cite{Chandrasekaran:2022cip,Jensen:2023yxy} the clocks are first treated as ideal, and only after deriving the relevant density operators are projectors included to enforce the energy bounds of the observer Hamiltonian (in their case, reducing the spectra from the full real line to values on a half-line).  The perspective-neutral approach reviewed in Sec.~\ref{Section: quantum clocks as quantum reference frames}, by contrast, directly begins with a kinematical Hilbert space on which the clock frames already have their final energy restrictions and thus may already be non-ideal (having limited spectral ranges), and then  implements the constraint by mapping states and operators to the physical Hilbert space and/or gauge-fixed `reduced' Hilbert spaces corresponding to the internal frame perspectives.  Relative to us, we therefore say that \cite{Chandrasekaran:2022cip,Jensen:2023yxy} consider an `extended' kinematical Hilbert space and constraint operator to start with, which we indicate with the label `ext':
\begin{equation}\begin{split}\label{eq: extended kinematical H}
    \mathcal{H}_{\text{ext}}:= \mathcal{H}_{S}\otimes L^2(\mathbb{R})_1 \otimes L^2(\mathbb{R})_{2},
\end{split}\end{equation}
\begin{equation}\begin{split} H^{\text{ext}}:=H_{S}+H^{\text{ext}}_{1}-H^{\text{ext}}_{2}.\label{extcon}
\end{split}\end{equation}
The relative minus sign again accounts for the different relative direction of time evolution in $\mathcal{U},\mathcal{U}'$, assuming clock $C_2$ lives in $\mathcal{U}'$.
$H^{\text{ext}}_{1}$ and $H^{\text{ext}}_{2}$ are therefore unbounded ideal clock frame Hamiltonians. The final output of both routes coincides, as we now explain, see Fig.~\ref{fig: commutative constraints}. 

The relation to our Hamiltonians is given by
\begin{equation}
H_i=\Pi_i\,H_i^{\rm ext}
\end{equation}
with projector onto the ultimately permitted energy intervals of clock $C_i$
\begin{equation}\label{piext}
\Pi_i=\sum_\alpha\Theta(\epsilon_{\rm max}^{i,\alpha}-H_i)\Theta(-\epsilon^{i,\alpha}_{\rm min}+H_i)\,,
\end{equation}
where $\alpha$ sums over the different possible non-overlapping intervals (in \cite{Chandrasekaran:2022cip} only a single interval was used). Similarly, we have
\begin{equation}
    \mathcal{H}_{\rm kin}=\Pi_1\,\Pi_2\left(\mathcal{H}_{\rm ext}\right)=\mathcal{H}_S\otimes\mathcal{H}_1\otimes\mathcal{H}_2
\end{equation}
with $\mathcal{H}_i=\bigoplus_\alpha L^2(\sigma_i^\alpha)$, where $\sigma^\alpha_i=[\epsilon_{\rm min}^{i,\alpha},\epsilon_{\rm max}^{i,\alpha}]$, and for the clock states
\begin{equation}\label{tprojection}
    \ket{t}_i=\Pi_i\,\ket{t}_i^{\rm ext}\,.
\end{equation}
Note that the projection does not affect the covariance properties in Eq.~\eqref{clockcov}, but is responsible for the clock state fuzziness in Eqs.~\eqref{eq:chi} and~\eqref{fuzzy}.

In the case of  ideal frames, the  Hilbert space of coinvariants, as discussed in \cite[Sec.~4.2]{Chandrasekaran:2022cip} is isomorphic to a simple tensor product of the QFT Hilbert space and one frame factor, say that of clock $C_1$; it thus removes clock $C_2$ in the complementary patch $\mathcal{U}'$.  This is implemented via a certain $T$ map
from (extended) kinematical states to a representation of coinvariants. Let us directly generalise it to non-ideal clocks and show that it is equivalent to Page-Wootters reduction. The map is
\begin{equation}\label{eq: T map}
T_2: \mathcal{H}_{\rm kin}\longrightarrow
\mathcal{H}_{S}\otimes \mathcal{H}_1\,,
\end{equation}
reading  in our notation and conventions \cite[Eq.~(4.10)]{Chandrasekaran:2022cip}
\begin{eqnarray}\label{eq:isom}
    T_2\ket{\psi}=\int_\mathbb{R}\dd{t}e^{-it(H_S+H_1)}\ket{\psi(t)}\,,
\end{eqnarray}
where $\ket{\psi}\in\mathcal{H}_{\rm kin}$ and $\ket{\psi(t)}=\left(\bra{t}_2\otimes \mathds{1}_{S1}\right)\ket{\psi}$.
Clearly, using the clock covariance in Eq.~\eqref{clockcov}
\begin{eqnarray}\label{PWred2}
    T_2\ket{\psi}=\mathcal{R}'_2(0)\,\Pi_{\rm phys}\,\ket{\psi}\underset{\eqref{MR1}}{=}\mathcal{R}_2(0)M\ket{\psi}\,,
\end{eqnarray}
where $\Pi_{\rm phys}$ is the coherent group averaging map Eq.~\eqref{piphys} (updated in the obvious way to accommodate two clocks) and $\mathcal{R}'_2(0)$ is the Page-Wootters reduction map as in Eq.~\eqref{R1}, however, for clock $C_2$ reading $\tau_2=0$. The ideal case of CLPW is recovered by setting $\Pi_i=\mathds{1}_i$. Hence, the $T$ map from the (extended) kinematical Hilbert space to a coinvariant representation invoked by CLPW \cite[Eq.~(4.10)]{Chandrasekaran:2022cip} is equivalent to first going perspective-neutral via group averaging (or $M$) and then ``jumping into the clock $C_2$'s perspective when it reads $\tau_2=0$'' via a Page-Wootters reduction.

Written this way, it is clear that the action of $T_2$ implements the constraint Eq.~\eqref{2clockconstraint2} in the form ${H=H_S+H_1-H_2}$. Nevertheless, it is  somewhat instructive to double check that this also holds for non-ideal clocks,
\begin{equation}
    T_2H\ket{\psi}=0\,,\qquad \forall\,\ket{\psi}\in\mathcal{H}_{\rm kin}\,,
\end{equation}
which we briefly do in App.~\ref{app_doublecheck}. Hence,
\begin{eqnarray}
    T_2H_2\ket{\psi} = T_2(H_S+H_1)\ket{\psi} = (H_S+H_1)T_2\ket{\psi}\,.\label{tihi}
\end{eqnarray}
Now $\Pi_2=\mathds{1}$ on $\mathcal{H}_{\rm kin}$. However, by Eq.~\eqref{tihi}, this turns into a non-trivial projector for the internal frame perspective
\begin{eqnarray}\label{TP}
T_2\Pi_2\ket{\psi}=\Pi_{|2}T_2\ket{\psi}\,,
\end{eqnarray}
where we recover Eqs.~\eqref{pi1} and~\eqref{eq:reducedProjector} (with the appropriate minus signs for $H=H_S+H_1-H_2$)
\begin{eqnarray}\label{eq:nontrivialProj}
    \Pi_{|2}=\sum_\alpha\Theta(\epsilon_{\rm max}^{2,\alpha}-H_S-H_1)\Theta(-\epsilon^{2,\alpha}_{\rm min}+H_S+H_1)\,.
\end{eqnarray}
In other words, we recover the reduced Hilbert space in $C_2$-perspective of Sec.~\ref{ssec_QRFchanges}
\begin{eqnarray}\label{eq:reducedHilbertSpace}
   \mathcal{H}_{|2}\equiv T_2\left(\mathcal{H}_{\rm kin}\right) = \mathcal{R}_2(\tau)\left(\mathcal{H}_{\rm phys}\right) = \Pi_{|2}\left(\mathcal{H}_{S} \otimes \mathcal{H}_{1}\right)\,,
\end{eqnarray}
i.e.\ a subspace of $\mathcal{H}_S\otimes\mathcal{H}_1$ when $C_2$ is non-ideal and so $\Pi_{|2}$ is a non-trivial projector. By contrast, when $C_2$ is ideal, $\Pi_{|2}=\mathds{1}_{S1}$, and we recover the \emph{initial} result of \cite[Sec.~4.2]{Chandrasekaran:2022cip}. 

In particular, had we started like CLPW with $\mathcal{H}_{\rm ext}$, first implemented the constraint $H^{\rm ext}$ via $T^{\rm ext}_2$ (defined in the obvious way) and \emph{then} imposed the restrictions $\Pi_i$ on the clock energy spectra, we would find in place of Eq.~\eqref{TP}
\begin{equation}\label{text1}
    T_2^{\rm ext}\Pi_1\Pi_2\ket{\psi}_{\rm ext}=\Pi_1\Pi_{|2}T_2^{\rm ext}\ket{\psi}_{\rm ext}\,,
\end{equation}
 where $\Pi_i$ is now a non-trivial projector on $\mathcal{H}_{\rm ext}$. However, thanks to Eq.~\eqref{tprojection},
 \begin{equation}\label{text2}
     T_2^{\rm ext}\Pi_1\Pi_2\ket{\psi}_{\rm ext}=T_2\ket{\psi}\,,
 \end{equation}
with $\ket{\psi}\in\mathcal{H}_{\rm kin}$, and so it is clear that the two routes in Fig.~\ref{fig: commutative constraints} give the same end result. Thus, CLPW's and JSS's observers are nothing but clock QRFs according to the perspective-neutral formalism.

\subsubsection{Reduction of observable algebras} \label{sect:algebraReduction}

Let us briefly consider the reduction of the crossed observable algebras $\mathcal{A}^H_{SC_1}=\left(\mathcal{A}_S\otimes\mathcal{B}(\mathcal{H}_1)\otimes \mathds{1}_2\right)^H$ and $\mathcal{A}^H_{S'C_2}:=\left(\mathcal{A}_S'\otimes \mathds{1}_1\otimes\mathcal{B}(\mathcal{H}_2)\right)^H=(\mathcal{A}^H_{SC_1})'$, corresponding to $\mathcal U$ and $\mathcal U'$, respectively, which are commutants within $\mathcal{A}_{\rm kin}^H$, and check that the procedure by CLPW agrees with the procedure of the perspective-neutral formalism. To determine the representation on the space of coinvariants, CLPW set \cite[Sec.~4.2]{Chandrasekaran:2022cip}
\begin{equation}
    \mathcal{A}T_2\ket{\psi}:=T_2\mathcal{A}_{\rm kin}^H\ket{\psi}\,,
\end{equation}
which we have adapted to our conventions and notation and extended to clocks that are not necessarily ideal. Using Eq.~\eqref{coinvrep}, we thus have
\begin{equation}
\begin{split}
    \mathcal{A}T_2\ket{\psi}&=\mathcal{R}_2(0)M\mathcal{A}_{\rm kin}^H\ket{\psi}=\mathcal{R}_2(0){r}(\mathcal{A}^H_{\rm kin})\Ket{\psi}\\
    &=\mathcal{R}_2(0){r}(\mathcal{A}^H_{\rm kin})\mathcal{R}_2^\dag(0)\mathcal{R}_2(0)\Ket{\psi}\,,
    \end{split}
\end{equation}
so that
\begin{equation}
    \mathcal{A}=\mathcal{R}_2(0){r}(\mathcal{A}^H_{\rm kin})\mathcal{R}^\dag_2(0)\,
\end{equation}
agrees with the Page-Wootters reduced algebras as in \cite[Thm.~3]{Hoehn:2019fsy}. Hence, by Eqs.~\eqref{wrong2}:
\begin{equation}\label{eq: T2 reduce SC1}
\begin{split}
    \mathcal{A}_{SC_1|2}&:=\mathcal{R}_2(0){r}(\mathcal{A}^H_{SC_1})\mathcal{R}_2^\dag(0)\\
    &=\Pi_{|2}\mathcal{A}^H_{SC_1}\\
    &=\Pi_{|2}\Big\langle O_{C_1}^{\tau_1}(a), e^{isH_1}\,\Big|\,a\in\mathcal{A}_S,s\in\mathbb{R}\Big\rangle''
    \end{split}
\end{equation}
and by Eq.~\eqref{projalgeb}:
\begin{equation}\label{eq: T2 reduce SC2}
\begin{split}
    \mathcal{A}_{S'C_2|2}&:=\mathcal{R}_2(0){r}(\mathcal{A}^H_{S'C_2})\mathcal{R}_2^\dag(0)\\
    &=\Pi_{|2}\Big\langle a',e^{is(H_S+H_1)}\,\Big|\,a'\in\mathcal{A}_S',s\in\mathbb{R}\Big\rangle''\,\Pi_{|2}\,.
    \end{split}
\end{equation}
Both algebras now act on the reduced space in $C_2$-perspective $\mathcal{H}_{|2}$. From the arguments in App.~\ref{app_ginvalg} and Sec.~\ref{ssec_vN} it follows that both $\mathcal{A}_{SC_1|2}$ and $\mathcal{A}_{S'C_2|2}$ are Type $\rm{II}$ von Neumann factors; they are of Type $\rm{II}_1$ if the respective clock has an energy bound from below, otherwise they are Type $\rm{II}_\infty$, see \cite{Chandrasekaran:2022cip} and Sec.~\ref{Section: arbitrary number of observers}. Furthermore, Haag duality (including the clocks) is preserved by reduction: on $\mathcal{H}_{|2}$ we have
\begin{equation}
    \mathcal{A}_{SC_1|2}=(\mathcal{A}_{S'C_2|2})'
\end{equation}
because the commutant of $\mathcal{A}^H_{SC_1}$ on $\mathcal{H}_S\otimes\mathcal{H}_1$ is $\big\langle a',e^{is(H_S+H_1)}\,\big|
\,a'\in\mathcal{A}'_S,s\in\mathbb{R}\big\rangle''$ and these commutant relations are preserved by projection with $\Pi_{|2}$, yielding $\mathcal{H}_{|2}$ via Eq.~\eqref{eq:reducedHilbertSpace} \cite[EP7, p.~21]{Joneslec}.

Thanks to Eqs.~\eqref{text1} and~\eqref{text2} it can be easily checked that the reduced algebras $\mathcal{A}_{SC_1|2}$ and $\mathcal{A}_{S'C_2|2}$ coincide with the reduced algebras in CLPW upon implementing the clock energy restrictions \cite[Sec.~4.2]{Chandrasekaran:2022cip}.\footnote{In fact, CLPW perform another unitary conjugation in $C_2$-perspective to put $A_{SC_1|2}$ into a standard crossed product form, a step we refrain from here.} 

We will now move on to generalise the construction by CLPW in \cite{Chandrasekaran:2022cip} and JSS in \cite{Jensen:2023yxy} to arbitrarily many observers with clocks using the perspective-neutral framework. We will later demonstrate in Sec.~\ref{Section: comparison with previous}  how to recover the density operators and entropies obtained in these references from our more general formalism.

\section{Subregion density operators with non-degenerate clocks}\label{Section: arbitrary number of observers}

Let us now consider a system consisting of quantum fields living in a Hilbert space $\mathcal{H}_S$, and some arbitrary number $n\in\mathbb{N}$ of observers carrying clocks $C_i$, $i=1,\dots,n$, each with non-degenerate continuous spectra $\sigma_i\subset\RR$, and so Hilbert spaces $\mathcal{H}_i = L^2(\sigma_i)$. We focus initially on the case where the clocks are non-degenerate. Thus the total kinematical Hilbert space is 
\begin{equation}
    \mathcal{H}_{\text{kin}} = \mathcal{H}_S\otimes\bigotimes_{i=1}^n\mathcal{H}_i.
\end{equation}
From here onward, we will now mostly invoke the language of co-invariants when describing constrained states and observables, but, as described in Sec.~\ref{Section: quantum clocks as quantum reference frames}, using RAQ would lead to an equivalent formulation.

We will assume in this section that all the observers have access to field observables in some fixed spacetime subregion $\mathcal{U}$ or its complement $\mathcal{U}'$. Heuristically, one may imagine that the observers are travelling along some worldlines in $\mathcal{U}$ or $\mathcal{U}'$, and that they are somehow able to measure the fields in the vicinities of their worldlines. The timelike tube theorem \cite{Borchers1961,Araki1963AGO,Witten:2023qsv,Strohmaier:2023opz} is then what permits them to measure the fields in the rest of $\mathcal{U}$ or $\mathcal{U}'$ (under certain conditions), as briefly mentioned in Subsec.~\ref{sect:linearisationConditions}.

Each observer will also be assumed to have some amount of access to the degrees of freedom of the clocks of the other observers. At the present level of abstraction, exactly how much access is essentially just an operational specification. But we can use physical and heuristic principles to motivate this specification (although note that the following is non-rigorous and should be taken with a grain of salt). Indeed, suppose we continue to use the picture of observers travelling along worldlines. If an observer Alice can act with the fields in the vicinity of another observer Bob's worldline (invoking the timelike tube theorem), then she also ought to be able to act on Bob with his Hamiltonian -- because the local gravitational constraints allow us to rewrite this Hamiltonian as a field operator supported in Bob's neighborhood. Also, Alice ought to be able to measure time differences of Bob's clock -- because such time differences can be written in terms of the metric proper time along his worldline (again ultimately due to the gravitational constraints). At the kinematical level, we can just allow her to measure the absolute time of his clock, which will reduce to time differences after imposing gauge invariance. So, Alice can act with all operators on Bob's clock's Hilbert space. More generally, we will assume each observer can act with the full set of operators on some subset $R$ of the available clocks (including their own).\footnote{It should be noted that this line of reasoning will \emph{not} necessarily lead to the same structure in the case of \emph{degenerate} clocks, which we explore further in Sec.~\ref{sect:rhoDeg}.} Despite the heuristic reasoning given here, in the interest of generality we will not make any assumptions about whether the clocks in $R$ are located inside or outside $\mathcal{U}$.

We assume that there is a gauge constraint of the form $H=H_S+\sum_{i=1}^n H_i$, where $H_S$ is the generator of some diffeomorphism acting on the fields and preserving $\mathcal{U}$ and $\mathcal{U}'$, and $H_i$ is the Hamiltonian of clock $C_i$. Overall $H$ should be understood as the generator of the diffeomorphism acting on the fields and the clocks simultaneously. While $H_S$ throughout Sect.~\ref{Section: quantum clocks as quantum reference frames} could have been arbitrary, in particular here we will take it to be the generator of a boost $\xi$ as described in Sect.~\ref{sect:CLPW}, which we will take to correspond to the modular flow of some state of the fields in $\mathcal{U}$; since the algebra of the fields is Type III, this is necessarily an outer automorphism, so $H_S$ cannot be expressed in terms of an operator localised to $\mathcal{U}$.

Let $R\subset \{C_i|i=1,\dots,n\}$ be some subset of the available clocks (which could be for example the set of clocks contained in the region $\mathcal{U}$), and let $R^c = \{C_i|i=1,\dots,n\}\setminus R$ be its complement. We are interested in the observables accessible with the clocks $C_i\in R$. As in the previous sections, the algebra of such observables may be formulated as the gauge-invariant subalgebra of kinematical operators acting on the clocks and the fields in $\mathcal{U}$: 
\begin{equation}
    \mathcal{A}_{SR}^H = (\mathcal{A}_S\otimes \mathcal{B}(\mathcal{H}_R))^H,
\end{equation}
where $\mathcal{H}_R=\bigotimes_{C_i\in R}\mathcal{H}_i$, and $\mathcal{A}_S$ is the algebra of field observables with support in $\mathcal{U}$. Picking any particular $C_i\in R$, we can decompose this algebra as
\begin{equation}
    \mathcal{A}_{SR}^H = \{O^\tau_i(a),U_i(t)\mid a\in \mathcal{A}_S\otimes\mathcal{B}(\mathcal{H}_{R_{\bar{i}}}), \tau\in\RR, t\in \RR\}'',
\end{equation}
where $R_{\bar{i}}=R\setminus C_i$, $\mathcal{H}_{R_{\bar{i}}}=\bigotimes_{C_j\in R_{\bar{i}}}\mathcal{H}_j$, and
\begin{equation}
    O^\tau_i(a) = \int_{-\infty}^\infty\dd{t} e^{-iHt} (a \otimes \dyad{\tau}_i ) e^{iHt},\qquad U_i(t) = e^{-iH_i t},
\end{equation}
with $\ket{\tau}_i$ being the clock states for $C_i$, and we are omitting identity factors and often rearranging tensor factors to simplify the notation.

The corresponding algebra of physical observables is $r(\mathcal{A}^H_{SR})$, acting on the physical Hilbert space $\mathcal{H}_{\text{phys}}$, obtained by imposing the constraint $H=0$ as described previously.

We will assume $H_S$ generates the modular flow with respect to $\mathcal{A}_S$ of some cyclic and separating QFT state $\ket{\psi_S}\in\mathcal{H}_S$; more precisely, we assume its modular operator is of the form $\Delta_{\psi_S} = e^{-\beta H_S}$, for some inverse temperature $\beta$. We will also take $R^c$ to be non-empty. As described previously in Sec.~\ref{ssec_vN} (whose argument extends to arbitrarily many clocks), these assumptions guarantee that
\begin{equation}
    r:\mathcal{A}^H_{SR} \to r(\mathcal{A}^H_{SR})
    \label{eq:physicalRep}
\end{equation}
is a faithful representation, and that $r(\mathcal{A}^H_{SR})$ is a Type II factor which we will confirm later. Our aim in this section is to compute the density operator in the algebra $r(\mathcal{A}^H_{SR})$ corresponding to a physical state $\Ket{\phi}\in\mathcal{H}_{\text{phys}}$. In the interest of readability, some key details are reserved for Appendix~\ref{Appendix: density operator details}.

\subsection{Trace on the observable algebra}
\label{Section: trace}

Our goal is to create a trace using the basic property of a modular operator $\Delta_\Psi$, namely 
\begin{equation}
    \expval{a b}{\Psi} = \expval{b \Delta_\Psi a}{\Psi}.
\end{equation}
If we can find a $\ket{\Psi}$ for which $\Delta_\Psi$ acts as an identity on a, possibly projected, Hilbert space, then we propose the trace to be the expectation value of any operator in this state $\Tr(.):= \expval{.}{\Psi}$. This will define a good trace for the type II algebra $r(\mathcal{A}^H_{SR})$ due to faithfulness of the representation \eqref{eq:physicalRep} and the uniqueness of the trace up to scaling. To make the extension to multiple clocks, it will be necessary to introduce auxiliary Hilbert spaces $\mathcal{H}_j^*$ for every $C_j \in R_{\bar{i}}$ to be able to build $\ket{\Psi}$. Afterwards, we can map back from this extended Hilbert space to the physical Hilbert space by making use of the relative modular operator.

The first step is to write down a trace on $\mathcal{A}^H_{SR}$. To this end, let us define a `thermofield double' for the clocks in $R_{\bar{i}}$:
\begin{equation}
    \ket*{\psi_{R_{\bar{i}}}} = \bigotimes_{C_j\in R_{\bar{i}}}\int_{\sigma_j} \dd{\epsilon_j} e^{-\beta \epsilon_j/2} \ket{\epsilon_j}\otimes \bra{\epsilon_j} \in \bigotimes_{C_j\in R_{\bar{i}}} \mathcal{H}_j\otimes \mathcal{H}_j^*.
    \label{Equation: thermofield double}
\end{equation}
This is a (non-normalisable -- so technically speaking it is an abuse of notation to write $\ket*{\psi_{R_{\bar{i}}}}\in\bigotimes_{C_j\in R_{\bar{i}}} \mathcal{H}_j\otimes \mathcal{H}_j^*$) state for two copies of each of the clocks in $R_{\bar{i}}$. It is cyclic and separating for $\mathcal{B}(\mathcal{H}_{R_{\bar{i}}})$  (unless otherwise stated we always take operators to act on the first copy of the Hilbert space $\mathcal{H}_j$, rather than $\mathcal{H}_j^*$), with Tomita operator given by (as confirmed in App.~\ref{Appendix: density operator details})
\begin{equation}
    S_{\psi_{R_{\bar{i}}}} (\ket{\varphi}\otimes\bra{\zeta}) = e^{\beta H_{R_{\bar{i}}}/2} \ket{\zeta} \otimes \bra{\varphi}e^{-\beta H_{R_{\bar{i}}}/2}.
\end{equation}
One then finds the modular operator
\begin{equation}
    \Delta_{\psi_{R_{\bar{i}}}} = S_{\psi_{R_{\bar{i}}}}^\dagger S_{\psi_{R_{\bar{i}}}} = \exp(-\beta(H_{R_{\bar{i}}}-H_{R_{\bar{i}}^*})),
\end{equation}
where the ${}^*$ denotes that the operator acts on the second copy of the clocks. We then set $\ket*{\psi_{SR_{\bar{i}}}}=\ket{\psi_S}\otimes\ket*{\psi_{R_{\bar{i}}}}$. The state $\ket*{\psi_{SR_{\bar{i}}}}$ is cyclic and separating for $\mathcal{A}_S\otimes \mathcal{B}(\mathcal{H}_{R_{\bar{i}}})$. It has a Tomita operator
\begin{equation}
    S_{\psi_{SR_{\bar{i}}}} = S_{\psi_S}\otimes S_{R_{\bar{i}}},
\end{equation}
where $S_{\psi_S}$ is the Tomita operator associated with the QFT state $\ket{\psi_S}$, and modular operator
\begin{equation}
    \Delta_{\psi_{SR_{\bar{i}}}} = \Delta_{\psi_S}\otimes\Delta_{R_{\bar{i}}} = \exp(-\beta(H_S+H_{R_{\bar{i}}}-H_{R_{\bar{i}}^*})).
\end{equation}

Consider the state
\begin{equation}
    \ket{\Psi} = \ket*{\psi_{SR_{\bar{i}}}}\otimes e^{-\beta H_i/2}\ket{0}_i \in \mathcal{H}_S \otimes  \bigotimes_{C_j\in R_{\bar{i}}} \big(\mathcal{H}_j\otimes \mathcal{H}_j^* \big)\otimes \mathcal{H}_i.
    \label{Equation: Psi tracial state}
\end{equation}
$\mathcal{A}_{SR}^H$ acts on the Hilbert space containing this state. 

As shown in Appendix~\ref{Appendix: density operator details}, $\ket{\Psi}$ is separating for this action, and it is cyclic over a subspace
\begin{equation}
    \widetilde{\mathcal{H}}_{SR_{\bar{i}}R_{\bar{i}}^*C_i} = \widetilde{\Pi}(\mathcal{H}_S \otimes  \bigotimes_{C_j\in R_{\bar{i}}} \big(\mathcal{H}_j\otimes \mathcal{H}_j^* \big)\otimes \mathcal{H}_i),
\end{equation}
where 
\begin{equation}
    \widetilde{\Pi} = \int_{-\infty}^\infty\dd{\tau} \exp(-i(H_S + H_{R_{\bar{i}}} - H_{R_{\bar{i}}^*} + H_i)\tau) \braket{\tau}{0}_i \equiv \tilde{\Pi}[H_S + H_{R_{\bar{i}}} - H_{R^*_{\bar{i}}} + H_i,\sigma_i].
    \label{Equation: trace projection widetilde Pi}
\end{equation}
This is a projector similar to the Page--Wootters one \eqref{eq:reducedProjector}; it enforces the condition that when acting on a state, the value of $H_S + H_{R_{\bar{i}}} - H_{R^*_{\bar{i}}} + H_i$ must lie in the spectrum $\sigma_i$ of clock $C_i$.

Due to these properties, there is a Tomita operator for $\ket{\Psi}$ and $\mathcal{A}_{SR}^H$, defined on a dense subspace of $\widetilde{\mathcal{H}}_{SR_{\bar{i}}R_{\bar{i}}^*C_i}$; it may be written explicitly as
\begin{equation}
    S_\Psi = \int_{-\infty}^\infty\dd{t} e^{\beta H_i/2} \ket{t}_i\exp(-i(H_S + H_{R_{\bar{i}}} - H_{R_{\bar{i}}^*})t)S_{\psi_{SR_{\bar{i}}}} \bra{-t}_i e^{-\beta H_i/2}.
    \label{Equation: Psi tracial Tomita}
\end{equation}
From this, the modular operator of $\ket{\Psi}$ may be derived:
\begin{equation}
    \Delta_\Psi = S_\Psi^\dagger S_\Psi = \widetilde{\Pi}.
\end{equation}
Upon restriction to $\widetilde{\mathcal{H}}_{SR_{\bar{i}}R_{\bar{i}}^*R_i}$, one observes that $\Delta_\Psi$ is just the identity operator. A direct confirmation of these formulae may be found in Appendix~\ref{Appendix: density operator details}.

These properties suffice to show that 
\begin{equation}
    \Tr(A) = e^{S_{0,R}}\mel{\Psi}{A}{\Psi}
    \label{Equation: trace}
\end{equation}
defines a trace on $\mathcal{A}_{SR}^H$, where $S_{0,R}$ is some normalisation constant (the motivation for this notation will hopefully become clear later). Indeed, this functional is clearly normal, since it is the expectation value in a Hilbert space state. It is faithful because $\ket{\Psi}$ is separating. It obeys the cyclicity property because $\Delta_\Psi$ is the identity on the subspace over which $\ket{\Psi}$ is cyclic:
\begin{equation}
    \Tr(AB) = e^{S_{0,R}}\mel{\Psi}{AB}{\Psi} = e^{S_{0,R}}\mel{\Psi}{A\Delta_\Psi B}{\Psi} = e^{S_{0,R}}\mel{\Psi}{BA}{\Psi} = \Tr(BA).
\end{equation}
Finally, it is semifinite; for an example of a finite trace operator, consider
\begin{equation}
    A = \int_{-\infty}^\infty\dd{t}\bigotimes_{C_j\in R}\theta(H_j)\ket{t}_j\bra{t}_j\theta(H_j)\in\mathcal{A}_{SR}^H.
\end{equation}
Then one may show that
\begin{equation}
    \Tr(A) = e^{S_{0,R}}\prod_{C_j\in R}\int_{\sigma_j \cap \RR_+} \dd{\epsilon_j}e^{-\beta \epsilon_j},
\end{equation}
and all the integrals converge.  Given $\Tr$, we can define also a trace $\Tr_r$ on $r(\mathcal{A}_{SR}^H)$ by using the fact that the representation $r$ restricted to $\mathcal{A}_{SR}^H$ is faithful (cf.\ Sec.~\ref{ssec_vN}):
\begin{equation}
    \Tr_r(r(A)) = \Tr(A).
\end{equation}

The trace of the identity is given by 
\begin{equation}
    \Tr(1) = \Tr_r(1) = e^{S_{0,R}} \braket{\Psi}{\Psi} = e^{S_{0,R}}\braket*{\psi_{R_{\bar{i}}}}{\psi_{R_{\bar{i}}}} \bra{0}_ie^{-\beta H_i}\ket{0}_i.
    \label{eq:traceId}
\end{equation}
If $R$ includes more than one frame, then this is infinite, because the thermofield double state is not normalisable. However, if $R$ contains only a single frame, then we have
\begin{equation}
    \Tr(1) = e^{S_{0,R}} \bra{0}_ie^{-\beta H_i}\ket{0}_i = \frac1{2\pi}e^{S_{0,R}}\int_{\sigma_i}\dd{\epsilon_i} e^{-\beta \epsilon_i}.
\end{equation}
We see that $\Tr(1)$ is then finite if and only if $\sigma_i$ is bounded below. It is convenient in this case to set
\begin{equation}
    S_{0,R} = -\log(\frac1{2\pi}\int_{\sigma_i}\dd{\epsilon_i} e^{-\beta \epsilon_i}),
    \label{Equation: type II 1 normalisation}
\end{equation}
so that $\Tr(1) =1$.

\subsection{Density operator}
\label{subsec: density operator}

Next, given a physical state $\Ket{\phi}$, we wish to find the density matrix $\rho_\phi$ in the algebra $r(\mathcal{A}_{SR}^H)$. The density operator is defined to obey
\begin{equation}
    e^{S_{0,R}}\mel{\Psi}{r^{-1}(\rho_\phi)A}{\Psi} = \Tr(r^{-1}(\rho_\phi) A) = \Tr_r(\rho_\phi r(A)) = \Mel{\phi}{r(A)}{\phi}.
    \label{Equation: definition of density operator}
\end{equation}
In fact, the density operator is given by
\begin{equation}
    r^{-1}(\rho_\phi) = e^{-S_{0,R}}\Delta_{\phi|\Psi},
\end{equation}
i.e.\ the relative modular operator from $\ket{\Psi}$ to $\Ket{\phi}$, intertwined through the representation $r$, scaled by $e^{-S_{0,R}}$. Indeed, one may easily verify that such an operator suffices for equality between the left hand and right hand sides in~\eqref{Equation: definition of density operator}. Moreover, it is generally true that for any two states $\phi,\Psi$ with $\Psi$ cyclic and separating for an algebra, $\Delta_\Psi^{-1/2}\Delta_{\phi|\Psi}\Delta_\Psi^{-1/2}$ is an element of the algebra --- in our case $\Delta_\Psi$ is the identity, which confirms that the $r^{-1}(\rho_\phi)$ given above is an element of $\mathcal{A}^H_{SR}$. 

So to derive the density operator, let us first give the relative Tomita operator; it is
\begin{equation}
  S_{\phi|\Psi} = \Pi_{\text{phys}} \int_{-\infty}^\infty\dd{t}\ket{t}_i S_{\phi_{|i}(0)|\psi_{SR_{\bar{i}}}}\bra{0}_i \exp(-i(H_S+H_{R_{\bar{i}}}-H_{R_{\bar{i}}^*}+H_i)(t+i\beta/2)),
  \label{Equation: phi Psi relative tomita}
\end{equation}
where $\ket*{\phi_{|i}(\tau)} = \mathcal{R}_i(\tau)\Ket{\phi}$ is the reduced state when clock $C_i$ reads $\tau$, and $S_{\phi_{|i}(0)|\psi_{SR_{\bar{i}}}}$ is a relative Tomita operator for the algebra $\mathcal{A}_S\otimes\mathcal{B}(\mathcal{H}_{R_{\bar{i}}})$. One may employ this formula to obtain
\begin{equation}
    \Delta_{\phi|\Psi} = S^\dagger_{\phi|\Psi} S_{\phi|\Psi} = e^{\beta H_i/2}\int_{-\infty}^\infty \dd{t} e^{-iH_it} O_i^\tau(\Delta_{\psi_{SR_{\bar{i}}}}^{-1/2} S_{\phi_{|i}(\tau+t)|\psi_{SR_{\bar{i}}}}^\dagger S_{\phi_{|i}(\tau)|\psi_{SR_{\bar{i}}}}\Delta_{\psi_{SR_{\bar{i}}}}^{-1/2}) e^{\beta H_i/2},
\end{equation}
which holds for any $\tau$. A full derivation is given in Appendix~\ref{Appendix: density operator details}. The density operator is thus given by
\begin{equation}
    \rho_\phi = e^{-S_{0,R}}r(\Delta_{\phi|\Psi}) = e^{-S_{0,R}}V_i(i\beta/2)\int_{-\infty}^\infty \dd{t} V_i(t) \mathcal{O}_i^\tau(\Delta_{\psi_{SR_{\bar{i}}}}^{-1/2} S_{\phi_{|i}(\tau+t)|\psi_{SR_{\bar{i}}}}^\dagger S_{\phi_{|i}(\tau)|\psi_{SR_{\bar{i}}}}\Delta_{\psi_{SR_{\bar{i}}}}^{-1/2}) V_i(i\beta/2).
    \label{Equation: perspective neutral density operator}
\end{equation}
Note that the unusual combination of Tomita and modular operators appearing inside $\mathcal{O}^\tau_i(\dots)$ is in fact an element of $\mathcal{A}_S\otimes\mathcal{B}(\mathcal{H}_{R_{\bar{i}}})$, which is also shown in Appendix~\ref{Appendix: density operator details}.

The $\rho_\phi$ we have just found is the density operator at the perspective-neutral level. To find the density operator in the perspective of a particular $C_i\in R$, we can simply conjugate $\rho_\phi$ with the reduction map $\mathcal{R}_i(\tau)$. To be precise, because $\mathcal{R}_i(\tau)$ is an isometry, we can define a trace on the algebra $\mathcal{A}_{R_{\bar{i}}S|C_i}$ via
\begin{equation}
    \Tr_{|i}(\cdot) = \Tr_r(\mathcal{R}_i(\tau)^\dagger(\cdot)\mathcal{R}_i(\tau)),
\end{equation}
and it is easy to verify that the density operator of the perspective-reduced state $\ket*{\phi_{|i}(\tau)}$ with respect to this trace is given by $\rho_{\phi|i}(\tau) = \mathcal{R}_i(\tau) \rho_\phi \mathcal{R}_i(\tau)^\dagger$. In Appendix~\ref{Appendix: density operator details}, it is shown that
\begin{equation}
    \rho_{\phi|i}(\tau) = \Pi_{|i} e^{-S_{0,R}-\beta (H_{R^c}+H_{R_{\bar{i}}^*})} \int_{-\infty}^\infty\dd{t} e^{i(H_S+\sum_{ j\ne i }H_j)t} S_{\phi_{|i}(\tau+t)|\psi_{SR_{\bar{i}}}}^\dagger S_{\phi_{|i}(\tau)|\psi_{SR_{\bar{i}}}}.
    \label{eq:reducedDensityIntermed}
\end{equation}
Note that the $H_{R_{\bar{i}}^*}$ that appears in this formula will be cancelled out by contributions from the relative Tomita operators, and there will be no overall action on the second copy of the clocks in $R_{\bar{i}}$. To confirm this let us decompose the reduced state $\ket*{\phi_{|i}(\tau)}$ as
\begin{equation}
    \ket*{\phi_{|i}(\tau)} = \sum_I \ket*{\phi^I_S(\tau)}\otimes \ket*{\tilde\phi^I(\tau)},
    \label{Equation: reduced perspective bipartite decomposition}
\end{equation}
where $I$ is some continuous or discrete label, $\ket*{\phi^I_S(\tau)}\in\mathcal{H}_S$, and $\ket*{\tilde\phi^I(\tau)}\in\bigotimes_{i\ne j}\mathcal{H}_j$. Then one has
\begin{equation}
    \rho_{\phi|i}(\tau) = \Pi_{|i} e^{-S_{0,R}-\beta H_{R^c}} \int_{-\infty}^\infty\dd{t} e^{i(H_S+\sum_{ j\ne i}H_j)t} \sum_{I,J} S_{\phi_S^I(\tau+t)|\psi_S}^\dagger S_{\phi_S^J(\tau)|\psi_S}\otimes \tr_{R^c}(\ket*{\tilde\phi^I(\tau+t)}\bra*{\tilde\phi^J(\tau)}).
    \label{eq:reducedDensityOperator}
\end{equation}
This is explicitly shown in Appendix~\ref{Appendix: density operator details}. As claimed, the $H_{R_{\bar{i}}^*}$ has dropped out.

\subsection{Partial trace over a clock}
\label{Subsection: partial trace}

Suppose the set $R$ contains at least two clocks. We can imagine removing one particular clock $C_i\in R$ from this set, and asking how this affects our observations. We clearly have $r(\mathcal{A}_{S\,R\setminus C_i}^H)\subset r(\mathcal{A}_{SR}^H)$, so the set of operators we are allowed to act with is reduced -- since we can no longer dress to $C_i$, or reorient it. 

A good way to understand the effect of removing $C_i$ is to consider the consequences for the density operator. One may note that the density operator $\rho^R\in r(\mathcal{A}_{SR}^H)$ uniquely determines the density operator $\rho^{R\setminus C_i}\in r(\mathcal{A}_{S\,R\setminus C_i}^H)$ for the set of clocks $R\setminus C_i$, via the condition that the expectation values agree. Indeed, let $\Tr^R$ and $\Tr^{R\setminus C_i}$ denote the traces for the two sets of clocks; then we have
\begin{equation}
    \Tr^R(\rho^R a) = \Tr^{R\setminus C_i}(\rho^{R\setminus C_i} a) \qq{for all} a\in r(\mathcal{A}_{S\,R\setminus C_i}^H).
\end{equation}
This equation determines a map from density matrices in $r(\mathcal{A}_{SR}^H)$ to the corresponding density matrices in $r(\mathcal{A}_{S\,R\setminus C_i}^H)$. This map should be understood as a generalised partial trace (in the same way in which the traces constructed above are generalised matrix traces), which integrates out the degrees of freedom of $C_i$.

We may explicitly write down the partial trace. Indeed, pick a $C_j\in R$ with $C_i\ne C_j$, and let
\begin{equation}
    \rho^R = r(\tilde\rho^R), \qquad \rho^{R\setminus C_i} = r(\tilde\rho^{R\setminus C_i})
\end{equation}
denote the density operators for $R$ and $R\setminus C_i$ respectively (note that since $r$ is faithful on each algebra, each of $\tilde\rho^R,\tilde\rho^{R\setminus C_i}$ are uniquely determined by $\rho^R,\rho^{R\setminus C_i}$ respectively). Then for $a = r(\tilde a)\in r(\mathcal{A}_{S\,R\setminus C_i}^H)$ we have
\begin{align}
    \Tr^R(\rho^R a) &= e^{S_{0,R}}\qty\big(\bra*{\psi_{SR_{\bar{j}}}}\otimes\bra{0}_je^{-\beta H_j/2}) \tilde\rho^R \tilde a\qty\big(\ket*{\psi_{SR_{\bar{j}}}}\otimes e^{-\beta H_j/2}\ket{0}_j)\\
    &= e^{S_{0,R}}\qty\big(\bra*{\psi_{SR_{\overline{ij}}}}\otimes\bra{0}_je^{-\beta H_j/2}) \tr_i(e^{-\beta H_i}\tilde\rho^R) \tilde a\qty\big(\ket*{\psi_{SR_{\overline{ij}}}}\otimes e^{-\beta H_j/2}\ket{0}_j)\\
    &= e^{S_{0,R}-S_{0,R\setminus C_i}} \Tr^{R\setminus C_i}\qty(r(\tr_i(e^{-\beta H_i}\tilde\rho^R))a),
\end{align}
where $R_{\overline{ij}}=R\setminus \{C_i,C_j\}$, and $\tr_i$ denotes the ordinary Hilbert space trace for $\mathcal{H}_i$. Therefore, the partial trace over clock $C_i$ is given by
\begin{equation}
    \Tr^R_{C_i}: \rho^R \mapsto e^{S_{0,R}-S_{0,R\setminus C_i}} r\qty(\tr_i(e^{-\beta H_i}r^{-1}(\rho^R))).
\end{equation}
One observes that the constants $S_{0,R}$ and $S_{0,R\setminus C_i}$ are related to the overall normalisation of the partial trace. In some circumstances, we can use this relationship to constrain these constants in a canonical way, by considering the partial traces of certain operators. 

Consider, for example, the operator $\mathcal{O}^{\tau_j}_j(\dyad{\tau_i}_i)$, which conditions on the difference in times of clock $C_i$ and $C_j$ being $\tau_j-\tau_i$. Then one may show that
\begin{equation}
    \Tr^R_{C_i}(\mathcal{O}^{\tau_j}_j(\dyad{\tau_i}_i)) = \frac1{2\pi} e^{S_{0,R}-S_{0,R\setminus C_i}} \mathds{1} \int_{\sigma_i} \dd{\epsilon_i}e^{-\beta \epsilon_i}.
\end{equation}
If the spectrum of $H_i$ is bounded below, we see that this is finite and proportional to the identity. This is independent of the particular time difference $\tau_j-\tau_i$ that we condition on, which, physically speaking, makes intuitive sense: if we trace out the clock $C_i$, then it should not matter how we condition on its time relative to the other clocks. If we pick $S_{0,R}$ and $S_{0,R\setminus C_i}$ such that
\begin{equation}
    S_{0,R} - S_{0,R\setminus C_i} = -\log(\frac1{2\pi}\int_{\sigma_i}\dd{\epsilon_i} e^{-\beta \epsilon_i}),
\end{equation}
then we in fact have
\begin{equation}
    \Tr^R_{C_i}(\mathcal{O}^{\tau_j}_j(\dyad{\tau_i}_i)) = \mathds{1}.
    \label{Equation: partial trace canonical normalisation}
\end{equation}
This provides a canonical way to pick the constants $S_{0,R}$ and $S_{0,R\setminus C_i}$. 

In particular, suppose all the clocks in $R$ have spectrum bounded below, and let us enforce the following conditions: the trace on the algebra $r(\mathcal{A}_{SC_i}^H)$ obeys $\Tr^{C_i}(\mathds{1})=1$ for all $C_i\in R$; and the partial traces all obey~\eqref{Equation: partial trace canonical normalisation}. Then this uniquely determines
\begin{equation}
    S_{0,R} = -\sum_{C_i\in R}\log(\frac1{2\pi}\int_{\sigma_i}\dd{\epsilon_i} e^{-\beta \epsilon_i}).
\end{equation}

\subsection{Comparison with previous work}
\label{Section: comparison with previous}

We now pause to compare our density operator expressions \eqref{eq:reducedDensityOperator} to those obtained in \cite{Chandrasekaran:2022cip} (CLPW) and \cite{Jensen:2023yxy} (JSS) in similar settings.  To some extent this amounts to a mere translation between conventions, and where the treatments overlap we use this as a consistency check. However we also comment on some differences of methodology and generalisations available in our construction (aside from the primary generalisation to include an arbitrary numbers of reference frames, which we here eschew for sake of comparison).  Some details are placed in appendix \ref{app: comparison to prior work}.

We wish to compare our density operator expressions to the those of CLPW and JSS in the case that there are only two frames, which we denote $C_1$ and $C_2$.  We will consider the kinematical algebra $\mathcal{A}^H_{SC_1}$ and its physical or gauge-fixed representations. The first major methodological difference in our approach was noted already in section \ref{sect:PWCLPW} and illustrated in figure \ref{fig: commutative constraints}.  This is that both JSS and CLPW always begin with ideal clocks (having unbounded spectrum).  After computing density operators for semiclassical states, they then implement projections to limit the spectra of one or both clocks to the positive real line, effectively turning these from ideal to non-ideal clocks. Our approach, by contrast, has been to begin with a kinematical Hilbert space on which the clocks are already non-ideal, and then implement the gauge-constraint by mapping states and operators to the physical Hilbert space and/or gauge-fixed `reduced' Hilbert spaces.

Another hindrance to direct comparison of our density operator expressions is that we have written ours, by default, in the perspective of a frame whose reorientations are \textit{part of} the algebra under consideration, which in this case means taking the perspective of clock $C_1$ and not $C_2$.  This can be thought of as an alternate choice of gauge-fixing, via a map that is unitarily related to the $T_2$ map of \eqref{eq: T map} (which was introduced in Eq.~4.10 of \cite{Chandrasekaran:2022cip}).  In our language, the $T_2$ map is a reduction `to the perspective' of frame $C_2$ in orientation $\ket{\tau_2=0}$, which leaves the algebra $\mathcal{A}^H_{SC_1}$ intact in its usual form, apart from stripping the identity factor on $\mathcal{H}_2$ and, in the case of non-ideal $C_2$, implementing an additional projection for compatibility with the bounds of the $C_2$ sprectrum (compare equations \eqref{eq: T2 reduce SC1} and \eqref{eq: T2 reduce SC2}).   Note that regardless of which clock perspective we adopt, we are presently considering the algebra $\mathcal{A}^H_{SC_1}$ and its representation under different gauge fixings, as opposed to alternative choices of algebra (e.g. $\mathcal{A}^H_{SC_1}$ versus $\mathcal{A}^H_{SC_2}$ or $(\mathcal{A}^H_{SC_1})'$). Both CLPW and JSS consider states of the form 
\begin{equation}\label{eq: JSS product state}
    \ket*{\hat{\Phi}}_{|C_2}=\ket{\Phi}\otimes\ket{f}_{C_1},
\end{equation}
which are completely unentangled.\footnote{More specifically, we would say such states are unentangled between factors $S$ and $C_1$, as seen from the perspective of clock $C_2$, at least up to the effect of the projector $\Pi_{|2}$ which must ultimately be implemented if $C_2$ is non-ideal.}
We therefore proceed by mapping states of this form to the perspective of frame $C_1$, at which point we can apply our density operator expressions from section \ref{subsec: density operator}. Another frame change map then returns this density operator to the perspective of clock $C_2$, at which point we will have expression directly analogous to those of CLPW and JSS. The details of this computation are shown in appendix \ref{app: comparison to prior work}. The result (equation \eqref{eq: final density op C' perspective, appendix}) can be written
\begin{equation}\label{eq: final density op C2 perspective, main text}
  \rho_{\hat{\Phi}|C_2}
  =2\pi N  f(H_{C_1})\Pi_{|C_2}\Pi^{(\mathbb{R}\rightarrow\sigma_{C_1})}_{C_1}e^{-iH_ST^{\text{ext}}_{C_1}}
  S^\dagger_{\phi_S|\psi_{S}}  
  \Pi^{\text{ext}}_{|C_{2}}
  S_{\phi_S|\psi_{S}}
  e^{iH_ST^{\text{ext}}_{C_1}}\Pi^{(\sigma_{C_1}\rightarrow\mathbb{R})}_{C_1}f^*(H_{C_1})e^{\beta (H_S+H_{C_1})}.
\end{equation}
The operators in the center of this expression all act on the extended/ideal Hilbert space for clock $C_1$, and the operators $\Pi_{C_1}^{\mathbb{R}\rightarrow \sigma_{C_1}}:= \int_{\sigma_{C_1}}\dd\epsilon \ket{\epsilon}\bra{\epsilon}^{\text{ext}}$ and $\Pi_{C_1}^{\sigma_{C_1}\rightarrow \mathbb{R}}:= \int_{\sigma_{C_1}}\dd\epsilon \ket{\epsilon}^{\text{ext}}\bra{\epsilon}$ return this extended operator to the non-ideal clock $C_1$ Hilbert space.  In the case that clock $C_2$ is ideal, \ref{eq: final density op C2 perspective, main text} can be written
\begin{multline}\label{eq: density op when C2 ideal, main text}
  \rho^{(\sigma_{C_2}= \mathbb{R})}_{\hat{\Phi}|C_2}
  =2\pi N \\
  \times\Pi^{(\mathbb{R}\rightarrow\sigma_{C_1})}_{C_1} e^{\beta H^{\text{ext}}_{C_1}/2}f(H^{\text{ext}}_{C_1})e^{-i H_S T_{C_1}^{\text{ext}}}  \Delta_{\Psi_S}^{-1/2}\Delta_{\phi_S|\psi_{S}}\Delta_{\Psi_S}^{-1/2}e^{i H_S T_{C_1}^{\text{ext}}}f^*(H^{\text{ext}}_{C_1})e^{\beta H^{\text{ext}}_{C_1}/2}\Pi^{(\sigma_{C_1}\rightarrow\mathbb{R})}_{C_1}.
\end{multline}
Up to conventional normalisation constant, and swapping $H^{\text{ext}}_{C_1}\rightarrow \hat{q}$ and $T_{C_1}^{\text{ext}}\rightarrow-\hat{p}$\footnote{The minus sign on $p$ here comes from the fact that when $\hat{q}$ is the frame contribution to the constraint instead of $H_{C_1}$, algebra elements are dressed as $e^{i\hat{p} H_S}a e^{-i\hat{p}H_S}$ instead of $e^{-iT^{\text{ext}}_{C_1} H_S}a e^{iT^{\text{ext}}_{C_1} H_S}$.}, this is seen to be equivalent to JSS~\cite[equation 5.13]{Jensen:2023yxy}.  On the other hand, when $C_2$ is not ideal, equation \eqref{eq: final density op C2 perspective, main text} represents a nontrivial generalisation.  An important asymmetry between the clock frames is that when only clock $1$ is non-ideal, the density operator for the algebra $\mathcal{A}_{SC_1}^{H}$ can be obtained directly by projection (with $\Pi_{C_1}$) from the ideal case.  This is the method employed by both JSS and CLPW, and it is essentially what equation \eqref{eq: density op when C2 ideal, main text} expresses.  By contrast, when $C_2$ is non-ideal, direct projection with $\Pi_{|C_2}$ would not result in the correct density matrix.  This stems from the fact that this projector is an element of the commutant algebra rather than the algebra itself, and the projected algebra is not a subalgebra of the ideal algebra. This is discussed in more detail in appendix \ref{app: alternate method}, where we further verify expression \eqref{eq: final density op C2 perspective, main text} by giving an alternate derivation of the density operator, essentially modifying the method employed by JSS in~\cite[appendix E]{Jensen:2023yxy}.

While both JSS and CLPW discuss the possibility of non-ideal complementary frames, they only report explicit density operator expressions for states that are approximate product states, meaning they satisfy ($\Pi_{|C_2}\ket{\Phi_S}\otimes \ket{f}_{C_1} \approx \ket{\Phi_S}\otimes \ket{f}_{C_1}$).  Thus their results are not incorrect, but our expressions are more general in the case of non-ideal complementary frame.

\subsection{The need for a complementary frame}
\label{subsec: need for complementary frame}

In \cite[Sec.~4.3]{Chandrasekaran:2022cip}, when considering the algebra of observables in a static patch of de Sitter, it was noted that there is a stark difference between the case of including only a single observer to dress the patch algebra, versus including an additional observer associated with the complement patch.  In either case, the original subsystem algebra is a type III$_1$ factor.  When both observers are included, the dressed algebra becomes a type II von Neumann factor on the physical Hilbert space.  If the complement observer is absent, this transition to type II occurs for the gauge-invariant algebra at the kinematical level, but the representation of this algebra on the physical Hilbert space fails altogether to be von Neumann. Our technical summary of the reason for this was given already in Sec.~\ref{ssec_vN}, but we here review that discussion in the present context, in view of our generalisations to an arbitrary number of reference frames.  

All of the kinematical gauge-invariant algebras $\mathcal{A}_{SC_1\dots C_n}^{H}$ are von Neumann by construction.  In particular, they are closed under a bicommutant on the kinematical Hilbert space.  
To discern whether they remain von Neumann in their representation on the physical Hilbert space, it is useful to consider their bicommutant with respect to the set of gauge-invariant kinematical operators.\footnote{The von Neumann nature of these kinematical algebras refers explicitly to their representation on the kinematical Hilbert space, with the relevant commutant being with respect to the full set of bounded operators $\mathcal{B}(\mathcal{H}_{\text{kin}})$.  However we are free to consider other commutant operations, such as that with respect to a restricted set of operators like the gauge-invariant kinematical operators $\mathcal{B}(\mathcal{H}_{\text{kin}})^H$.} None of these algebras are closed under this restricted bicommutant, because at least the bounded functions of the constraint itself are appended by the operation.  If these are the \textit{only} operators which this bicommutant appends, then the representation on the physical Hilbert space will remain von Neumann, since the bounded functions of the constraint map trivially to the identity component ($c$-numbers) on the physical Hilbert space.
When the action of $H_S$ is ergodic on $\mathcal{A}'$, as is true in the case of de Sitter patches \cite{Chandrasekaran:2022cip,Chen:2024rpx}, the gauge invariant commutant of $\mathcal{A}_{SC_1\dots C_n}^{H}$ will contain none of the elements $a'\in \mathcal{A}'$, or dressed versions thereof, unless at least one clock $C_{n+1}$ is left over.  With an additional clock present, elements of the form $\mathcal{O}^\tau_{C_{n+1}}(a'), a' \in \mathcal{A}'$ are part of the gauge-invariant commutant, so that dressed elements of $\mathcal{A}'$ are excluded from the gauge-invariant bicommutant.  In the absence of this additional clock, the bicommutant always adds elements of the form $\mathcal{O}^\tau_{C_1}(a')$ (for example).  Thus it is apparent that algebras of the form $\mathcal{A}_{SC_1\dots C_n}^{H}$ will never be von Neuman when represented on the physical Hilbert space unless there is at least one complementary clock.  It has been observed \cite{Kaplan:2024xyk} that the need for a complement observer is likely connected to the failure of convergence of all single-particle states in a de Sitter group-averaging inner product \cite{Marolf:2008hg}.  Both are the result of pathologically `unbalanced' de Sitter configurations, which at the classical level manifest as a linearisation instability \cite{DeVuyst:2024grw}.

It is perhaps not surprising that the properties of a local algebra for an observer depends on the contents of the rest of the universe in gauge theory -- after all the gauge constraints are non-local conditions on the theory (although we note that the authors of~\cite{Chandrasekaran:2022cip} took an alternative viewpoint, and a complete interpretation of the discussion above remains open).

\section{Subregion density operators with degenerate clocks}
\label{sect:rhoDeg}

So far, we have been dealing with clocks whose Hamiltonian is non-degenerate when computing the density operators in Sec.~\ref{Section: arbitrary number of observers}. Let us now generalise our construction to the case where the clock Hamiltonians have degenerate spectra, in order to investigate the consequences of this degeneracy for density operators and later entropies. The case that is of most physical importance is when $H_C \propto p^2$ as it arises for free particles or in relativistic dispersion relations, but for now we will take the degeneracy to be more general. 

\subsection{Degenerate quantum clocks as QRFs}\label{sssec_degclock1}

We start by `replicating' Sec.~\ref{Section: quantum clocks as quantum reference frames} for degenerate clocks, following some of the discussion in \cite{Hoehn:2020epv}, before constructing density operators in their presence in Sec.~\ref{ssec_degdensity}.

\subsubsection{Covariant POVMs for degenerate clocks}\label{sssec_degpovm}

Let us summarise the necessary modifications of Sec.~\ref{sect:POVMs} when $H_C$ has a degenerate continuous spectrum, but the degeneracy does not depend on the energy (except possibly for a set of measure zero). Clock states \eqref{clockstates} now read
\begin{equation}\label{degclockstate}
    \ket{t,\lambda}:=\frac{1}{\sqrt{2\pi}}\int_{\sigma_C}\dd{\epsilon}e^{ig(\epsilon)}e^{-it\epsilon}\ket{\epsilon,\lambda},
\end{equation}
where $\lambda$ labels the degeneracy sectors. Since these sectors are orthogonal, we further have
\begin{eqnarray}
    \braket{t,\lambda}{t',\lambda'}=\delta_{\lambda\lambda'}\chi(t-t'),
\end{eqnarray}
where $\chi(t-t')$ is the overlap distribution in Eq.~\eqref{fuzzy}. Clearly, per $\lambda$-sector, all the previous properties apply and the clock states give rise to a covariant clock POVM per $\lambda$-sector. In particular, the normalisation takes the form
\begin{eqnarray}
    \int_\mathbb{R}\dd{t}\ket{t,\lambda}\!\bra{t,\lambda}=\Pi_\lambda\,,
\end{eqnarray}
with $\Pi_\lambda$ the projector onto the $\lambda$-degeneracy sector of $\mathcal{H}_C$ and $\sum_\lambda\Pi_\lambda=\mathds{1}_C$. A covariant clock POVM for the full Hilbert space $\mathcal{H}_C$ is thus obtained by summing over the degeneracy sectors. \\
\indent A relevant example is a clock Hamiltonian of the form $H_C=sp^2/2m$ with $s=\pm1$, as they appear in many relativistic and non-relativistic examples. Its spectrum $\sigma_C=s[0,\infty)$ is doubly degenerate, except for its zero eigenvalue, and clock states can be written as \cite{Hoehn:2020epv,Braunstein:1995jb}
\begin{eqnarray}
    \ket{t,\lambda}=\int_\mathbb{R}\dd{p}\sqrt{|p|}\Theta(-\lambda p)e^{ig(p)}e^{-itsp^2/2m}\ket{p},
\end{eqnarray}
where $\lambda=\pm1$ labels positive and negative frequency (or left and right moving) sectors. We have $\Pi_\pm=\Theta(\mp p)$ and, since $\ket{t,\lambda}$ has zero support on $p=0$, also $\Pi_\pm\ket{t,\mp}=0$. It can be shown \cite{holevoProbabilisticStatisticalAspects1982,Hoehn:2020epv,Busch1994} that the first moment operator of the corresponding POVM is a symmetric quantisation of the classically conjugate time variable $T=s\frac{mq}{2p}$,  which thus is monotonic along the flow generated by $H_C$ regardless of the frequency sector, i.e. 
\begin{eqnarray}
    T^{(1)}=\sum_{\lambda=\pm1} T^{(1)}_\lambda=\sum_{\lambda=\pm1}\int_\mathbb{R}\dd{t}t\ket{t,\lambda}\!\bra{t,\lambda}=s\frac{m}{4}\left(qp^{-1}+p^{-1}q\right).
\end{eqnarray}

\subsubsection{Decomposition of Hilbert spaces and observable algebras}\label{sssec_degclock2}

Let us analyze the equivalent of Secs.~\ref{sseHphys} and \ref{ssec_Hphys} for the case when the clock spectrum is degenerate in an energy-independent way (except possibly for a set of measure zero). This discussion will somewhat generalise the one in \cite{Hoehn:2020epv}, which was restricted to quadratic Hamiltonians. 
From the above, we know that if clock $C$ has degenerate spectrum, then its Hilbert space can be decomposed into a direct sum over the degeneracy sectors labeled by $\lambda$
\begin{equation}
    \label{eq:degClockHilbertspace}
    \mathcal{H}_C = \bigoplus_{\lambda} \mathcal{H}^\lambda_C,
\end{equation}
where each single $\mathcal{H}^\lambda_C$ is a copy of $L^2(\sigma_C)$. This means that the total kinematical Hilbert space decomposes as (where we collect all remaining degrees of freedom in a single system $S$)
\begin{equation}
    \label{eq:kinematicalDeg}
    \mathcal{H}_{\text{kin}} = \mathcal{H}_S\otimes \bigoplus_\lambda \mathcal{H}_C^{\lambda}.
\end{equation}
Alternatively, we can write the clock Hilbert spaces as 
\begin{equation}
    \label{eq:degClockH}
    \mathcal{H}_C = L^2(\sigma_C) \otimes \mathbb{C}^{m},
\end{equation}
where $m$ is the number of degeneracy sectors, i.e.\ the multiplicity. The first tensor factor above gives the energy wavefunction, while the latter accounts for the different sectors. We then have
\begin{equation}
    \mathcal{H}_{\text{kin}} = \mathcal{H}_S\otimes L^2(\sigma_C) \otimes \mathbb{C}^{m}
\end{equation}
The factor $\mathbb{C}^{m}$ may be viewed as a Hilbert space for a set of internal degrees of freedom of clock $C$.

The decomposition \eqref{eq:kinematicalDeg} carries over to the physical Hilbert space \cite{Hoehn:2020epv} as well 
\begin{equation}\label{hphysdecomp}
    \mathcal{H}_{\rm phys}=\bigoplus_\lambda\mathcal{H}_\lambda.
\end{equation}
One can see this by defining the charge operator $Q:=\sum_\lambda c_\lambda\Pi_\lambda\in\mathcal{A}_{\rm inv}$ for some $c_\lambda\in\mathbb{R}$ with $c_\lambda\neq c_{\lambda'}$, whose eigenspace with eigenvalue $c_\lambda$ corresponds to the $\lambda$-degeneracy sector.
Then one observes that $\Pi_\lambda$ and the  charge $Q$ commute with $H$, leading to the decomposition.

As for the algebra, on the kinematical Hilbert space we have the following algebra of observables prior to imposing gauge-invariance
\begin{equation}
    \mathcal{A}_{SC} = \mathcal{A}_S\otimes \mathcal{B}(L^2(\sigma_C))\otimes \mathcal{B}(\mathbb{C}^{m}).
\end{equation}
Compared to the non-degenerate case, we now also have access to operators which swap between the different degeneracy sectors of the clocks. These are contained in the tensor factor $\mathcal{B}(\mathbb{C}^{m})$. The gauge-invariant subalgebra may then be written as (rearranging tensor factors for convenience)
\begin{equation}
    \label{eq:AinvDegenerate}
    \mathcal{A}_\text{inv} = \Big(\mathcal{A}_S\otimes  \mathcal{B}(L^2(\sigma_C)) \Big)^H \otimes  \mathcal{B}(\mathbb{C}^{m}).
\end{equation}
The first factor is equivalent to the algebra in the non-degenerate case, consisting of the relational observables $O_{C}^\tau(a)$ \eqref{relobs} and clock reorientations $U_C(t)$, while the latter factor is just a matrix algebra and encodes the sector dependency. Thus, when $\mathcal{A}_S$ is von Neumann, then so is $\mathcal{A}_{\rm inv}$; the latter will again be a Type $\rm{II}$ factor when $\mathcal{A}_S$ is a Type $\rm{III}_{1}$ factor and $H$ a modular Hamiltonian.

\subsubsection{Operational superselection}
\label{sect:degSuperselection}

Suppose the system $S$ contains multiple degenerate clocks. There are then various subalgebras of $\mathcal{A}_\text{inv}$ one can consider, corresponding to subtly different subsystems. To see this, let us recall the motivations we gave at the beginning of Sec.~\ref{Section: arbitrary number of observers} for the degrees of freedom accessible to a certain observer travelling along a worldline. The basic assumption is that the observer can act on all of the degrees of freedom in the vicinity of the worldline. By the timelike tube theorem, it can then also act on the fields in the timelike envelope of the worldline, which (under certain assumptions \cite{Witten:2023qsv}) may be identified with the region $\mathcal{U}$. Using the gravitational constraints we then argued that the observer may have access to the times and energies of some subset of other clocks $C_i$. Thus, the observer may act with arbitrary operators on the $L^2(\sigma_i)$ factor of the clock Hilbert space. But it is not necessarily true that the observer can act on the internal degrees of freedom i.e.\ the $\mathbb{C}^{m_i}$ factor -- unless such degrees of freedom can be related to the fields via the gravitational constraints. Whether this is true in practice changes depending on the exact nature of the clock, and at the present level of abstraction it should be viewed as operational input. If an observer \emph{cannot} act on the internal degrees of freedom, then from the observer's perspective clock $C_i$ undergoes superselection -- since any of the operators the observer uses to act on $C_i$ must be block-diagonal in the degeneracy sectors.

Another way in which the subsystem can change is purely internal to the observer itself. There may be a good physical reason why an observer would be restricted to acting within some subset of its own degeneracy sectors. For example, for the quadratic Hamiltonian observer described in Sec.~\ref{sssec_degpovm}, the two degeneracy sectors correspond to positive and negative frequency modes of the clock, and it may be desirable to restrict to the positive frequency mode sector (see for example~\cite{Susskind:2023rxm}).

In any case, it is clear that there are various scenarios one can consider, and one needs operational input to choose between them. What should be clear is that, depending on the particular scenario, we will get different algebras of observables, and hence differing manifestations of the \observerdependence{} of entropy (see Sec.~\ref{sec:degClockExample}).

In Sec.~\ref{sssec_degclock2}, we have assumed there is no superselection and therefore we need to take the sector-swapping operators into account. Here, we discuss what happens if we take the gauge-invariant algebra to be superselected across the $\lambda$-sectors. 

Recalling from Sec.~\ref{sssec_degpovm} that the covariant clock POVM decomposes across the $\lambda$-degeneracy sectors of clock $C$, 
$E_C(X)=\sum_\lambda\int_{X\subset\mathbb{R}}\dd{t}\ket{t,\lambda}\!\bra{t,\lambda}$, we see that the relational observables in Eq.~\eqref{relobs} decompose across the $\lambda$-sectors:
\begin{equation}
O^\tau_C(a)=\sum_\lambda O^\tau_{C,\lambda}(a)=\sum_\lambda\int_\mathbb{R}\dd{t}U_{SC}(t)\left(a \otimes \ket{\tau,\lambda}\!\bra{\tau,\lambda}\right)U_{SC}^\dag(t),
\end{equation}
where $O^\tau_{C,\lambda}(a)=(\mathds{1}_S \otimes \Pi_\lambda) O_C^\tau(a)=O_C^\tau(a)(\mathds{1}_S \otimes \Pi_\lambda)$. Clearly, the clock reorientations decompose similarly and thereby give rise to a reducible representation of the translation group on $\mathcal{H}_C$,
\begin{equation}
    U_{C}(\tau)\otimes \mathds{1}_S=\sum_\lambda U_{C,\lambda}(\tau)\otimes \mathds{1}_S, \qquad U_{C,\lambda}(\tau):=\exp(-i\tau\Pi_\lambda H_C).
\end{equation}
In other words, the entire gauge-invariant algebra \eqref{gaugeinvalg} for the superselected situation decomposes across the degeneracy sectors
\begin{equation}
    \mathcal{A}^{ss}_{\rm inv}=\bigoplus_\lambda\mathcal{A}_{\rm inv}^\lambda=\bigoplus_\lambda\Big\langle O^\tau_{C,\lambda}(a),\mathds{1}_S \otimes U_{C,\lambda}(t)\,\big|\,a\in\mathcal{A}_S,\,t\in\mathbb{R}\Big\rangle.
\end{equation}
In particular, note that $\mathds{1}_S \otimes \Pi_\lambda\in\mathcal{A}^{ss}_{\rm inv}$, so $\mathcal{A}_{\rm inv}^\lambda$ is a proper subalgebra of $\mathcal{A}^{ss}_{\rm inv}$.

The gauge-invariant algebra is superselected across the $\lambda$-sectors and thus reducibly represented on $\mathcal{H}_{\rm kin}$. Indeed, since the charge $Q = \sum_\lambda c_\lambda\Pi_\lambda$ commutes with every element in $\mathcal{A}^{ss}_{\rm inv}$, it is part of its center. Hence, no observable in $\mathcal{A}^{ss}_{\rm inv}$ can map between the $\lambda$-sectors, leading to the superselection rule 
\begin{equation}
O_C^\tau(a)=\oplus_\lambda O_{C,\lambda}^\tau(a),\qquad\qquad \mathds{1}_S \otimes U_C(\tau) = \mathds{1}_S \otimes \oplus_\lambda U_{C,\lambda}(\tau).
\end{equation}
Interactions  will ruin this superselection rule if they do not commute with the projectors $\Pi_\lambda$.

For instance, in the case of a quadratic clock Hamiltonian $H_C=-p^2$, as briefly discussed in Sec.~\ref{sssec_degpovm} and as they appear in relativistic dispersion relations $H=-p_t^2+\abs{\vec{p}}^2+m^2$ and minisuperspace models, the observables will get superselected across the positive and negative frequency sectors \cite{Hoehn:2020epv}. Interactions can lift this superselection; e.g.\ in a closed FRW model with massive scalar field, positive and negative frequency modes mix \cite{Hohn:2011us}.

Invoking the argument in appendix~\ref{app_ginvalg} per $\lambda$-sector entails that, if $\mathcal{A}_S$ is von Neumann, then so are $\mathcal{A}^{ss}_{\rm inv}$ and each of the $\mathcal{A}_{\rm inv}^\lambda$. The individual $\mathcal{A}_{\rm inv}^\lambda$ may be factors (when considered as algebras acting within their corresponding $\lambda$-sectors), depending on some conditions. For example, they are when $\mathcal{A}_S$ is a type $\rm{III}_1$ factor and $H_S$ is a modular Hamiltonian as in the case of perturbative quantum gravity below.

Moreover, due to the decomposition \eqref{hphysdecomp}, we obtain a reducible physical representation $r$ of the gauge-invariant algebra (here expressed in co-invariant language)
\begin{equation}
    r(\mathcal{A}^{ss}_{\rm inv})=\bigoplus_\lambda r\left(\mathcal{A}_{\rm inv}^\lambda\right).
\end{equation}
Hence, coherent superpositions across $Q$ charge sectors in the physical Hilbert space are indistinguishable from corresponding classical mixtures  when probed with observables from $r(\mathcal{A}_{\rm inv})$.

While we started with one clock $C$, we can think of the different charge sectors as corresponding to different quantum clocks $C_\lambda$. The covariant POVM $E_{C,\lambda}(X)=\Pi_\lambda E_C(X)$ models $C_\lambda$ on the clock Hilbert space $\mathcal{H}_{C}^{\lambda}$. The $\lambda$-sector $\mathcal{H}_\lambda$ of the physical Hilbert space can thus be viewed as the sector of the theory in which this clock $C_\lambda$ exists.  A given observer may only have access to a clock with fixed charge $\lambda$, e.g.\ a clock with only positive frequencies in relativity. Such an observer can then only probe the $\lambda$-sector and will be oblivious to what happens outside it.

\subsubsection{Page-Wootters formalism for degenerate clocks}

When the clock spectrum is degenerate, a reduction map like Eq.~\eqref{PWred} can be defined making use of the decomposition of the physical Hilbert space in its degeneracy sectors \eqref{hphysdecomp} through $\mathcal{R}_C(\tau):\mathcal{H}_{\rm phys}\rightarrow\bigoplus_\lambda\mathcal{H}_{|C,\lambda}$, where $\mathcal{H}_{|C,\lambda}$ are the reduced Hilbert spaces of the $\lambda$-sector and
\begin{equation}
\label{eq:PWreddegSum}
    \mathcal{R}_C(\tau) := {\oplus_\lambda} \mathcal{R}_{C,\lambda}(\tau), \qq\qquad {\mathcal{R}_C^\dag(\tau):=\oplus_\lambda} \mathcal{R}_{C,\lambda}^\dag(\tau) \,.
\end{equation}
Sector-wise everything works as before and so 
\begin{equation}\label{jklv}
    \mathcal{R}_C(\tau)\mathcal{R}_C^\dag(\tau)=\oplus_\lambda\mathcal{R}_{C,\lambda}(\tau)\mathcal{R}_{C,\lambda}^\dag(\tau)=\oplus_\lambda\Pi_{|C,\lambda}\,,\quad\quad\mathcal{R}^\dag_C(\tau)\mathcal{R}_C(\tau)=\oplus_\lambda\mathcal{R}_{C,\lambda}^\dag(\tau)\mathcal{R}_{C,\lambda}(\tau)=\mathds{1}_{\rm phys}\,,
\end{equation}
where  $\Pi_{|C,\lambda}=\bra{\tau,\lambda}\Pi_{\rm phys}\ket{\tau,\lambda}=\Pi_{|C}$ for all $\lambda$, as all reduced spaces are isomorphic copies. So far, we have invoked the setting of co-invariants. Using Eq.~\eqref{MR1}, it is related to the PW-reduction in RAQ language
\begin{equation}
    \label{PWreddeg}
    \mathcal{R}_{C,\lambda}(\tau)=\mathcal{R}'_{C,\lambda}(\tau)M^\dag\,,\qquad\text{with}\qquad\mathcal{R}'_{C,\lambda}(\tau):=\bra{\tau,\lambda}\otimes \mathds{1}_S\,.
\end{equation}
Thus, only the reduction maps per $\lambda$-superselection sector 
$\mathcal{R}_{C,\lambda}(\tau)$ are invertible \cite{Hoehn:2020epv}.
Specifically, the reduction theorems in Eqs.~\eqref{obsredthm} and~\eqref{redalg} only hold per $\lambda$-sector,
\begin{equation}
    \mathcal{R}_{C,\lambda}(\tau)\,r(\mathcal{A}^\lambda_{\rm inv})\mathcal{R}^\dag_{C,\lambda}(\tau) =\Pi_{|C}\big\langle a,U_S(t)\,\big|\,a\in\mathcal{A}_S,\,t\in\mathbb{R}\big\rangle\Pi_{|C}.
\end{equation}
Since the clock states \eqref{degclockstate} satisfy the same covariance property \eqref{clockcov} as in the non-degenerate case, the reduced form of this algebra will be the same for each superselection sector.

Every $\lambda$-sector also has its own relational Schr\"odinger equation Eq.~\eqref{Schrod}, but again it will take the same form in each sector. For example, for quadratic Hamiltonians, $\tau$-time will run forward in \emph{both} the positive and negative frequency sectors \cite{Hoehn:2020epv}.

The reduction maps also act on the sector swaps, i.e.\ 
elements of $\mathcal{B}(\mathbb{C}^{m})$ in $\eqref{eq:AinvDegenerate}$. Taking a single sector swap as {$S^C_{\lambda'\lambda}$ of the same clock, we have that due to orthogonality when we reduce with respect to the same clock that 
\begin{equation}
    \label{eq:reductionSectorSwap}
    R_{C, \lambda_1}(\tau) S^C_{\lambda_2\lambda_3}R_{C, \lambda_4}^\dag(\tau) = \delta_{\lambda_1 \lambda_2} \delta_{\lambda_3\lambda_4}\Pi_{|C}.
\end{equation}
Hence, }sector swaps of the clock we reduce to vanish under the reduction to a single degeneracy sector  due to them being {$\lambda$-block} off-diagonal. Nevertheless, they will generally not vanish when we perform a global reduction as in \eqref{eq:PWreddegSum}{, as this can lead to sector-wise off-diagonal conjugation with the reduction maps}. In this case, the reduced algebra will still have swap operators between sectors of the clock we reduce to.

In contrast, if we reduce a sector swap of some other clock $C'$ for instance, then
\begin{equation}
    \label{eq:reductionOtherSectorSwap}
    R_{C, \lambda}(\tau) S^{C'}_{\sigma'\sigma}R_{C, \lambda}^\dag(\tau) 
    = S^{C'}_{\sigma'\sigma}\Pi_{\vert C}.
\end{equation}
Sector swaps of other clocks than the one we reduce to do not vanish, regardless of whether we perform a sector-wise or global reduction.

\subsubsection{Changing degenerate clocks}\label{sssec_degclockchange}

Let us discuss how we can perform QRF transformations, as in Sec.~\ref{ssec_QRFchanges}, between degenerate clocks. Since the reduction maps \eqref{eq:PWreddegSum} are  invertible per sector, QRF transformations can be defined on a combined sector $(\lambda_1, \lambda_2)$ of the two degenerate clocks. We will restrict the coming discussion to this situation, which can be thought of as a superselection case of Sec.~\ref{sect:degSuperselection}. 
One can define a larger QRF transformation by taking direct sums of the ones we will discuss. This larger transformation is invertible because they are sector-wise invertible, in the same way as \eqref{jklv}.

Suppose Alice has access to the superselection sector $\lambda_1$ of clock $C_1$, while Bob only has access to superselection sector $\lambda_2$ of $C_2$. They can then only compare observables in the subalgebra
\begin{eqnarray}
    \label{eq:algebraDegTwo}
    \mathcal{A}_{\rm inv}^{\lambda_1,\lambda_1}&:=&\left(\mathcal{A}_S \otimes \mathcal{B}(\mathcal{H}_{1,\lambda_1})\otimes\mathcal{B}(\mathcal{H}_{2,\lambda_2})\right)^{H}\\
    &=&\Big\langle O^{\tau_i}_{C_i,\lambda_i}(a),\mathds{1}_S \otimes U_{i,\lambda_i}(t)\otimes \mathds{1}_{j}\,\Big|\,a\in\mathcal{A}_S\otimes\mathcal{B}(\mathcal{H}_{j,\lambda_j}),\,t\in\mathbb{R}\Big\rangle\,\qquad i,j=1,2, \,\,i\neq j,\nonumber
\end{eqnarray}
as it contains all those observables which can be written as relational observables and reorientations relative to either of the clocks $C_{1,\lambda_1}$ and $C_{2,\lambda_2}$. This is a subalgebra of both $\mathcal{A}_{\rm inv}^{\lambda_1}$ and $\mathcal{A}_{\rm inv}^{\lambda_2}$, in fact, it is their overlap $\mathcal{A}_{\rm inv}^{\lambda_1,\lambda_2}=\mathcal{A}_{\rm inv}^{\lambda_1}\cap\mathcal{A}_{\rm inv}^{\lambda_2}$. This will become important later when Alice and Bob with different clocks cannot compare the full density operators they have respectively access to, but only the part residing in the overlap of the algebras.

The physical Hilbert space splits into superselection sectors \eqref{hphysdecomp} relative to both clocks \cite{Hoehn:2020epv}
\begin{eqnarray}
    \mathcal{H}_{\rm phys}=\bigoplus_{\mu_1}\mathcal{H}_{\mu_1}=\bigoplus_{\mu_2}\mathcal{H}_{\mu_2},
\end{eqnarray}
where we use $\mu_i$ as the dummy summation variable over the superselection sectors of clock $C_i$, to distinguish it from the specific sectors $\lambda_i$. The domain of the physical representation $r(\mathcal{A}_{\rm inv}^{\lambda_1,\lambda_2})$ is the overlap of the two sectors $\mathcal{H}_{\lambda_1,\lambda_2}:=\mathcal{H}_{\lambda_1}\cap\mathcal{H}_{\lambda_2}$, i.e.\ the subsector where both clocks $C_{1,\lambda_1}$ and $C_{2,\lambda_2}$ exist and we can thus compare and relate their corresponding relational descriptions. 

As the ``quantum coordinate maps'' in Eq.~\eqref{PWreddeg} are now defined per superselection sector for each clock, we also have that the two observers can only compare states in the sector overlap $\mathcal{H}_{\lambda_1,\lambda_2}$, i.e.\ clock changes are defined here \cite{Hoehn:2020epv} (see also \cite{Hohn:2018toe,Hohn:2018iwn}):
\begin{equation*}
    \begin{tikzcd}[column sep=8em, row sep=3em]
        & \mathcal{H}_{\lambda_1,\lambda_2} \arrow[swap]{dl}{\mathcal{R}_{1,\lambda_1}(\tau_1)} \arrow{dr}{\mathcal{R}_{2,\lambda_2}(\tau_2)} & \\
        \mathcal{H}_{\lambda_2|1}=\Pi_{|1}\left(\mathcal{H}_S\otimes \mathcal{H}_{2,\lambda_2}\right) \arrow[swap]{rr}{V_{1,\lambda_1 \to 2,\lambda_2}(\tau_1,\tau_2) = \mathcal{R}_{2,\lambda_2}(\tau_2) \circ \mathcal{R}_{1,\lambda_1}^\dag(\tau_1)} & & \mathcal{H}_{\lambda_1|2}=\Pi_{|2}\left(\mathcal{H}_S\otimes\mathcal{H}_{1,\lambda_1}\right)
    \end{tikzcd}
\end{equation*}
The doubly superselected clock change transformation \eqref{clockchange} thus becomes
\begin{eqnarray}
V_{1,\lambda_1\to2,\lambda_2}(\tau_1,\tau_2):=\mathcal{R}_{2,\lambda_2}(\tau_2)\circ\mathcal{R}_{1,\lambda_1}^\dag(\tau_1)=\int_\mathbb{R}\dd{t}\ket{t+\tau_1,\lambda_1}_1\otimes\bra{\tau_2-t,\lambda_2}_2\otimes U_S(t).
\end{eqnarray}
As before, it is a controlled, hence non-local unitary.

{By combining sectors in the reduction maps appropriately, one can also transform sector swaps from one perspective to another. Furthermore, a full QRF transformation could be carried out by summing over all sector-wise ones.}

\subsubsection{Subsystem relativity for degenerate clocks} \label{sssec_degrelativity}

Suppose again that Alice has access to superselection sector $\lambda_1$ of clock $C_1$, while Bob only has access to superselection sector $\lambda_2$ of $C_2$. As they can only compare observables in the subalgebra $\mathcal{A}_{\rm inv}^{\lambda_1,\lambda_2}$ \eqref{eq:algebraDegTwo}, we consider subsystem relativity within it. Defining for $i=1,2$, $i\neq j$,
\begin{equation}
    \mathcal{A}^H_{SC_i,\lambda_i}:=\left(\mathcal{A}_S\otimes\mathcal{B}(\mathcal{H}_{i,\lambda_i})\otimes \mathds{1}_{j,\lambda_j}\right)^H
\end{equation}
with $\mathds{1}_{j,\lambda_j}$ the identity on the subspace $\mathcal{H}_{j,\lambda_j}$, we clearly have 
\begin{equation}
   \mathcal{A}^H_{SC_1,\lambda_1}\cap\mathcal{A}^H_{SC_2,\lambda_2}= \left(\mathcal{A}_S \otimes \mathds{1}_{1,\lambda_1}\otimes \mathds{1}_{2,\lambda_2}\right)^H.
\end{equation}
That is, Alice and Bob can again only compare internal relational observables of $S$ and, as before, when $\mathcal{A}_S$ is a Type $\rm{III}_1$ von Neumann factor {on which the modular Hamiltonian $H_S$ acts ergodically (as in de Sitter space \cite{Chandrasekaran:2022cip,Chen:2024rpx})}, then this overlap contains only $c$-numbers. It is also clear that this conclusion carries over to physical representations of these algebras, similarly to Eq.~\eqref{physoverlap}.

We noted in Sec.~\ref{ssec_subsystem relativity} that internal QRF perspectives $\mathcal{R}_i(\tau)$ are (direct sums/integrals of) TPSs on the physical Hilbert space $\mathcal{H}_{\rm phys}$ and that changing QRF changes such a TPS. Perspective maps $\mathcal{R}_{i,\lambda_i}$ of degenerate clocks are similarly (direct sums/integrals of) TPSs, however, only on the corresponding superselection sector $\mathcal{H}_{\lambda_i}\subset\mathcal{H}_{\rm phys}$. Since changes from Alice's to Bob's clock are defined on  intersections $\mathcal{H}_{\lambda_1,\lambda_2}=\mathcal{H}_{\lambda_1}\cap\mathcal{H}_{\lambda_2}$, the clock change means that we switch from understanding the subsystem structure of $\mathcal{H}_{\lambda_1,\lambda_2}$ in terms of the TPS of its ambient space $\mathcal{H}_{\lambda_1}$ to understanding it in terms of the TPS of the distinct ambient space $\mathcal{H}_{\lambda_2}$.

\subsection{Density operators}\label{ssec_degdensity}

In comparison to Sec.~\ref{Section: arbitrary number of observers}, each clock $C_i$ is now assigned a label $\lambda_i$ for its degeneracy sectors. Similarly to Sec.~\ref{sssec_degclock2}, we can extend those results to include multiple degenerate clocks. For instance,
the total kinematical Hilbert space decomposes as 
\begin{equation}
    \mathcal{H}_{\text{kin}} = \mathcal{H}_S\otimes \bigoplus_\Lambda \bigotimes_{i=1}^n\mathcal{H}_i^{\lambda_i}
\end{equation}
where $\Lambda$ denotes the collection of degeneracy sector labels $\{\lambda_i\}$ for all clocks. Through \eqref{eq:degClockH} per clock this can alternatively be written as 
\begin{equation}
    \mathcal{H}_{\text{kin}} = \mathcal{H}_S\otimes \bigotimes_{i=1}^nL^2(\sigma_i) \otimes \mathbb{C}^{m_i}
\end{equation}
Again, each factor $\mathbb{C}^{m_i}$ may be viewed as a Hilbert space for a set of internal degrees of freedom of clock $C_i$.

As explained in Sec.~\ref{sect:degSuperselection}, there are multiple subalgebras one can consider. In the rest of this subsection we will explore what happens in the different cases. First, we will treat the case where there is no superselection whatsoever. Then we will see what happens when all the clocks are subjected to superselection. There are of course a multitude of intermediate cases where only some clocks are superselected, but to avoid overcomplicating the narrative we will not explicitly study these. It is straightforward to see how these intermediate cases mix the properties of two cases we study. Finally, we will see what happens when we additionally \emph{restrict} the sector of a particular clock.

\subsubsection{No superselection}

Let us first consider the case where there is no superselection. Then the algebra of observables for a set of clocks $R$ may then be written (without yet imposing gauge-invariance)
\begin{equation}
    \mathcal{A}_{SR} = \mathcal{A}_S\otimes \bigotimes_{C_j\in R} \mathcal{B}(L^2(\sigma_j))\otimes \mathcal{B}(\mathbb{C}^{m_j}).
\end{equation}
The gauge-invariant subalgebra may then be written (rearranging tensor factors for convenience)
\begin{equation}
    \mathcal{A}^H_{SR} = \qty\Big(\mathcal{A}_S\otimes \bigotimes_{C_j\in R} \mathcal{B}(L^2(\sigma_j)))^H\otimes \bigotimes_{C_j\in R}  \mathcal{B}(\mathbb{C}^{m_j}).
    \label{eq:degClockAlgebra}
\end{equation}
The first factor is equivalent to the algebra in the non-degenerate case, while the latter factor is just a matrix algebra. The trace on the full $\mathcal{A}^H_{SR}$ may be constructed as the tensor product of the traces on the two tensor factors. Since the latter tensor factor is a finite Type I algebra, the Type of $\mathcal{A}^H_{SR}$ is the same as in the non-degenerate case, i.e.\ if $R$ contains only a single clock of energy bounded below, then $\mathcal{A}^H_{SR}$ is Type II${}_1$, otherwise it is Type II${}_\infty$.

As previously, the trace may be written as the expectation value in a state $\ket{\Psi}$. This state may be constructed like~\eqref{Equation: Psi tracial state}, except we have to additionally include maximally mixed states over two copies of the multiplicity Hilbert spaces $\mathbb{C}^{m_j}$. Thus,
\begin{equation}
    \ket{\Psi} = \ket{\psi_S} \otimes\ket*{\psi_{R_{\bar{i}}}}\otimes e^{-\beta H_i/2}\ket{0}_i \otimes \bigotimes_{C_j\in R} \sum_{\lambda_j=1}^{m_j} \ket{\lambda_j}_j\otimes\ket{\lambda_j}_j,
\end{equation}
where $\ket{\lambda_j}_j$, $a\lambda_j=1,\dots,m_j$ is an orthonormal basis of $\mathbb{C}^{m_j}$ labelling the degeneracy sectors. Let us now set
\begin{equation}
    \ket*{\psi_{SR_{\bar{i}}}} = \ket{\psi_S} \otimes\ket*{\psi_{R_{\bar{i}}}} \otimes \bigotimes_{C_j\in R_{\bar{i}}} \sum_{\lambda_j=1}^{m_j} \ket{\lambda_j}_j\otimes\ket{\lambda_j}_j,
\end{equation}
so that
\begin{equation}
    \ket{\Psi} = \ket*{\psi_{SR_{\bar{i}}}}\otimes e^{-\beta H_i/2}\ket{0}_i \otimes \sum_{\lambda_i=1}^{m_i} \ket{\lambda_i}_i\otimes\ket{\lambda_i}_i.
\end{equation}

One then finds that a relative Tomita operator for $\mathcal{A}_{SR}^H$ from $\ket{\Psi}$ to a physical state $\Ket{\phi}$ may be written
\begin{multline}
  S_{\phi|\Psi} = \Pi_{\text{phys}} \sum_{\lambda_i,\lambda'_i}\int_{-\infty}^\infty\dd{t}\ket{t}_i\otimes \ket{\lambda_i}_i S_{\phi_{|i}(0,\lambda'_i)|\psi_{SR_{\bar{i}}}}\bra{0}_i\otimes\bra{\lambda'_i}_i\otimes\bra{\lambda_i}_i \\
  \exp(-i(H_S+H_{R_{\bar{i}}}-H_{R_{\bar{i}}^*}+H_i)(t+i\beta/2)),
  \label{Equation: phi Psi relative tomita degenerate no superselection}
\end{multline}
Here, $S_{\phi_{|i}(0,\lambda'_i)|\psi_{SR_{\bar{i}}}}$ is a relative Tomita operator for 
\begin{equation}
    \mathcal{A}_{R_{\bar{i}}S} = \mathcal{A}_S\otimes \bigotimes_{C_j\in R_{\bar{i}}} \mathcal{B}(L^2(\sigma_j))\otimes \mathcal{B}(\mathbb{C}^{m_j}),
\end{equation}
from $\ket*{\psi_{SR_{\bar{i}}}}$ to
\begin{equation}
    \ket*{\phi_{|i}(0,\lambda'_i)} = \bra{\lambda'_i}_i \mathcal{R}_i(0)\Ket{\phi} = \mathcal{R}_{i,\lambda'_i}(0)\Ket{\phi}.
\end{equation}
As before, we may use~\eqref{Equation: phi Psi relative tomita degenerate no superselection} to deduce the density operator. One finds
\begin{equation}
    \rho_\phi = \sum_{\lambda_i,\lambda_i'} e^{-S_{0,R}}V_i(i\beta/2)\int_{-\infty}^\infty \dd{t} V_i(t) \mathcal{O}_i^\tau(\Delta_{\psi_{SR_{\bar{i}}}}^{-1/2} S_{\phi_{|i}(\tau+t,\lambda'_i)|\psi_{SR_{\bar{i}}}}^\dagger S_{\phi_{|i}(\tau,\lambda_i)|\psi_{SR_{\bar{i}}}})  V_i(i\beta/2)\otimes r\qty(\ket{\lambda_i'}\bra{\lambda_i}).
    \label{Equation: degenerate density operator without superselection}
\end{equation}
In computing the entropy of such a density operator, one would have to account for the correlations between the different degeneracy sectors.

\subsubsection{Superselection for all clocks}
\label{sect:superAll}

Suppose now that we are in a situation where superselection occurs for all the clocks, just like in Sec.~\ref{sect:degSuperselection}. This means that we are not allowed to act with operators which change the degeneracy sectors of the clocks.

In this case the gauge-invariant algebra of the QFT and a set of clocks $R$ decomposes:
\begin{equation}\label{superselectedalgebra}
    \mathcal{A}^H_{SR} = \bigoplus_{\Lambda_R} \mathcal{A}^H_{RS,\Lambda_R},
\end{equation}
where $\Lambda_R = \{\lambda_i\mid C_i\in R\}$. Each $\mathcal{A}^H_{RS,\Lambda_R}$ is the gauge-invariant algebra within a degeneracy sector of the clocks in $R$. Each $\mathcal{A}^H_{RS,\Lambda_R}$ is structurally equivalent to the algebra $\mathcal{A}^H_{SR}$ for non-degenerate clocks.

Due to the degeneracy, $\mathcal{A}^H_{SR}$ is not a factor. However, each $\mathcal{A}^H_{RS,\Lambda_R}$ is a Type II factor, and thus has a unique trace up to a constant factor. For simplicity, we will choose this constant factor to be the same in each sector, which amounts to imposing that the constant $S_{0,R}$ in~\eqref{Equation: trace} is sector-independent. Then~\eqref{Equation: trace} defines the trace on the full algebra $\mathcal{A}^H_{SR}$.

A direct sum of algebras of the same Type, also has that Type. Thus if $R$ contains only a single clock of energy bounded below, then $\mathcal{A}^H_{SR}$ is Type II${}_1$, otherwise it is Type II${}_\infty$.

The direct sum structure descends also to the physical Hilbert space:
\begin{equation}
    \mathcal{H}_{\text{phys}} = \bigoplus_{\Lambda_R} \mathcal{H}_{\text{phys}}^{\Lambda_R},
\end{equation}
Here for convenience we are choosing only to decompose relative to the degeneracy sector labels of $R$, but each $\mathcal{H}_{\text{phys}}^{\lambda_R}$ may of course be further decomposed relative to the degeneracy sector labels of $R^c$. Each $\mathcal{A}^H_{RS,\lambda_R}$ has a physical representation $r(\mathcal{A}^H_{RS,\lambda_R})$ acting on $\mathcal{H}_{\text{phys}}^{\Lambda_R}$.

The point now is that we can exploit this direct sum structure to compute the density operator sector by sector. In particular, let us decompose the physical state as
\begin{equation}
    \label{eq:physicalStateDeg}
    \Ket{\phi} = \sum_{\Lambda_R} \sqrt{p^{\Lambda_R}} \Ket*{\phi^{\Lambda_R}},
\end{equation}
where $p^{\Lambda_R}\ge 0$ is chosen such that $\Ket*{\phi^{\Lambda_R}}\in\mathcal{H}_{\text{phys}}^{\Lambda_R}$ is normalised. Explicitly, we have
\begin{equation}
    p^{\Lambda_R} = \Bra{\phi}\Pi^{\Lambda_R}\Ket{\phi}, \qq{and} \Ket*{\phi^{\Lambda_R}} = \frac1{\sqrt{p^{\Lambda_R}}} \Pi^{\Lambda_R}\Ket{\phi}
    \label{eq:degReducedStates}
\end{equation}
where $\Pi^{\Lambda_R}=\bigotimes_{C_i\in R}\Pi_{\lambda_i}$ is the projector onto the $\Lambda_R$ degeneracy sector. Then, working sector by sector, the density matrix corresponding to $\Ket{\phi}$ is given by
\begin{equation}
    \rho_\phi = \sum_{\Lambda_R} p^{\Lambda_R} \Pi^{\Lambda_R} \rho_\phi^{\Lambda_R},
    \label{Equation: degenerate density operator}
\end{equation}
where
\begin{equation}
    \rho_\phi^{\Lambda_R} = e^{-S_{0,R}}V_i(i\beta/2)\int_{-\infty}^\infty\dd{t} V_i(t) \mathcal{O}_i^\tau(\Delta_{\psi_{SR_{\bar{i}}}}^{-1/2} S_{\phi^{\Lambda_R}_{|i}(\tau+t)|\psi_{SR_{\bar{i}}}}^\dagger S_{\phi^{\Lambda_R}_{|i}(\tau)|\psi_{SR_{\bar{i}}}}\Delta_{\psi_{SR_{\bar{i}}}}^{-1/2}) V_i(i\beta/2).
    \label{Equation: degenerate density operator with superselection}
\end{equation}
Here,
\begin{equation}
    \label{eq:PWDeg}
    \ket*{\phi^{\Lambda_R}_{|i}(\tau)} = \mathcal{R}_i(\tau) \Ket*{\Phi^{\Lambda_R}}
\end{equation}
is the state in the perspective of $C_i$, in the $\Lambda_R$ sector.

The entropy may now similarly be decomposed sector by sector. Indeed, we have
\begin{equation}
    \log\rho_\phi = \sum_{\Lambda_R} \Pi^{\Lambda_R} \qty(\log p^{\Lambda_R} + \log\rho_\phi^{\Lambda_R}),
\end{equation}
which implies that
\begin{equation}
    S[\phi] = -\Bra{\phi}\log\rho_\phi\Ket{\phi} = -\sum_{\Lambda_R} p^{\Lambda_R} \log p^{\Lambda_R} + \sum_{\Lambda_R} p^{\Lambda_R} S^{\Lambda_R}[\phi],
    \label{Equation: degenerate entropy}
\end{equation}
where $S^{\Lambda_R}[\phi] = -\Bra*{\phi^{\Lambda_R}}\log \rho_\phi^{\Lambda_R}\Ket*{\phi^{\Lambda_R}}$ is the entropy in the $\Lambda_R$ sector. The first term on the right hand side above is a Shannon entropy for the probability distribution of the sectors, while the second term is the entropy \emph{averaged over} the sectors.

To go further with the computation of the entropy amounts to evaluating $S^{\Lambda_R}[\phi]$. However, within each sector this is just equivalent to evaluating the entropy in the non-degenerate case. Thus, the entropy in the superselecting case may be computed by applying non-degenerate arguments sector by sector, and then applying~\eqref{Equation: degenerate entropy}.

\subsubsection{Fixing the sector for one clock}

Let us suppose we fix the sector of a single clock $C_i$ to $\lambda_i$. This amounts to restricting to the Hilbert space
\begin{equation}
    \Pi_{\lambda_i}\mathcal{H}_{\text{kin}} = \mathcal{H}_S\otimes \mathcal{H}_i^{\lambda_i}\otimes\bigotimes_{j\ne i} \mathcal{H}_j.
\end{equation}
Similarly, the algebra of operators (in both the no superselection and superselection cases) is restricted to $\Pi_{\lambda_i}\mathcal{A}_{SR}^H\Pi_{\lambda_i}$. The trace on this algebra may be simply constructed as the restriction of the trace on the original larger algebra $\mathcal{A}_{SR}^H$. Similarly, if $\rho$ is the density operator on the larger algebra then $\Pi_{\lambda_i}\rho\Pi_{\lambda_i}$ is the density operator on the restricted algebra.

Thus, we can straightforwardly write down the density operator in either of the two cases, using the previous results~\eqref{Equation: degenerate density operator without superselection} and~\eqref{Equation: degenerate density operator}. In the case of no superselection for the other clocks, we obtain:
\begin{equation}
    \rho_\phi = e^{-S_{0,R}}\Pi_{\lambda_i}V_i(i\beta/2)\int_{-\infty}^\infty \dd{t} V_i(t) \mathcal{O}_i^\tau(\Delta_{\psi_{SR_{\bar{i}}}}^{-1/2} S_{\phi_{|i}(\tau+t,\lambda_i)|\psi_{SR_{\bar{i}}}}^\dagger S_{\phi_{|i}(\tau,\lambda_i)|\psi_{SR_{\bar{i}}}}) V_i(i\beta/2).
\end{equation}

In the case of superselection, we simply restrict the sum in~\eqref{Equation: degenerate density operator} to those $\Lambda_R$ consistent with $\lambda_i$. The $\rho_\phi^{\Lambda_R}$ are still given by~\eqref{Equation: degenerate density operator with superselection}.

\section{Semiclassical regime}\label{sec_entropy}

In the preceding sections we have worked out explicit expressions for density operators associated with a variety of gauge-invariant subalgebras in the presence of a variety of types of clock frames. We now turn to the computation of entanglement entropies associated with such algebras.
For a general state $\Ket{\phi}$, it does not seem possible to go further than the implicit expression 
\begin{equation}
    S[\phi] = -\bra*{\phi_{|i}(\tau)}\log\rho_{\phi|i}(\tau)\ket*{\phi_{|i}(\tau)} 
\end{equation}
for the entropy. To make progress we will first consider a restricted (but physically important) class of states: those falling within a semiclassical regime defined by the following two assumptions. 
(We will focus on the case of non-degenerate clocks, but the results of this section may be straightforwardly generalised to the density operators for degenerate clocks given in Section~\ref{sect:rhoDeg}.)
\begin{itemize}
    \item First, we assume that there is an $\epsilon>0$ such that for any $a\in\mathcal{A}_{S}\otimes\bigotimes_{C_j\in R_{\bar{i}}}\mathcal{B}(\mathcal{H}_{C_j})$ 
    \begin{equation}
        \abs{\Mel{\phi}{\mathcal{O}_{C_i}^\tau(a) V_i(t)}{\phi}}\ll \abs{\Mel{\phi}{\mathcal{O}_{C_i}^\tau(a)}{\phi}} \qq{if} \abs{t} > \order{\epsilon}.
        \label{Equation: semiclassical condition 1}
    \end{equation}
    In other words, the expectation values of dressed observables are peaked when no overall reorientation of $C_i$ is being done. One can roughly speaking think of this as a condition that ``the state $\Ket{\phi}$ is peaked around a particular orientation for $C_i$''. Of course, unlike the above equation, this na\"ive statement is not gauge-invariant. The correct and more precise statement is that there is some degree of freedom (or collection of degrees of freedom) in the state $\Ket{\phi}$ which is capable of keeping track of reorientations of $C_i$, and which is peaked around `no reorientation'. This is easiest to see if we set $a=1$; the above equation then means that reorienting the clock by a little bit $\epsilon$ results in an almost orthogonal state. Allowing $a$ to be general means that this happens independently of the action of any dressed operators.
    \item Second, consider the operator $r(e^{iH_St})$, i.e.\ the physical representation of evolution by the QFT Hamiltonian.
    Then we assume
    \begin{equation}
        r(e^{iH_St})\Ket{\phi}\approx e^{i\mathcal{E}t}\Ket{\phi} \qq{if} \abs{t} <\order{\epsilon},
        \label{Equation: semiclassical condition 2}
    \end{equation}
    for some constant 
    \begin{equation}
        \mathcal{E} = \Expval{r(H_S)}{\phi}.
    \end{equation}
    Thus, within times of $\order{\epsilon}$, the QFT degrees of freedom only evolve by an overall phase. Put another way, the system is in an approximate eigenstate of $H_S$ (of course, it could also be an exact eigenstate of $H_S$, such as the vacuum).
\end{itemize}
In combination, by~\eqref{Equation: semiclassical condition 1} there is an effectively fixed time (the orientation of $C_i$), which, by~\eqref{Equation: semiclassical condition 2}, acts as a \emph{classical} parameter on which the \emph{quantum} state of the fields depends. This is why we call this a \emph{semiclassical} regime. 

An alternative formulation of these assumptions is as follows:
\begin{equation}
    (\Delta H_S)^2_\phi \ll (\Delta H_i)^2_\phi,
    \label{Equation: semiclassical fluctuation hierarchy}
\end{equation}
i.e.\ the variance of the QFT Hamiltonian is suppressed relative to the variance of the clock Hamiltonian. Indeed, by the first assumption above we have $(\Delta H_i)^2_\phi\sim 1/\epsilon^2$, and if we define 
\begin{equation}
    \eta = \norm{\epsilon(r(H_S)-\mathcal{E})\Ket{\phi}} = (\Delta H_i)^{-1}_\phi (\Delta H_S)_\phi,
\end{equation}
then the second assumption amounts to taking $\eta\ll 1$, which reduces to~\eqref{Equation: semiclassical fluctuation hierarchy}. Another equivalent way to state these assumptions is that the orientation of $C_i$ must be sufficiently peaked that fluctuations in the QFT Hamiltonian are suppressed.

With these assumptions in place, we are basically allowed to ignore (in a controlled manner) the QFT Hamiltonian $H_S$ in certain places, when computing the entropy $S[\phi]$. To heuristically set the stage for justifying this statement (and indeed for explaining what it means), consider the algebra whose entropy we are computing:
\begin{equation}
    \mathcal{A}_{R_{\bar{i}}S|C_i} = \{a, H_S+\sum_{j\ne i}H_j\mid a\in \mathcal{A}_{S}\otimes\mathcal{B}(\mathcal{H}_{R_{\bar{i}}})\}.
\end{equation}
Suppose we were to remove $H_S$ from the right hand side. Then what we would end up with would be a rather different algebra, which (since the modified reorientation operator would now commute with field operators, and since the Hamiltonians of the frames in $R_{\bar{i}}$ may be absorbed into the $\mathcal{B}(\mathcal{H}_{R_{\bar{i}}})$ factor) can be written as
\begin{equation}
    \widehat{\mathcal{A}} = \{a, \sum_{j\ne i}H_j\mid a\in \mathcal{A}_{S}\otimes\mathcal{B}(\mathcal{H}_{R_{\bar{i}}})\}
    = \mathcal{A}_{S}\otimes \mathcal{B}(\mathcal{H}_{R_{\bar{i}}}) \otimes \{H_{R^c}\}''.
    \label{Equation: semiclassical algebra}
\end{equation}
The approximation for the entropy $S[\phi]$ that we will give will have the interpretation of a kind of entropy with respect to this algebra (but not a von Neumann entropy -- indeed the algebra is Type III, being the tensor product of the Type III algebra $\mathcal{A}_{S}$ with some other algebra).

Suppose we were to also take the formula~\eqref{eq:reducedDensityIntermed} for the density operator in the perspective of $C_i$, and remove $H_S$ wherever it appears. We would end up with\footnote{Here, in writing $\hat{\rho}_\phi(\tau)$ we leave off any remnant of the projector $\Pi_{|i}$, because for reasons discussed in appendix \ref{Appendix: semiclassical approximation details}, we can ignore this in the following discussion of the semiclassical entropy.}
\begin{equation}
    \hat\rho_{\phi}(\tau) = e^{-S_{0,R}-\beta (H_{R^c}+H_{R_{\bar{i}}^*})} \int_{-\infty}^\infty\dd{t} e^{i\sum_{j\ne i}H_jt} S_{\hat\phi(\tau+t)|\psi_{SR_{\bar{i}}}}^\dagger S_{\hat\phi(\tau)|\psi_{SR_{\bar{i}}}},
    \label{Equation: rho hat}
\end{equation}
where
\begin{equation}
    \ket*{\hat\phi(\tau+t)} = e^{-i\sum_{j\ne i}H_jt}\ket*{\phi_{|i}(\tau)}
\end{equation}
is the state we would get by evolving the reduced state $\ket*{\phi_{|i}(\tau)}$ with the `modified reorientation operator' $\sum_{j\ne i} H_j$.  Under the semiclassical assumptions made above, it turns out that the entropy may be approximated as follows (a derivation is given in Appendix~\ref{Appendix: semiclassical approximation details}):
\begin{equation}
    S[\phi] = -\bra*{\phi_{|i}(\tau)}\log\rho_{\phi|i}(\tau)\ket*{\phi_{|i}(\tau)} 
    \approx-\bra*{\hat\phi(\tau)}\log\hat\rho_{\phi}(\tau)\ket*{\hat\phi(\tau)}.
    \label{Equation: hat entropy}
\end{equation}
As alluded to above, the QFT Hamiltonian $H_S$ has more or less been eliminated from the right hand side. 

Let us now consider the quantum information theoretic meaning of this approximate result. To this end, consider the Hilbert space
\begin{equation}
    \widehat{\mathcal{H}} = \mathcal{H}_S\otimes\mathcal{H}_{R_{\bar{i}}} \otimes \mathcal{H}_{R_{\bar{i}}}^*\otimes L^2(\sigma_{R^c}).
\end{equation}
This includes field degrees of freedom in $\mathcal{H}_S$, and degrees of freedom of two copies of the frames in $R_{\bar{i}}$ --- but it also contains an additional degree of freedom for the total energy of the frames in $R^c$. Any element of $\widehat{\mathcal{A}}$ may be written as a $\mathcal{A}_S \otimes \mathcal{B}(\mathcal{H}_{R_{\bar{i}}})$ valued function of $H_{R^c}$. Writing such an element as $a(H_{R^c})$ where $a:\sigma_{R^c}\to \mathcal{A}_S\otimes \mathcal{B}(\mathcal{H}_{R_{\bar{i}}})$, we can define a faithful, normal representation of $\widehat{\mathcal{A}}$ on $\widehat{\mathcal{H}}$ as follows:
\begin{equation}
    a(H_{R^c}) (\ket{\varphi}\otimes \ket{E}) = a(E)\ket{\varphi} \otimes \ket{E},
\end{equation}
where $\ket{\varphi}\in\mathcal{H}_S\otimes\mathcal{H}_{R_{\bar{i}}}\otimes\mathcal{H}_{R_{\bar{i}}}^*$, operators are taken to act on the first copy of the frames in $\mathcal{H}_{R_{\bar{i}}}$ and $\ket{E}$ is an energy $E$ eigenstate in $L^2(\sigma_{R^c})$. It is easy to verify that the following (non-normalisable) state is cyclic/separating for this representation:
\begin{equation}
    \ket*{\widehat{\Psi}} = \ket{\psi_S} \otimes \bigotimes_{C_j\in R_{\bar{i}}} \int_{\sigma_j} \dd{E_j} \ket{E_j}\otimes\bra{E_j} \otimes \frac1{\sqrt{2\pi}}\int_{\sigma_{R^c}} \dd{E} \ket{E}.
    \label{Equation: semiclassical Psi}
\end{equation}
The expectation value in this state gives a faithful normal weight on the algebra:
\begin{equation}
    \widehat\Psi:\widehat{\mathcal{A}} \to\CC, \qquad a(H_{R^c}) \mapsto \frac1{2\pi}\int_{\sigma_{R^c}}\dd{E} \bra*{\psi_S} \tr_{R_{\bar{i}}}\qty(a(E)) \ket*{\psi_S}.
\end{equation}
This weight may be thought of as one in which the fields are in the KMS `vacuum' $\ket{\psi_S}$, the frames in $R_{\bar{i}}$ are maximally mixed, and the energy of the complementary frames $R^c$ is uniformly distributed.

In Appendix~\ref{Appendix: semiclassical approximation details}, it is shown that the modular operator of $\ket*{\widehat\Psi}$ is $\Delta_{\widehat\Psi}=e^{-\beta H_S}$, and that moreover the relative modular operator from $\ket*{\widehat\Psi}$ to $\ket*{\hat\phi(\tau)}$ is related to~\eqref{Equation: rho hat} by
\begin{equation}
    \Delta_{\hat\phi(\tau)|\widehat\Psi} = e^{S_{0,R}+\beta H_{R^c}}\hat\rho_\phi(\tau).
    \label{Equation: semiclassical relative modular}
\end{equation}
Using these results in the formula for the entropy, we find
\begin{align}
    S[\phi] &\approx S_{0,R} + \beta \bra*{\hat\phi(\tau)}H_{R^c}\ket*{\hat\phi(\tau)} - \bra*{\hat\phi(\tau)}\log \Delta_{\hat\phi(\tau)|\widehat\Psi} \ket*{\hat\phi(\tau)} \label{Equation: entropy decomp (not separating)}\\
     &= S_{0,R} + \beta \bra*{\hat\phi(\tau)}\qty(H_S+H_{R^c})\ket*{\hat\phi(\tau)} + \bra*{\hat\phi(\tau)}\log \Delta_{\widehat\Psi|\hat\phi(\tau)} \ket*{\hat\phi(\tau)},\label{Equation: entropy decomp}
\end{align}
where the second line holds when $\ket*{\hat\phi(\tau)}$ is separating for the algebra above, and then follows from $\log\Delta_{\hat\phi(\tau)|\widehat\Psi} = \log\Delta_{\widehat\Psi} + \log\Delta_{\hat\phi(\tau)} - \log\Delta_{\widehat\Psi|\hat\phi(\tau)}$.  The first term in this entropy formula reflects an overall state-independent ambiguity, coming from the choice of normalisation in the trace. The second term may be rewritten as the expectation of $-\beta H_R$ in the state $\Ket{\phi}$. The final term may be recognised as a relative entropy between $\ket*{\widehat\Psi}$ and $\ket*{\hat\phi(\tau)}$ with respect to the algebra above. Overall, the entropy in the semiclassical regime may be written
\begin{equation}
    S[\phi] \approx S_{0,R} - \beta \expval{H_R}_\phi - S_{\text{rel}}(\hat\phi(\tau)||\widehat\Psi). 
    \label{Equation: semiclassical entropy}
\end{equation}

\subsection{Shannon entropy contribution from energy of complementary clocks}
\label{Subsection: complementary energy}

The `effective' algebra $\widehat{\mathcal{A}}$ represented in equation \eqref{Equation: semiclassical algebra} has a non-trivial center generated by $H_{R^c}$, the total Hamiltonian of the complementary clocks. It turns out that the relative modular operator $\Delta_{\hat\phi(\tau)|\widehat\Psi}$ commutes with $H_{R^c}$, and thus may be decomposed as a direct integral over the eigenspaces of $H_{R^c}$. This results in a contribution to the entropy $S[\phi]$ that takes the form of a classical Shannon entropy for the probability distribution corresponding to measurements of $H_{R^c}$.

More precisely, the projection-valued measure for measurements of $H_{R^c}$ is $P(E)\dd{E}$, where
\begin{equation}
    P(E) = \delta\qty\big(E- \sum_{\mathclap{C_j\in R^c}} H_j) = \frac1{2\pi} \int_{-\infty}^\infty e^{i(E-\sum_{C_j\in R^c} H_j)t}\dd{t}.
    \label{Equation: complementary energy pvm}
\end{equation}
From this, we may write down the Born probability distribution for measurements of $H_{R^c}$:
\begin{equation}
    p_E = \bra*{\hat\phi(\tau)}P(E)\ket*{\hat\phi(\tau)}.
\end{equation}
Furthermore, the expectation values of operators $a\in \mathcal{A}_S\otimes \mathcal{B}(H_{R_{\bar{i}}})$ may be written
\begin{equation}
    \bra*{\hat\phi(\tau)}a\ket*{\hat\phi(\tau)} = \int_{\sigma_{R^c}}\dd{E} p_E \phi_E(a),
\end{equation}
where\footnote{Here we are assuming for simplicity that $p_E\ne 0$ for all $E\in \sigma_{R^c}$. It is straightforward to account for the case where this does not hold, by simply excluding values of $E$ for which $p_E=0$ in the following arguments.}
\begin{equation}
    p_E\phi_E(a) = \bra*{\hat\phi(\tau)}P(E)a\ket*{\hat\phi(\tau)}
\end{equation}
This functional $\phi_E$ defines a state on $\mathcal{A}_S\otimes\mathcal{B}(H_{R_{\bar{i}}})$, and it may be straightforwardly confirmed that the relative modular operator from 
\begin{equation}
    \ket*{\bar\psi} = \ket{\psi_S} \otimes \bigotimes_{C_j\in R_{\bar{i}}} \int_{\sigma_j} \dd{E_j} \ket{E_j}\otimes\bra{E_j}
\end{equation}
to $\phi_E$ takes the form
\begin{equation}
    \Delta_{\phi_E|\bar\psi} = \frac1{p_E}e^{-\beta H_{R_{\bar{i}}^*}} S_{\hat\phi(\tau)|\psi_{SR_{\bar{i}}}}^\dagger P(E)S_{\hat\phi(\tau)|\psi_{SR_{\bar{i}}}}.
    \label{Equation: energy conditioned relative modular}
\end{equation}
Note that $\bar\psi$ is the tensor product of the QFT KMS state $\ket{\psi_S}$ with a maximally mixed state for the frames in $R_{\bar{i}}$.

Using the identity
\begin{equation}
    S_{\hat\phi(\tau+t)|\psi_{SR_{\bar{i}}}} = e^{-iH_{R^c}t}S_{\hat\phi(\tau)|\psi_{SR_{\bar{i}}}}e^{iH_{R_{\bar{i}}}t},
\end{equation}
we may also write
\begin{align}
    \Delta_{\hat\phi(\tau)|\widehat\Psi} &= 
    e^{-\beta H_{R_{\bar{i}}^*}} \int_{-\infty}^\infty\dd{t} e^{i\sum_{j\ne i}H_jt} S_{\hat\phi(\tau+t)|\psi_{SR_{\bar{i}}}}^\dagger S_{\hat\phi(\tau)|\psi_{SR_{\bar{i}}}}\\
    &=e^{-\beta H_{R_{\bar{i}}^*}} \int_{-\infty}^\infty\dd{t} e^{iH_{R^c}t} S_{\hat\phi(\tau)|\psi_{SR_{\bar{i}}}}^\dagger e^{iH_{R^c}t}S_{\hat\phi(\tau)|\psi_{SR_{\bar{i}}}}\\
    &=2\pi \int_{\sigma_{R^c}}\dd{E} P(E) e^{-\beta H_{R_{\bar{i}^*}}}S_{\hat\phi(\tau)|\psi_{SR_{\bar{i}}}}^\dagger P(E)S_{\hat\phi(\tau)|\psi_{SR_{\bar{i}}}}.
\end{align}
Recognising the integrand as $P(E) p_E \Delta_{\phi_E|\bar\psi}$, and using the facts that the integrand commutes with $H_{R^c}$, and $P(E)P(E') = \delta(E-E')P(E)$, we may write
\begin{align}
    \log \Delta_{\hat\phi(\tau)|\widehat\Psi} &= \int \dd{E} P(E) \log(2\pi p_E \Delta_{\phi_E|\bar\psi})\label{Equation: relative modular E integral}\\
    &= \log(2\pi) + \int \dd{E} P(E) \qty(\log p_E + \log \Delta_{\phi_E|\bar\psi}).
\end{align}
Substituting this into the entropy formula, we find
\begin{equation}
    S[\phi] \approx S_{0,R} - \log(2\pi) + \beta\expval*{H_{R^c}}{\hat\phi(\tau)} - \int_{\sigma_{R^c}} \dd{E} p_E \log p_E - \int_{\sigma_{R^c}} \dd{E} p_E \phi_E(\log\Delta_{\phi_E|\bar\psi}).
\end{equation}
The third term appearing above is the promised Shannon entropy of $H_{R^c}$. In the case that $\phi_E$ is separating for $\mathcal{A}_S\otimes\mathcal{B}(H_{R_{\bar{i}}})$, we may rewrite the final term in terms of relative entropies as we have done previously. The result is
\begin{equation}
    S[\phi]\approx S_{0,R}-\log (2\pi) - \beta \expval{H_R}_\phi - \int_{\sigma_{R^c}} \dd{E} p_E \log p_E - \int_{\sigma_{R^c}} \dd{E} p_E S_{\text{rel}}(\phi_E||\bar\psi).
\end{equation}
The final term now readily can be identified as the relative entropy from $\bar\psi$ of the state $\phi_E$, \emph{averaged over} $E$, according to the probability distribution $p_E$. The structure of this formula is reminiscent of the entropies in~\cite{Donnelly_2012, Donnelly_2015}, as is to be expected for algebras with non-trivial center.

\subsection{Generalised entropy}
\label{Subsection: generalised entropy}

By arguing as in~\cite{Jensen:2023yxy}, one may view~\eqref{Equation: semiclassical entropy} as a kind of gravitational generalised entropy, with an expected area term. However, our formula holds more generally than has been previously considered. First, we are allowing the state $\ket*{\phi_{|i}(\tau)}$ to be arbitrarily entangled between the QFT and the clocks (insofar as this is compatible with the semiclassical assumptions above), whereas for example~\cite{Chandrasekaran:2022cip, Jensen:2023yxy} considered only the case where $\ket*{\phi_{|i}(\tau)}$ factorises as a product state between the QFT and the clocks. Second, we are computing the entropy one gets with respect to arbitrary subsets $R$ of clocks (whereas previously only a single clock has been involved). The resulting entropy then in fact differs from the generalised entropy, depending on which clocks $R$ we use. To see this, let us proceed with a rough outline of the arguments of~\cite{Jensen:2023yxy}; for further details of the assumptions involved we refer the reader to that paper. We introduce some kind of UV regulator on the fields, such that the QFT algebra $\mathcal{A}_S$ is no longer Type III. Then $\widehat{\mathcal{A}}$ itself is no longer Type III, which means that it has well-defined density operators for states (and weights). In terms of these density operators, the modular and relative modular operators factorise as follows:
\begin{equation}
    \Delta_{\widehat\Psi} = \rho_{\widehat\Psi} (\rho'_{\widehat\Psi})^{-1}, \qquad \Delta_{\hat\phi(\tau)|\widehat\Psi} = \rho (\rho'_{\widehat\Psi})^{-1}.
\end{equation}
Here $\rho_{\widehat\Psi}$ and $\rho$ are the UV regulated density operators of $\widehat\Psi$ and $\hat\phi$ respectively, while $\rho'_{\widehat\Psi}$ is the UV regulated density operator of $\widehat\Psi$ with respect to the commutant of $\widehat{\mathcal{A}}$. Since $\Delta_{\widehat\Psi} = e^{-\beta H_S}$ we can determine these operators in terms of $H_S$. As specified at the beginning of Section~\ref{Section: arbitrary number of observers}, $H_S$ is the generator of a diffeomorphism acting on the fields, and as such may be written in terms of the energy momentum tensor of the fields $T^{ab}$ as follows:
\begin{equation}
    H_S = H_\xi - H'_\xi, \qq{where} H_\xi = \int_\Sigma \dd{\Sigma_a}\xi_b T^{ab} \qq{and} H'_\xi = -\int_{\Sigma'} \dd{\Sigma_a}\xi_b T^{ab}.
\end{equation}
Here, $\Sigma,\Sigma'$ are Cauchy surfaces for $\mathcal{U}$ and its causal complement $\mathcal{U}'$, and $\xi$ is the vector field corresponding to the diffeomorphism generated by $H_S$. Thus we may write
\begin{equation}
    \rho_{\widehat\Psi} = \frac{e^{-\beta H_\xi}}{Z_\xi},\qquad
    \rho'_{\widehat\Psi} = \frac{e^{-\beta H'_\xi}}{Z_\xi},
\end{equation}
where we have introduced a normalisation constant $Z_\xi$. We thus have
\begin{equation}
    \log\Delta_{\hat\phi(\tau)|\Psi} = \log \rho + \beta H'_\xi + \log Z_\xi.
\end{equation}
Using this in~\eqref{Equation: entropy decomp (not separating)}, one finds
\begin{equation}
    S[\phi] \approx S_{0,R} - \log Z_\xi + \beta \bra*{\hat\phi(\tau)}(H_{R^c}-H'_\xi)\ket*{\hat\phi(\tau)} - \bra*{\hat\phi(\tau)}\log\rho\ket*{\hat\phi(\tau)}.
\end{equation}
We may use the freedom in the factor $e^{-{S_{0,R}}}$ appearing in the trace~\eqref{Equation: trace} to set $S_{0,R} = \log Z_\xi$, thus eliminating the first two terms. The last term is the (UV regulated) von Neumann entropy $S_{\widehat{\mathcal{A}}}[\hat\phi(\tau)]$ of the state $\ket*{\hat\phi(\tau)}$ with respect to the algebra $\widehat{\mathcal{A}}$, i.e.\ of the fields in $\mathcal{U}$, the frames $R_{\bar{i}}$, and the complementary frame energy $H_{R^c}$. To address the middle term, we use the conjecture of~\cite{Jensen:2023yxy} that $\xi$ essentially behaves as a boost near the boundary of $\mathcal{U}$, from which it follows that
\begin{equation}
    \beta H_\xi' = \beta H_{R_{\text{out}}} -\frac{A}{4G_N}
\end{equation}
is a constraint in the gravitational theory. Here, $H_{R_{\text{out}}}=\sum_{C_j\in R_{\text{out}}} H_j$ is the total Hamiltonian of all the frames $R_{\text{out}}$ outside $\mathcal{U}$, while $A$ is the area of the boundary of $\Sigma$, i.e.\ the codimension 2 corner of $\mathcal{U}$, and $G_N$ is Newton's constant. Note that under our present assumptions, $R^c$ and $R_{\text{out}}$ need not be the same set of clocks. Putting this all together (and using also the fact that $H_{R^c}-H_{R_{\text{out}}} = H_{R_{\text{in}}} - H_R$, where $H_{R_{\text{in}}}$ is the total Hamiltonian of the frames \emph{inside} of $\mathcal{U}$), we get
\begin{equation}
    S[\phi] \approx \expval{\frac{A}{4G_N}}_{\hat\phi(\tau)} + S_{\widehat{\mathcal{A}}}[\hat\phi(\tau)] - \beta\expval{H_{R_{\text{in}}}-H_R}_\phi.
    \label{Equation: generalised entropy}
\end{equation}
Thus, we obtain something very much like a gravitational generalised entropy, with a couple of frame-dependent caveats. The most obvious caveat is in the final term, which gives the difference in energies between the frames located in $\mathcal{U}$, and the frames making up the set $R$. This term disappears in the case $R=R_{\text{in}}$, although of course the above formula holds more generally than this. The other caveat is that the second term is not just an entropy of the QFT, but also of the frames. As such, it is sensitive to the way in which the frames are entangled with each other, and with the QFT.

\subsection{Special case: separable QFT and frames}
\label{Subsection: separable state}

A special case occurs when the state $\ket*{\hat\phi(\tau)}$ is approximately separable between the fields and the frames:
\begin{equation}
    \ket*{\hat\phi(\tau)} \approx \ket*{\phi_S(\tau)}\otimes\ket*{\tilde\phi(\tau)}, \qq{where} \ket*{\phi_S(\tau)}\in\mathcal{H}_S, \, \ket*{\tilde{\phi}(\tau)}\in\mathcal{H}_{R_{\bar{i}}}\otimes\mathcal{H}_{R^c}.
    \label{Equation: approximate product state}
\end{equation}

Our results apply in the presence of arbitrary entanglement, but previous works such as~\cite{Chandrasekaran:2022cip,Jensen:2023yxy} exclusively employed states of this product form in the semiclassical limit. However, even though the semiclassical assumptions restrict the kind of entanglement allowed, such product states~\eqref{Equation: approximate product state} only make up a small subset of the full set of semiclassical states, for which generically the fields and the frames will be non-trivially entangled. In the interest of making contact with these earlier papers, let us show how our results specialize to the case~\eqref{Equation: approximate product state}.

In a product state, the relative entropy term in the entropy may be further decomposed into contributions from $\ket*{\phi_S(\tau)}$ and $\ket*{\tilde\phi(\tau)}$. Indeed, the state $\ket*{\widehat\Psi}$ is also separable, which allows us to decompose the relative modular operator as
\begin{equation}
    \Delta_{\hat{\phi}(\tau)|\widehat\Psi} \approx \Delta_{\phi_S(\tau)|\psi_S}\otimes\Delta_{\tilde\phi(\tau)|\widetilde\Psi},
\end{equation}
where
\begin{equation}
    \ket*{\widetilde\Psi} = \bigotimes_{C_j\in R_{\bar{i}}} \int_{\sigma_j} \dd{E_j} \ket{E_j}\otimes\bra{E_j} \otimes \frac1{\sqrt{2\pi}}\int_{\sigma_{R^c}} \dd{E} \ket{E}.
\end{equation}
We can explicitly confirm this by using a decomposition as in~\eqref{eq:reducedDensityOperator} to write
\begin{align}
    \Delta_{\hat\phi(\tau)|\widehat\Psi} &= e^{S_{0,R}+\beta H_{R^c}} \hat\rho_\phi(\tau) \\
    &\approx \Delta_{\phi_S(\tau)|\psi_S} \otimes \int_{-\infty}^\infty\dd{t} e^{i\sum_{j\ne i}H_j t} \tr_{R^c}(\ket*{\tilde\phi(\tau+t)}\bra*{\tilde\phi(\tau)}),
\end{align}
and we may identify the latter factor as $\Delta_{\tilde\phi(\tau)|\widetilde\Psi}$. Actually, it may be confirmed that taking expectation values in $\ket*{\widetilde\Psi}$ gives a trace on $\mathcal{B}(\mathcal{H}_{R_{\bar{i}}})\otimes\{H_{R^c}\}''$ (although certainly not the only one since this is not a factor). Since $\Delta_{\tilde\phi(\tau)|\widetilde\Psi}$ is an element of this algebra, it is thus actually the density operator of the state $\ket*{\hat\phi(\tau)}$ with respect to this trace and algebra. We write this as $\rho_{\phi,R_{\bar{i}}}(\tau)=\Delta_{\tilde\phi(\tau)|\widetilde\Psi}$ to denote that it is the density operator of the frames. Using these results in the formula for the entropy~\eqref{Equation: semiclassical entropy}, one finds
\begin{align}
    S[\phi] &\approx S_{0,R} + \beta \bra*{\hat\phi(\tau)}H_{R^c}\ket*{\hat\phi(\tau)} - \bra*{\hat\phi(\tau)}\log \Delta_{\phi_S(\tau)|\psi_S} \ket*{\hat\phi(\tau)} + S_R[\phi] \\
     &= S_{0,R} - \beta \expval{H_R}_\phi - S_{\text{rel}}(\phi_S(\tau)||\psi_S) + S_R[\phi],
     \label{Equation: separable semiclassical entropy}
\end{align}
where the second line holds if $\ket*{\phi_S(\tau)}$ is separating with respect to $\mathcal{A}_S$, and
\begin{equation}\label{eq: frame entropy contribution}
    S_R[\phi] = -\bra*{\tilde\phi(\tau)}\log\rho_{\phi,R_{\bar{i}}}(\tau)\ket*{\tilde\phi(\tau)}
\end{equation}
is the entropy of the frames. 

It should be noted that $S_R[\phi]$, as written above, appears to depend on a choice of $C_i\in R$ -- but it is actually independent of this choice, to leading order in the semiclassical approximation. To see this, consider the (Type I) algebra
\begin{equation}
    r(\mathcal{B}(\mathcal{H}_R)^H) = \{\mathcal{O}^\tau_{C_i}(a), V_i(t) \mid a\in \mathcal{B}(\mathcal{H}_{R_{\bar{i}}}), t\in \RR\}'',
\end{equation}
which acts at the perspective-neutral level. On the right-hand side, we have decomposed this algebra with respect to a particular $C_i\in R$, but clearly the algebra itself is independent of the choice of $C_i$. By the semiclassical and separability assumptions, the expectation value of a general element of this algebra may be approximated as:
\begin{equation}
    \Bra{\phi} \mathcal{O}^\tau_{C_i}(a) V_i(t) \Ket{\phi} = \bra*{\phi_{|i}(\tau)} a e^{iH_St}e^{i\sum_{j\ne i}H_jt}\ket*{\phi_{|i}(\tau)} \approx \bra*{\phi_{|i}(\tau)} a e^{i\mathcal{E}t}e^{i\sum_{j\ne i}H_jt}\ket*{\phi_{|i}(\tau)}
\end{equation}
The operators $a e^{i\sum_{j\ne i}H_jt}$ span the algebra $\mathcal{B}(\mathcal{H}_{R_{\bar{i}}})\otimes\{H_{R^c}\}''$. Thus the density operator for $r(\mathcal{B}(\mathcal{H}_R)^H)$ may be obtained by pulling back the density operator $\rho_{\phi,R_{\bar{i}}}(\tau)$ for $\mathcal{B}(\mathcal{H}_{R_{\bar{i}}})\otimes\{H_{R^c}\}''$ through the homomorphism
\begin{equation}
    \mathcal{O}^\tau_{C_i}(a) V_i(t) \leftrightarrow a e^{i\mathcal{E}t}e^{i\sum_{j\ne i}H_jt}.
\end{equation}
One obtains
\begin{equation}
    \rho_{\phi,R} = \int_{-\infty}^\infty \dd{t} e^{-i\mathcal{E}t} r(e^{-iH_Rt}) \mathcal{O}_{C_i}^\tau\qty(\tr_{SR^c}(e^{-i H_{R^c}t}\ket*{\phi_{|i}(\tau)}\bra*{\phi_{|i}(\tau)})).
\end{equation}
Moreover, the von Neumann entropies of the density operators $\rho_{\phi,R}$ and $\rho_{\phi,R_{\bar{i}}}(\tau)$ will be approximately equal. Now, 
\begin{equation}
    \mathcal{O}_{C_i}^\tau\qty(\tr_{SR^c}(e^{-i H_{R^c}t}\ket*{\phi_{|i}(\tau)}\bra*{\phi_{|i}(\tau)})) = \tr_{SR^c}(\mathcal{O}_{C_i}^\tau\qty(e^{-i H_{R^c}t}\ket*{\phi_{|i}(\tau)}\bra*{\phi_{|i}(\tau)})),
\end{equation}
and
\begin{align}
    \mathcal{O}_{C_i}^\tau\qty(e^{-i H_{R^c}t}\ket*{\phi_{|i}(\tau)}\bra*{\phi_{|i}(\tau)})
    &=\mathcal{O}_{C_i}^\tau\qty(e^{-i H_{R^c}t}V_{j\to i}(\tau,\tau)\ket*{\phi_{|j}(\tau)}\bra*{\phi_{|j}(\tau)}V_{j\to i}(\tau,\tau)^\dagger) \\
    &=\mathcal{O}_{C_i}^\tau\qty(V_{j\to i}(\tau,\tau)e^{-i H_{R^c}t}\ket*{\phi_{|j}(\tau)}\bra*{\phi_{|j}(\tau)}V_{j\to i}(\tau,\tau)^\dagger) \\
    &=\mathcal{O}_{C_j}^\tau\qty(e^{-i H_{R^c}t}\ket*{\phi_{|j}(\tau)}\bra*{\phi_{|j}(\tau)})
\end{align}
for any $C_j\in R$, so $\rho_{\phi,R}$ is independent of the choice of $C_i$. Therefore, $S_R[\phi]$, which was above formulated as the von Neumann entropy of $\rho_{\phi,R_{\bar{i}}}[\phi]$ in \eqref{eq: frame entropy contribution}, may be alternatively formulated as the von Neumann entropy of $\rho_{\phi,R}$, which is independent of the choice of $C_i$.

Note also that the Shannon entropy of $H_{R^c}$ is contained within the frame entropy $S_R[\phi]$. Indeed, one may decompose the frame density operator as follows:
\begin{align}
    \rho_{\phi,R_{\bar{i}}}(\tau) &= \int_{-\infty}^\infty\dd{t} e^{i\sum_{j\ne i}H_j t} \tr_{R^c}(\ket*{\tilde\phi(\tau+t)}\bra*{\tilde\phi(\tau)}) \\
    &= \int_{-\infty}^\infty\dd{t} e^{i H_{R^c} t} \tr_{R^c}(e^{-i H_{R^c}t}\ket*{\tilde\phi(\tau)}\bra*{\tilde\phi(\tau)}) \\
    &\approx 2\pi \int_{H_{R^c}}\dd{E} P(E) p_E \rho_{R_{\bar{i}},E},
\end{align}
where
\begin{equation}
    p_E = \ev*{P(E)}{\hat\phi(\tau)} \approx \ev*{P(E)}{\tilde\phi(\tau)}
\end{equation}
is the probability distribution of $H_{R^c}$, and
\begin{equation}
    p_E\rho_{R_{\bar{i}},E} = \tr_{R^c}(P(E)\ket*{\tilde\phi(\tau)}\bra*{\tilde\phi(\tau)}).
\end{equation}
Note that $\rho_{R_{\bar{i}},E}$ is the reduced density operator of the frames in $R_{\bar{i}}$, conditioned on the energy of the frames in $R^c$ being $E$. Using this decomposition, one finds
\begin{align}
    S_R[\phi] &= -\int_{\sigma_{R^c}}\dd{E}\bra*{\tilde\phi(\tau)}P(E)\log(2\pi p_E \rho_{R,E})\ket*{\tilde\phi(\tau)} \\
    &= -\log(2\pi) - \int_{\sigma_{R^c}}\dd{E}p_E\log p_E + \int_{\sigma_{R^c}}\dd{E} p_E S_{R_{\bar{i}},E}[\phi],
    \label{Equation: separable shannon}
\end{align}
where 
\begin{equation}
    S_{R_{\bar{i}},E}[\phi] = -\tr_{R_{\bar{i}}}\qty(\rho_{R_{\bar{i}},E}\log\rho_{R_{\bar{i}},E})
\end{equation}
is the entanglement entropy of the frames in $R_{\bar{i}}$, conditioned on $H_{R^c}=E$. Thus the total frame entropy $S_R[\phi]$ contains a term corresponding to the Shannon entropy of $H_{R^c}$, as well as an average of $H_{R^c}$-conditioned entanglement entropies of the frames in $R_{\bar{i}}$.

After UV regularisation, an approximate product state of the form~\eqref{Equation: approximate product state} has an approximate product density operator $\rho=\rho_S\otimes \rho_{\phi,R_{\bar{i}}}(\tau)$, where $\rho_S$ is the density operator of the fields in $\mathcal{U}$, and $\rho_{\phi,R}(\tau)$ is the density operator of the frames. The term $S_{\widehat{\mathcal{A}}}[\hat\phi(\tau)]$ in the generalised entropy~\eqref{Equation: generalised entropy} then may be written as the sum of the von Neumann entropies of these two density operators, and one finds
\begin{equation}
    S[\phi] \approx \expval{\frac{A}{4G_N}}_{\hat\phi(\tau)} + S_{\text{QFT}}[\hat\phi(\tau)] + S_R[\phi]- \beta\expval{H_{R_{\text{in}}}-H_R}_\phi.
    \label{Equation: generalised entropy product}
\end{equation}
Here $S_{\text{QFT}}[\hat\phi(\tau)]$ is the von Neumann entropy of $\rho_S$, i.e. the entanglement entropy of the fields in $\mathcal{U}$. One may observe that the frame dependence of this formula has been entirely relegated to the final two terms. 

\subsection{Linear order correction and entanglement}
\label{Section: semiclassical linear correction}

All of the formulas for the entropy discussed above hold to leading order in the semiclassical approximation; at higher orders in $\eta$ there are corrections to these formulas. In Appendix~\ref{Appendix: linear corrections}, it is shown that, to $\order{\eta}$, the entropy receives corrections of the following form:
\begin{equation}
    S[\phi] = S_0[\phi] - 2\pi \int_{\sigma_{R^c}} \dd{E} \int_0^\infty\dd{\lambda} \phi_E\qty(\frac1{\lambda+\Delta_{\phi_E|\bar\psi}}(H_S-\mathcal{E}) \partial_E\qty(p_E \Delta_{\phi_E|\bar\psi})\frac1{\lambda+\Delta_{\phi_E|\bar\psi}}) + \order{\eta^2},
    \label{Equation: linear correction}
\end{equation}
where $S_0[\phi]$ is the leading order contribution discussed above, and $p_E,\phi_E,\bar\psi$ were defined in Section~\ref{Subsection: complementary energy}. 

The nature of this linear order correction is a little difficult to decipher from this expression. To get something a bit easier to digest, let us introduce again a UV regulator on the fields as in Subsec.~\ref{Subsection: generalised entropy}, such that all algebras have well-defined density operators. Then we can write 
\begin{equation}
    \Delta_{\phi_E|\bar\psi} = \rho_E (\rho'_{\bar\psi})^{-1},
\end{equation}
where $\rho_E$ is the density operator for $\phi_E$ in the algebra $\mathcal{A}_S\otimes\mathcal{B}(\mathcal{H}_{R_{\bar{i}}})$, and $\rho'_{\bar\psi}$ is the density operator for $\bar\psi$ in its complement; as in Subsec.~\ref{Subsection: generalised entropy}, we have $\rho'_{\bar\psi} = e^{-\beta H_\xi'}/Z_\xi$. With this decomposition, it is shown in Appendix~\ref{Appendix: linear corrections} that one may perform the $\lambda$ integral in~\eqref{Equation: linear correction} to obtain
\begin{equation}
  S[\phi] = S_0[\phi] - 2\pi\int_{\sigma_{R^c}}\dd{E}\qty[\partial_E\qty\Big(p_E \qty\big(\phi_E\qty(H_\xi)-\mathcal{E})) - p_E\phi_E\qty(H_\xi'\partial_E \log\qty(p_E\rho_E))] + \order{\eta^2}.
    \label{Equation: linear correction UV regulated}
\end{equation}
Thus, the linear order correction depends in a non-trivial way on the correlations between the energies of the QFT degrees of freedom and the complementary clocks, and other statistical properties of the state.

In the special case in Subsec.~\ref{Subsection: separable state}, the state is assumed to exactly factorise between the QFT and clocks, as in~\eqref{Equation: approximate product state}. We then have that $\phi_E$ factorises as $\phi_E=\phi_S\otimes \tilde\phi_E$, where $\phi_S$ is the state on $\mathcal{A}_S$ given by
\begin{equation}
    \phi_S (a) = \bra*{\phi_S(\tau)}a\ket*{\phi_S(\tau)},
\end{equation}
and $\tilde\phi_E$ is the frames part of the state, conditioned on the complementary frame energy being $E$. Similarly, with a UV cutoff in place, the density operator factorises $\rho_E = \rho_S\otimes \tilde\rho_E$, with $\rho_S\in\mathcal{A}_S$ and $\tilde\rho_E\in\mathcal{B}(\mathcal{H}_{R_{\bar{i}}})$. The integrand in~\eqref{Equation: linear correction UV regulated} may then be written\footnote{To show this one needs to use $\tilde\phi_E(\partial_E\log \tilde\rho_E) = 0$.}
\begin{equation}
    \phi_S(H_S-\mathcal{E}) \,\partial_Ep_E.
\end{equation}
But the first factor on the right hand side vanishes by the definition of $\mathcal{E}=\phi(H_S)=\phi_S(H_S)$. So the linear order correction to the entropy vanishes, if we use a UV regulator. Because the QFT algebra is hyperfinite, this continues to be true even without the UV regulator.\footnote{Indeed, the von Neumann entropy is a relative entropy with respect to the trace, and the relative entropy on a hyperfinite algebra is the limit of the relative entropies on its approximating algebras~\cite{Araki:1976zv} (which are the UV regulated algebras in this case).} Thus there are no linear order corrections for a separable state of the form considered in Subsec.~\ref{Subsection: separable state}. This is in agreement with~\cite{Faulkner:2024gst}.

On the other hand, if the state is non-trivially entangled, there will generically be a non-zero linear correction. To make this a bit more obvious, let us assume that $p_E$ vanishes at the boundary of $\sigma_{R^c}$ (this is compatible with the semiclassical assumptions; indeed, if $\sigma_{R^c}=\RR$, then a normalisable state requires this assumption to hold). Then the first term in~\eqref{Equation: linear correction UV regulated} can be integrated and seen to vanish. Let's also assume the state is such that $\rho_E$ is independent of $E$. Then, integrating by parts, one may verify that
\begin{equation}
  S[\phi] = S_0[\phi] - 2\pi\int_{\sigma_{R^c}}\dd{E}p_E\partial_E \phi_E(H_\xi') + \order{\eta^2}.
    \label{Equation: linear correction UV regulated special case}
\end{equation}
Now note that $\phi_E(H_\xi')$ can now depend arbitrarily on $E$, so long as 
\begin{equation}
  \mathcal{E} = \int_{\sigma_{R^c}}\dd{E}p_E\phi_E(H_\xi-H_{\xi}').
\end{equation}
The semiclassical approximation remains valid if $\partial_E \phi_E(H_\xi')=\order{\eta}$ (so that the second term in~\eqref{Equation: linear correction UV regulated special case} is $\order{\eta}$). Clearly, these conditions allow for a non-zero contribution at linear order.

One may thus in principle use the linear order part of the entropy as a diagnostic for whether there is entanglement between the QFT and clocks. It should also be noted that for non-ideal clocks, exactly factorizing states are much harder to come by, due to the non-trivial projection operator $\Pi_{|i}$ defining the reduced Hilbert space. In any case we should typically expect linear corrections to the entropy.

\subsection{Comparison with previous work}

At this stage, we may directly specialise our entropy formula to the states studied in~\cite{Chandrasekaran:2022cip,Jensen:2023yxy}. As described in Section~\ref{Section: comparison with previous}, in those works there were only two clocks $C_1,C_2$ present. In the perspective of one of the clocks (without loss of generality $C_1$) a product state of the form
\begin{equation}
    \ket*{\phi_{|1}(\tau)} = \ket*{\phi_S}\otimes\ket{f_2}
\end{equation}
was assumed, where $\ket*{\phi_S}\in\mathcal{H}_S$ and $\ket{f}\in\mathcal{H}_2$. The entropies of the algebras $r(\mathcal{A}^H_{SC_1})$ and $r(\mathcal{A}^H_{SC_2})$ where then studied.

It was also assumed in~\cite{Chandrasekaran:2022cip,Jensen:2023yxy} that the energy wavefunction $f_2(E) =\bra{E}_2\ket{f_2}$ of the clock $C_2$ is slowly varying as a function of $E$. This implies that its time wavefunction $\tilde{f}_2(t) = \bra{t}_2\ket{f_2}$ is rapidly varying. Changing to the perspective of clock $C_2$, we have
\begin{equation}
    \ket*{\phi_{|2}(\tau)} = \int_{-\infty}^\infty\dd{t} \tilde{f}_2(t) e^{iH_St}\ket*{\phi_S} \otimes \ket{-t}_1.
\end{equation}
Another assumption of~\cite{Chandrasekaran:2022cip,Jensen:2023yxy} was that $H_S$ approximately commutes with the relative modular Hamiltonian from $\psi_S$ to $\phi_S$. For us, this translates to our second semiclassical assumption, that $\ket*{\phi_S}$ is in an approximate eigenstate of $H_S$, i.e.\ $e^{iH_St}\ket*{\phi_S}\approx e^{i\mathcal{E}t}\ket*{\phi_S}$. To be a bit more precise about the nature of the wavefunction $\tilde{f}_2(t)$, we are assuming that it is rapidly varying enough that this approximate $H_S$ eigenstate assumption is reliable in integrals of the above form. Alternatively expressed in terms of energies, the assumption is that the variance of $H_2$ is much larger than the variance of $H_S$. We can then write
\begin{equation}
    \ket*{\phi_{|2}(\tau)} \approx\ket*{\phi_S} \otimes \ket{f_1},\qq{where} \ket{f_1}= \int_{-\infty}^\infty\dd{t} \tilde{f}_2(t) e^{i\mathcal{E}t} \ket{-t}_1.
\end{equation}
Thus, we see that the assumption of approximate separability applies to the state in the perspective of both $C_1$ and $C_2$. Note that the energy wavefunctions of the two clocks are related by
\begin{equation}
    f_1(E) = \bra{E}_1\ket{f_1}= \int_{-\infty}^\infty\dd{t} \tilde{f}_2(t) e^{i(\mathcal{E}+E)t} = f_2(E+\mathcal{E})
    \label{Equation: f1 vs f2}
\end{equation}
(we set $f_{1,2}(E)=0$ if $E\not\in\sigma_{1,2}$). So $f_1(E)$ is also slowly varying.

The first part of our semiclassical assumption can now be seen to apply to either clock. Indeed, for any $a\in\mathcal{A}_S$,
\begin{equation}
    \Bra{\phi}\mathcal{O}^\tau_{C_i}(a)V_i(t)\Ket{\phi} \approx \bra{\phi_S}ae^{iH_St}\ket{\phi_S} \bra{f_j}{e^{iH_jt}}\ket{f_j}\approx \bra{\phi_S}a\ket{\phi_S} \int_{\sigma_{C_j}}\dd{E} \abs{f_j(E)}^2 e^{i(E+\mathcal{E})t},
\end{equation}
where $(i,j)\in\{(1,2),(2,1)\}$. Due to $f_j(E)$ being slowly varying, the right hand side is sharply peaked at $t=0$, as required. Thus, both of our semiclassical assumptions are satisfied, and we can apply the semiclassical entropy formulae.

In particular, the entropy of the algebra $r(\mathcal{A}_{SC_i}^H)$ may be written (invoking~\eqref{Equation: separable semiclassical entropy} and~\eqref{Equation: separable shannon}, using $H_{R^c}=H_j$ so $p_E=\abs{f_j(E)}^2$, and noting $R_{\bar{i}}=\emptyset$)
\begin{equation}
    S[\phi] \approx S_{0,C_i} - \beta \expval{H_i}_\phi - S_{\text{rel}}(\phi_S||\psi_S) -\log(2\pi) + S_{\text{obs}},
\end{equation}
where
\begin{equation}
    S_{\text{obs}} = - \int_{\sigma_i}\dd{E} \abs{f_i(E)}^2\log\abs{f_i(E)}^2 = - \int_{\sigma_j}\dd{E} \abs{f_j(E)}^2\log\abs{f_j(E)}^2
\end{equation}
(the second equality holds because of~\eqref{Equation: f1 vs f2}). In the case where the spectrum of $C_i$ is bounded below, the algebra is Type II$_{1}$, and we may normalise the trace such that $\Tr(\mathds{1})=1$; as described in Section~\ref{Section: trace}, this amounts to fixing the constant $S_{0,C_i}$ such that it obeys~\eqref{Equation: type II 1 normalisation}. In the particular case that the spectrum is of the form $\sigma_i = [0,\infty)$, we have $S_{0,C_i} = \log\beta + \log(2\pi)$. Writing $H_i=H_{\text{obs}}$, we then have
\begin{equation}
    S[\phi] \approx - \beta \expval{H_{\text{obs}}}_\phi - S_{\text{rel}}(\phi_S||\psi_S) + S_{\text{obs}} + \log \beta,
\end{equation}
which agrees with, for example,~\cite[Eq.\ (5.30)]{Jensen:2023yxy}.

\section{Antisemiclassical regime}
\label{sect:antiSemiclassical}

In the previous section, we described a semiclassical regime in which the fluctuations of a given clock $C_i$ are much smaller than the fluctuations of the QFT. In this regime, the von Neumann entropy for the algebra of $r(\mathcal{A}_{SR}^H)$ was shown to agree with the expected gravitational generalised entropy formula (after imposing a UV cutoff).

However, this semiclassical regime only concerns a very small corner of the Hilbert space. For most physical states, the clocks will fluctuate to roughly the same degree as the fields. The generalised entropy formula~\eqref{Equation: generalised entropy} should not be expected to hold non-semiclassically, but it is interesting to ask how to extrapolate it away from this regime. In Subsec.~\ref{Section: semiclassical linear correction}, we described how the entropy changes at linear order away from the semiclassical regime.

In contrast, in this section we will describe a regime which is very far away from the semiclassical one. In particular, we will consider states for which the fluctuations of the time of $C_i$ are much \emph{larger} than the fluctuations of the other degrees of freedom. This is precisely the opposite of the semiclassical case, and for this reason we call this the `antisemiclassical' regime. Since the time of the clock is fluctuating so much, it turns out to be of rather little use for making observations of the other degrees of freedom (we make this more precise below). This enables a simplification of the entropy formula -- but a different simplification to the semiclassical one.

Thus, in this paper we will have described two regimes, semiclassical and antisemiclassical. It would be very useful to understand how to interpolate between them, a task we leave to future work. 

As in the case of the semiclassical regime, we can more precisely define the antisemiclassical regime by examining expectation values of operators in the algebra. In particular, let $a\in\mathcal{A}_{\mathcal{U}}\otimes\bigotimes_{C_j\in R_{\bar{i}}}\mathcal{B}(\mathcal{H}_{C_j})$, and consider the expectation value of $\mathcal{O}_{C_i}^\tau(a)V_i(t)$. The antisemiclassical assumption is that this expectation value is well-approximated by taking $\Ket{\phi}$ to be an approximate eigenstate of the clock Hamiltonian $r(H_i)$, with energy $\bar{E}$:
\begin{equation}
    \Mel{\phi}{\mathcal{O}_{C_i}^\tau(a) V_i(t)}{\phi}\approx e^{-i\bar{E} t} F(t)\Mel{\phi}{\mathcal{O}_{C_i}^\tau(a) }{\phi}.
    \label{Equation: antisemiclassical assumption}
\end{equation}
Here, $F(t)$ is some real (at leading order we can absorb any phase of $F(t)$ into the $e^{-i\bar{E}t}$ term) slowly varying function of $t$, reflecting the fact that $\Ket{\phi}$ is not an exact eigenstate. Since $\Ket{\phi}$ is in an approximate eigenstate of $r(H_i)$, the time read by the clock has very large fluctuations, as required. Note that if we set $a=1$, and consider the complex conjugate of both sides in combination with $t\to -t$, we can observe that $F(t)=F(-t)$.

There has to be a timescale involved in the above assumption; it turns out to be given by $\beta$. In particular, we assume that~\eqref{Equation: antisemiclassical assumption} holds for $t < \beta$, and that for $t>\beta$ the left hand side of ~\eqref{Equation: antisemiclassical assumption} is very close to vanishing. This timescale will be required to obtain a simplified entropy formula. 

Recall that the semiclassical regime can be understood as applying when the fluctuations of the QFT Hamiltonian are sufficiently suppressed relative to those of the clock Hamiltonian, as in~\eqref{Equation: semiclassical fluctuation hierarchy}. The antisemiclassical regime will hold when instead
\begin{equation}
    (\Delta H_i)^2_\phi \ll \beta^{-2}.
\end{equation}
Thus, the fluctuations of the clock Hamiltonian should be less than thermal vacuum fluctuations of the QFT Hamiltonian.

The assumption~\eqref{Equation: antisemiclassical assumption} then implies that 
\begin{equation}
    S_{\phi_{|i}(\tau+t)|\psi_{SR_{\bar{i}}}}^\dagger S_{\phi_{|i}(\tau)|\psi_{SR_{\bar{i}}}} \approx e^{i\bar{E} t} F(t)
    \Delta_{\phi_{|i}(\tau)|\psi_{SR_{\bar{i}}}},
    \label{Equation: antisemiclassical consequence}
\end{equation}
as can be confirmed by considering matrix elements of both sides in between states of the form $a$ for $a\in\mathcal{A}_{\mathcal{U}}\otimes\bigotimes_{C_j\in R_{\bar{i}}}\mathcal{B}(\mathcal{H}_{C_j})$, and using the fact that $\ket*{\psi_{SR_{\bar{i}}}}$ is cyclic.

Using this in~\eqref{eq:reducedDensityIntermed}, we find
\begin{equation}
    \rho_{\phi|i}(\tau) \approx \Pi_{|i} e^{-S_{0,R}-\beta (H_{R^c}+H_{R_{\bar{i}}^*})} W \Delta_{\phi_{|i}(\tau)|\psi_{SR_{\bar{i}}}},
\end{equation}
where
\begin{equation}
    W = \int_{-\infty}^\infty\dd{t} e^{i(H_S+\sum_{j\ne i}H_j+\bar{E})t} F(t) =\int_{-\infty}^\infty\dd{t} e^{-i(H_S+\sum_{j\ne i}H_j)t} \Mel{\phi}{V_i(t)}{\phi}.
\end{equation}
Due to the slowly varying nature of $F(t)$, the support of $W$ is sharply peaked around states for which $H_S+\sum_{j\ne i}H_j+\bar{E} = 0$. 

Note that $\Delta_{\phi_{|i}(\tau)|\psi_{SR_{\bar{i}}}}$ is approximately gauge-invariant; indeed
\begin{align}
    e^{i(H_S+\sum_{j\ne i}H_j+\bar{E})t} F(t)\Delta_{\phi_{|i}(\tau)|\psi_{SR_{\bar{i}}}} 
    &\approx e^{i(H_S+\sum_{j\ne i}H_j)t}S_{\phi_{|i}(\tau+t)|\psi_{SR_{\bar{i}}}}^\dagger S_{\phi_{|i}(\tau)|\psi_{SR_{\bar{i}}}}\\
    &= S_{\phi_{|i}(\tau)|\psi_{SR_{\bar{i}}}}^\dagger S_{\phi_{|i}(\tau-t)|\psi_{SR_{\bar{i}}}} e^{i(H_S+\sum_{j\ne i}H_j)t}\\
    &\approx \Delta_{\phi_{|i}(\tau)|\psi_{SR_{\bar{i}}}}e^{i(H_S+\sum_{j\ne i}H_j+\bar{E})t} F(t).
    \label{Equation: antisemiclassical commutation}
\end{align}
where to get the last line we used the Hermitian conjugate of~\eqref{Equation: antisemiclassical consequence} with $t\to -t$. Thus, mapping back to the perspective-neutral level, the density operator is given by $\rho^R_\phi = \mathcal{R}_i(\tau)^\dagger\rho_{\phi|i}(\tau)\mathcal{R}_i(\tau)=r(\tilde\rho^R_\phi)$, where
\begin{equation}
    \tilde\rho^R_\phi \approx e^{-S_{0,R}+\beta(H_S+H_R-H_{R_{\bar{i}}^*})} \int_{-\infty}^\infty\dd{t} e^{iH_it}\Mel{\phi}{V_i(t)}{\phi} \Delta_{\phi_{|i}(\tau)|\psi_{SR_{\bar{i}}}}.
    \label{Equation: antisemiclassical R rho}
\end{equation}
This is the density operator of the system relative to a set of clocks $R$. 

At this stage let us consider two cases. Suppose first that $R$ contains at least one other clock than $C_i$, and recall from Subsec.~\ref{Subsection: partial trace} that the corresponding density operator $\rho^{R\setminus C_i}_\phi$ of the system relative to the set of clocks $R\setminus C_i$ may be found as the partial trace of $\rho^R_\phi$ as follows:
\begin{align}
    \rho^{R\setminus C_i}_\phi &= e^{S_{0,R}-S_{0,R\setminus C_i}}r\qty\Big[\tr_i(e^{-\beta H_i}\tilde\rho^R_\phi)] \\
    &\approx e^{-S_{0,R\setminus C}+\beta(H_S+H_{R_{\bar{i}}}-H_{R_{\bar{i}}^*})} \tr_i\qty(\int_{-\infty}^\infty\dd{t} e^{iH_it}\Mel{\phi}{V_i(t)}{\phi}) \Delta_{\psi_{|i}(\tau)|\psi_{SR_{\bar{i}}}}.
    \label{Equation: antisemiclassical R minus C_i rho}
\end{align}
Noting that
\begin{equation}
    \tr_i\qty(\int_{-\infty}^\infty\dd{t} e^{iH_it}\Mel{\phi}{V_i(t)}{\phi})=2\pi\int_{\sigma_i}\dd{E} \Mel{\phi}{\delta(r(H_i)-E))}{\phi} = 2\pi,
\end{equation}
we can substitute~\eqref{Equation: antisemiclassical R minus C_i rho} back into~\eqref{Equation: antisemiclassical R rho} to obtain
\begin{equation}
    \rho^R_\phi \approx \frac1{2\pi} e^{S_{0,R\setminus C_i}-S_{0,R}+\beta H_i} \rho_\phi^{R\setminus C_i}\int_{-\infty}^\infty\dd{t}V_i(-t)\Mel{\phi}{V_i(t)}{\phi} . 
\end{equation}
In the other case, where $R$ contains only a single clock, we have that $\Delta_\psi^{-1/2}\Delta_{\psi_{|i}(\tau)|\psi_S}\Delta_\psi^{-1/2}$ is an approximately gauge-invariant element of $\mathcal{A}_S$, and therefore (under the assumption that gauge transformations act ergodically on $\mathcal{A}_S$) at leading order a multiple of the identity. Moreover, its expectation value in the state $\ket{\psi_S}$ is $1$ -- so we must have $\Delta_\psi^{-1/2}\Delta_{\psi_{|i}(\tau)|\psi_S}\Delta_\psi^{-1/2}\approx\mathds{1}$. Substituting this in to~\eqref{Equation: antisemiclassical R rho} yields
\begin{equation}
    \rho^R_\phi \approx \frac1{2\pi} e^{-S_{0,R}+\beta H_i} \int_{-\infty}^\infty\dd{t}V_i(-t)\Mel{\phi}{V_i(t)}{\phi}. 
\end{equation}
The two cases are rather similar, and can be treated on equal grounds if we set $S_{0,\emptyset}=0$ and $\rho_\phi^{\emptyset}=\mathds{1}$. Let us do so in the following.

Note that
\begin{equation}
    \frac1{2\pi}\int_{-\infty}^\infty\dd{t}V_i(-t)\Mel{\phi}{V_i(t)}{\phi} = r\qty\Big[\int_{\sigma_i}\dd{E} \dyad{E}_i p_i(E)], 
\end{equation}
where $\ket{E}_i$ are energy eigenstates for $C_i$, and $p_i(E) = \Mel{\phi}{r(\dyad{E}_i)}{\phi}$ is the probability density for the energy of $C_i$ in the state $\Ket{\phi}$. Thus, the density operator may be written
\begin{equation}
    \rho^R_\phi \approx e^{S_{0,R\setminus C_i}-S_{0,R}+\beta H_i}r\qty\Big[\int_{\sigma_i}\dd{E} \dyad{E}_i p_i(E)]\rho_\phi^{R\setminus C_i},
\end{equation}
Thus, the density operator for the system relative to $R$ can essentially be computed in terms of the density operator for the system relative to $R\setminus C_i$, and the probability density for the energy of the clock $C_i$. The clock $C_i$ and the rest of the degrees of freedom are separable in this density operator; indeed, all but the last factor above make up an operator acting on $C_i$. This means there are no additional correlations between the system and $C_i$ which need to be taken into account.

This is a direct way to see that the antisemiclassical clock $C_i$ is fairly useless for observing the system; the state of the system relative to $R$ is basically determined by the state of the system relative to $R\setminus C_i$, so we do not gain any new information by using $C_i$ (other than its own energy's probability density).

The entropy of $\rho_\phi^R$ has a similar decomposition. Indeed the logarithm of the density operator may be written (here we use the assumption that~\eqref{Equation: antisemiclassical assumption} holds on timescales less than $\beta$, so that $e^{\beta H_i}$ approximately commutes with the other contributions to the density operator)
\begin{equation}
    \log \rho_\phi^R \approx S_{0,R\setminus C_i}-S_{0,R} + \beta H_i + r\qty\Big[\int_{\sigma_i}\dd{E}\dyad{E}_i \log p_i(E)] +\log \rho_\phi^{R\setminus C_i},
\end{equation}
and with this, we can compute the entropy as
\begin{equation}
    S^R[\phi] = -\Mel{\phi}{\log \rho_\phi^R}{\phi} \approx S_{0,R}-S_{0,R\setminus C_i}-\beta \bar{E} - \int_{\sigma_i}\dd{E} p_i(E) \log p_i(E) + S^{R\setminus C_i}[\phi]
    \label{Equation: antisemiclassical entropy}
\end{equation}
(note that $S^{\emptyset}[\phi]=0$). Thus, the entropy decomposes as the sum of $-\beta\bar{E}$, a Shannon entropy for the energy of the clock $C_i$ and the entropy of the system relative to $R\setminus C_i$, plus some constants.

In the particular case where $R=\{C_i\}$ contains only one clock, the formula simplifies to 
\begin{equation}
    S^R[\phi] \approx S_{0,R}-\beta \bar{E} - \int_{\sigma_i}\dd{E} p_i(E) \log p_i(E).
    \label{Equation: antisemiclassical entropy one clock}
\end{equation}
This (leading order contribution to the) entropy does not depend on the state of the fields at all. If $\sigma_i$ is bounded below such that the algebra of operators is Type II$_1$ as described earlier, then it is interesting to comment on what happens if we substitute in the canonical normalisation~\eqref{Equation: type II 1 normalisation}. One finds 
\begin{equation}
    S^R[\phi] \approx - \int_{\sigma_i}\dd{E} p_i(E) (\log p_i(E)-\log s_i(E)),
    \label{Equation: antisemiclassical entropy one clock Boltzmann}
\end{equation}
where
\begin{equation}
    s_i(E) = e^{-\beta E}\Big/\int_{\sigma_i}\dd{E}'e^{-\beta E'}
\end{equation}
is the Boltzmann distribution for the clock at inverse temperature $\beta$. We may recognise $S^R[\phi]$ as (minus) the classical relative entropy between this Boltzmann distribution and the actual probability distribution of the clock's energies.

\section{\Observerdependence{} of gravitational entropy}
\label{sect:observerDependence}

We have described how to use clock QRFs to define gravitational subregion algebras with well-defined von Neumann entropies. But depending on which set of clocks $R$ one uses, one gets different von Neumann algebras, and therefore different entropies. Thus, entropy is \observerdependent{}.  These differences are ultimately manifestations of the subsystem relativity discussed on Sect.~\ref{ssec_subsystem relativity}.

Any comparison of entropies between different algebras will be sensitive to how we normalise the traces of these algebras. For the algebras discussed in this paper, the normalisations are parametrised by the numbers $S_{0,R}$, and these contribute overall state-independent constants to the entropies, as previously discussed. Thus, in order to technically speaking have a full comparison of the entropies for different choices of $R$, one needs to have fixed these constants. In special cases there are canonical ways to do so; for example, when $R$ contains a single clock whose energy spectrum is bounded below, we may pick $S_{0,R}$ such that the trace of the identity is $1$, as discussed at~\eqref{Equation: type II 1 normalisation}. One may also in certain cases use the partial trace to relate the trace normalisations of different algebras, as described in Sec.~\ref{Subsection: partial trace}. But in the general case, there does not seem to be any canonical way to pick these normalisations.

One might therefore be concerned that the ambiguity in the constants $S_{0,R}$ will muddy our proposed picture of \observerdependence{} in the entropy. The reason this is not the case is that these constants are necessarily \emph{state-independent}. Thus, even if one does not have a canonical way to fix these constants, the entropy functionals one gets for different sets of clocks $R$ will depend on the state in fundamentally different ways, and this dependence cannot be absorbed into the constants. Thus, the state-dependence of the entropies is unambiguous, and it is strictly speaking this state-dependence that we will now discuss.

In this section, we will explore several examples of this phenomenon: a physically motivated setup involving a gravitational interferometer, a scenario involving superselection and degenerate clocks, and a situation where only some clocks satisfy the conditions necessary for a semiclassical approximation. We will also comment on what happens to entropies when we have \emph{periodic} clocks~\cite{Chataignier:2024eil}, and how these compare to the monotonic clocks we have so far discussed in this paper.
 
\subsection{Case study: gravitational interferometer}
\label{sect:Shapiro}
We first consider an example which we dub the ``gravitational interferometer'' for reasons explained below, which was also discussed in section 4.2 of \cite{DeVuyst:2024pop}.  Technical details such as the density matrix computations will be relegated to App. \ref{app: grav interferometer}.
With this case study we want to explicitly show a feasible setup in which we can identify a manifestation of subsystem relativity, as explained in Sect.~\ref{ssec_subsystem relativity}, namely the \observerdependence{} of gravitational entropy. This relativity refers to what different observers would attribute to the QFT degrees of freedom alone. Concretely, we will work with two clocks $C_1, C_2$ in the same subregion $\mathcal{U}$ and one clock $C_3$ in the complementary region $\mathcal{U}'$.  These regions could be complementary static patches of a de Sitter space, the left and right exteriors of a maximally extended Schwarzschild spacetime, or possibly the more general regions as suggested by JSS \cite{Jensen:2023yxy}.  With this in mind, we will use the words ``patch'' and ``subregion'' interchangeably, while acknowledging that the latter scenario is on less rigorous footing regarding assumptions about the constraint and QFT modular flow \cite{DeVuyst:2024grw}.  For purposes of this discussion we take for granted that the necessary ingredients are in place.

We will consider the algebra of field operators in subregion $\mathcal{U}$, dressed with respect to one or the other clock.  On the perspective-neutral level, these are $r(\mathcal{A}^H_{\mathcal{U}C_1})$ and $r(\mathcal{A}^H_{\mathcal{U}C_2})$. These algebras encode the gauge-invariant degrees of freedom that each observer ascribes to the QFT degrees of freedom alone. The entropy associated with these will thus correspond to the entropy each observer assigns to the QFT (including graviton) degrees of freedom within $\mathcal{U}$. It may be that they each only have access to these degrees of freedom. This is the minimum guaranteed by the timelike tube theorem \cite{Borchers1961,Araki1963AGO,Strohmaier:2023hhy,Strohmaier:2023opz}; it depends on operational assumptions whether they may also have access to other clocks in the same patch. Even if they do, these algebras account for what they each assign to the QFT alone.  Crucially, although these pertain to the fields in the same subregion, they are inequivalent gauge-invariant algebras
\begin{equation}\label{eq: unequal subsystem algebras}
    r(\mathcal{A}^H_{\mathcal{U}C_1}) \neq r(\mathcal{A}^H_{\mathcal{U}C_2}) 
\end{equation}
whose overlap is described by Eq.~\eqref{physoverlap} and may be trivial as in Eq.~\eqref{trivial} when the modular Hamiltonian $H_S$ has an ergodic action on the regional QFT algebra.  Consequently, since the density operators are defined with respect to distinct algebras, so will the entropy turn out to be unequal under the same global state. At the reduced level, this means that operators, in particular the density operators, cannot be transformed into each other by conjugation with a unitary QRF transformation \eqref{clockchange}.

As emphasised in the Introduction, the usual \emph{horizon entropy} contribution that arises in a decomposition of the entropy in semiclassical states is associated with the full dressed algebra of all fields and all clocks within the same patch.  In this context, this means $r(\mathcal{A}^H_{SC_1C_2})$ rather than either of the two subalgebras \eqref{eq: unequal subsystem algebras}.  Thus we emphasise that the relativity of gravitational entropy discussed here is not a statement about relativity of horizon area or horizon entropy.  Though the reduced density operators associated to the full patch algebra $r(\mathcal{A}^H_{SC_1C_2})$ would take different expressions in the perspectives of $C_1$ and $C_2$, these are always unitarily related.  Thus the \textit{total} generalised entropy of the same patch (or other region) is not frame-dependent. To be more explicit about the differences that \textit{do} occur between the algebras \eqref{eq: unequal subsystem algebras}, we will design our setup to highlight two possible causes of disparity between these dressed QFT entropies: first, when the clocks are not isomorphic, i.e.~their energy spectra differ, and second, when the entanglement structure differs.

\begin{figure}
    \centering
    \begin{tikzpicture}[scale=0.5]
        \draw[fill=black!10] (-1,0) circle (3) node {\parbox{2cm}{\centering\large massive \\ object}};
        
        \draw[thick,blue,-{Stealth[scale=1.1,angle'=45]}] (6,-4) -- (6,-3)
            .. controls (6,-2) and (7,-2) .. (7,-1)
            -- (7,0.3);
        \draw[thick,blue] (7,0) -- (7,1)
            .. controls (7,2) and (6,2) .. (6,3)
            -- (6,4);
        
        \draw[thick,blue,-{Stealth[scale=1.1,angle'=45]}] (6,-4) -- (6,-3)
            .. controls (6,-2) and (5,-2) .. (5,-1)
            -- (5,0.3);
        \draw[thick,blue] (5,0) -- (5,1)
            .. controls (5,2) and (6,2) .. (6,3)
            -- (6,4);

        \draw[thick,blue,-{Stealth[scale=1.1,angle'=45]}] (13,-4) -- (13,0.3);
        \draw[thick,blue] (13,0) -- (13,4);

        \node[below] at (6,-4.3) {\large $C_2$};
        \node[below] at (13,-4.3) {\large $C_1$};
        \node[right] at (6,-3.5) {$\ket{0}_2$};
        \node[right] at (6,+3.5) {$\ket{\tau_2}_2+\ket{\tau_2+\Delta\tau}$};

        \draw [semithick,decorate,decoration={brace,amplitude=5pt,raise=-4,mirror}]
  (7.6,-1.7) -- (7.6,1.7);
        \node[right] at (7.6,0) {$\tau_2+\Delta\tau$};
        \draw [semithick,decorate,decoration={brace,amplitude=5pt,raise=-4}]
  (4.4,-1.7) -- (4.4,1.7);
        \node[left] at (4.4,0) {$\tau_2$};
    \end{tikzpicture}
    \caption{
        Clock $C_2$ starts at time reading 0 and is then split into two branches where one branch makes a flyby encounter with a massive object while the other is kept far away (or we accelerate them by different amounts). Due to time dilation, the flyby branch will lack behind in time by an amount $\Delta \tau$ w.r.t.~the other branch. After recombination of both branches, we thus achieve a superposition of two clock states. Meanwhile, clock $C_1$ just evolves as is without splitting into multiple branches.
    }
    \label{Figure: interferometer}
\end{figure}
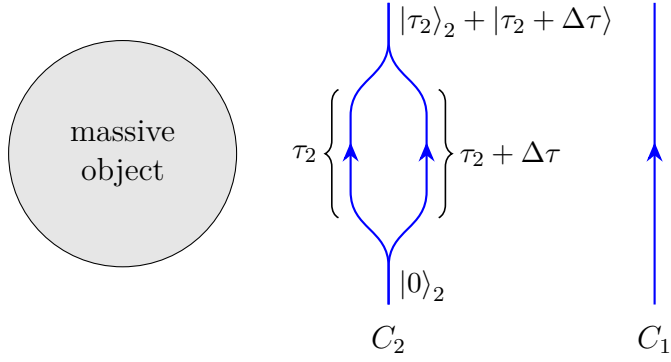

We now describe the setup.  We take the three clocks to be non-ideal, such that individual clock states are normalisable.  We will first consider a situation where, from the perspective of clock $C_1$, the state on the other factors is entirely unentangled (i.e.~a product state).  This will decidedly \textit{not} be the case from the perspective of clock $C_2$, so with this class of wavefunctions it will be easy to emphasise the role of differing entanglement structure between the two perspectives. We will then consider a more specific state within this class in which the second clock is in a simple superposition of clock orientation states when seen from the perspective of the first clock. The third, complementary clock will also be taken to be a simple clock state.  Regarding the configuration of states on $C_1$ and $C_2$, this situation can be achieved by a gravitational interferometer experiment where we have a clock split into two branches where in one branch the clock has a flyby to a massive object. Due to the stronger gravitational field close to the massive object, the time reading of the flyby clock will be dilated. Bringing both branches together then results in the desired superposition (see Fig.~\ref{Figure: interferometer}). In effect, this is just a quantum version of Shapiro time delay \cite{PhysRevLett.13.789}.  

We start with a state which from $C_1$-perspective is in a full product state between all components
\begin{equation}
    \ket*{\phi_{\vert 1}} = \sqrt{N} \Pi_{\vert 1} (\ket{\phi_S} \otimes \ket{f}_2 \otimes \ket{g}_3).
    \label{eq:interferometer}
\end{equation}
In this expression, $\ket{f}_2 \in \mathcal{H}_2$ and $\ket{g}_3 \in \mathcal{H}_3$ are both normalised on their respective factors, with overall normalisation constant $N$ such that the perspectival state is also normalised.  Furthermore, without loss of generality we will set the gauge parameters to zero: $\tau_i = 0 = \tau_j$. Eventually we will take $\ket{f}_2$ to be the superposition and $\ket{g}_3$ to be a pure clock state, but we will leave it general for now. We also let $f$ and $g$ represent the energy-basis wavefunctions on their respective factors.  In App. \ref{app:  approximate product states} it is shown that a frame change \eqref{clockchange} to the perspective of clock $C_2$ results in the reduced state
\begin{equation}
    \ket*{\phi_{\vert 2}} := V_{1 \to 2}(0,0) \ket*{\phi_{\vert 1}} = \sqrt{2\pi N} f(-H_S-H_1-H_3) \Pi_{\vert 2} (\ket{\phi_S} \otimes \ket{0}_1 \otimes \ket{g}_3).
    \label{eq:interferometerChange}
\end{equation}
Using these two expressions for the same global state, the reduced density operators for the algebras $\mathcal{A}_{SC_1}$ and $\mathcal{A}_{SC_2}$ are likewise computed in App.~\ref{app: grav interferometer}. The results are reported together in equation \eqref{eq: rho1 and rho2}, in the perspectives of $C_1$ and $C_2$, respectively.  We report these here with the simplifying assumption that $\phi_S$ lies in the canonical cone of $\psi_S$, such that consequently $J_{\phi_S \vert \psi_S} = J_{\psi_S}$ and we obtain the polar decomposition $S_{\phi_S \vert \psi_S} = J_{\psi_S} \Delta_{\phi_S \vert \psi_S}^{1/2}$.\footnote{This choice turns out to be not that restricting, as one can apply a unitary from the commutant algebra to such a state to reach any other cyclic-separating state.} With these choices, we obtain the reduced density operators 
\begin{align}
    \label{eq: rho1 before approx}
    \hspace{-5pt}\rho_{\phi \vert 1} &= 2 \pi  N e^{-S_{0,1} -\beta(H_2+H_3)} \Pi_{|1} \Delta^{1/2}_{\phi_S \vert \psi_S}
    \int_{\sigma_2}\dd \epsilon_2 \Pi(-\bar{H}_1-\epsilon_2,\sigma_3)
    |f(\epsilon_2)|^2|g(-\bar{H}_1-\epsilon_2)|^2  
    \Delta^{1/2}_{\phi_S \vert \psi_S}\Pi_{|1},\\
    \label{eq: rho2 before approx}
    \hspace{-5pt}\rho_{\phi \vert 2} &= 2 \pi N e^{-S_{0,2}-\beta(H_1+H_3)} f(\bar{H}_2)
    \Pi_{|2}\Delta^{1/2}_{\phi_S \vert \psi_S}
    \int_{\sigma_1} \dd \epsilon_1 \Pi(-\bar{H}_2-\epsilon_1,\sigma_3)
    |g(-\bar{H}_2-\epsilon_1)|^2
    \Delta^{1/2}_{\phi_S \vert \psi_S}\Pi_{|2} f^*(\bar{H}_2).
\end{align}
In the expressions above, $\bar{H}_i := H_i - H$ denotes the constraint-equivalent of $H_i$ on $\mathcal{H}_{|i}$, and as usual the non-trivial projectors $\Pi(-\bar{H}_i - \eps_j, \sigma_k)$ restrict $-\bar{H}_i - \eps_j$ to eigenspaces in the spectral range $\sigma_k$.

The two non-trivial projectors within the integrals will generally render entropy computations difficult. Therefore, to have both of them act approximately trivially we take clock $C_3$ to be effectively ideal.  There are situations in which an ideal clock in the complement does appear naturally. For instance, when the spacetime has a boundary in the complement, then $H_3$ could be the second-order ADM boundary Hamiltonian which is unbounded, cf.~Sec.~\ref{sect:linearisationConditions}. We also now take the wave function of this clock to be a simple clock state $\ket{g}_3 = \sqrt{2\pi / \norm{\sigma_3}} \ket{\tau_3}$, with $\norm{\sigma_3} := \int_{\sigma_i} \dd{\eps_i}$ the spectral range. Then the modulus just becomes a constant: $\abs*{g(E_3)}^2 = \norm{\sigma_3}^{-1}$. With these simplifications, the integral between the relative modular operators in \eqref{eq: rho1 before approx} gives $\norm{\sigma_3}^{-1}$, while that in \eqref{eq: rho2 before approx} gives $\norm{\sigma_1}\norm{\sigma_3}^{-1}$. We acquire the two density operators 
\begin{align}
    \label{eq:density1}
    \rho_{\phi \vert 1} &\approx 2 \pi N \norm{\sigma_3}^{-1} e^{-S_{0,1}-\beta(H_2+H_3)} \Pi_{|1} \Delta_{\phi_S \vert \psi_S}  \Pi_{|1},  \\
    \rho_{\phi \vert 2} &\approx 2\pi  N\norm{\sigma_1} \norm{\sigma_3}^{-1} e^{-S_{0,2}-\beta(H_1 + H_3)} 
    f(\bar{H}_2) \Pi_{|2} \Delta_{\phi_S \vert \psi_S} \Pi_{|2} f^*(\bar{H}_2).
    \label{eq:density2}
\end{align}
These density operators \eqref{eq:density1} and \eqref{eq:density2} are very similar, but still indicate some aspects of subsystem relativity. The wavefunction $f$ has disappeared from \eqref{eq:density1} altogether while it is still present in \eqref{eq:density2}. Therefore properties such as entanglement entropy of the algebra $\mathcal{A}_{SC_2}$ will depend on $f$, while the same is not true for $\mathcal{A}_{SC_1}$ even in the same physical state. This is appropriate considering that the former algebra contains (for example) reorientations of clock $C_2$ while the latter excludes these.
An additional manifestation of subsystem relativity is evident in the role of the two projectors $\Pi_{|1}$ and $\Pi_{|2}$.  These indicate different `coarse-grainings' of the dressed QFT degrees of freedom, which are essentially represented in the density operators through the relative modular operator $\Delta_{\phi_S|\psi_S}$. Taking these algebras to represent what can be learnt about the QFT state on the subregion `from the perspective' of each clock, we see that the differences can be stark and physically significant.

At this point, one might object that taking the limit that $C_3$ becomes ideal leads to unreasonable clock states as these are not normalisable. However, this limit does not cause trouble because the reduced states $\ket*{\phi_{\vert 1}}$ and $\ket*{\phi_{\vert 2}}$ are normalised as states on their respective reduced Hilbert spaces (not necessarily on each individual kinematical factor). The projectors $\Pi_{\vert 1}$ and $\Pi_{\vert 2}$ keep the unbounded $\sigma_3$ integral in check and the normalisation constant $N$ can be adapted accordingly.

To compute the entropy beyond semiclassical order, we now consider the specific example of Shapiro time delay under consideration. We thus let $\ket{f}_2$ be a superposition of two clock states, i.e.
\begin{equation}
    \ket{f}_2 = \sqrt{N_2} \qty\big(\ket{\tau_2}_2 + \ket{\tau_2 + \Delta \tau}_2), \quad f(E) = \sqrt{\frac{N_2}{2\pi}} e^{-i \tau_2 E} (1 + e^{-i \alpha E / \norm{\sigma_2}}),
    \label{eq:superposf}
\end{equation}
with the dimensionless parameter $\alpha := \Delta \tau \norm{\sigma_2}$ and the $\alpha$-dependent normalisation constant
\begin{equation}
    N_2 :=\frac{\pi}{\norm{\sigma_2}\left(1+\frac{\sin(\alpha\epsilon_2^{\text{max}}/\norm{\sigma_2})}{\alpha}-\frac{\sin(\alpha\epsilon_2^{\text{min}}/\norm{\sigma_2})}{\alpha}\right)},
\end{equation}
which reduces to $\pi/ (2\norm{\sigma_2}) $ when $\alpha \to 0$. We also note down the expression for the state in the other perspective \eqref{eq:interferometerChange}
\begin{equation}
    \label{eq:stateV}
    \ket*{\phi_{\vert 2}} \propto \Pi_{|2} e^{iH_S \tau_2} \qty(\ket{\phi_S}\otimes\ket{-\tau_2}_1\otimes\ket*{\tau_3-\tau_2}_3 + e^{iH_S \Delta\tau}\ket{\phi_S}\otimes\ket{-\tau_2-\Delta\tau}_1\otimes\ket*{\tau_3-\tau_2-\Delta\tau}_3)
\end{equation}
to show that superposition in one perspective ($C_1$) generally leads to entanglement under a QRF transformation (to $C_2$ perspective) \cite{Hoehn:2019fsy,Hoehn:2020epv,Giacomini:2017zju,Castro-Ruiz:2019nnl,AliAhmad:2021adn,Hoehn:2023ehz,delaHamette:2020dyi}. The density operator associated to this state is \eqref{eq:density2}
\begin{align}
    \rho_{\phi \vert 2} &\approx \frac{N N_2 \norm{\sigma_1}}{\norm{\sigma_3}} e^{-S_{0,2}-\beta(H_1 + H_3)} 
    e^{-i\tau_2\bar{H}_2} (1+e^{-i\alpha\bar{H}_2 / \norm{\sigma_2}}) \Pi_{|2} \Delta_{\phi_S \vert \psi_S} \Pi_{|2}  (1+e^{i\alpha\bar{H}_2 / \norm{\sigma_2}}) e^{i\tau_2\bar{H}_2}.
    \label{eq:density2Shapiro}
\end{align}
While the entropy for \eqref{eq:density1} can be computed straightforwardly, namely the $f$-independent
\begin{equation}
    \label{eq:entropy1}
    S[\phi_{\vert 1}] = S_{0,1} - \log(2\pi N \norm{\sigma_3}^{-1}) + \beta \expval*{H_2 + H_3}_{\phi_{\vert 1}} - \expval*{\log(\Pi_{|1} \Delta_{\phi_S \vert \psi_S}\Pi_{|1})}_{\phi_{\vert 1}},
\end{equation}
this is not true for \eqref{eq:density2Shapiro} owing to the non-commutativity between $H_S$ and $\Delta_{\phi_S \vert \psi_S}$. Nevertheless, the non-commutative part can be written in the form $e^{Y} e^X e^{Y^\dagger}$ on which we need to apply the BCH-formula. In general, this leads to an infinite series of terms unless we have more specific information about the commutator. Luckily, there is a way to  expand this into a small parameter if such one appears in $Y$, which is here the case when we take $\alpha \ll 1$, representing the superimposed clock approximately be a single clock state. The details are worked out in App.~\ref{sect:BCH}. We thus find 
\begin{equation}
    \rho_{\phi \vert 2} \approx \frac{2\pi N \norm{\sigma_1}}{\norm{\sigma_2} \norm{\sigma_3}}  e^{-S_{0,2}-\beta(H_1 + H_3)} e^{-i\tau_2\bar{H}_2} 
    \left(1 - i \frac{\alpha}{2 \norm{\sigma_2}} \bar{H}_2 \right) \Pi_{|2} \Delta_{\phi_S \vert \psi_S} \Pi_{|2}  \left(1 + i \frac{\alpha}{2 \norm{\sigma_2} } \bar{H}_2 \right) e^{i\tau_2\bar{H}_2}.
\end{equation}
We can now make use of formula \eqref{eq:linearBCH} by making the identification 
\begin{equation}
    X \equiv \log(\Pi_{|2} \Delta_{\phi_S \vert \psi_S} \Pi_{|2}), \quad Y \equiv \log(1 - i \frac{\alpha}{2 \norm{\sigma_2}} \bar{H}_2) = - i \frac{\alpha}{2 \norm{\sigma_2}} \bar{H}_2 + \mathcal{O}(\alpha^2).
\end{equation}
This specific piece up to first order in $\alpha$ is hence determined by
\begin{align}
    \log(e^Y e^X e^{Y^\dagger}) 
    &= \log(\Pi_{|2} \Delta_{\phi_S \vert \psi_S} \Pi_{|2}) + i \frac{\alpha}{2 \norm{\sigma_2}} \comm{H_S}{\log(\Pi_{|2} \Delta_{\phi_S \vert \psi_S} \Pi_{|2})} + \mathcal{O}(\alpha^2) \nonumber \\
    &= \log(\Pi_{|2} \Delta_{\phi_S \vert \psi_S} \Pi_{|2}) -  
    \frac{1}{2} \Delta \tau \dv{}{t} \log(\Pi_{|2} \Delta_{\phi_S(t) \vert \psi_S} \Pi_{|2}) \eval_{t = 0} \nonumber \\
    &= \frac{1}{2} \left(\log(\Pi_{|2} \Delta_{\phi_S \vert \psi_S} \Pi_{|2}) + \log(\Pi_{|2} \Delta_{\phi_S(-\Delta \tau) \vert \psi_S} \Pi_{|2}) \right) + \mathcal{O}(\alpha^2).
    \label{eq:entropyBCH}
\end{align}
Technically, this does not capture all the non-commutativity since there is also the conjugation by the unitary $e^{i\tau_2\bar{H}_2}$, but this can simply be taken outside the $\log$ as $\log(U (\cdot)U^\dagger) = U \log(\cdot) U^\dagger$. This leads to the entropy
\begin{align}
    \label{eq:entropy2}
    S[\phi_{\vert 2}] = S_{0,2}
    &- \log(2\pi N \norm{\sigma_1}\norm{\sigma_2}^{-1} \norm{\sigma_3}^{-1}) + \beta \expval*{H_1 + H_3}_{\phi_{\vert 2}}
    \nonumber \\
    &- \frac{1}{2} \left(\expval*{\log(\Pi_{|2}\Delta_{\phi_S(-\tau_2) \vert \psi_S}\Pi_{|2})}_{\phi_{\vert 2}} + \expval*{\log(\Pi_{|2}\Delta_{\phi_S(-\tau_2 - \Delta \tau) \vert \psi_S}\Pi_{|2})}_{\phi_{\vert 2}} \right) + \mathcal{O}(\alpha^2).
\end{align}
The $f$-dependence in this expression creeps in through the dependence on $\Delta \tau$. Also note that by taking the expectation value in the state \eqref{eq:stateV}, the result is in fact not dependent on $\tau_2$.

Given expressions \eqref{eq:entropy1} and \eqref{eq:entropy2} we can discuss the two sources of entropy relativity in more detail:
\begin{enumerate}
    \item $\mathbf{\alpha \to 0:}$ in the limit that $\alpha \to 0$, with $\Delta \tau \to 0$ and $\norm{\sigma_2}$ fixed, we have that $\ket{f}_2$ \eqref{eq:superposf} becomes a pure clock state with constant modulus. The density operator $\rho_{\phi \vert 2}$ \eqref{eq:density2} then becomes of the same form as $\rho_{\phi \vert 1}$ \eqref{eq:density1}, and the same holds for their respective entropies \eqref{eq:entropy2} and \eqref{eq:entropy1}. The only difference in entropy in this case then stems from the different fuzzinesses of the clocks. In the case the clocks are non-isomorphic, i.e.~$\sigma_1 \neq \sigma_2$, the terms 
    $\log(\Pi_{\vert 1} \Delta_{\phi_S \vert \psi_S} \Pi_{\vert 1} )$ and $\log(\Pi_{\vert 2} \Delta_{\phi_S \vert \psi_S} \Pi_{\vert 2} )$ functionally differ in the sense that the QFT degrees of freedom are coarse-grained in different ways;
    \item $\mathbf{\sigma_1 = \sigma_2:}$ when the clocks are isomorphic, then the projected modular operators will be functionally the same. When we let $\ket{f}_2$ be a superposition $\alpha \neq 0$, the entropy associated to $\ket*{\phi_{\vert 2}}$ \eqref{eq:stateV} has an extra term proportional to a commutator $i \alpha \expval*{\comm*{\log(\Pi_{\vert 2} \Delta_{\phi_S \vert \psi_S} \Pi_{\vert 2} )}{H_S}}_{\phi_{\vert 2}}$  \eqref{eq:entropyBCH} as a consequence of the different entanglement structure from the product state $\ket*{\phi_{\vert 1}}$ \eqref{eq:interferometer}.
\end{enumerate}

In conclusion, by studying the quantum version of Shapiro time delay we exemplified two possible sources for entropy relativity. The first one is the case when the clocks are non-isomorphic, which also shows up at the semiclassical level. For the second one with different entanglement structures in the reduced states, we had to go beyond the semiclassical level and looked at an expansion to first order in a small parameter to see the different entropy contributions.

\subsection{Degenerate clocks}
\label{sec:degClockExample}

Let us now discuss the case in which the two observers, Alice and Bob, occupy the same spacetime region and carry each a degenerate clock described by a Hamiltonian $H_C\propto p^2$, as also considered in \cite{Hoehn:2020epv}. We will illustrate that in this case  subsystem relativity also applies across degeneracy sectors when we consider the full patch algebra, i.e.~including \emph{all} clocks in their subregion -- provided we make an operational restriction as described in Sec.~\ref{sect:degSuperselection}. We thus have
a constraint of the form
\begin{equation}
    H = H_S + \frac{p_A^2}{2m_A} + \frac{p_B^2}{2m_B} - H_3,
\end{equation}
where $H_3$ is a non-degenerate, non-ideal clock and the minus sign reflects the opposite time direction in the complementary region, in which we assume clock $C_3$ to be located. 

As explained in Secs.~\ref{sssec_degclock2} and~\ref{sssec_degclockchange}, the physical Hilbert space decomposes into a direct sum of degeneracy sectors. Since Alice's clock $C_A$ and Bob's clock $C_B$ are both doubly degenerate, having each a positive and negative frequency sector, the physical Hilbert space consists of four sectors
\begin{equation}
    \mathcal{H}_\text{phys} = \mathcal{H}_{++} \oplus \mathcal{H}_{+-} \oplus \mathcal{H}_{-+} \oplus \mathcal{H}_{--},
\end{equation}
where the first label refers to the sector of $C_A$ and the second one to that of $C_B$. Starting with a physical state $\Ket{\Phi}$, this can be reduced to either observer's perspective in one of the four sectors through the PW reduction maps $\mathcal{R}_i^{\lambda \lambda'}(\tau)$ leading to four states $\ket*{\phi_{\vert i}^{\lambda \lambda'}}$. QRF transformations are then performed within one sector overlap, i.e.~$V_{i, \lambda \to j, \lambda'}(\tau_i,\tau_j)$ within $\Lambda_R = \lambda \lambda'$  (cf.\ Sec.~\ref{sssec_degclockchange}).

For the algebras, we will consider the case where $R = \{C_A, C_B\}$ and $R^c=\{C_3\}$ where each observer only has access to the full +-sector of their own clock $R_i$, $i=A,B$, which means they have access to the parts of both sectors of the other clock $R_{\bar{i}}$ that overlap with their own $+$-sector. Thus, we will operationally restrict the gauge-invariant algebras associated to each observer and it is quite common to only permit one frequency sector each. Namely, we impose that each observer only has access to their clock's positive frequency/right moving modes and does not have access to the degeneracy sector swap operators in $ \bigotimes_{C_j \in R} \mathcal{B}(\mathbb{C}^{m_j})$ for neither of the clocks; this restriction thus entails an operational superselection that may be justified in different ways, as we described in Sec.~\ref{sect:degSuperselection}. We emphasise that this is not a necessary restriction, but  we want to see what the repercussions for entropy relativity are.

In this case, the operationally restricted gauge-invariant algebras can be written as a direct sum over the superselection sectors, as in Eq.~\eqref{superselectedalgebra}. They are of the form
\begin{equation}
    \mathcal{A}_A = \mathcal{A}^H_{SR,++} \oplus \mathcal{A}^H_{SR,+-}, \quad 
    \mathcal{A}_B = \mathcal{A}^H_{SR,++} \oplus \mathcal{A}^H_{SR,-+}\,.
\end{equation}
With this decomposition of the algebra, the entropy can then be calculated according to \eqref{Equation: degenerate entropy} by looking at the contributions per sector $S^{\Lambda_R}[\phi]$. Notice that we are dealing with an algebra of type II$_\infty$ because we have more than one clock in $R$ \eqref{eq:traceId} despite the clock spectra being bounded from below $\sigma_A = \sigma_B = \mathbb{R}_+$. Our above assumption also means that Alice and Bob do not have access to the identity component $\mathbb{C}\mathds{1}_{\lambda\lambda'}$ on the remaining sectors, as otherwise they could rescale states and observables there. This has important consequences for the probabilities $p^{\Lambda_R}$ in \eqref{Equation: degenerate entropy} which will have to be reinterpreted as conditional probabilities, as we explain below.

The algebras on the physical level are then obtained by acting with $r(.)$ which is ensured to be faithful since $R^c$ is non-empty, containing clock $C_3$, cf.~Sec.~\ref{ssec_vN}. Density matrices and entropies are to be computed per sector, as explained in Sec.~\ref{sect:superAll}, in the same way as the nondegenerate case. Note that the nontrivial overlap between both only consists of one overlap sector
\begin{equation}
    r(\mathcal{A}_A) \cap r(\mathcal{A}_B) = r(\mathcal{A}^H_{SR,++}).
\end{equation}
This situation thus differs from the gravitational interferometer example in Sec.~\ref{sect:Shapiro}, where $R$ contained only one clock for each observer's perspective and the overlap between the two happened to be trivial (when $H_S$ has an ergodic action, see Eq.~\eqref{trivial}), unlike here.

As a consequence, Alice and Bob can only meaningfully compare states/observables on this overlap. The density matrix $\rho_\phi$ will have a part $\rho_\phi^{++}$ that lies in this algebra and can be probed by both Alice and Bob. While the two will use distinct observables from $r(\mathcal{A}^H_{SR,++})$ to probe this state, their description of it will be related by a unitary QRF transformation $V_{A,+ \to B,+}(\tau_A,\tau_B)$, and so the entropy contribution of the $++$-sector will be the same for both observers
\begin{equation}
    S^{++}_A[\phi] =  S^{++}_B[\phi].
\end{equation}

The situation changes on the other halves of their respective algebras. Alice will be able to access the $\rho_\phi^{+-}$ part of the full density matrix whilst Bob will have access to $\rho_\phi^{-+}$. There is no a priori reason for these density matrices to be the same or even lead to the same entropies. It is easy to engineer setups in which this is indeed true, such that 
\begin{equation}
    S^{+-}_A[\phi] \neq S^{-+}_B[\phi].
\end{equation}
Thence, this case of \observerdependence{} is sourced by what we may call subsystem relativity across the degeneracy sectors, cf.~Sec.~\ref{sssec_degrelativity}.
The easiest example of such an engineered setup is if one chooses $C_3$ to be nearly ideal, just as we did for the interferometer Eqs.~(\ref{eq:density1},~\ref{eq:density2}), and picks reduced states of the form
\begin{equation}\label{degclockstateex}
    \ket*{\phi^{\Lambda_R}_{\vert i}} \propto \Pi_{\vert i} ( \ket*{\phi_S^{\Lambda_R}} \otimes \ket{\tau_{\bar{i}}}\otimes \ket{\tau_3}), \qquad \Lambda_R \in \{+-, -+ \}.
\end{equation}
Recall that an ideal clock $C_3$ is not necessarily an entirely unreasonable assumption, as the ADM boundary Hamiltonian (for second-order metric perturbations) is the Hamiltonian of an ideal clock as  mentioned in Sec.~\ref{sect:linearisationConditions}. As $C_3$ is, by assumption, in the causal complement of the current region of interest, we may thus simply assume that the complementary region is one with an asymptotic boundary such that $C_3$ is an ADM boundary clock.

The states \eqref{degclockstateex} can be mapped back to the gauge-invariant level through the inverse PW-reduction maps \eqref{eq:PWDeg}. For this simple state it can be shown that $S^{\Lambda_R}[\phi]$ only depends on the chosen QFT-state $\ket*{\phi_S^{\Lambda_R}}$. To have a different entropy one just chooses a QFT-state with differing properties in the distinct sector overlaps, $\ket*{\phi_S^{+-}} \neq \ket*{\phi_S^{-+}}$.

Finally, let us consider the coefficients $p^{\Lambda_R}$ in \eqref{Equation: degenerate entropy}. Since Alice and Bob both only have access to two out of four sectors, they can only access two out four coefficients through \eqref{eq:degReducedStates}. These probabilities can only be probed `relative to their existence', so they will rather be conditional probabilities $p^{\Lambda_R}_i$ instead. By this we mean that, for example, Alice needs to renormalise as
\begin{equation}
    p^{\Lambda_R}_A = \frac{p^{\Lambda_R}}{p^{++} + p^{+-}}, \qquad \Lambda_R \in \{++,+-\},
\end{equation}
with the $p^{\Lambda_R}$ from \eqref{eq:degReducedStates}.
These are the conditional probabilities relative to Alice and Bob and should  be thought of as what they would measure, given what they have access to by assumption.
They will generally differ per observer: $\{p^{++}_A, p^{+-}_A \}$ for Alice and  $\{p^{++}_B, p^{-+}_B \}$ for Bob. Even though they both have full access to the $++$-sector, their conditional probabilities need not be equal $p^{++}_A \neq p^{++}_B$ since the weights given by the other sector need not be equal $p^{+-}_A \neq p^{-+}_B$ either, and for most global states will not be. This leads to additional operational differences in the entropy  \eqref{Equation: degenerate entropy}: the conditional Shannon entropy, the first contribution, will be different because their conditional probability distributions will generally differ. There will also be a difference in the second piece of \eqref{Equation: degenerate entropy}, where the entropy contributions per sector are multiplied by these conditional probabilities. Hence, the last piece can be reinterpreted as a conditional expectation value of the overall von Neumann entropy of the state accessible to either observer.

This examples serves to illustrate that there can be several sources of entropy relativity, depending on the exact operational assumptions to which we subject the different observers with clocks.

\subsection{Semiclassical vs.\ non-semiclassical clocks}
\label{Subsection: semiclassical vs nonsemiclassical}

It is entirely possible in a given state for a semiclassical regime to apply to one clock but not to another; indeed it could be that an antisemiclassical regime applies to the other clock. One would have to apply two different entropy formulas for the two clocks (one would be the generalised entropy~\eqref{Equation: generalised entropy}, and the other would be~\eqref{Equation: antisemiclassical entropy}), and the two formulas would not agree with each other. Indeed, the generalised entropy clearly depends on the state of the fields, whereas~\eqref{Equation: antisemiclassical entropy one clock} does not. 

Let us now give a slightly more explicit demonstration of this phenomenon. Consider a system with three ideal clocks $C_1,C_2,C_3$, and suppose the state $\Ket{\phi}$ in the perspective of clock $C_3$ is a product state of a QFT state $\ket{\phi_S}\in\mathcal{H}_S$, and clock states $\ket{\phi_1}\in\mathcal{H}_1$ and $\ket{\phi_2}\in\mathcal{H}_2$:
\begin{equation}
    \ket*{\phi_{|3}(0)} = \ket{\phi_S}\otimes\ket{\phi_1}\otimes\ket{\phi_2}.
\end{equation}
The variances of the Hamiltonians $H_S,H_1,H_2$ may be computed in this state as the variances in the respective factors $\ket{\phi_S},\ket{\phi_1},\ket{\phi_2}$:
\begin{align}
    (\Delta H_S)^2 &= \expval{(H_S-\expval{H_S}_{\phi_S})^2}_{\phi_S},\\
    (\Delta H_1)^2 &= \expval{(H_1-\expval{H_1}_{\phi_1})^2}_{\phi_1},\\
    (\Delta H_2)^2 &= \expval{(H_2-\expval{H_2}_{\phi_2})^2}_{\phi_2}.
\end{align}
Now, let us choose these states such that
\begin{equation}
    (\Delta H_2)^2 \ll \beta^{-2},\qquad (\Delta H_S)^2 \ll (\Delta H_1)^2.
\end{equation}
This is of course possible, because the spectra of each of $H_S,H_1,H_2$ are unbounded above and below. Moreover, these imply that a semiclassical regime applies to clock $C_1$, but an antisemiclassical regime applies to clock $C_2$, as required.

\subsection{Periodic vs.\ monotonic clocks}
\label{Subsection: periodic vs monotonic}

In this paper we have so far considered only \emph{monotonic} clocks, i.e.\ those which record distinct times for any value of $\RR$. We shall now briefly consider instead \emph{periodic} clocks, i.e.\ those with times cyclically evolving through some finite interval $[0,t_{\text{max}})$. Such clocks were thoroughly investigated in~\cite{Chataignier:2024eil}, and below we shall quickly review some of the relevant details.

A periodic clock $C_{\text{per}}$ is not a complete reference frame for a full group $\RR$ of time evolutions. Rather, it can only be used to fix the gauge up to some discrete subgroup $\ZZ$ generated by evolution in time by the amount $t_{\text{max}}$. As such, the periodic clock is a QRF for the quotient group $\rm{U}(1)=\RR/\ZZ$. Its Hamiltonian $H_{C_{\text{per}}}$ is the quantisation of the generator of a representation of $\rm{U}(1)$, and so must have a discrete spectrum of eigenvalues, which may be written in the form
\begin{equation}\label{periodicspec}
    \sigma_{C_{\text{per}}} = \qty{\omega \qty(n_\mu+\frac{\varphi}{2\pi})\Bigm\vert n_\mu\in\ZZ},
\end{equation}
where $\varphi\in\RR$, $\omega = \frac{2\pi}{t_{\text{max}}}$, and $\mu$ ranges over some finite or infinite set of labels. A non-trivial phase $\varphi$ means that the clock transforms under a \emph{projective} unitary representation of $\rm{U}(1)$, since we have $e^{-iH_{C_{\text{per}}}t_{\text{max}}} = e^{-i\varphi}$. For simplicity, let us restrict to the case where $\varphi=0$, so that the representation is genuinely unitary (this restriction may be relaxed as described in~\cite{Chataignier:2024eil}). We shall also assume without loss of generality that $t=t_{\text{max}}$ is the smallest non-zero value for which $e^{-iH_{C_{\text{per}}}t}=1$, which in terms of the spectrum means that the $\{n_\mu\}$ are coprime. As a further simplifying assumption, we take the spectrum to be non-degenerate, so that the Hilbert space of $C_{\text{per}}$ is $\mathcal{H}_{C_{\text{per}}}=L^2(\sigma_{C_{\text{per}}})$.

As in the monotonic case, one may then define clock states as follows (setting the arbitrary phase, as in Eq.~\eqref{clockstates}, to $g(\epsilon_\mu)=0$):
\begin{equation}
    \ket{t}_{\text{per}} = \sum_{\varepsilon_\mu\in \sigma_{C_{\text{per}}}} e^{-i\varepsilon_\mu t}\ket{\varepsilon_\mu},
\end{equation}
where $\ket{\varepsilon_\mu}$ is the $H_{C_{\text{per}}}=\varepsilon_\mu$ eigenstate. We have $e^{-iH_{C_{\text{per}}}t'}\ket{t}_{\text{per}} = \ket{t+t'}_{\text{per}}$ and  $\ket{t}_{\text{per}}=\ket{t+t_{\text{max}}}_{\text{per}}$. These states form a resolution of the identity
\begin{equation}
    \mathds{1} = \frac1{t_{\text{max}}}\int_0^{t_{\text{max}}} \dd{t}\dyad{t}_{\text{per}}\,,
\end{equation}
thus furnishing a $\rm{U}(1)$-covariant clock POVM. A general operator acting on the fields and the periodic clock, i.e.\ an $a\in\mathcal{A}_S\otimes \mathcal{B}(\mathcal{H}_{C_{\text{per}}})$, may be written in the form
\begin{equation}
    a = \frac1{t_{\text{max}}}\int_0^{t_{\text{max}}}\dd{t}a(t,t') \otimes \dyad{t'}_{\text{per}}, \qquad a(t,t')\in \mathcal{A}_S.    
\end{equation}
Much of the discussion then proceeds similarly to the case of the monotonic clock; we refer the reader to~\cite{Chataignier:2024eil} for more details.

Let us in particular give an outline of some relevant properties of the invariant subalgebra $(\mathcal{A}_S\otimes \mathcal{B}(\mathcal{H}_{C_{\text{per}}}))^{H_S+H_{C_{\text{per}}}}$. For now $\mathcal{A}_S$ will refer abstractly to the algebra of `system' operators; we will shortly specialise to the case where the system includes the quantum fields in a subregion.

It is instructive to note that we can construct this subalgebra in three steps. First, we can impose invariance under the group $\ZZ$ of evolutions by integer multiples of $t_{\text{max}}$. Since the clock is invariant under this group, we can restrict our attention to the system, for which we go to the subalgebra
\begin{equation}
    \mathcal{A}_S^{t_{\text{max}}} := (\mathcal{A}_S)^{e^{-iH_S t_{\text{max}}}} \subset \mathcal{A}_S,
\end{equation}
As a second step, we can take the crossed product of $\mathcal{A}^{t_{\text{max}}}_S$ with its modular flow. To this end, let us for the moment assume that $C_{\rm per}$ is an \emph{ideal} periodic clock, which means that we allow all $n_\mu\in\mathbb{Z}$ in Eq.~\eqref{periodicspec} with the consequence that $\braket{t}{t'}_{\rm per}\propto\delta(t-t')$ and $\mathcal{H}_{\rm C_{\rm per}}\simeq L^2(\rm{U}(1))$, as appropriate for a $\rm{U}(1)$-crossed product. Indeed, note that now modular flow is a representation of the leftover $\rm{U}(1)=\RR/\ZZ$ of period $t_{\text{max}}$, so this crossed product may be written $\mathcal{A}_S^{t_{\text{max}}}\rtimes \rm{U}(1)$. This algebra is guaranteed to have a trace, because the crossed product makes modular flow inner.

This crossed product would thus correspond to employing a periodic clock whose spectrum is given by $\omega\ZZ$. The third step in the construction of $(\mathcal{A}_S\otimes \mathcal{B}(\mathcal{H}_{C_{\text{per}}}))^{H_S+H_{C_{\text{per}}}}$ is therefore to project the spectrum of the clock down to $\sigma_{C_{\text{per}}}\subseteq \omega\ZZ$ (analogous to the projection onto the spectrum employed when considering non-ideal monotonic clocks). 

One should then also represent this on a physical Hilbert space (which may be constructed as demonstrated in~\cite{Chataignier:2024eil}). For reasons that are essentially analogous to the case of a monotonic clock explained earlier (and so whose details we skip), one would end up with a normal and faithful (so long as there is at least one other clock in the full system) representation of a Type II${}_1$ (if $\sigma_{C_{\text{per}}}$ is bounded below) or Type II${}_\infty$ (if not) von Neumann algebra $\mathcal{A}_{SC_{\text{per}}}^H$ (assuming the invariant subalgebra is itself Type II). As such, one can construct density operators and traces by similar formulae to those we have given in this paper for the monotonic case.

Having explained how to generalise the results of the paper to periodic clocks, let us now qualitatively and briefly explain how the entropies will differ when using a periodic clock vs a monotonic clock. The essential point is that in the intermediate algebra $\mathcal{A}_S^{t_{\text{max}}}\rtimes \rm{U}(1)$ used in the construction of $\mathcal{A}_{SC_{\text{per}}}^H$, we must restrict to periodic system observables before ever involving the clock. This will clearly change the way in which the entropy functional for a periodic clock depends on the state, relative to that of a monotonic clock. This is a clear source of entropy relativity (see further discussion in~\cite{periodic}). 

Let us now consider what happens when $\mathcal{A}_S$ is the algebra of QFT operators in a subregion. In this case, the above construction typically simplifies significantly, because the spectrum of the vacuum modular flow is purely continuous, and hence $\mathcal{A}_S^{t_{\text{max}}}=\mathbb{C}\mathds{1}$ is trivial. Consider, for example, the vacuum modular Hamiltonian of a Rindler wedge, i.e.\ the boost generator. In any QFT there are no bounded operators that evolve periodically under a boost (of course there are operators which create particles of a certain fixed boost charge -- but such operators are \emph{unbounded} and thus not elements of $\mathcal{A}_S$). In this case, the invariant subalgebra simplifies to $\mathcal{B}(\mathcal{H}_{C_{\text{per}}})^{H_{\text{per}}}$. Therefore, the periodic clock cannot be used \emph{by itself} to measure anything in the QFT. In this sense, a periodic clock is even worse than an antisemiclassical monotonic one for observing the QFT, and in this case the only contributions to the entropy associated with the periodic clock arise from its own internal degrees of freedom.

The situation changes when we allow the system algebra $\mathcal{A}_S$ to contain more than just operators acting on the QFT. For example, it could be $\mathcal{A}_S=\mathcal{A}_{\mathcal{U}}\otimes\mathcal{B}(\mathcal{H}_C)$, where $\mathcal{A}_{\mathcal{U}}$ is the algebra of the QFT in a subregion $\mathcal{U}$, while $\mathcal{H}_C$ is the Hilbert space of another clock, which we take to be ideal and \emph{monotonic}. Note that now $H_S$ is the modular Hamiltonian of the QFT plus the energy of clock $C$. Then clearly $\mathcal{A}_S^{t_{\text{max}}}$ is non-trivial: it contains as a subalgebra the crossed product of $\mathcal{A}_{\mathcal{U}}$ with the clock $C$.\footnote{In fact, $\mathcal{A}_S^{t_{\text{max}}}$ is in this case Type III$_{\lambda}$, where $\lambda = e^{-2\pi/T}$. This is because the modular flow of this algebra is outer, but it acts periodically with period $T$, which is exactly the definition of Type III$_\lambda$. This is consistent with the following: it may be confirmed that $\mathcal{A}_S^{t_{\text{max}}}$ contains as a subalgebra the crossed product of $\mathcal{A}_{\mathcal{U}}$ with periodic modular flow, which is Type III${}_\lambda$~\cite[Lemma XVIII.4.17]{takesaki2013theory}.}
Then the invariant algebra $(\mathcal{A}_S\otimes\mathcal{H}_{C_{\text{per}}})^{H_S+H_{C_{\text{per}}}}$ contains more than just clock observables, and so the corresponding entropy has contributions from the QFT. On the other hand, this algebra is equivalent to the crossed product of $\mathcal{A}_{\mathcal{U}}\otimes\mathcal{B}(\mathcal{H}_{C_{\text{per}}})$ with the monotonic clock $C$, which we already know is Type II$_\infty$. The entropy would be equivalent to the one seen by the monotonic clock $C$, so there would be no meaningful consequences of the periodicity of $C$.

\section{Conclusion}
\label{Section: Conclusion}

The basic idea underlying this paper has been that the apparent importance of QRFs in a rigorous definition of entropy in quantum gravity~\cite{leutheusser2023causal,leutheusser2023emergent,Witten:2021unn,Chandrasekaran:2022cip,Chandrasekaran_2023,Kudler-Flam:2024psh,Jensen:2023yxy,Ali:2024jkx,Klinger:2023tgi,Faulkner:2024gst,AliAhmad:2023etg,KirklinGSL,Kudler-Flam:2023qfl,Witten:2023xze, Kudler-Flam:2023hkl,Gesteau:2023hbq,Balasubramanian:2023dpj,Soni:2023fke,AliAhmad:2023etg,Aguilar-Gutierrez:2023odp,Gomez:2023wrq,Gomez:2023upk,Klinger:2023tgi,Klinger:2023auu} leads one to an inevitable conclusion: gravitational entropy is \observerdependent{}. Our goal has been to explore this phenomenon. We have built upon and given a much more detailed and technical exposition of the methods used in our shorter companion paper~\cite{DeVuyst:2024pop}. We argued that many parts of the framework~\cite{Chandrasekaran:2022cip} introduced coincide with the perspective-neutral framework for QRFs \cite{DeVuyst:2024pop,delaHamette:2021oex,Hoehn:2023ehz,Hoehn:2019fsy,Hoehn:2020epv,Chataignier:2024eil,AliAhmad:2021adn,Hoehn:2021flk,Giacomini:2021gei,Castro-Ruiz:2019nnl,delaHamette:2021piz,Vanrietvelde:2018dit,Vanrietvelde:2018pgb,Hoehn:2021flk,Suleymanov:2023wio,Carrozza:2024smc}, which generalises the Page-Wootters formalism \cite{Page:1983uc,1984IJTP...23..701W}, and we summarised this coincidence with the informal equation ``$\text{PW} = \text{CLPW}$''. We considered observers carrying clocks of a more general nature than previously considered, in the sense that they can have degenerate energies, they can be periodic or monotonic, and they can be arbitrarily entangled with other degrees of freedom. Furthermore, we showed how the entropies of a subregion relative to arbitrarily many clocks reproduces the generalised entropy formula in a semiclassical regime, and we also considered corrections to this regime. We also explored its opposite, the antisemiclassical regime. Finally, we investigated several explicit examples of the phenomenon of \observerdependent{} entropies, explaining how it can manifest from using different kinds of clocks, but also from using clocks whose states have different properties (such as the degree of their entanglement with other degrees of freedom, or the magnitude of their energy fluctuations).

Our main tools were the perspective-neutral QRF formalism and the mathematical theory of von Neumann algebras. To that end, we have explained how these two frameworks fit together. We have given full, detailed derivations of algebras, traces, density operators, and entropies for the gravitational subsystems defined using quantum clocks.

Before ending, let us speculate on three other possible sources of \observerdependence{} in the entropy. 

First, in this paper we have made the simplifying assumption that the clocks and fields are non-interacting (since we are taking the total Hamiltonian to decompose as a sum of QFT Hamiltonian and clock Hamiltonians). More realistic clocks would have to backreact non-negligibly on the fields; depending on the kinds of interactions, one would thus have modified notions of physical subsystems and thus different entropies. Interacting QRFs have not been studied in much detail in the literature (see~\cite{Hohn:2011us,Marolf:2009wp,Hoehn:2023axh,Smith2019quantizingtime,Smith:2019imm,Castro-Ruiz:2019nnl} for some exceptions). It would be interesting to see whether much headway can be made in the gravitational setting, though there are challenges~\cite{Hohn:2011us,Marolf:2009wp,Hoehn:2023axh,Dittrich:2016hvj}.

Next, when considering degenerate clocks, we have for simplicity restricted to the case where the multiplicity of each energy level is independent of the energy. This again will not be the case for more realistic clocks. The first law of thermodynamics relates the energy-dependence of this multiplicity (via the canonical ensemble entropy of the clock) to its \emph{temperature}. Thus, allowing for a non-trivial dependence of this kind could allow one to derive an \observerdependence{} of the entropy arising from the temperature of the clocks. Again, this would be interesting to investigate.

Finally, in this paper we considered QRFs transforming only under a single gauge transformation -- the modular boost of the subregion. We moreover only implemented the single constraint corresponding to this particular gauge symmetry. We emphasise that we did not give a concrete justification for why this makes sense, although we did discuss it briefly in Sec.~\ref{sect:linearisationConditions}. Of course, the gravitational gauge group is \emph{much} larger than one-dimensional; it is the full spacetime diffeomorphism group. Moreover, observers will typically not just carry clocks -- they will also carry other objects such as (for example) rods and Lorentz frames. These will transform non-trivially under other diffeomorphisms, and are required for dressing more complicated field observables. Thus, at some point one will need to address these other gauge symmetries and their corresponding constraints. An intermediate step to the full diffeomorphism group would be to impose constraints generating a larger subgroup than the boost $\RR$ considered in this paper. An obvious question is how the setup we have explored can be generalised to crossed products and QRFs for more general locally compact groups, and for more general higher-dimensional symmetry groups as investigated in \cite{delaHamette:2021oex,AliAhmad:2024eun,Fewster:2024pur}. See~\cite{KirklinGSL} for an application of the perspective-neutral formalism to perturbative quantum gravity with a \emph{two}-dimensional group of diffeomorphisms, which allowed for an extension of the generalised second law beyond the semiclassical regime.

If one is able to carry out this generalisation, then a natural expectation would be that the entropy of a subsystem relative to different observers would now continue to be sensitive to the nature of the QRFs carried by those observers. In general, such QRFs transform in a representation of a subgroup of diffeomorphisms. A basic example would be a QRF transforming under $\operatorname{SO}(3)$ or its double cover $\operatorname{SU}(2)$ (which is natural for example in the case of the de Sitter static patch, whose symmetry group is $\RR\times \operatorname{SO}(3)$, with $\RR$ being the boost). Representations of $\operatorname{SU}(2)$ are labeled by their spin $j$, and it is natural to expect that the entropy depends on the value of the observer spin. 

Somewhat related to this, the trajectory of the observer carrying the quantum clock is usually taken as classical in most works on gravitational crossed product algebras. There is an interesting recent development \cite{Kolchmeyer:2024fly}, where this setting was generalised to also describing the location of the observers in a quantum manner in two-dimensional de Sitter space. This allows their associated subregions to overlap. Such an expanded description of a quantum observer also constitutes a QRF (with additional degrees of freedom than just the clock) and it would be interesting to extend our observations to that case. We leave investigation of this and related questions to future work.

\phantomsection
\addcontentsline{toc}{section}{\numberline{}Acknowledgements}
\section*{Acknowledgments}

We thank Gon\c{c}alo Ara\'ujo Regado, Wissam Chemissany, \r{A}smund Folkestad, Laurent Freidel, Elliot Gesteau, Ted Jacobson, Fabio Mele, Federico Piazza, Gautam Satishchandran and Antony Speranza for helpful discussions.
This work was supported in part by funding from the Okinawa Institute of Science and Technology Graduate University.~It was also made possible through the support of the ID\# 62312 grant from the John Templeton Foundation, as part of the \href{https://www.templeton.org/grant/the-quantum-information-structure-of-spacetime-qiss-second-phase}{\textit{`The Quantum Information Structure of Spacetime'} Project (QISS)} from the John Templeton Foundation.~The opinions expressed in this project/publication are those of the author(s) and do not necessarily reflect the views of the John Templeton Foundation. Research at Perimeter Institute is supported in part by the Government of Canada through the Department of Innovation, Science and Economic Development and by the Province of Ontario through the Ministry of Colleges and Universities.

\appendix

\section{Additional derivations concerning non-ideal clocks}
\label{app_additionalnon-ideal}
\subsection{Fourier transform for non-ideal clocks}\label{app_fourier}

While the clock states in Eq.~\eqref{clockstates} need not be orthogonal and the clock energy may not range over the full reals,  a Fourier transform still exists \cite{Hoehn:2019fsy}. Indeed, 
\begin{eqnarray}
    \ket{\epsilon}=\frac{1}{\sqrt{2\pi}}\int_\mathbb{R}\dd{t}e^{-ig(\epsilon)}e^{it\epsilon}\ket{t}\,
\end{eqnarray}
with overlap
\begin{eqnarray}
   \braket{\epsilon}{t}=\frac{1}{\sqrt{2\pi}} e^{ig(\epsilon)}e^{-it\epsilon} \,.
\end{eqnarray}
Hence, for a state $\ket{\psi}$ with wave function $\psi(t)$
\begin{eqnarray}
    \ket{\psi}=\int\dd{t}\psi(t)\ket{t}
\end{eqnarray}
by making use of (\ref{eq:chi}) we get that
\begin{eqnarray}\label{fourier1}
    \tilde\psi(\epsilon):=\braket{\epsilon}{\psi}=\frac{1}{\sqrt{2\pi}}\int_\mathbb{R}\dd{t}e^{ig(\epsilon)}e^{-it\epsilon}\psi(t).
\end{eqnarray}
Conversely, the inverse Fourier transformation returns a ``filtered version'' of the function $\psi(t)$, with energy support only on the spectrum, which describes an equivalent state on the clock Hilbert space.
\begin{equation}\begin{split}\label{invfourierfiltered}
    &\bar{\psi}(t):=\frac{1}{\sqrt{2\pi}}\int_{\sigma_C}\dd{\epsilon} e^{-ig(\epsilon)}e^{it\epsilon}\tilde\psi(\epsilon)=\int\dd t'\psi(t')\braket{t}{t'}\,,\\
    &\int \dd t~ \bar{\psi}(t)\ket{t} = \int \dd t~ \psi(t)\ket{t}\,.
\end{split}\end{equation}

\subsection{Algebra of gauge-invariant clock-system observables and type conversion}\label{app_ginvalg}

Here, we prove Eq.~\eqref{gaugeinvalg} from Sec.~\ref{sseHphys}, i.e.\ that the algebra of gauge-invariant clock-system observables is comprised of relational observables, describing $S$ relative to $C$, and reorientations of $C$:
\begin{eqnarray}
    \left(\mathcal{A}_S \otimes \mathcal{B}(\mathcal{H}_C)\right)^{H}=\Big\langle O^\tau_C(a),\mathds{1}_S \otimes U_C(t)\,|\,a\in\mathcal{A}_S,\,t\in\mathbb{R}\Big\rangle.
\end{eqnarray}
\indent To this end, consider any $O\in\left(\mathcal{A}_S \otimes \mathcal{B}(\mathcal{H}_C)\right)^{H}$ and consider its `partial matrix elements' in the clock states Eq.~\eqref{clockstates}
\begin{equation}
    O(t,t'):=\langle t|O|t'\rangle.
\end{equation}
Gauge invariance, i.e.\ $[U_{SC}(t),O]=0$ for all $t$, implies
\begin{eqnarray}
    O(t,t')=U_S(t)O(0,t'-t)U_S^\dag(t).
\end{eqnarray}
Since any such $O(0,t'-t)$ can be written as a linear combination (or integral) of objects of the form $f(t'-t)a$, where $f(t)$ is some function and $a\in\mathcal{A}_S$, we have that a basis for such partial matrix elements is given by
\begin{equation}
    \{O_{\tilde t,a}(t,t')=\chi(t+\tilde t-t')U_S(t)\,a\,U_S^\dag(t)\,|\,\tilde t\in\mathbb{R},\,a\in\mathcal{A}_S\},
\end{equation}
where $\chi(t)$ is the clock state overlap distribution in Eq.~\eqref{fuzzy}. To see this, consider an arbitrary function $f(t)$ and recall the Fourier transforms in appendix~\ref{app_fourier}, setting $g(\epsilon)=0$ for simplicity,
\begin{eqnarray}
    f(t'-t)&=&\frac{1}{\sqrt{2\pi}}\int_{\sigma_C}\dd{\epsilon}e^{i(t'-t)\epsilon}\tilde f(\epsilon)=\frac{1}{2\pi}\int_{\sigma_C}\dd{\epsilon}\int_\mathbb{R}\dd{\tilde t}e^{i(t'-t-\tilde t)\epsilon}f(\tilde t)\nonumber\\
    &\underset{\eqref{eq:chi}}{=}&\int_\mathbb{R}\dd{\tilde t}\chi(t'-t-\tilde t) f(\tilde t).
\end{eqnarray}
\indent Now integrating the basis elements and recalling Eq.~\eqref{eq:chi},
\begin{eqnarray}
    \int_\mathbb{R}\dd{t}\dd{t'}O_{\tilde t,a}(t,t')\ket{t}\!\bra{t'}&=&\int_\mathbb{R}\dd{t}\dd{t'}\braket{t+\tilde t}{t'}\ket{t}\!\bra{t'}\otimes U_S(t)\,a\,U_S^\dag(t)\nonumber\\
    &=&\int_\mathbb{R}\dd{t}\ket{t}\!\bra{t+\tilde t}\otimes U_S(t)\,a\,U_S^\dag(t) \left(\int_\mathbb{R}\dd{t'}\ket{t'}\!\bra{t'}\otimes \mathds{1}_S\right)\nonumber\\
    &\underset{\eqref{relobs}}{=}&O_C^{\tau=0}(a) \left(U_C^\dag(\tilde t)\otimes \mathds{1}_S\right),
\end{eqnarray}
where we made use of the clock state covariance \eqref{clockcov} in the last line. 
This suffices for the claim since 
\begin{equation}
    O_C^\tau(a)=\left(\mathds{1}_S \otimes U_C(\tau)\right)\,O_C^{0}(a)\,\left(U_C^\dag(\tau)\otimes \mathds{1}_S\right),
\end{equation}
i.e.\ clock reorientations change the $\tau$ label at which the relational observable is evaluated.

Next, we prove the statement in Sec.~\ref{sseHphys} that if $\mathcal{A}_S$ is a von Neumann algebra, then so is the gauge-invariant clock-system algebra $\left(\mathcal{A}_S \otimes \mathcal{B}(\mathcal{H}_C)\right)^{H}$. We begin with the case that $H_C$ features a non-degenerate spectrum. We can then extend the clock to an auxiliary ideal clock if it is not already ideal. We extend the Hilbert space $\mathcal{H}_C$ to a new Hilbert space $\tilde{\mathcal{H}}_C=\mathcal{H}_C\oplus \mathcal{H}_{\rm add}$ by adding all the missing energy eigenstates to extend $\sigma_C$ over the full real line. In particular, by spectral decomposition, we extend the clock Hamiltonian to a new operator $\tilde H_C$ on $\tilde{\mathcal{H}}_C$ which has a nondegenerate spectrum covering all of $\mathbb{R}$. At this stage the clock is ideal and we have $\tilde{\mathcal{H}}_C\simeq L^2(\mathbb{R})$. 

Extending the constraint to the extended kinematical Hilbert space $\tilde{\mathcal{H}}_{\rm kin}:=\tilde{\mathcal{H}}_C\otimes\mathcal{H}_S$, $H\to \tilde H_C+H_S$, the gauge-invariant algebra becomes a crossed product of $\mathcal{A}_S$ by the translation group generated by the system Hamiltonian $H_S$:
\begin{equation}\label{crossedproductapp}
   \left(\mathcal{A}_S \otimes\mathcal{B}(L^2(\mathbb{R})_C)\right)^{\tilde H_C+H_S} =\mathcal{A}_S\rtimes_{U_S}\mathbb{R}=\Big\langle e^{-i\tilde T H_S } \,a\, e^{i\tilde T  H_S },e^{-it\tilde H_C}\,\big|\,a\in\mathcal{A}_S,\,t\in\mathbb{R}\Big\rangle'',
\end{equation}
where $\tilde T$ is the self-adjoint operator conjugate to $\tilde H_C$ on $\tilde{\mathcal{H}}_C$. This is a von Neumann algebra when $\mathcal{A}_S$ is one (e.g., see the appendices of \cite{Chandrasekaran:2022cip,Jensen:2023yxy} for a summary of crossed products). The crossed product is not necessarily a factor, even if $\mathcal{A}_S$ is. However, it is one when $\mathcal{A}_S$ is a type $\rm{III}_1$ factor and $H_S$ is the modular Hamiltonian associated with a cyclic and separating vector $\ket{\psi_S}$ in $\mathcal{H}_S$ for $\mathcal{A}_S$. In this case, the crossed product is a von Neumann factor of type $\rm{II}_\infty$.

Some intuition regarding why such a type transition must occur can be garnered by first noting that type $\rm{III}_1$ algebras have only outer modular flows, meaning that $H_S$ itself cannot be an element of $\mathcal{A}_S$. However, a related state can easily be identified on $\tilde{\mathcal{H}}_C\otimes\mathcal{H}_S$ for which $H_S$ continues to act as the modular Hamiltonian, now for the full crossed product algebra.  In particular, the (non-normalizable) state $\ket{\tilde t_C=0}\otimes \ket{\psi_S}$ has this property, as can be easily checked.  A modular flow on elements $\hat{a}\in \mathcal{A}_S\rtimes_{U_S}\mathbb{R}$ of the crossed product algebra is thus given by
\begin{equation}
    e^{-iu H_S}\hat{a}e^{iu H_S} 
    = e^{iu \tilde H_C}e^{-iu (H_S+\tilde H_C)}\hat{a}e^{iu (H_S+\tilde H_C)}e^{-iu \tilde H_C}
    = e^{iu \tilde H_C}\hat{a}e^{-iu \tilde H_C},
    \quad u\in \mathbb{R}\,,
\end{equation}
where we have used the fact that the crossed product algebra commutes with $H_S+\tilde H_C$.  We see that $\tilde H_C$ itself generates a modular flow for the crossed product algebra, and since bounded functions of $\tilde H_C$ are part of the algebra, this is now an inner modular flow.  The crossed product algebra therefore cannot be of type $\rm{III}$.  In fact, Takesaki \cite{Takesaki1979} further showed how to leverage this inner modular flow to identify a trace on the crossed product algebra, demonstrating that it is of type $\rm{II}_\infty$ (see \cite{Sorce:2023fdx} for an accessible review of von Neumann type classifications, and appendix B of \cite{Jensen:2023yxy} for a related summary in this context). Thus, it is the addition of the clock reorientations, given by $\tilde H_C$, to the algebra that is responsible for turning the modular flow inner and thereby for the type conversion.

To show that the gauge-invariant clock-system algebra for the original (i.e. non-extended) clock Hamiltonian is also a von Neumann algebra, we first note that $H_C=\tilde\Pi_{|C}\tilde{H}_C$, where $\tilde\Pi_{|C}$ is the orthogonal projector on $\tilde{\mathcal{H}}_C$ onto its subspace $\mathcal{H}_C$. When $\sigma_C$ is just an interval, it is given by $\tilde\Pi_{|C}:=\Theta(\epsilon_{\rm max}-\tilde{H}_C)\Theta(\tilde{H}_C-\epsilon_{\rm min})$, where $\epsilon_{\rm max/min}$ are the spectral limits of the original $H_C$; otherwise, it is given by a sum over such products of theta functions over the respective intervals. Since the extended clock reorientations $e^{-it\tilde H_C}$ constitute a basis of bounded functions of $\tilde H_C$, we have that $\tilde\Pi_{|C}$ is contained in the crossed product algebra in Eq.~\eqref{crossedproductapp}.

Next, we note that the clock states of the two Hamiltonians according to Eq.~\eqref{clockstates} are related by $\ket{\tau}=\tilde\Pi_{|C}\ket{\tilde{\tau}}$. Hence,
\begin{equation}\label{tilderelobs}
    O^{0}_C(a)=\tilde\Pi_{|C}\left(e^{-i\tilde T H_S } \,a\, e^{i\tilde T  H_S }\right)\tilde\Pi_{|C}.
\end{equation}
For the original gauge-invariant clock-system algebra, we thus have that it is a subalgebra of the crossed product
\begin{equation}
    \left(\mathcal{A}_S \otimes \mathcal{B}(\mathcal{H}_C)\right)^{H}=\tilde\Pi_{|C}\left(\mathcal{A}_S\rtimes_{U_S}\mathbb{R}\right)\tilde\Pi_{|C}.
\end{equation}
In general, $\Pi\mathcal{A}\Pi$ is a von Neumann algebra if $\mathcal{A}$ is a von Neumann algebra containing the projection $\Pi$ \cite[statement EP7, p.~21]{Joneslec}; $\Pi\mathcal{A}\Pi$ is further a factor, if $\mathcal{A}$ is \cite[footnote 7]{Chandrasekaran:2022cip}. This proves the claim. In particular, if $\mathcal{A}_S$ is a type $\rm{III}_1$ factor and $H_S$ an associated modular Hamiltonian, then $ \left(\mathcal{A}_S \otimes \mathcal{B}(\mathcal{H}_C)\right)^{H}$ is a type II factor. We emphasise that the von Neumann structure is at this stage defined in terms of the kinematical Hilbert space. In Sec.~\ref{ssec_vN}, we explore the von Neumann nature of the action of these algebras on the physical Hilbert space.

The case when $H_C$ has a degenerate spectrum, but the degeneracy is energy-independent (except possibly in a set of measure zero) follows from this and is discussed in Sec.~\ref{sect:degSuperselection}; when there is also an operational superselection, $ \left(\mathcal{A}_S \otimes \mathcal{B}(\mathcal{H}_C)\right)^{H}$ is then still a von Neumann algebra when $\mathcal{A}_S$ is, but decomposes across degeneracy sectors and can thus not be a factor. 

\subsection{Bounded physical operators from bounded kinematical ones}\label{app_bounded}

In this appendix, we use the reduction theorem
\begin{equation}
    \mathcal{R}_C(\tau)\mathcal{O}_C^\tau(a)\mathcal{R}_C^{-1}(\tau)=\Pi_{|C}\,a\,\Pi_{|C},\quad\qquad a\in\mathcal{L}(\mathcal{H}_S)
\end{equation}
from \cite{Hoehn:2019fsy} in Eqs.~\eqref{obsredthm} and~\eqref{equivred} to briefly prove the statement in Sec.~\ref{ssec_obsred} that every element in $\mathcal{B}(\mathcal{H}_{\rm phys})$ can be obtained from an element in $\mathcal{B}(\mathcal{H}_{\rm kin})$ via the relational observable construction in Eq.~\eqref{relobs} and that this relational observable is also bounded, hence an element of $\left(\mathcal{B}(\mathcal{H}_{\rm kin})\right)^{H}$. Relational observables are thus complete on the physical Hilbert space.

We begin with the case that $H_C$ is non-degenerate. Since $\mathcal{R}_C(\tau)$ is unitary and $\Pi_{|C}$ is bounded, we have that $r({O}_C^\tau(a))\in\mathcal{B}(\mathcal{H}_{\rm phys})$ if and only if $a\in\mathcal{B}(\mathcal{H}_S)$, where we recall that $r$ is the representation of $\mathcal{A}_{\rm inv}$ on $\mathcal{H}_{\rm phys}$. Similarly, $\mathds{1}_C\otimes a\in\mathcal{B}(\mathcal{H}_{\rm kin})$ if and only if $a\in\mathcal{B}(\mathcal{H}_S)$. We will now also demonstrate that $O_C^\tau(a)\in\left(\mathcal{B}(\mathcal{H}_{\rm kin})\right)^{H}$ if and only if $a\in\mathcal{B}(\mathcal{H}_S)$. To see this, we again extend $C$ to an auxiliary ideal clock if it is not already ideal, as in appendix~\ref{app_ginvalg}. On the extended kinematical Hilbert space $\tilde{\mathcal{H}}_{\rm kin}$ we can then write, as in Eq.~\eqref{tilderelobs},
\begin{equation}
    O^{\tau}_C(a)=\tilde\Pi_{|C}\left(e^{-i(\tilde T-\tau) H_S } \,a\, e^{i(\tilde T-\tau)  H_S }\right)\tilde\Pi_{|C}.
\end{equation}
Since $e^{-i(\tilde T-\tau) H_S }$ is unitary on  $\tilde{\mathcal{H}}_{\rm kin}$ and $\tilde\Pi_{|C}$ is bounded, we see that $O_C^\tau(a)\in\mathcal{B}(\tilde{\mathcal{H}}_{\rm kin})$ if and only if $a$ is bounded. But if $O_C^\tau(a)$ is bounded on $\tilde{\mathcal{H}}_{\rm kin}$, it clearly is also bounded in the subspace $\mathcal{H}_{\rm kin}=\tilde\Pi_{|C}(\tilde{\mathcal{H}}_{\rm kin})$ on which it has support. Since the argument holds for any $a\in\mathcal{B}(\mathcal{H}_S)$ and since the unitarity of $\mathcal{R}_C$ entails $\mathcal{B}(\mathcal{H}_{\rm phys})\simeq\Pi_{|C}\mathcal{B}(\mathcal{H}_S)\Pi_{|C}$, this proves the claim.

In the case of a clock with energy-independent $H_C$-degeneracy (except possibly in a set of measure zero), this argument holds per $\lambda$-superselection sector, cf.~Sec.~\ref{sssec_degclock2}.

\subsection{Reduced Hilbert space decomposition for non-ideal clocks}\label{app_non-idealclockTPS}

Here, we explain the decomposition in Eq.~\eqref{non-idealdecompeq} of the main text, establishing that non-ideal clocks induce a direct sum/integral of TPSs on $\mathcal{H}_{\rm phys}$.

Consider the projector $\Pi_{|1}$ in Eq.~\eqref{pi1}. It restricts the eigenvalues of $H_2+H_S$ on the physical Hilbert space to
\begin{equation}\label{eq: eval constraint 1}
    -\epsilon^{\text{max}}_1 \leq E_S + E_2  \leq -\epsilon^{\text{min}}_1.
\end{equation}
Let us first investigate which system energies are compatible with \textit{any} eigenvalue of $H_2$.  Employing the extrema of $\sigma_2$ in the bounds \eqref{eq: eval constraint 1} implies that for system energies in the range
\begin{equation}\label{stringent2}
    -\epsilon^{\text{max}}_1 - \epsilon^{\text{min}}_2  \leq E_S \leq -\epsilon^{\text{min}}_1 - \epsilon^{\text{max}}_2 ,
\end{equation}
any eigenvalue of $H_2$ can occur and remain consistent the constraint.  This range is nonempty if $\sigma_1$ contains $\sigma_2$.
Denoting by $\Pi_S$ the projector onto this energy range, we have that the tensor product $\mathcal{H}_{C_2}\otimes\Pi_S(\mathcal{H}_S)$ is a subspace of $\mathcal{H}_{|1}$.

Next, let us consider the energy range
\begin{equation}\label{range2}
    -\epsilon^{\text{min}}_1 - \epsilon^{\text{max}}_2 <E_S\leq -\epsilon^{\text{min}}_1 - \epsilon^{\text{min}}_2 .
\end{equation}
System energies cannot be larger than the upper bound on the right while still satisfying $E_S+E_2\in-\sigma_1$ for \emph{some} $E_2\in\sigma_2$. For any fixed $E_S$, the compatible $C_2$ energies are 
\begin{equation}\label{range3}
    \epsilon^{\rm min}_2\leq E_2\leq -\epsilon^{\rm min}_1-E_S.
\end{equation}
For any $E_S$ in the range \eqref{stringent2}, this imposes no restriction on $E_2$. However, for $E_S$ in the range \eqref{range2} this is different. Parametrizing such system energies as $E_S=-\epsilon^{\text{min}}_1 - \epsilon^{\text{max}}_2 +\delta$ with $0<\delta\leq+\epsilon_2^{\rm max}-\epsilon_2^{\rm min}$, Eq.~\eqref{range3} entails
\begin{equation}
    \epsilon^{\rm min}_2\leq E_2\leq \epsilon_2^{\rm max}-\delta,
\end{equation}
and so the admissible $C_2$ energies becomes squashed from above in an $E_S$-dependent manner. Assuming $\sigma_S:=\rm{Spec}(H_S)$ to be purely continuous, we can thus write these contributions to $\mathcal{H}_{|1}$ as 
\begin{equation}
    \mathcal{H}_{SC_2}^+:=\int^{-\epsilon^{\rm min}_1-\epsilon_2^{\rm min}}_{-\epsilon_1^{\rm min}-\epsilon_2^{\rm max}}\dd{E}_S\Pi_{C_2}^{[\epsilon_2^{\rm min},-\epsilon_1^{\rm min}-E_S]}\left(\mathcal{H}_{C_2}\right)\otimes\mathcal{H}_S^{E_S},
\end{equation}
where $\Pi_{C_2}^{[\epsilon_2^{\rm min},-\epsilon_1^{\rm min}-E_S]}$ is the projector onto the subspace of $\mathcal{H}_{C_2}$ compatible with Eq.~\eqref{range3} and $\mathcal{H}_S^{E_S}$ is the improper subspace spanned by states with fixed energy $E_S$. When $H_S$ is degenerate, as will typically be the case for modular Hamiltonians of QFTs with multiple species, the tensor product under the integral will be nontrivial. Hence, $\mathcal{H}_{SC_2}^+$ is a direct integral of TPSs. When $\sigma_S$ is discrete, this will turn into a direct sum instead.

Similarly, one treats system energies in the range
\begin{eqnarray}
    -\epsilon_1^{\rm max}-\epsilon_2^{\rm max}\leq E_S<-\epsilon_1^{\rm max}-\epsilon_2^{\rm min},
\end{eqnarray}
resulting in another direct integral/sum of TPSs 
\begin{equation}
    \mathcal{H}_{SC_2}^-:=\int^{-\epsilon^{\rm max}_1-\epsilon_2^{\rm min}}_{-\epsilon_1^{\rm max}-\epsilon_2^{\rm max}}\dd{E}_S\Pi_{C_2}^{[-\epsilon_1^{\rm max}-E_S,\epsilon_2^{\rm max}]}\left(\mathcal{H}_{C_2}\right)\otimes\mathcal{H}_S^{E_S}.
\end{equation} 
This establishes the decomposition in Eq.~\eqref{non-idealdecompeq}.

\section{Constraint implementation via the \texorpdfstring{$T$}{T}-map}\label{app_doublecheck}

Here we briefly check that the $T$-map in Eq.~\eqref{eq:isom} implements the constraint Eq.~\eqref{2clockconstraint2} (with relative minus sign between the clock Hamiltonians as in Eq.~\eqref{extcon}) also for non-ideal clocks as used in the main body (\cite[Sec.~4.2]{Chandrasekaran:2022cip} used ideal clocks in their implementation). 

We have
\begin{eqnarray}
    T_2H\ket{\psi}&=&\int_\mathbb{R}\dd{t}e^{-it(H_S+H_1)}\ket{H\psi(t)}=\int_\mathbb{R}\dd{t}e^{-it(H_S+H_1)}\left((H_S+H_1)\ket{\psi(t)}-\ket{H_2\psi(t)}\right)\nonumber\\
    &=&-\int_\mathbb{R}\dd{t} e^{-it(H_S+H_1)}\left(i\partial_{t}\ket{\psi(t)}+\ket{H_2\psi(t)}\right)\,,
\end{eqnarray}
where we used partial integration and that $\lim_{t\to\pm\infty}\ket{\psi(t)}=0$, coming from $\ket{\psi}\in\mathcal{H}_{\rm kin}$ being square integrable. Now
\begin{eqnarray}
    \ket{H_2\psi(t)}=\bra{t}H_2\ket{\psi}=-i\partial_{t}\ket{\psi(t)}\,,
\end{eqnarray}
where we invoked Eq.~\eqref{clockstates}. Hence, reduced states solve the constraint in the sense that
\begin{eqnarray}
    T_2H\ket{\psi}=0\,,\qquad\forall\,\ket{\psi}\in\mathcal{H}_{\rm kin}\,.
\end{eqnarray}

\section{Details of derivation of density operator}
\label{Appendix: density operator details}

We first demonstrate that
\begin{equation}
    S_{\psi_{R_{\bar{i}}}} (\ket{\varphi}\otimes\bra{\zeta}) = e^{\beta H_{R_{\bar{i}}}/2} \ket{\zeta} \otimes \bra{\varphi}e^{-\beta H_{R_{\bar{i}}}/2}.
\end{equation}
is the Tomita operator of $\ket*{\psi_{R_{\bar{i}}}}$. For any $A\in \mathcal{B}(\mathcal{H}_{R_{\bar{i}}})$, one has
\begin{align}
    S_{\psi_{R_{\bar{i}}}} A \ket*{\psi_{R_{\bar{i}}}} &= S_{\psi_{R_{\bar{i}}}} A \bigotimes_{C_j\in R_{\bar{i}}}\int_{\sigma_j} \dd{\epsilon_j} \ket{\epsilon_j}\otimes \bra{\epsilon_j} e^{-\beta H_{R_{\bar{i}}}/2} \\
    &= \bigotimes_{C_j\in R_{\bar{i}}}\int_{\sigma_j} \dd{\epsilon_j} \ket{\epsilon_j}\otimes \bra{\epsilon_j} A^\dagger e^{-\beta H_{R_{\bar{i}}}/2}\\
    &= \bigotimes_{C_j\in R_{\bar{i}}}\int_{\sigma_j} \dd{\epsilon_j} \int_{\sigma_j} \dd{\epsilon_j'}\ket{\epsilon_j}\otimes \bra{\epsilon_j} A^\dagger \ket*{\epsilon_j'}\bra*{\epsilon_j'}e^{-\beta H_{R_{\bar{i}}}/2}\\
    &= \bigotimes_{C_j\in R_{\bar{i}}}\int_{\sigma_j} \dd{\epsilon_j} \int_{\sigma_j} \dd{\epsilon_j'}\ket{\epsilon_j} \bra{\epsilon_j} A^\dagger \ket*{\epsilon_j'}\otimes\bra*{\epsilon_j'}e^{-\beta H_{R_{\bar{i}}}/2}\\
    &= \bigotimes_{C_j\in R_{\bar{i}}} \int_{\sigma_j} \dd{\epsilon_j'}A^\dagger \ket*{\epsilon_j'}\otimes\bra*{\epsilon_j'}e^{-\beta H_{R_{\bar{i}}}/2}\\
    &= A^\dagger \ket*{\psi_{R_{\bar{i}}}} ,
\end{align}
as required.

Next, let us show that the state $\ket{\Psi}$ defined in~\eqref{Equation: Psi tracial state} is separating for the algebra $\mathcal{A}_{SR}^H$. Consider the map 
\begin{equation}
    \gamma: \mathcal{H}_S \otimes  \bigotimes_{C_j\in R_{\bar{i}}} \big(\mathcal{H}_j\otimes \mathcal{H}_j^* \big)\otimes \mathcal{H}_i \to \mathcal{H}_S \otimes  \bigotimes_{C_j\in R_{\bar{i}}} \big(\mathcal{H}_j\otimes \mathcal{H}_j^* \big)\otimes \mathcal{B}(\mathcal{H}_i) 
\end{equation}
defined by
\begin{equation}
    \gamma: \ket{\alpha} \mapsto \int_{-\infty}^\infty \dd{t} \exp(-i(H_S + H_{R_{\bar{i}}} - H_{R_{\bar{i}}^*} + H_i)t) \ket{\alpha} \bra{t}_i e^{\beta H_i/2},
\end{equation}
where $H_{R_{\bar{i}}} = \sum_{C_j\in R_{\bar{i}}} H_j$, and the ${}^*$ in $H_{R_{\bar{i}}^*}$ denotes that it is the same operator but acting on the second copy of the frames $R_{\bar{i}}$, i.e.\ on $\mathcal{H}_j^*$. Then one may confirm, using the resolution of the identity in terms of the clock states $\ket{t}_i$, and the fact that $(H_S + H_{R_{\bar{i}}} - H_{R_{\bar{i}}^*})\ket*{\psi_{SR_{\bar{i}}}} = 0$, 
\begin{equation}
    \gamma A \ket{\Psi} = A (\ket*{\psi_{SR_{\bar{i}}}}\otimes \mathds{1}_{C_i}),
\end{equation}
Since $\ket*{\psi_{SR_{\bar{i}}}}$ is separating, we can have $\gamma A\ket{\Psi}=0$ only if $A=0$, which implies $\ket{\Psi}$ is separating. 

Let us also show that the state $\ket{\Psi}$ is cyclic over the image $\widetilde{\mathcal{H}}_{SR_{\bar{i}}R_{\bar{i}}^*C_i}$ of the projection operator $\widetilde\Pi$ defined in~\eqref{Equation: trace projection widetilde Pi}. Consider a general operator in the algebra, which may be written in the form 
\begin{equation}
    A = \int_{-\infty}^\infty \dd{t} U_i(t) O_i^0(a(t)) 
e^{\beta H_i/2}\in \mathcal{A}_{SR}^H, 
\end{equation}
with $a:\RR\to \mathcal{A}_S\otimes\mathcal{B}(\mathcal{H}_{R_{\bar{i}}})$. Then we have 
\begin{align}
A\ket{\Psi}
&=\int_{-\infty}^{\infty} \dd\tau \int_{-\infty}^{\infty} \dd t~
e^{-iH\tau} a(t)\otimes \ket{t}_i\bra{0}_i e^{iH\tau}\ket*{\psi_{SR_{\bar{i}}}}\otimes\ket{0}_i\\
&=\int_{-\infty}^{\infty} \dd\tau \int_{-\infty}^{\infty} \dd t~
\exp(-i(H_S+H_{R_{\bar{i}}}-H_{R^*_{\bar{i}}}+H_i)\tau) \bra{\tau}\ket{0}_i a(t) \ket*{\psi_{SR_{\bar{i}}}}\otimes\ket{t}_i\\
&=\widetilde{\Pi}\int_{-\infty}^{\infty} \dd t ~ a(t) \ket*{\psi_{SR_{\bar{i}}}}\otimes\ket{t}_i,
\end{align}
where in the second line we used $\bra{0}_i e^{iH\tau}\ket*{\psi_{SR_{\bar{i}}}} = \bra{\tau}_i e^{i H_{R^*_{\bar{i}}}\tau} \ket*{\psi_{SR_{\bar{i}}}}$ since $(H_S+H_{R_{\bar{i}}}-H_{R^*_{\bar{i}}})\ket*{\psi_{SR_{\bar{i}}}}=0$.
Since $\ket*{\psi_{SR_{\bar{i}}}}$ is cyclic for $\mathcal{A}_S\otimes \mathcal{B}(\mathcal{H}_{R_{\bar{i}}})$, we can vary over all states in $\widetilde{\mathcal{H}}_{SR_{\bar{i}}R_{\bar{i}}^*C_i}$ in this way, so $\ket{\Psi}$ is cyclic over this subspace.

Next, we demonstrate that~\eqref{Equation: Psi tracial Tomita} is the Tomita operator of $\ket{\Psi}$: 
\begin{align}
    S_\Psi A\ket{\Psi} &= \int_{-\infty}^\infty\dd{t} e^{\beta H_i/2} \ket{t}_i\exp(-i(H_S + H_{R_{\bar{i}}} - H_{R_{\bar{i}}^*})t)S_{\psi_{SR_{\bar{i}}}} \bra{-t}_i e^{-\beta H_i/2} A e^{-\beta H_i/2}\ket{0}_i \ket*{\psi_{SR_{\bar{i}}}} \\
    &= \int_{-\infty}^\infty\dd{t} e^{\beta H_i/2} \ket{t}_i\exp(-i(H_S + H_{R_{\bar{i}}} - H_{R_{\bar{i}}^*})t)\bra{0}_i e^{-\beta H_i/2} A^\dagger e^{-\beta H_i/2}\ket{-t}_i \ket*{\psi_{SR_{\bar{i}}}} \\
    &= \int_{-\infty}^\infty\dd{t} e^{\beta H_i/2} \ket{t}_i\bra{t}_i e^{-\beta H_i/2} A^\dagger e^{-\beta H_i/2}\ket{0}_i \ket*{\psi_{SR_{\bar{i}}}} \\
    &= A^\dagger \ket{\Psi},
\end{align}
as required. We then have the modular operator
\begin{align}
    \Delta_\Psi &= S_\Psi^\dagger S_\Psi\\
    &=\begin{multlined}[t]
        \int_{-\infty}^\infty\dd{t}\int_{-\infty}^\infty\dd{t'} e^{-\beta H_i/2} \ket{-t}_i S_{\psi_{SR_{\bar{i}}}}^\dagger \\
        \exp(i(H_S + H_{R_{\bar{i}}} - H_{R_{\bar{i}}^*})(t-t')) \bra{t}_i e^{\beta H_i}\ket{t'}_i S_{\psi_{SR_{\bar{i}}}} \bra{-t'}_ie^{-\beta H_i/2}
    \end{multlined}\\
    &= \begin{multlined}[t]
        \int_{-\infty}^\infty\dd{t}\int_{-\infty}^\infty\dd{t'} e^{-\beta H_i/2} \ket{-t}_i \Delta_{\psi_{SR_{\bar{i}}}} \\
        \exp(i(H_S + H_{R_{\bar{i}}} - H_{R_{\bar{i}}^*})(t-t')) \bra{t'}_i e^{\beta H_i}\ket{t}_i \bra{-t'}_ie^{-\beta H_i/2}
    \end{multlined}\\
    &= \int_{-\infty}^\infty \dd{t}\int_{-\infty}^\infty\dd{t'} \exp(-i(H_S + H_{R_{\bar{i}}} - H_{R_{\bar{i}}^*}+H_i)t)e^{\beta H_i/2}\ket{-t'}_i\braket{t}{0}_i \bra{-t'}_i e^{-\beta H_i/2} \\
    &= \widetilde{\Pi}.
\end{align}
where in the third line we used that 
\begin{equation}
    \exp(i(H_S + H_{R_{\bar{i}}} - H_{R_{\bar{i}}^*})(t-t'))S_{\psi_{SR_{\bar{i}}}} \exp(-i(H_S + H_{R_{\bar{i}}} - H_{R_{\bar{i}}^*})(t-t')) = S_{\psi_{SR_{\bar{i}}}},
\end{equation}
and in the fourth line we changed variables $t \to  t'-t - i\beta$. Thus $\Delta_\Psi$ is the identity (on $\widetilde{\mathcal{H}}_{SR_{\bar{i}}R_{\bar{i}}^*C_i}$).

Next, we show that~\eqref{Equation: phi Psi relative tomita} is the relative Tomita operator from $\ket{\Psi}$ to a physical state $\Ket{\phi}$ with $\ket*{\phi_{|i}(\tau)} = \mathcal{R}_i(\tau)\Ket{\phi}$.
This may be confirmed as follows:
\begin{align}
    S_{\phi|\Psi}A\ket{\Psi} &= \Pi_{\text{phys}}\int_{-\infty}^\infty\dd{t}\ket{t}_i S_{\phi_{|i}(0)|\psi_{SR_{\bar{i}}}}\bra{0}_i A \ket{t}_i\ket*{\psi_{SR_{\bar{i}}}} \\
    &= \Pi_{\text{phys}} \int_{-\infty}^\infty\dd{t}\ket{t}_i \bra{t}_i A^\dagger \ket{0}_i\ket*{\phi_{|i}(0)} \\
    &= \Pi_{\text{phys}} A^\dagger \ket*{\phi_{|i}(0)}\ket{0}_i \\
    &= r(A)^\dagger \Pi_{\text{phys}} \ket*{\phi_{|i}(0)}\ket{0}_i = r(A)^\dagger \Ket{\phi}.
\end{align}
Using the identity
\begin{equation}
    \exp\Big(-i(H_S+\sum_{j\ne i} H_j)t\Big) S_{\phi_{|i}(0)|\psi_{SR_{\bar{i}}}} \exp(i(H_S+H_{R_{\bar{i}}}-H_{R_{\bar{i}}^*})t) = S_{\phi_{|i}(t)|\psi_{SR_{\bar{i}}}},
\end{equation}
we thus have the relative modular operator (na\"ively in computing this object we might get a $\Pi_{\text{phys}}^2$, but the target space of the relative Tomita operator is the physical Hilbert space, whose inner product we should use when contracting $S_{\phi|\Psi}^\dagger$ with $S_{\phi|\Psi}$ -- this removes one factor of $\Pi_{\text{phys}}$, as described in Sec.~\ref{sseHphys})
\begin{align}
    \Delta_{\phi|\Psi} &= S_{\phi|\Psi}^\dagger S_{\phi|\Psi} \\
    &= 
    \begin{multlined}[t]
        \int_{-\infty}^\infty\dd{t}\exp(i(H_S+H_{R_{\bar{i}}}-H_{R_{\bar{i}}^*}+H_i)(t-i\beta/2))\ket{0}_i S_{\phi_{|i}(0)|\psi_{SR_{\bar{i}}}}^\dagger\bra{t}_i \\
        \int_{-\infty}^\infty \dd{t'}\exp(-i(H_S+\sum_{j=1}^n H_{j})t') \\
        \int_{-\infty}^\infty\dd{t''}\ket{t''}_i S_{\phi_{|i}(0)|\psi_{SR_{\bar{i}}}}\bra{0}_i \exp(-i(H_S+H_{R_{\bar{i}}}-H_{R_{\bar{i}}^*}+H_i)(t''+i\beta/2))
    \end{multlined}\\
    &= 
    \begin{multlined}[t]
        e^{\beta H_i/2}\int_{-\infty}^\infty\dd{t}\int_{-\infty}^\infty\dd{t'}\int_{-\infty}^\infty\dd{t''}\exp(i(H_S+H_{R_{\bar{i}}}-H_{R_{\bar{i}}^*})(t-t'))\ket{-t}_i \bra{t''}\ket{t-t'}_i  \\
        \Delta_{\psi_{SR_{\bar{i}}}}^{-1/2} S_{\phi_{|i}(-t')|\psi_{SR_{\bar{i}}}}^\dagger S_{\phi_{|i}(0)|\psi_{SR_{\bar{i}}}}\Delta_{\psi_{SR_{\bar{i}}}}^{-1/2}\bra{-t''}_i \exp(-i(H_S+H_{R_{\bar{i}}}-H_{R_{\bar{i}}^*})t'')e^{\beta H_i/2}
    \end{multlined}\\
    &= 
    \begin{multlined}[t]
        e^{\beta H_i/2}\int_{-\infty}^\infty\dd{t}\exp(i(H_S+H_{R_{\bar{i}}}-H_{R_{\bar{i}}^*}+H_i)t)\bra{0}\ket{t}_i\int_{-\infty}^\infty\dd{t'}\int_{-\infty}^\infty\dd{t''}  \\
        \exp(-i(H_S+H_{R_{\bar{i}}}-H_{R_{\bar{i}}^*})t'')\Delta_{\psi_{SR_{\bar{i}}}}^{-1/2} S_{\phi_{|i}(t')|\psi_{SR_{\bar{i}}}}^\dagger S_{\phi_{|i}(0)|\psi_{SR_{\bar{i}}}}\Delta_{\psi_{SR_{\bar{i}}}}^{-1/2}\\
        \otimes e^{-iH_it'}\ket{t''}_i\bra{t''}_i \exp(i(H_S+H_{R_{\bar{i}}}-H_{R_{\bar{i}}^*})t'')e^{\beta H_i/2}
    \end{multlined}\\
    &= \widetilde{\Pi} e^{\beta H_i/2}\int \dd{t'} e^{-iH_it'} O_i^0(\Delta_{\psi_{SR_{\bar{i}}}}^{-1/2} S_{\phi_{|i}(t')|\psi_{SR_{\bar{i}}}}^\dagger S_{\phi_{|i}(0)|\psi_{SR_{\bar{i}}}}\Delta_{\psi_{SR_{\bar{i}}}}^{-1/2}) e^{\beta H_i/2}.
\end{align}
In the fourth line we changed variables $t\to t+t'+t''$, $t'\to -t'$, and $t''\to -t''$. Note that we can get rid of the $\widetilde{\Pi}$ at the front, if we treat this as an operator acting on $\widetilde{\mathcal{H}}_{SR_{\bar{i}}R_{\bar{i}}^*C_i}$. To summarise the above, we have
\begin{equation}
    \Delta_{\phi|\Psi} = e^{\beta H_i/2}\int_{-\infty}^\infty \dd{t} e^{-iH_it} O_i^\tau(\Delta_{\psi_{SR_{\bar{i}}}}^{-1/2} S_{\phi_{|i}(\tau+t)|\psi_{SR_{\bar{i}}}}^\dagger S_{\phi_{|i}(\tau)|\psi_{SR_{\bar{i}}}}\Delta_{\psi_{SR_{\bar{i}}}}^{-1/2}) e^{\beta H_i/2},
\end{equation}
where we have also used covariance of the dressed observable to include an additional $\tau$ parametrisation.

We have
\begin{equation}
    \Delta_{\psi_{SR_{\bar{i}}}}^{-1/2} S_{\phi_{|i}(t')|\psi_{SR_{\bar{i}}}}^\dagger S_{\phi_{|i}(0)|\psi_{SR_{\bar{i}}}}\Delta_{\psi_{SR_{\bar{i}}}}^{-1/2} \in \mathcal{A}_S\otimes\mathcal{B}(\mathcal{H}_{R_{\bar{i}}}),
    \label{Equation: modular combination in the algebra}
\end{equation}
which is what justifies writing the term in the integrand above as a dressed observable. To see this, suppose $a,b,c\in\mathcal{A}_S\otimes\mathcal{B}(\mathcal{H}_{R_{\bar{i}}})$. Then 
\begin{align}
    \MoveEqLeft\bra*{\psi_{SR_{\bar{i}}}} a S_{\psi_{SR_{\bar{i}}}}^\dagger S_{\phi_{|i}(t')|\psi_{SR_{\bar{i}}}}^\dagger S_{\phi_{|i}(0)|\psi_{SR_{\bar{i}}}}S_{\psi_{SR_{\bar{i}}}} bc \ket*{\psi_{SR_{\bar{i}}}} \\
    &= \bra*{\psi_{SR_{\bar{i}}}} bc S_{\phi_{|i}(0)|\psi_{SR_{\bar{i}}}}^\dagger S_{\phi_{|i}(t')|\psi_{SR_{\bar{i}}}}a\ket*{\psi_{SR_{\bar{i}}}} \\
    &= \bra*{\phi_{|i}(t')}abc\ket*{\phi_{|i}(0)} \\
    &= \bra*{\psi_{SR_{\bar{i}}}} ab S_{\psi_{SR_{\bar{i}}}}^\dagger S_{\phi_{|i}(t')|\psi_{SR_{\bar{i}}}}^\dagger S_{\phi_{|i}(0)|\psi_{SR_{\bar{i}}}}S_{\psi_{SR_{\bar{i}}}} c \ket*{\psi_{SR_{\bar{i}}}}.
\end{align}
Since $\ket*{\psi_{SR_{\bar{i}}}}$ is cyclic, we may conclude that 
\begin{equation}
    S_{\psi_{SR_{\bar{i}}}}^\dagger S_{\phi_{|i}(t')|\psi_{SR_{\bar{i}}}}^\dagger S_{\phi_{|i}(0)|\psi_{SR_{\bar{i}}}}S_{\psi_{SR_{\bar{i}}}} \in (\mathcal{A}\otimes\mathcal{B}(\mathcal{H}_{R_{\bar{i}}}))'.
\end{equation}
But we have
\begin{equation}
    \Delta_{\psi_{SR_{\bar{i}}}}^{-1/2} S_{\phi_{|i}(t')|\psi_{SR_{\bar{i}}}}^\dagger S_{\phi_{|i}(0)|\psi_{SR_{\bar{i}}}}\Delta_{\psi_{SR_{\bar{i}}}}^{-1/2} = J_{\psi_{SR_{\bar{i}}}}\qty\big(S_{\psi_{SR_{\bar{i}}}}^\dagger S_{\phi_{|i}(t')|\psi_{SR_{\bar{i}}}}^\dagger S_{\phi_{|i}(0)|\psi_{SR_{\bar{i}}}}S_{\psi_{SR_{\bar{i}}}})J_{\psi_{SR_{\bar{i}}}},
\end{equation}
where $J_{\psi_{SR_{\bar{i}}}}$ is the modular conjugation of $\ket*{\psi_{SR_{\bar{i}}}}$. Since the modular conjugation gives an isomorphism between the algebra and its commutant, we have~\eqref{Equation: modular combination in the algebra}.

Within the perspective of a clock $C_i\in R$, at time $\tau$, the density operator is given by~\eqref{eq:reducedDensityIntermed}. We can show this as follows:
\begin{align}
    \rho_{\phi|i}(\tau) &= \mathcal{R}_i(\tau)\rho_\phi \mathcal{R}_i(\tau)^\dagger  \\
    &= 
    \begin{multlined}[t]
    e^{-S_{0,R}}e^{-\beta(H_S+\sum_{j\ne i}H_j)/2} \int_{-\infty}^\infty \dd{t} e^{i(H_S+\sum_{j\ne i} H_j)t}\\
    \Pi_{|i}\Delta_{\psi_{SR_{\bar{i}}}}^{-1/2} S_{\phi_{|i}(\tau+t)|\psi_{SR_{\bar{i}}}}^\dagger S_{\phi_{|i}(\tau)|\psi_{SR_{\bar{i}}}}\Delta_{\psi_{SR_{\bar{i}}}}^{-1/2}\Pi_{|i} e^{-\beta(H_S+\sum_{j\ne i}H_j)/2} 
    \end{multlined}\\
    &= e^{-S_{0,R}-\beta (H_{R^c}+H_{R_{\bar{i}}^*})} \int_{-\infty}^\infty \dd{t} e^{i(H_S+\sum_{j\ne i}H_j)t} \Pi_{|i} S_{\phi_{|i}(\tau+t)|\psi_{SR_{\bar{i}}}}^\dagger S_{\phi_{|i}(\tau)|\psi_{SR_{\bar{i}}}}\Pi_{|i}.
\end{align}
Note that
\begin{align}
    S_{\phi_{|i}(\tau+t)|\psi_{SR_{\bar{i}}}}^\dagger S_{\phi_{|i}(\tau)|\psi_{SR_{\bar{i}}}}\Pi_{|i} &= \int_{-\infty}^\infty\dd{t'} S_{\phi_{|i}(\tau+t)|\psi_{SR_{\bar{i}}}}^\dagger S_{\phi_{|i}(\tau)|\psi_{SR_{\bar{i}}}} e^{-i(H_S+\sum_{j\ne i}H_j)t'} \braket{0}{t'}_i\\
    &= \int_{-\infty}^\infty\dd{t'} e^{-i(H_S+\sum_{j\ne i}H_j)t'}S_{\phi_{|i}(\tau+t-t')|\psi_{SR_{\bar{i}}}}^\dagger S_{\braket{t'}{0}_i \phi_{|i}(\tau-t')|\psi_{SR_{\bar{i}}}}.
\end{align}
Using this in the formula for $\rho_{\phi|i}(\tau)$, and changing variables $t \to t+t'$, we get
\begin{equation}
    \rho_{\phi|i}(\tau) = e^{-S_{0,R}-\beta (H_{R^c}+H_{R_{\bar{i}}^*})} \int_{-\infty}^\infty \dd{t} e^{i(H_S+\sum_{j\ne i}H_j)t} \Pi_{|i} S_{\phi_{|i}(\tau+t)|\psi_{SR_{\bar{i}}}}^\dagger 
    \int \dd{t'} S_{\braket{t'}{0}_i\phi_{|i}(\tau-t')|\psi_{SR_{\bar{i}}}}.
\end{equation}
But
\begin{equation}
    \int \dd{t'} S_{\braket{t'}{0}_i\phi_{|i}(\tau-t')|\psi_{SR_{\bar{i}}}} = S_{\Pi_{|i}\phi_{|i}(\tau)|\psi_{SR_{\bar{i}}}} = S_{\phi_{|i}(\tau)|\psi_{SR_{\bar{i}}}},
\end{equation}
since $\phi_i(\tau)$ is in the image of $\Pi_{|i}$. So
\begin{equation}
    \rho_{\phi|i}(\tau) = \Pi_{|i} e^{-S_{0,R}-\beta (H_{R^c}+H_{R_{\bar{i}}^*})} \int_{-\infty}^\infty\dd{t} e^{i(H_S+\sum_{j\ne i}H_j)t} S_{\phi_{|i}(\tau+t)|\psi_{SR_{\bar{i}}}}^\dagger S_{\phi_{|i}(\tau)|\psi_{SR_{\bar{i}}}},
\end{equation}
as required.

Finally, suppose we decompose the reduced state $\ket*{\phi_{|i}(\tau)}$ as in~\eqref{Equation: reduced perspective bipartite decomposition}. Then one has
\begin{equation}
    S_{\phi_{|i}(\tau)|\psi_{SR_{\bar{i}}}} = \sum_I S_{\phi_S^I(\tau)|\psi_S} \otimes S_{\tilde\phi^I(\tau)|\psi_{R_{\bar{i}}}},
\end{equation}
so
\begin{equation}
    S_{\phi_{|i}(\tau+t)|\psi_{SR_{\bar{i}}}}^\dagger S_{\phi_{|i}(\tau)|\psi_{SR_{\bar{i}}}} = \sum_{I,J} S_{\phi_S^I(\tau+t)|\psi_S}^\dagger S_{\phi_S^J(\tau)|\psi_S} \otimes S_{\tilde\phi^I(\tau+t)|\psi_{R_{\bar{i}}}}^\dagger S_{\tilde\phi^J(\tau)|\psi_{R_{\bar{i}}}} 
\end{equation}
Using
\begin{equation}
    \bra*{\psi_{R_{\bar{i}}}}a S_{\tilde\phi^I(\tau+t)|\psi_{R_{\bar{i}}}}^\dagger S_{\tilde\phi^J(\tau)|\psi_{R_{\bar{i}}}} b \ket*{\psi_{R_{\bar{i}}}} = \bra*{\tilde\phi^J(\tau)}ba\ket*{\tilde\phi^I(\tau+t)},
\end{equation}
for any $a,b\in\mathcal{B}(\mathcal{H}_{R_{\bar{i}}})$, we may identify
\begin{equation}
    S_{\tilde\phi^I(\tau+t)|\psi_{R_{\bar{i}}}}^\dagger S_{\tilde\phi^J(\tau)|\psi_{R_{\bar{i}}}} = e^{\beta H_{R_{\bar{i}}^*}} \tr_{R^c}(\ket*{\tilde\phi^I(\tau+t)}\bra*{\tilde\phi^J(\tau)}).
\end{equation}
Substituting this into the formula for the density matrix, one has
\begin{equation}
    \rho_{\phi|i}(\tau) = \Pi_{|i} e^{-S_{0,R}-\beta H_{R^c}} \int_{-\infty}^\infty\dd{t} e^{i(H_S+\sum_{j\ne i}H_j)t} \sum_{I,J} S_{\phi_S^I(\tau+t)|\psi_S}^\dagger S_{\phi_S^J(\tau)|\psi_S}\otimes \tr_{R^c}(\ket*{\tilde\phi^I(\tau+t)}\bra*{\tilde\phi^J(\tau)}),\label{eq: appendix density matrix}
\end{equation}
as reported in the main text, equation \eqref{eq:reducedDensityOperator}.  A related expression will prove useful when the state decomposition between QFT and frame factors is expressed in terms of a continuous parameter instead of a discrete sum: 
\begin{equation}\begin{split}\label{eq: continuous state decomp, appendix}
    \ket*{\phi_{|i}(t)}&:=\int_{-\infty}^{\infty} \dd{\mu} \ket*{\phi^{\mu}_S(t)}\otimes\ket*{\tilde{\phi}^{\mu}_{R_{\bar{i}}{R^c}}(t)} ~\in \mathcal{H}_{|R_i},
\end{split}\end{equation}  
The direct analogue of equation \eqref{eq: appendix density matrix} for this state is
\begin{equation}\begin{split}\label{eq: density op SR_i}
  \rho_{\phi|i}&
  \begin{multlined}[t]
      =\Pi_{|i}e^{-S_{0,R}-\beta H_{{R^c}}}\int_{-\infty}^\infty\dd{t}\dd{\mu}\dd{\nu}~  e^{i(H_S+H_{R_{\bar{i}}}+H_{R^c})t}
      S_{\phi^\nu_S(t)|\psi_{S}}^\dagger S_{\phi^\mu_S(0)|\psi_{S}}\\
      \otimes
      \Tr_{{R^c}}
        \left(
        \ket*{\tilde{\phi}^{\nu}_{R_{\bar{i}}{R^c}}(t)}
        \bra*{\tilde{\phi}^{\mu}_{R_{\bar{i}}{R^c}}(0)}
        \right)
  \end{multlined}
  \\
  &\begin{multlined}[t]
  =\Pi_{|i}e^{-S_{0,R}-\beta H_{{R^c}}}\int_{-\infty}^\infty\dd{\mu}\dd{\nu}~ 
  S_{\phi^\nu_S(0)|\psi_{S}}^\dagger \\
  \qquad\qquad\bigg(
\int_{-\infty}^\infty\dd{t}
  e^{i(H_S-H_{R_{\bar{i}}}-H_{R^c})t}
  \Tr_{{R^c}}
    \left(
    \ket*{\tilde{\phi}^{\mu}_{R_{\bar{i}}{R^c}}(0)}
    \bra*{\tilde{\phi}^{\nu}_{R_{\bar{i}}{R^c}}(t)}
    \right)
  \bigg)S_{\phi^\mu_S(0)|\psi_{S}}.
  \end{multlined}
\end{split}\end{equation}
In the second expression we have isolated the $t$-integral between the relative Tomita operators, which are taken to act as complex conjugation in the energy bases of the frames.  Finally, if we are interested in cases where the set $R_{\bar{i}}$ is empty (so that the algebra considered is just the QFT subregion algebra dressed by a single frame), then the trace in the above expression becomes an inner product, and we have
\begin{equation}\begin{split}\label{eq: rho without R_ibar}
  &\rho_{\phi|i}=
  \Pi_{|i}e^{-S_{0,R}-\beta H_{{R^c}}}\int_{-\infty}^\infty\dd{\mu}\dd{\nu}~ 
  S_{\phi^\nu_S(0)|\psi_{S}}^\dagger \bigg(
\int_{-\infty}^\infty\dd{t}
  e^{i(H_S-H_{R^c})t}
    \bra{\tilde{\phi}^{\nu}_{R^c}(t)}\ket{\tilde{\phi}^{\mu}_{R^c}(0)}
  \bigg)S_{\phi^\mu_S(0)|\psi_{S}}.
\end{split}\end{equation}
This expression will be referred to in several computations.

\section{Details of semiclassical approximations}
\label{Appendix: semiclassical approximation details}

Let us write $\rho_{\phi|i}(\tau)=\Pi_{|i}\tilde\rho_{\phi|i}(\tau)$, where
\begin{equation}
    \tilde\rho_{\phi|i}(\tau) = e^{-S_{0,R}-\beta (H_{R^c}+H_{R_{\bar{i}}^*})} \int_{-\infty}^\infty\dd{t} e^{i(H_S+\sum_{j\ne i}H_j)t} S_{\phi_{|i}(\tau+t)|\psi_{SR_{\bar{i}}}}^\dagger S_{\phi_{|i}(\tau)|\psi_{SR_{\bar{i}}}}.
    \label{Equation: rho tilde}
\end{equation}
One may confirm that $\tilde\rho_{\phi|i}(\tau)$ is Hermitian; since $\rho_{\phi|i}(\tau)$ is also Hermitian, we have
\begin{equation}
    \Pi_{|i}\tilde\rho_{\phi|i}(\tau) = \rho_{\phi|i}(\tau) = \rho_{\phi|i}(\tau)^\dagger = \tilde\rho_{\phi|i}(\tau)\Pi_{|i},
\end{equation}
so $\tilde\rho_{\phi|i}(\tau)$ commutes with $\Pi_{|i}$, and so we may write
\begin{equation}
    \log\rho_{\phi|i}(\tau) = \Pi_{|i}\log\tilde\rho_{\phi|i}(\tau).
\end{equation}
The entropy of the density matrix may therefore be written
\begin{equation}
    S[\phi] = -\Bra{\phi}\log\rho_\phi\Ket{\phi} = -\bra*{\phi_{|i}(\tau)}\log\rho_{\phi|i}(\tau)\ket*{\phi_{|i}(\tau)} = -\bra*{\phi_{|i}(\tau)}\log\tilde\rho_{\phi|i}(\tau)\ket*{\phi_{|i}(\tau)}.
    \label{Equation: tilde entropy}
\end{equation}
We will show now that~\eqref{Equation: hat entropy} holds in the semiclassical regime.  First, note that, for any $a,b\in\mathcal{A}_{\mathcal{U}}\otimes\mathcal{B}(\mathcal{H}_{R_{\bar{i}}})$, by the definition of the relative Tomita operators, we have
\begin{equation}
    \mel*{\psi_{SR_{\bar{i}}}}{b S_{\phi_{|i}(t+\tau)|\psi_{SR_{\bar{i}}}}^\dagger S_{\phi_{|i}(\tau)|\psi_{SR_{\bar{i}}}}a}{\psi_{SR_{\bar{i}}}} = \Mel{\phi}{\mathcal{O}_{C_i}^\tau(ab)V_i(t)^\dagger}{\phi},
\end{equation}
so~\eqref{Equation: semiclassical condition 1} implies
\begin{equation}
    \abs{\mel*{\psi_{SR_{\bar{i}}}}{b S_{\phi_{|i}(t+\tau)|\psi_{SR_{\bar{i}}}}^\dagger S_{\phi_{|i}(\tau)|\psi_{SR_{\bar{i}}}}a}{\psi_{SR_{\bar{i}}}}} \ll \abs{\mel*{\psi_{SR_{\bar{i}}}}{b S_{\phi_{|i}(\tau)|\psi_{SR_{\bar{i}}}}^\dagger S_{\phi_{|i}(\tau)|\psi_{SR_{\bar{i}}}}a}{\psi_{SR_{\bar{i}}}}} \qq{if} \abs{t}>\order{\epsilon}.
\end{equation}
Since $\ket*{\psi_{SR_{\bar{i}}}}$ is cyclic, it must be that the operator $S_{\phi(t+\tau)|\psi_{SR_{\bar{i}}}}^\dagger S_{\phi(\tau)|\psi_{SR_{\bar{i}}}}$ is itself sharply peaked within $\abs{t}<\order{\epsilon}$. Thus, one may ignore contributions outside of this window in the integral in~\eqref{Equation: rho tilde}. Consider 
\begin{multline}
    \bra*{\phi_{|i}(\tau)}(\tilde\rho_{\phi|i}(\tau))^n\ket*{\phi_{|i}(\tau)} \\
    =\bra*{\phi_{|i}(\tau)}\qty(e^{-S_{0,R}-\beta (H_{R^c}+H_{R_{\bar{i}}^*})} \int_{-\infty}^\infty\dd{t} e^{i(H_S+\sum_{j\ne i}H_j)t} S_{\phi_{|i}(\tau+t)|\psi_{SR_{\bar{i}}}}^\dagger S_{\phi_{|i}(\tau)|\psi_{SR_{\bar{i}}}})^n\ket*{\phi_{|i}(\tau)}.
    \label{Equation: rho tilde n}
\end{multline}
By~\eqref{Equation: semiclassical condition 2}, for $\abs{t'}< \order{\epsilon}$, we have
\begin{multline}
    e^{-iH_St'} S_{\phi_{|i}(\tau+t)|\psi_{SR_{\bar{i}}}}^\dagger S_{\phi_{|i}(\tau)|\psi_{SR_{\bar{i}}}}e^{iH_St'} \\
    = S_{e^{-iH_St'}\phi_{|i}(\tau+t)|\psi_{SR_{\bar{i}}}}^\dagger S_{e^{-iH_St'}\phi_{|i}(\tau)|\psi_{SR_{\bar{i}}}} \approx S_{\phi_{|i}(\tau+t)|\psi_{SR_{\bar{i}}}}^\dagger S_{\phi_{|i}(\tau)|\psi_{SR_{\bar{i}}}}.
\end{multline}
Since each of the $t$ integrals that appears in~\eqref{Equation: rho tilde n} is peaked around $\abs{t}< \order{\epsilon}$, we can use this to (approximately) commute all the $e^{iH_St}$ factors to the right, where we can then use $e^{iH_St}\ket*{\phi_{|i}(\tau)}\approx e^{i\mathcal{E}t}\ket*{\phi_{|i}(\tau)}$. We thus obtain
\begin{multline}
    \bra*{\phi_{|i}(\tau)}(\tilde\rho_{\phi|i}(\tau))^n\ket*{\phi_{|i}(\tau)} \\
    \approx\bra*{\phi_{|i}(\tau)}\qty(e^{-S_{0,R}-\beta (H_{R^c}+H_{R_{\bar{i}}^*})} \int_{-\infty}^\infty\dd{t} e^{i(\mathcal{E}+\sum_{j\ne i}H_j)t} S_{\phi_{|i}(\tau+t)|\psi_{SR_{\bar{i}}}}^\dagger S_{\phi_{|i}(\tau)|\psi_{SR_{\bar{i}}}})^n\ket*{\phi_{|i}(\tau)}.
    \label{Equation: rho tilde n 2}
\end{multline}
Moreover, by~\eqref{Equation: semiclassical condition 2}, in the window $\abs{t}<\order{\epsilon}$ we can write $\ket*{\phi_{|i}(\tau+t)}\approx e^{-i\mathcal{E}t}\ket*{\hat\phi(\tau+t)}$, where
\begin{equation}
    \ket*{\hat\phi(\tau+t)} = e^{-i\sum_{j\ne i}H_jt}\ket*{\phi_{|i}(\tau)}.
\end{equation}
One therefore has the approximation
\begin{equation}
    \bra*{\phi_{|i}(\tau)}(\tilde\rho_{\phi|i}(\tau))^n\ket*{\phi_{|i}(\tau)} 
    \approx\bra*{\hat\phi(\tau)}(\hat\rho_{\phi}(\tau))^n\ket*{\hat\phi(\tau)}
    \label{Equation: hat tilde n}
\end{equation}
where (using $S^\dagger_{\phi_{|i}(\tau+t)|\psi_{SR_{\bar{i}}}} \approx e^{-i\mathcal{E}t}S^\dagger_{\hat\phi(\tau+t)|\psi_{SR_{\bar{i}}}}$)
\begin{equation}
    \hat\rho_{\phi}(\tau) = e^{-S_{0,R}-\beta (H_{R^c}+H_{R_{\bar{i}}^*})} \int_{-\infty}^\infty\dd{t} e^{i\sum_{j\ne i}H_jt} S_{\hat\phi(\tau+t)|\psi_{SR_{\bar{i}}}}^\dagger S_{\hat\phi(\tau)|\psi_{SR_{\bar{i}}}}.
    \label{Equation: appendix hat rho}
\end{equation}
Finally, taking an $n$ derivative of~\eqref{Equation: hat tilde n}, and then setting $n=0$, one finds
\begin{equation}
    S[\phi] = -\bra*{\phi_{|i}(\tau)}\log\tilde\rho_{\phi|i}(\tau)\ket*{\phi_{|i}(\tau)} 
    \approx-\bra*{\hat\phi(\tau)}\log\hat\rho_{\phi}(\tau)\ket*{\hat\phi(\tau)}.
    \label{Equation: hat tilde entropy}
\end{equation}
In combination with~\eqref{Equation: tilde entropy}, this implies~\eqref{Equation: hat entropy}.

Consider the state $\ket*{\widehat\Psi}$ defined in~\eqref{Equation: semiclassical Psi}, and let  $a(H_{R^c}),b(H_{R^c})$ be any two elements of $\widehat{\mathcal{A}}$ (written as $\mathcal{A}_S\otimes\mathcal{B}(\mathcal{H}_{R_{\bar{i}}})$-valued functions of $H_{R^c}$). Then we have
\begin{align}
    \MoveEqLeft\bra*{\widehat\Psi}a(H_{R^c})e^{-\beta H_S} \hat\rho_\phi(\tau)b(H_{R^c})\ket*{\widehat\Psi} \\
    &= \frac1{2\pi}\int_{\sigma_{R^c}}\dd{E}\bra*{\psi_S}\tr_{R_{\bar{i}}}\qty(a(E) \Delta_\psi b(E))\ket*{\psi_S} \\
    &= \frac1{2\pi}\int_{\sigma_{R^c}}\dd{E}\bra*{\psi_S}\tr_{R_{\bar{i}}}\qty(b(E) a(E))\ket*{\psi_S} \\
    &= \bra*{\widehat\Psi}b(H_{R^c}) a(H_{R^c})\ket*{\widehat\Psi},
\end{align}
so $e^{-\beta H_S} = \Delta_{\widehat\Psi}$ is the modular operator of $\Psi$. Similarly, we have
\begin{align}
    \MoveEqLeft\bra*{\widehat\Psi}a(H_{R^c})e^{S_{0,R}+\beta H_{R^c}} \hat\rho_\phi(\tau)b(H_{R^c})\ket*{\widehat\Psi} \\
    &= \frac1{2\pi}\int_{\sigma_{R^c}}\dd{E}\bra*{\psi_S}\tr_{R_{\bar{i}}}\qty(a(E) e^{-\beta H_{R_{\bar{i}}^*}} \int_{-\infty}^\infty\dd{t} e^{i(H_{R_{\bar{i}}}+E)t} S_{\hat\phi(\tau+t)|\psi_{SR_{\bar{i}}}}^\dagger S_{\hat\phi(\tau)|\psi_{SR_{\bar{i}}}}b(E))\ket*{\psi_S} \\
    &= \frac1{2\pi}\int_{\sigma_{R^c}}\dd{E}\bra*{\psi_{SR_{\bar{i}}}}a(E) \int_{-\infty}^\infty\dd{t} e^{i(H_{R_{\bar{i}}}+E)t} S_{\hat\phi(\tau+t)|\psi_{SR_{\bar{i}}}}^\dagger S_{\hat\phi(\tau)|\psi_{SR_{\bar{i}}}}b(E)\ket*{\psi_{SR_{\bar{i}}}} \\
    &= \frac1{2\pi}\int_{\sigma_{R^c}}\dd{E}\int_{-\infty}^\infty\dd{t}\bra*{\psi_{SR_{\bar{i}}}}a(E)  e^{i(H_{R_{\bar{i}}}+E)t} S_{\hat\phi(\tau+t)|\psi_{SR_{\bar{i}}}}^\dagger S_{\hat\phi(\tau)|\psi_{SR_{\bar{i}}}}b(E)\ket*{\psi_{SR_{\bar{i}}}} \\
    &= \frac1{2\pi}\int_{\sigma_{R^c}}\dd{E}\int_{-\infty}^\infty\dd{t}\bra*{\hat\phi(\tau)} b(E)a(E)e^{i(H_{R_{\bar{i}}}+E)t} \ket*{\hat\phi(\tau+t)}\\
    &= \frac1{2\pi}\int_{\sigma_{R^c}}\dd{E}\int_{-\infty}^\infty\dd{t}\bra*{\hat\phi(\tau)} b(E)a(E)e^{i(E-H_{R^c})t} \ket*{\hat\phi(\tau)}\\
    &= \bra*{\hat\phi(\tau)} b(H_{R^c}) a(H_{R^c}) \ket*{\hat\phi(\tau)}
\end{align}
which shows that~\eqref{Equation: semiclassical relative modular} holds.

\subsection{Linear corrections}
\label{Appendix: linear corrections}

Let us now consider the corrections to~\eqref{Equation: hat tilde entropy}. We start by writing the exact density operator as
\begin{equation}
    \rho_{\phi|i}(\tau) = \Pi_{|i} e^{-S_{0,R}-\beta(H_{R^c}+H_{R_{\bar{i}}^*})} \int_{-\infty}^\infty\dd{t} e^{i(H_S+\sum_{j\ne i}H_i)t}\qty(e^{-i\mathcal{E}t}S^\dagger_{\hat\phi(\tau+t)|\psi_{SR_{\bar{i}}}}S_{\hat\phi(\tau)|\psi_{SR_{\bar{i}}}} + Y(t)),
\end{equation}
where
\begin{equation}
    Y(t) = S^\dagger_{\phi_{|i}(\tau+t)|\psi_{SR_{\bar{i}}}}S_{\phi_{|i}(\tau)|\psi_{SR_{\bar{i}}}} - S^\dagger_{e^{-i\mathcal{E}t}\hat\phi(\tau+t)|\psi_{SR_{\bar{i}}}}S_{\hat\phi(\tau)|\psi_{SR_{\bar{i}}}}.
\end{equation}
The small parameter in which we are expanding is $\eta = \norm*{\epsilon (H_S-\mathcal{E})\ket*{\phi_{|i}(\tau)}}$. By the semiclassical assumption, for $\abs{t}<\order{\epsilon}$, we have $Y(t) = \order{\eta}$, so we can write the exact density operator as a dominant part $\rho^{(0)}_{\phi|i}(\tau)$ plus an $\order{\eta}$ correction $\rho^{(Y)}_{\phi|i}(\tau)$:
\begin{equation}
    \rho_{\phi|i}(\tau) = \rho^{(0)} + \delta^{(Y)}\rho,
\end{equation}
where
\begin{align}
    \rho^{(0)} &= \Pi_{|i} e^{-S_{0,R}-\beta(H_{R^c}+H_{R_{\bar{i}}^*})} \int_{-\infty}^\infty\dd{t} e^{i(H_S-\mathcal{E}+\sum_{j\ne i}H_i)t}S^\dagger_{\hat\phi(\tau+t)|\psi_{SR_{\bar{i}}}}S_{\hat\phi(\tau)|\psi_{SR_{\bar{i}}}},\\
    \delta^{(Y)}\rho &= \Pi_{|i} e^{-S_{0,R}-\beta(H_{R^c}+H_{R_{\bar{i}}^*})} \int_{-\infty}^\infty\dd{t} e^{i(H_S+\sum_{j\ne i}H_i)t} Y(t).
\end{align}
We may further expand $e^{i(H_S-\mathcal{E})t}=1 + i(H_S-\mathcal{E})te^{i(H_S-\mathcal{E})t} + \dots$ in $\rho^{(0)}_{\phi|i}(\tau)$ to obtain
\begin{multline}
    \rho^{(0)} = \Pi_{|i}\hat\rho_\phi(\tau) \\
    + \Pi_{|i} e^{-S_{0,R}-\beta(H_{R^c}+H_{R_{\bar{i}}^*})} \int_{-\infty}^\infty\dd{t} i e^{i(H_S-\mathcal{E}+\sum_{j\ne i}H_i)t}(H_S-\mathcal{E})tS^\dagger_{\hat\phi(\tau+t)|\psi_{SR_{\bar{i}}}}S_{\hat\phi(\tau)|\psi_{SR_{\bar{i}}}} + \dots.
\end{multline}
The second term is $\order{\eta}$ (because the integral is dominated by $\abs{t}<\order{\epsilon}$), and the remaining terms in the ellipsis are $\order{\eta^2}$ (here these approximations hold when one computes expectation values of functions of these operators, in the state $\ket*{\phi_{|i}(\tau)}$). 

Using the integral formula
\begin{equation}
    \dv{s}\log X(s) = \int_0^\infty\dd{\lambda} \frac1{\lambda+X(s)} \dv{X(s)}{s}  \frac1{\lambda+ X(s)},
    \label{Equation: logarithm derivative integral}
\end{equation}
we may then Taylor expand to obtain
\begin{multline}
   \log \rho_{\phi|i}^{(0)}(\tau) = \Pi_{|i}\Big[\log\hat\rho_\phi(\tau) \\
   + \int_{-\infty}^\infty\dd{t}\int_0^\infty\dd{\lambda}\frac1{\lambda+\Delta_{\hat\phi(\tau)|\widehat\Psi}}i(H_S-\mathcal{E})te^{i\sum_{j\ne i}H_it}S^\dagger_{\hat\phi(\tau+t)|\psi_{SR_{\bar{i}}}}S_{\hat\phi(\tau)|\psi_{SR_{\bar{i}}}}\frac1{\lambda+\Delta_{\hat\phi(\tau)|\widehat\Psi}}\Big] + \order{\eta^2}.
\end{multline}
By inserting an integral over the projection-valued measure for the energy of the complementary frames defined in~\eqref{Equation: complementary energy pvm}, and using~\eqref{Equation: energy conditioned relative modular}, with similar techniques to those used in Section~\ref{Subsection: complementary energy} this may be rewritten as
\begin{multline}
   \log \rho_{\phi|i}^{(0)}(\tau) = \Pi_{|i}\Big[\log\hat\rho_\phi(\tau) \\
   + 2\pi\int_{\sigma_{R^c}}\dd{E}P(E)\int_0^\infty\dd{\lambda}\frac1{\lambda+p_E\Delta_{\phi_E|\bar\psi}}(H_S-\mathcal{E}) \partial_E\qty(p_E \Delta_{\phi_E|\bar\psi})\frac1{\lambda+p_E\Delta_{\phi_E|\bar\psi}}\Big] + \order{\eta^2}.
\end{multline}
After a change of variables $\lambda\to p_E \lambda$, this takes the form
\begin{multline}
   \log \rho_{\phi|i}^{(0)}(\tau) = \Pi_{|i}\Big[\log\hat\rho_\phi(\tau) \\
   + 2\pi \int_{\sigma_{R^c}}\dd{E}P(E)\frac1{p_E}\int_0^\infty\dd{\lambda}\frac1{\lambda+\Delta_{\phi_E|\bar\psi}}(H_S-\mathcal{E}) \partial_E\qty(p_E \Delta_{\phi_E|\bar\psi})\frac1{\lambda+\Delta_{\phi_E|\bar\psi}}\Big] + \order{\eta^2}.
\end{multline}
Taking an expectation value in the state $\ket*{\phi_{|i}(\tau)}=\ket*{\hat\phi(\tau)}$ thus yields the entropy
\begin{equation}
    S[\phi] = S_0[\phi] - 2\pi \int_{\sigma_{R^c}} \dd{E} \int_0^\infty\dd{\lambda} \phi_E\qty(\frac1{\lambda+\Delta_{\phi_E|\bar\psi}}(H_S-\mathcal{E}) \partial_E\qty(p_E \Delta_{\phi_E|\bar\psi})\frac1{\lambda+\Delta_{\phi_E|\bar\psi}}) + \order{\eta^2},
\end{equation}
where $S_0[\phi]$ is the leading order semiclassical contribution.

To make progress, we will now impose a UV cutoff on the fields, such that we may decompose $H_S=H_\xi-H'_\xi$, and all algebras have well-defined density operators. Then we can write
\begin{equation}
    \Delta_{\phi_E|\bar\psi} = \rho_E (\rho'_{\bar\psi})^{-1},
\end{equation}
where $\rho_E$ is the density operator for $\phi_E$ in the algebra $\mathcal{A}_S\otimes\mathcal{B}(\mathcal{H}_{R_{\bar{i}}})$, and $\rho'_{\bar\psi}$ is the density operator for $\bar\psi$ in its complement; as in~\ref{Subsection: generalised entropy}, we have $\rho'_{\bar\psi} = e^{-\beta H_\xi'}/Z_\xi$. With this, we can decompose the entropy as
\begin{equation}
    S[\phi] = S_0[\phi] - \delta S +\delta'S + \order{\eta^2},
\end{equation}
where
\begin{align}
  \delta S &= 2\pi \int_{\sigma_{R^c}} \dd{E} \int_0^\infty\dd{\lambda} \phi_E\qty(\frac1{\lambda+\rho_E(\rho'_{\bar\psi})^{-1}}(H_\xi-\mathcal{E}) \partial_E\qty(p_E \rho_E(\rho'_{\bar\psi})^{-1})\frac1{\lambda+\rho_E(\rho'_{\bar\psi})^{-1}}),\\
  \delta' S &= 2\pi \int_{\sigma_{R^c}} \dd{E} \int_0^\infty\dd{\lambda} \phi_E\qty(\frac1{\lambda+\rho_E(\rho'_{\bar\psi})^{-1}}H_\xi' \partial_E\qty(p_E \rho_E(\rho'_{\bar\psi})^{-1})\frac1{\lambda+\rho_E(\rho'_{\bar\psi})^{-1}}).
\end{align}

Recognising the $\dd{E}$ integrand in $\delta S$ as 
\begin{equation}
  \dv{s}\log(\qty(\rho_E+s(H_\xi-\mathcal{E}) \partial_E(p_E\rho_E))(\rho'_{\bar\psi})^{-1}) =  \dv{s}\log(\rho_E+s(H_\xi-\mathcal{E}) \partial_E(p_E\rho_E)), 
\end{equation}
we can write it as
\begin{equation}
  \delta S = 2\pi \int_{\sigma_{R^c}} \dd{E} \int_0^\infty\dd{\lambda} \phi_E\qty(\frac1{\lambda+\rho_E}(H_\xi-\mathcal{E}) \partial_E\qty(p_E \rho_E)\frac1{\lambda+\rho_E}).
\end{equation}
Every operator that appears here is now an element of $\mathcal{S}_S\otimes\mathcal{B}(\mathcal{H}_{R_{\bar{i}}})$, so we can write $\phi_E(\cdot)$ as $\Tr(\rho_E(\cdot))$, where $\Tr$ is the trace on $\mathcal{A}_S\otimes\mathcal{B}(\mathcal{H}_{R_{\bar{i}}})$. Then, using the cyclic property of the trace we have
\begin{align}
  \delta S &= 2\pi \int_{\sigma_{R^c}} \dd{E} \int_0^\infty\dd{\lambda} \Tr\qty(\frac1{\lambda+\rho_E}\rho_E\frac1{\lambda+\rho_E}(H_\xi-\mathcal{E}) \partial_E\qty(p_E \rho_E))\\
           &= 2\pi \int_{\sigma_{R^c}} \dd{E} \Tr\qty((H_\xi-\mathcal{E}) \partial_E\qty(p_E \rho_E))\\
           &= 2\pi \int_{\sigma_{R^c}} \dd{E} \partial_E \qty(p_E (\phi_E\qty(H_\xi)-\mathcal{E})).
\end{align}
In the second line, we just did the $\lambda$ integral.

For $\delta'S$, we note that $H'_\xi$ commutes with $\rho_E(\rho'_{\bar\psi})^{-1}$, so
\begin{align}
  \delta' S &= 2\pi \int_{\sigma_{R^c}} \dd{E} \int_0^\infty\dd{\lambda} \phi_E\qty(H_\xi' \frac1{\lambda+\rho_E(\rho'_{\bar\psi})^{-1}}\partial_E\qty(p_E \rho_E(\rho'_{\bar\psi})^{-1})\frac1{\lambda+\rho_E(\rho'_{\bar\psi})^{-1}})\\
  &= 2\pi \int_{\sigma_{R^c}} \dd{E} \phi_E\qty(H_\xi' \qty(\partial_E p_E + p_E\partial_E\log \rho_E))\\
  &= 2\pi \int_{\sigma_{R^c}} \dd{E} p_E\phi_E\qty(H_\xi'\partial_E \log\qty(p_E\rho_E)).
\end{align}

So overall we have
\begin{equation}
  S[\phi] = S_0[\phi] - 2\pi\int_{\sigma_{R^c}}\dd{E}\qty[\partial_E\qty\Big(p_E \qty\big(\phi_E\qty(H_\xi)-\mathcal{E})) - p_E\phi_E\qty(H_\xi'\partial_E \log\qty(p_E\rho_E))] + \order{\eta^2}.
\end{equation}

\section{Comparing density operators to prior work}
\label{app: comparison to prior work} 

For the sake of comparing to works \cite{Chandrasekaran:2022cip} (CLPW) and \cite{Jensen:2023yxy} (JSS), we consider the case of two clock frames, which we label $C$ and $C'$, both possibly non-ideal.  Suppose that in reduced perspective of clock $C'$ at orientation $\tau_{C'}=0$, the state is unentangled (up to the affect of the projector) between the QFT and clock $C$ factors.  A frame change map $V_{C'\rightarrow C}:=\int \dd t e^{-itH_S}\ket{t}_{C}\bra{-t}_{C'}$ takes this state to the perspective of frame $C$ at orientation $\tau_C=0$, which we write in terms of an integral decomposition between QFT and frame $C'$ factor:
\begin{align}
    \label{eq: C' perspective}
    &\ket*{\hat{\Phi}}_{|C'}:=\sqrt{N} \Pi_{|C'}\ket*{\Phi}\otimes \ket{f}_C\\
    \label{eq: C perspective}
    &\ket*{\hat{\Phi}}_{|C}=V_{C'\rightarrow C}\ket*{\hat{\Phi}}_{|C'}
    =\int \dd t \left(\ket*{\phi_S^\mu(t)}\right)\otimes\left( \ket*{\tilde{\phi}^\mu_{C'}(t)}\right)\\
    \label{eq: S state}
    &\ket*{\phi_S^\mu(t)}:=\ket{\phi_S(t+\mu)},\\
    \label{eq: C' state}
    &\ket*{\tilde{\phi}^\mu_{C'}(t)}:=\sqrt{N}\braket{-\mu}{f}_{C}\ket{\mu+t}_{C'}
\end{align}
In \eqref{eq: C' perspective} we have included a projector, $\Pi_{|C'}$, which is necessary if clock $C'$ is non-ideal, and a normalisation constant, $N:=\left(\bra{\Phi}\otimes\bra{f}_C\Pi_{|C'} \ket{\Phi}\otimes\ket{f}_C\right)^{-1}$.  In terms of the decomposition \eqref{eq: C perspective}, the general density matrix for the algebra $\mathcal{A}^H_{SC}$ in the perspective of clock $C$ is (see equation \eqref{eq: rho without R_ibar})
\begin{equation}\begin{split}\label{eq: density op with time integral}
  \rho_{\hat{\Phi}|C} 
  &=\Pi_{|C}e^{-S_{0,C}-\beta H_{C'}}\int_{-\infty}^\infty\dd{\mu}\dd{\nu}~  
  S_{\phi^\nu_S(0)|\psi_{S}}^\dagger \bigg(\int_{\infty}^{\infty} \dd t e^{i(H_S-H_{C'})t} 
    \braket{\tilde{\phi}^{\nu}_{C'}(t)}{\tilde{\phi}^{\mu}_{C'}(0)}
    \bigg)S_{\phi^\mu_S(0)|\psi_{S}}.
\end{split}\end{equation}
Plugging in states \eqref{eq: C' state} to the central $t$-integral gives
\begin{equation}\begin{split}\label{eq: just time integral}
    \int_{-\infty}^\infty\dd{t}&e^{i(H_S-H_{C'})t}
    \braket{\tilde{\phi}^{\nu}_{C'}(t)}{\tilde{\phi}^{\mu}_{C'}(0)}
    =N\int_{-\infty}^\infty\dd{t}e^{i(H_S-H_{C'})t}
    \braket{f}{-\nu}_{C}\braket{-\mu}{f}_{C}\braket{\nu+t}{\mu}_{C'}\\
    &=\frac{N}{2\pi}
    \left(\int_{\sigma'_{C}}\dd\epsilon'_{C}
    e^{-i\nu (H_S-H_{C'}-\epsilon'_{C})}f^*(\epsilon'_{C})\right)
    \Pi(H_S-H_{C'},-\sigma_{C'})
    \left(\int_{\sigma_{C}}\dd\epsilon_{C} e^{i\mu (H_S-H_{C'}-\epsilon_{C})}f(\epsilon_{C})\right).
\end{split}\end{equation}
Recall that $\Pi(A,\sigma):=\Theta(A-\sigma_{\text{min}})\Theta(\sigma_{\text{max}}-A))$ denotes a projector that limits the operator $A$ to the spectral range $\sigma$.
The density operator thus becomes
\begin{equation}\begin{split}\label{eq: density op C perspective}
  \rho_{\hat{\Phi}|C}
  &=\frac{N}{2\pi} \Pi_{|C} \left(\int_{-\infty}^\infty\dd{\nu}\int_{\sigma'_{C}}\dd\epsilon'_{C}e^{-i\nu (H_S+H_{C'}+\epsilon'_{C})}f(\epsilon'_{C})\right)S_{\phi_S(0)|\psi_{S}}^\dagger \Pi(H_S-H_{C'},-\sigma_{C'})S_{\phi_S(0)|\psi_{S}}\\
  &\qquad\quad\times\left(\int_{-\infty}^\infty\dd{\mu} \int_{\sigma_{C}}\dd\epsilon_{C} e^{i\mu (H_S+H_{C'}+\epsilon_{C})}f^*(\epsilon_{C})\right)\Pi_{|C}e^{-S_{0,C}-\beta H_{C'}}\\
  &=2\pi N \Pi_{|C} f(-H_S-H_{C'})
  S_{\phi_S(0)|\psi_{S}}^\dagger \Pi(H_S-H_{C'},-\sigma_{C'})S_{\phi_S(0)|\psi_{S}} f^*(-H_S-H_{C'})\Pi_{|C}e^{-S_{0,C}-\beta H_{C'}}.\\
\end{split}\end{equation}
In writing the first line we used the convenient fact that the continuous index $\mu$ on the QFT contribution \eqref{eq: S state} can be expressed written as a time translation (because we started with a simple product state): $\ket{\phi_S^\mu(t)}=\ket{\phi_S(t+\mu)}$.  
Equation \eqref{eq: density op C perspective} gives the density operator in the perspective of frame $C$.  Conjugating with the inverse of the above frame change map returns from the perspective of clock $C$ at orientation $\tau_{C}=0$ to that of clock $C'$ at orientation $\tau_{C'}=0$, providing the density operator relevant for comparison to CLPW and JSS: 
\begin{equation}
    \rho_{\Phi|C'}(\tau'=0):=V_{C \rightarrow C'}~\rho_{\Phi|C}(\tau=0)~V_{C' \rightarrow C}.
\end{equation}
We use the fact that 
\begin{equation}
    V_{C\rightarrow C'}\mathcal{F}(H_S+H_{C'}) \Pi_{|C}\Pi_{C'}= \mathcal{F}(-H_C)\Pi_{C}\Pi_{|C'}V_{C\rightarrow C'},
\end{equation}
and
\begin{equation}
    V_{C\rightarrow C'}\mathcal{F}(H_{C'})\Pi_{|C}\Pi_{C'} = \mathcal{F}(-H_S-H_C)\Pi_{C}\Pi_{|C'}V_{C\rightarrow C'},
\end{equation}
where $\mathcal{F}$ is any bounded function of its argument, and similar identities for $V_{C'\rightarrow C}$.  This leads to the following expression for the $C$-perspective density operator:
\begin{multline}\label{eq: density op C' perspective, appendix}
  \rho_{\hat{\Phi}|C'}
  =2\pi N \Pi_{|C'} f(H_{C})V_{C\rightarrow C'}
  S_{\phi_S(0)|\psi_{S}}^\dagger \Pi(H_S-H_{C'},-\sigma_{C'})S_{\phi_S(0)|\psi_{S}}\\
  V_{C'\rightarrow C} f^*(H_C)\Pi_{|C'}e^{-S_{0,C}+\beta (H_S+H_{C})}.
\end{multline}
This looks rather different from the expressions of JSS and CLPW.  We can find an alternative expression by manipulating the central terms as follows:
\begin{align}
  \nonumber\MoveEqLeft V_{C\rightarrow C'} S_{\phi_S|\psi_{S}}^\dagger \Pi(H_S-H_{C'},-\sigma_{C'})S_{\phi_S|\psi_{S}}V_{C'\rightarrow C}\\
  &\label{1st}
  \begin{multlined}[t]
      =\left(\int_{-\infty}^{\infty} \dd t e^{-it H_S}\ket{t}_C\bra{-t}_{C'}\right)
      S_{\phi_S|\psi_{S}}^\dagger \left(\int_{-\infty}^{\infty} \dd t' e^{-it'(H_S-H_{C'})}\bra{0}\ket{t'}_{C'}\right)S_{\phi_S|\psi_{S}}\\
      \left(\int_{-\infty}^{\infty} \dd t'' e^{it'' H_S}\ket{-t''}_{C'}\bra{t''}_{C}\right)
  \end{multlined}\\
  &\label{2nd}=\int_{-\infty}^{\infty} \dd t \dd t' \dd t'' e^{-it H_S}\ket{t}_C
  \Delta^{\frac{1}{2}}_{\phi_S|\psi_{S}}  
  e^{-it'\tilde{H}_S}\bra{0}\ket{t'}_{C'}
  \braket{t''+t'}{t}_{C'}
  \Delta^{\frac{1}{2}}_{\phi_S|\psi_{S}}
  e^{it'' H_S}\bra{t''}_{C}\\
  &\label{3rd}
  \begin{multlined}[t]
      =\int_{-\infty}^{\infty} \dd t \dd t' \dd t'' \dd \alpha~
      e^{-it H_S}\ket{t}_C
      \Delta^{\frac{1}{2}}_{\phi_S|\psi_{S}}  
      e^{-it'\tilde{H}_S}\braket{0}{t'}_{C'}
      \braket{\alpha}{t}_{C'}\\
      \delta(t''+t'-\alpha)
      \Delta^{\frac{1}{2}}_{\phi_S|\psi_{S}}
  e^{it'' H_S}\bra{t''}_{C}
  \end{multlined}\\
  &\label{4th}
  \begin{multlined}[t]
      =\int_{-\infty}^{\infty} \dd t \dd t' \dd t'' \dd \alpha~
      e^{-i(t+\alpha) H_S}\ket{t+\alpha}_C \braket{0}{t}_{C'}
      \Delta^{\frac{1}{2}}_{\phi_S|\psi_{S}}  
      e^{-it'\tilde{H}_S}\braket{0}{t'}_{C'}\\
      \delta(t''+t'-\alpha)
      \Delta^{\frac{1}{2}}_{\phi_S|\psi_{S}}
      e^{it'' H_S}\bra{t''}_{C}
  \end{multlined}\\
  &\label{5th}=\Pi_{|C'}\int_{-\infty}^{\infty} \dd \alpha \dd t' \dd t''~
  e^{-i \alpha H_S}\ket{\alpha}_C 
  \Delta^{\frac{1}{2}}_{\phi_S|\psi_{S}}  
  e^{-it'\tilde{H}_S}\braket{0}{t'}_{C'}
  \braket{\alpha}{t''+t'}^{\text{ext}}_C
  \Delta^{\frac{1}{2}}_{\phi_S|\psi_{S}}
  e^{it'' H_S}\bra{t''}_{C}\\
  &\label{6th}=\Pi_{|C'}\Pi^{(\mathbb{R}\rightarrow\sigma_C)}_Ce^{-iH_S T^{\text{ext}}_C }
  \Delta^{\frac{1}{2}}_{\phi_S|\psi_{S}}  
  \int_{-\infty}^{\infty} \dd t' e^{-it'(\tilde{H}_S+H^{\text{ext}}_C)}\braket{0}{t'}_{C'}
  \Delta^{\frac{1}{2}}_{\phi_S|\psi_{S}}
  e^{iH_S T^{\text{ext}}_C}\Pi^{(\sigma_C\rightarrow\mathbb{R})}_C\\
  &\label{7th}=\Pi_{|C'}\Pi^{(\mathbb{R}\rightarrow\sigma_C)}_{C}e^{-iH_ST^{\text{ext}}_C}
  S^\dagger_{\phi_S|\psi_{S}}  
  \Pi^{\text{ext}}_{|C'}
  S_{\phi_S|\psi_{S}}
  e^{iH_ST^{\text{ext}}_C}\Pi^{(\sigma_C\rightarrow\mathbb{R})}_C.
\end{align}
In \eqref{1st} we have just written out integrals explicitly.
In \eqref{2nd} we replaced the relative Tomitas with relative modular operators using $S_{\phi_S|\psi_S} = J_{\phi|\psi}\Delta^{\frac{1}{2}}_{\phi|\psi}$ and letting the relative conjugation operators act via complex conjugation in the energy basis on the intermediate frame factors.  We also introduced the notation $\tilde{H}_S:=J_{\phi|\psi}H_S J_{\phi|\psi}$. In \eqref{3rd}, we introduced an additional integral over $\alpha$ and a delta function $\delta(t''+t'-\alpha)$.  In \eqref{4th}, we shifted $t\rightarrow t+\alpha$ and then isolated $t$-dependent terms to the left.  In line \eqref{5th}, further isolating the $t$-dependence allows us to replace the $t$-integral with the projector $\Pi_{|C'}$ on the left.  We also write the delta function as if it arises from a bra and ket on an extended/ideal clock $C$ Hilbert space $\braket{\alpha}{t''+t'}_C^{\text{ext}}$.  This makes the following steps more clear.  In line \eqref{6th} we rewrite the $\alpha$ and $t''$ integrals in terms of $e^{\pm H_S T^{\text{ext}}_C}$, with extended clock operators acting on acting an extended/ideal frame $C$ Hilbert space.  This requires the inclusion of the projectors $\Pi^{(\mathbb{R}\rightarrow \sigma_C)}_C:=\int_{\sigma_C}\dd\epsilon\ket{\epsilon}_C\bra{\epsilon}^{\text{ext}}_{C}$ and $\Pi^{(\sigma_C\rightarrow \mathbb{R})}_C:=\int_{\sigma_C}\dd\epsilon\ket{\epsilon}^{\text{ext}}_C\bra{\epsilon}_{C}$ on the outer sides of these operators, mapping back to the non-ideal clock $C$ Hilbert space.  In line \eqref{7th} we again replace the relative modular operators with relative Tomitas, which results in the $t'$ integral giving the projector $\Pi^{\text{ext}}_{|C'}:=\Pi(H_S+H_C^{\text{ext}},-\sigma')$. Finally, employing this expression in \eqref{eq: density op C' perspective, appendix}, we arrive at an expression for the density operator which is more amenable to comparison with the JSS and CLPW expressions:
\begin{equation}\begin{split}\label{eq: final density op C' perspective, appendix}
  \rho_{\hat{\Phi}|C'}
  &=2\pi N  f(H_{C})\Pi_{|C'}\Pi^{(\mathbb{R}\rightarrow\sigma_C)}_{C}e^{-iH_ST^{\text{ext}}_C}
  S^\dagger_{\phi_S|\psi_{S}}  
  \Pi^{\text{ext}}_{|C'}
  S_{\phi_S|\psi_{S}}
  e^{iH_ST^{\text{ext}}_C}\Pi^{(\sigma_C\rightarrow\mathbb{R})}_Cf^*(H_C)e^{-S_{0,C}+\beta (H_S+H_{C})}.\\
\end{split}\end{equation}

\subsection*{Alternative derivation}
\label{app: alternate method}
We now give an alternate derivation of the density operator \eqref{eq: final density op C' perspective, appendix}.  We first recall the method of JSS for deriving density operators in the case of two frames that are both ideal.  In this case, the relevant algebra is
\begin{equation}
    \mathcal{A}^{H^{\text{ext}}}_{SC}=\{e^{-i H_S T^{\text{ext}}_C} a e^{iH_S T^{\text{ext}}_C}, e^{is H^{\text{ext}}_C}\}'' \text{ for } a\in \mathcal{A}_S, s\in \mathbb{R},\\
\end{equation}
where the superscript ``ext" again indicates the extension to ideal clock Hilbert spaces and operators.  For any state $\ket*{\hat{\Phi}} \in \mathcal{H}_{\text{QFT}}\otimes L^2(\mathbb{R})$ that is cyclic and separating for this algebra, the density operator may be found by a two step process.  First, the modular operator for this state and algebra, $\Delta^{\text{ext}}_{\hat{\Phi}}$, is identified by `solving' its defining equation:
\begin{equation}\label{eq: mod op defining equation, appendix}
    \bra*{\hat{\Phi}}\hat{a}\hat{b}\ket*{\hat{\Phi}}=
    \bra*{\hat{\Phi}}\hat{b}\Delta^{\text{ext}}_{\hat{\Phi}}\hat{a}\ket*{\hat{\Phi}}, \qquad \hat{a}, \hat{b} \in \mathcal{A}^{H^{\text{ext}}}_{SC}.
\end{equation}
For von Neumann factors of type I and II, the modular operator can be written as a product of elements affiliated with $\mathcal{A}^{H^{\text{ext}}}_{SC}$ and elements affiliated with $(\mathcal{A}^{H^{\text{ext}}}_{SC})'$.  The former piece is (up to normalisation) the density operator $\rho^{\text{ext}}_{\hat{\Phi}}$ for the algebra:
\begin{equation}\begin{split}
    \Delta^{\text{ext}}_{\hat{\Phi}} &= \rho^{\text{ext}}_{\hat{\Phi}}(\rho'^{\text{ext}}_{\hat{\Phi}})^{-1},
    \quad \rho^{\text{ext}}_{\hat{\Phi}}\in \mathcal{A}^{H^{\text{ext}}}_{SC}, 
    ~\rho'^{\text{ext}}_{\hat{\Phi}}\in (\mathcal{A}^{H^{\text{ext}}}_{SC})',
    \\
    \Tr\left(\rho^{\text{ext}}_{\hat{\Phi}}\hat{a}\right) &= \bra*{\hat{\Phi}}\hat{a}\ket*{\hat{\Phi}},
    \quad \forall \hat{a}\in \mathcal{A}^{H^{\text{ext}}}_{SC}.
\end{split}\end{equation}

If we now consider a non-ideal clock $C$, with $C'$ still ideal, the relevant algebra is then $\Pi_{C}\mathcal{A}^{H^{\text{ext}}}_{SC}\Pi_{C}$, where $\Pi_C$ is the projector onto the finite spectral range of the non-ideal clock Hamiltonian $H_C$.  Since $\Pi_{C} \in \mathcal{A}^{H^{\text{ext}}}_{SC}$, this is a proper subalgebra of $\mathcal{A}^{H^{\text{ext}}}_{SC}$.  This implies that $\Delta^{\text{ext}}_{\hat{\Phi}}$ continues to satisfy equation \eqref{eq: mod op defining equation, appendix} and $\rho^{\text{ext}}_{\hat{\Phi}}$ continues to give the correct expectation values for all $\hat{a}, \hat{b}\in \Pi_{C}\mathcal{A}^{H^{\text{ext}}}_{SC}\Pi_{C}$.  Projecting the latter as $\Pi_{C}\rho^{\text{ext}}_{\hat{\Phi}}\Pi_{C}$ renders it an element of  $\Pi_{C}\mathcal{A}^{H^{\text{ext}}}_{SC}\Pi_{C}$, which then serves as the density operator for this projected algebra.

If we instead take the case that $C'$ is non-ideal while $C$ is ideal, the relevant algebra is then $\Pi_{|C'}\mathcal{A}^{H^{\text{ext}}}_{SC}$.   The projector $\Pi_{|C'}$ restricts $-(H_S+H_{C})$ to the spectral range of $H_{C'}$, since this is the constraint-equivalent of $H_{C'}$ acting on $\mathcal{H}_{|C'}$.  Since $\Pi_{|C'}\in (\mathcal{A}^{H^{\text{ext}}}_{SC})'$, this projected algebra is not a subalgebra of $\mathcal{A}^{H^{\text{ext}}}_{SC}$, and the same modular operator $\Delta^{\text{ext}}_{\hat{\Phi}}$ is not guaranteed to satisfy \eqref{eq: mod op defining equation, appendix} for elements of this modified algebra.  We might instead try to seek a new modular operator $\tilde{\Delta}_{\hat{\Phi}}$ for this projected algebra.  This would entail satisfying
\begin{equation}\label{eq: mod op projected state?}
    \bra*{\hat{\Phi}}\hat{a}\Pi_{|C'}\hat{b}\ket*{\hat{\Phi}}=
    \bra*{\hat{\Phi}}\hat{b}\Pi_{|C'}\tilde{\Delta}_{\hat{\Phi}}\Pi_{|C'}\hat{a}\ket*{\hat{\Phi}}, \qquad \hat{a}, \hat{b} \in \mathcal{A}^{H^{\text{ext}}}_{SC}.
\end{equation}
However the state $\ket*{\hat{\Phi}}$ cannot be separating for a projected algebra of this form.  In the absence of a cyclic-separating state, we cannot hope to find the modular operator $\tilde{\Delta}_{\hat{\Phi}}$.  Instead, we determine to find a relative modular operator $\Delta_{\Pi_{|C'}\hat{\Phi}|\hat{\Phi}}$ for the unprojected algebra.  This would entail satisfying
\begin{equation}\label{eq: relative mod op to projected state, appendix}
    \bra*{\hat{\Phi}}\hat{a}\Pi_{|C'}\hat{b}\ket*{\hat{\Phi}}=
    \bra*{\hat{\Phi}}\hat{b}\Delta_{\Pi_{|C'}\hat{\Phi}|\hat{\Phi}}\hat{a}\ket*{\hat{\Phi}}, \qquad \hat{a}, \hat{b} \in \mathcal{A}^{H^{\text{ext}}}_{SC}
\end{equation}
This relative modular operator will be decomposable as
\begin{equation}\begin{split}
    \Delta_{\Pi_{|C'}\hat{\Phi}|\hat{\Phi}} &= \rho_{\Pi_{|C'}\hat{\Phi}}(\rho'_{\hat{\Phi}})^{-1},
    \quad \rho_{\Pi_{|C'}\hat{\Phi}}\in \mathcal{A}^{H^{\text{ext}}}_{SC}, 
    \quad\rho'_{\hat{\Phi}}\in (\mathcal{A}^{H^{\text{ext}}}_{SC})',
    \\
    \Tr\left(\rho_{\Pi_{|C'}\hat{\Phi}}\hat{a}\right) &= \bra*{\hat{\Phi}}\Pi_{|C'}\hat{a}\ket*{\hat{\Phi}},
    \quad \forall \hat{a}\in \mathcal{A}^{H^{\text{ext}}}_{SC}
    \text{ or } \Pi_{|C'}\mathcal{A}^{H^{\text{ext}}}_{SC}.
\end{split}\end{equation}
The projected operator $\Pi_{|C'}\rho_{\Pi_{|C'}\hat{\Phi}}\in \Pi_{|C'}\mathcal{A}^{H^{\text{ext}}}_{SC}$ will then serve as the density matrix for the projected algebra, continuing to give the right expectation values for all its elements.  To find $\rho_{\Pi_{|C'}\hat{\Phi}}$ we therefore first find $\Delta_{\Pi_{|C'}\hat{\Phi}|\hat{\Phi}}$ satisfying equation \eqref{eq: relative mod op to projected state, appendix}.
Using the notation $a_x:=e^{-ixH_S}ae^{ixH_S}$, we let
\begin{equation}\begin{split}\label{eq: a and b}
    \hat{a}&=a_{T^{\text{ext}}_C}e^{iuH^{\text{ext}}_C},\\
    \hat{b}&=a_{T^{\text{ext}}_C}e^{ivH^{\text{ext}}_C},
\end{split}\end{equation}
with $a, b\in \mathcal{A}_S$ and $u, v\in \mathbb{R}$, so that $\hat{a}$ and $\hat{b}$ are elements of the unprojected/ideal algebra.  For readability we will hereafter drop the superscript `ext' in the remainer of this section, but all $H_C$ and $T_C$ below should be understood as the ideal operators.  Elements of the form $\Pi_{|C'}\hat{a}$ and $\Pi_{|C'}\hat{b}$ additively generate the algebra $\Pi_{|C'}\mathcal{A}_{SC}$, so we only need to seek $\Delta_{\Pi_{|C'}\hat{\Phi}|\hat{\Phi}}$ satisfying equation \eqref{eq: relative mod op to projected state, appendix} for arbitrary such elements.  We first consider the left hand side of that equation.  For the frame, we employ bras and kets in the clock basis unless otherwise specified, and we let $f$ denote the energy-basis wavefunction while $\tilde{f}$ denotes the time-basis wavefunction.
\begin{equation}\begin{split}\label{eq: LHS appendix}
    &\bra*{\hat{\Phi}}\hat{a}\Pi_{|C'}\hat{b}\ket*{\hat{\Phi}}
    =\int \dd s' \dd s \dd t \chi_{C'}(-t) \tilde{f}^*(s')\tilde{f}(s)
    \bra{\Phi}\bra{s'}_Ca_{T_C}e^{iuH_C}e^{-i t(H_S+H_C)}b_{T_C}e^{ivH_C}\ket{\Phi}\ket{s}_C\\
    &=\int \dd s' \dd s \dd t \chi_{C'}(-t) \tilde{f}^*(s')\tilde{f}(s)\bra{s'+u}\ket{s-v+t}_C\bra{\Phi}a_{s'}e^{-i t H_S}b_{s-v}\ket{\Phi}\\
    &=\int \dd s' \dd s \dd t \chi_{C'}(-t) \tilde{f}^*(s')\tilde{f}(s)\braket{s'+u}{s-v+t}_C\bra{\Psi}b_{s-v}S^\dagger_{\Phi|\Psi}e^{i t H_S}S_{\Phi|\Psi}a_{s'}\ket{\Psi}\\
    &=\int \dd s' \dd s \dd t \chi_{C'}(-t) \tilde{f}^*(s')\tilde{f}(s)\bra{\Psi}\bra{-s+v}_C b_{-v}e^{isH_S}S^{\dagger}_{\Phi|\Psi}e^{i t (H_S+H_C)}S_{\Phi|\Psi}e^{-is'H_S}a\ket{-s'-u}_C\ket{\Psi}\\
    &= \bra{\Psi,v} b_{-v}\left(\int \dd s\tilde{f}(s) e^{is(H_S-H_C)}\right)S^\dagger_{\Phi|\Psi}\\
    &\qquad\qquad\left(\int \dd t e^{i t (H_S+H_{C})}\chi_{C'}(t)\right)S_{\Phi|\Psi}\left(\int \dd s' \dd t\tilde{f}^*(s')e^{-is'(H_S-H_C)}\right)a\ket{\Psi,-u}\\
    &=2\pi \bra{\Psi,v} b e^{-iT_CH_S}f(H_C-H_S)S^\dagger_{\Phi|\Psi}\Pi_{|C'}S_{\Phi|\Psi}f^*(H_C-H_S)a\ket{\Psi,-u}.
\end{split}\end{equation}
In going from the second to third line, we used $\bra{\Phi}a_{s'}e^{-i t H_S}b_{s-v}\ket{\Phi}=\bra{\Psi}b_{s-v}S^\dagger_{\Phi|\Psi}e^{i t H_S}S_{\Phi|\Psi}a_{s'}\ket{\Psi}$.  Next we turn to the right hand side of equation \eqref{eq: relative mod op to projected state, appendix}:
\begin{equation}\begin{split}\label{eq: RHS appendix}
    &\bra*{\hat{\Phi}}\hat{b}\Delta_{\Pi_{|C'}\hat{\Phi}|\hat{\Phi}}\hat{a}\ket*{\hat{\Phi}}=\int \dd s' \dd s \tilde{f}^*(s')\tilde{f}(s)\bra*{\Phi}\bra*{s'}_C b_{T_C}e^{ivH_C}\Delta_{\Pi_{|C'}\hat{\Phi}|\hat{\Phi}}
    a_{T_C}e^{iuH_C}\ket*{\Phi}\ket{s}_C\\
    &=\int \dd s' \dd s \tilde{f}^*(s')\tilde{f}(s)\bra*{\Phi}\bra*{s'+v}_C b_{s'}\Delta_{\Pi_{|C'}\hat{\Phi}|\hat{\Phi}}a_{s-u}\ket*{\Phi}\ket{s-u}_C\\
    &=\int \dd s' \dd s \tilde{f}^*(s')\tilde{f}(s)\bra*{\Psi}\bra*{s'+v}_C S_{\Psi}^\dagger S^\dagger_{\Phi|\Psi} b_{s'}\Delta_{\Pi_{|C'}\hat{\Phi}|\hat{\Phi}}a_{s-u}S_{\Phi|\Psi}S_{\Psi}\ket*{\Psi}\ket*{s-u}_C\\
    &=\int \dd s' \dd s \tilde{f}^*(s')\tilde{f}(s)\bra*{\Psi}\bra*{s'+v}_C b_{s'}S_{\Psi}^\dagger S^\dagger_{\Phi|\Psi}\Delta_{\Pi_{|C'}\hat{\Phi}|\hat{\Phi}}S_{\Phi|\Psi}S_{\Psi}a_{s-u}\ket*{\Psi}\ket*{s-u}_C\\
    &=\int \dd s' \dd s \tilde{f}^*(s')\tilde{f}(s)\bra*{\Psi,v}be^{is'(H_C+H_S)}S_{\Psi}^\dagger S^\dagger_{\Phi|\Psi}\Delta_{\Pi_{|C'}\hat{\Phi}|\hat{\Phi}}S_{\Phi|\Psi}S_{\Psi}e^{-is(H_C+H_S)}a_{-u}\ket*{\Psi,-u}\\
    &=2\pi \bra*{\Psi,v} b f^*(H_C+H_S)S_{\Psi}^\dagger S^\dagger_{\Phi|\Psi}\Delta_{\Pi_{|C'}\hat{\Phi}|\hat{\Phi}}S_{\Phi|\Psi}S_{\Psi}f(H_C+H_S)e^{-iT_CH_S}a\ket*{\Psi,-u}\\
\end{split}\end{equation}
We have used that $S_{\Phi|\Psi}S_\Psi$ is an element affiliated with $\mathcal{A}'_S$.  Putting together equations \eqref{eq: LHS appendix} and \eqref{eq: RHS appendix} allows us write $\Delta_{\Pi_{|C'}\hat{\Phi}|\hat{\Phi}}$ as
\begin{equation}\begin{split}\label{eq: combined}
    &\Delta_{\Pi_{|C'}\hat{\Phi}|\hat{\Phi}}=
    S^\dagger_{\Psi|\Phi}S^\dagger_{\Psi}\frac{1}{f^*(H_C+H_S)}
    e^{-iT_CH_S}f(H_C-H_S)S^\dagger_{\Phi|\Psi}\\
    &\qquad\qquad\Pi_{|C'}S_{\Phi|\Psi}f^*(H_C-H_S)e^{iT_CH_S}\frac{1}{f(H_C+H_S)}S_\Psi S_{\Psi|\Phi}\\
    &=
    S^\dagger_{\Psi|\Phi}S^\dagger_{\Psi}\frac{e^{-\frac{\beta}{2}(H_C+H_S)}}{f^*(H_C+H_S)}
    \bigg[e^{\frac{\beta}{2}H_C}f(H_C)e^{-iT_CH_S}\Delta^{-1/2}_{\Psi}\Delta^{1/2}_{\Phi|\Psi}J^\dagger_{\Phi|\Psi}\\
    &\qquad\qquad\Pi_{|C'}J_{\Phi|\Psi}\Delta^{1/2}_{\Phi|\Psi}\Delta^{-1/2}_{\Psi}e^{iT_CH_S}f^*(H_C)e^{\frac{\beta}{2}H_C}\bigg]\frac{e^{-\frac{\beta}{2}(H_C+H_S)}}{f(H_C+H_S)}S_\Psi S_{\Psi|\Phi}\\
\end{split}\end{equation}
The second expression is arranged so that terms inside brackets are affiliated with $\mathcal{A}_{SC}$ while those outside the brackets are affiliated with $\mathcal{A}'_{SC}$.  This is straightforward to see for the terms outside the brackets, by recalling that $S_\Psi S_{\Psi|\Phi}$ is affiliated with $\mathcal{A}'_{SC}$, as are any functions of the combination $H_C+H_S$. To see that the terms inside the brackets are affiliated with $\mathcal{A}_{SC}$, note that although $\Pi_{|C'}$ is itself affiliated with $\mathcal{A}'_{SC}$, it can be written as
\begin{equation}\begin{split}
    \Pi(H_S+H_C,-\sigma_{C'})
    =e^{-iT_CH_S}\Pi(H_C,-\sigma_{C'})e^{iT_CH_S}
    =J_\Psi e^{iT_CH_S}\Pi(H_C,-\sigma_{C'})e^{-iT_CH_S} J_\Psi
\end{split}\end{equation}
The terms in brackets in \eqref{eq: combined} may then be written as
\begin{equation}\begin{split}\label{eq: affiliated with A}
    &e^{\frac{\beta}{2}H_C}f(H_C)e^{-iT_CH_S}\Delta^{-1/2}_{\Psi}\Delta^{1/2}_{\Phi|\Psi}J_{\Psi|\Phi}J_\Psi e^{iT_C H_S}\\
    &~\times\Pi(H_C,-\sigma_{C'}) e^{-iT_C H_S} J_\Psi J_{\Phi|\Psi}\Delta^{1/2}_{\Phi|\Psi}\Delta^{-1/2}_{\Psi}e^{iT_CH_S}f^*(H_C)e^{\frac{\beta}{2}H_C}.
\end{split}\end{equation}
The combinations $J_\Psi J_{\Phi|\Psi}\Delta^{1/2}_{\Phi|\Psi}\Delta^{-1/2}_\Psi$ and $\Delta^{-1/2}_{\Psi}\Delta^{1/2}_{\Phi|\Psi}J_{\Psi|\Phi}J_\Psi$ are both affiliated with $\mathcal{A}$ (see, e.g. \cite{Jensen:2023yxy} appendix C), so conjugating them with $e^{\pm iT_C H_S}$ results in operators affiliated $\mathcal{A}^H_{SC}$.  All the other terms, including the projector in the middle, are now functions of $H_C$ and are therefore affiliated with $\mathcal{A}_{SC}$.  Given that the full term in brackets in \eqref{eq: combined} is affiliated with $\mathcal{A}_{SC}$, the relative modular operator may be completely factorised between $\mathcal{A}_{SC}$ and $\mathcal{A}'_{SC}$:
\begin{equation}\begin{split}
    \label{eq: rel mod op result, appendix}&\Delta_{\Pi_{|C'}\hat{\Phi}|\hat{\Phi}}=S^\dagger_{\Psi|\Phi}S^\dagger_{\Psi}\frac{e^{-\beta(H_C+H_S)}}{|f(H_C+H_S)|^2}S_\Psi S_{\Psi|\Phi}\\
    &\qquad \times e^{\frac{\beta}{2}(H_S+H_C)}f(H_C)e^{-iT_CH_S}S^\dagger_{\Phi|\Psi}\Pi_{|C'}S_{\Phi|\Psi}e^{iT_CH_S}f^*(H_C)e^{\frac{\beta}{2}(H_S+H_C)}.
\end{split}\end{equation}

After some minor simplifications we therefore find
\begin{align}
    \rho'_{\hat{\Phi}}&=
    S_{\Phi|\Psi} J_\Psi |f(H_C+H_S)|^2e^{\beta H_C} J_\Psi S^\dagger_{\Phi|\Psi}
\end{align}
as the density matrix for the commutant algebra, which equivalent to JSS equation 5.21. The second line of \eqref{eq: rel mod op result, appendix} gives the density operator for $\mathcal{A}^{H^{\text{ext}}}_{SC}$ in the state $\Pi_{|C'}\ket*{\hat{\Phi}}$, which upon projection with $\Pi_{|C'}$ acts as the density matrix for the algebra $\Pi_{|C'}\mathcal{A}^{H^{\text{ext}}}_{SC}$:
\begin{align}
    \rho_{\Pi_{|C'}\hat{\Phi}}&=\Pi_{|C'}e^{\frac{\beta}{2}(H_S+H_C)}f(H_C)e^{-iT_CH_S}S^\dagger_{\Phi|\Psi}\Pi_{|C'}S_{\Phi|\Psi}e^{iT_CH_S}f^*(H_C)e^{\frac{\beta}{2}(H_S+H_C)}.
\end{align}
Recalling that all operators in this expression are extended/ideal operators, conjugating this density operator with projectors $\Pi_C$, $\Pi_{|C'}$, to the appropriate reduced Hilbert space $\mathcal{H}_{|C'}$ then gives an expression equivalent to equation \eqref{eq: final density op C' perspective, appendix} from the previous subsection.

\section{Details of the gravitational interferometer example}
\label{app: grav interferometer}

In this appendix we will give a more detailed account of the calculations involved in section \ref{sect:Shapiro} concerning the ``gravitational interferometer'' example.  Before considering that particular state, we will consider density operators associated with a broader (but still very special) class of states, namely those which appear completely or approximately unentangled from the perspective of a specified frame. Despite being very special, these states provide a paradigmatic example of the perspective-dependence of certain properties, particularly those related to entanglement.   We will first consider the density operators associated with such states (for two different gauge invariant algebras) without imposing any semiclassicality conditions.  Then, in order to compute entropies, we will need to further specify a state and make several approximations.  Here we will make assumptions that run broadly parallel to the semiclassical assumptions of section \ref{sec_entropy}, but are slightly different.  Finally, we will finally restrict to the state used in the ``gravitational interferometer'' example of section \ref{sect:Shapiro}, which is one such approximate product state.

\subsection*{Approximate product states: $C_1$ perspective versus $C_2$ perspective}
\label{app:  approximate product states}
Consider the case of three frames, $\{C_1, C_2, C_3\}$.  We stipulate that in the perspective of frame $C_1$, the state exhibits no explicit entanglement across factors:
\begin{equation}\label{eq: unentangled state}
    \ket*{\phi_{|1}(0)}=\sqrt{N}\Pi_{|1} \ket{\phi_S}\ket{f}_2\ket{g}_3.
\end{equation}
Here $N = \left(\bra{\phi_S}\bra{f}_2\bra{g}_3 \Pi_{|1}\ket{\phi_S}\ket{f}_2\ket{g}_3\right)$ is an overall normalisation factor, while we take the states on each factor to be individually normalised: $\braket{\phi_S}{\phi_S} = \braket{f}{f}_2 = \braket{g}{g}_3 =1$. We will also let $f$ and $g$ represent the energy basis wavefunctions on the frame factors, for example $\ket{f}_2=\int_{\sigma_2}\dd\epsilon f(\epsilon)\ket{\epsilon}_2$.  We say that the state \eqref{eq: unentangled state} exhibits no ``explicit" entanglement because if clock $1$ is non-ideal, the projector $\Pi_{|1}$ is nontrivial, and the reduced Hilbert space does not have simple tensor product structure.  This inhibits our ability to discuss entanglement across factors because the state space does not fully factorise (see appendix \ref{app_non-idealclockTPS} for discussion of how a projected Hilbert space can be decomposed as a direct integral or sum of tensor product factorizations).  Nevertheless, considering their structure on the extended (prior to projection) Hilbert space, these states are unentangled ``up to'' the effect of the projector, and becomes completely unentangled in the limit of an ideal frame $C_1$. 

A state which is approximately unentangled in the perspective of one frame will generically be highly entangled from the perspective of another frame.
The frame-change map \eqref{clockchange} allows us to map the state \eqref{eq: unentangled state} to the perspective of another frame (we consider frame $C_2$), meaning from the reduced Hilbert space $\mathcal{H}_{|1}$ to the reduced Hilbert space $\mathcal{H}_{|2}$:
\begin{align}
    \ket*{\phi_{|2}(0)}
    &=V_{1\rightarrow 2}^{0,0}\ket*{\phi_{|1}(0)}\nonumber\\
    &=\sqrt{N} \bra{0}_2 \int_{-\infty}^{\infty} \dd t e^{-it(H_S+H_1+H_2+H_3)}\ket{0}_1 \int_{-\infty}^{\infty}\dd t' e^{-it'(H_S+H_2+H_3)}\braket{0}{t'}_1\ket{\phi_S}\ket{f}_2\ket{g}_3\nonumber\\
    &=\sqrt{N} \bra{0}_2 \int_{-\infty}^{\infty} \dd t \int_{-\infty}^{\infty} \dd t' e^{-it(H_S+H_2+H_3)}\ket{t-t'}_1\braket{-t'}{0}_1\ket{\phi_S}\ket{f}_2\ket{g}_3\nonumber\\
    &=\sqrt{N} \int_{-\infty}^{\infty} \dd t e^{-it(H_S+H_3)}\ket{\phi_S} \ket{t}_1 \braket{-t}{f}_2\ket{g}_3\label{eq: unentangled state, change perspective}\\
    &=\sqrt{2\pi N}\Pi_{|2}f(-H_S-H_1-H_3)\ket{\phi_S}\ket{0}_1 \ket{g}_3\nonumber
\end{align}
The last two lines give alternate expressions that may both be useful. These make apparent that the state is generically entangled between the QFT and remaining frame factors.  The states \eqref{eq: unentangled state} and $\eqref{eq: unentangled state, change perspective}$ descend from the same physical state under different gauge fixings. The statement that these represent ``unentangled" and ``entangled" states, respectively, refers to the tensor product structure of the extended (unprojected) reduced Hilbert spaces. 

\subsection*{Density operators: exact expressions}
We will now consider the reduced density operators associated with two different gauge invariant algebras, those of the QFT subregion operators dressed with respect to frame $C_1$ or frame $C_2$, along with reorientations of the same frame.  We have elsewhere denoted these algebras $\mathcal{A}_{SC_1}^H$ and $\mathcal{A}_{SC_2}^H$.  
In section \ref{Section: arbitrary number of observers} (and appendix \ref{Appendix: density operator details}), we derived a general density matrix expression for the algebra $\mathcal{A}^H_{SR}$ (or rather its representation on the physical Hilbert space, $r(\mathcal{A}^H_{SR})$) in the perspective of a frame $R_i$, where $R=R_i\cup R_{\bar{i}}$ is any subset of the frames and $R_i$ is any one of its members.  In the case that there is only one frame in $R$ (so that $R_{\bar{i}}$ is empty) the density operator is \eqref{eq: rho without R_ibar}, expressed in terms of the state decomposition $\eqref{eq: continuous state decomp, appendix}$.  These expressions cover the cases of interest here (algebras $r(\mathcal{A}^H_{SC_1})$ and $r(\mathcal{A}^H_{SC_2})$), under different interpretations of $R$ and $R^c$. 

We first focus on $r(\mathcal{A}^H_{SC_1})$.  This means taking $R=\{C_1\}$ and $R^c=\{C_2,C_3\}$. To proceed we must write the state from the perspective of $C_1$, equation \eqref{eq: unentangled state}, in terms of the decomposition \eqref{eq: continuous state decomp, appendix}.  Explicitly writing out the projector $\Pi_{|1}$ on this state leads to:
 \begin{align}
    \label{eq: full state C1}
    &\ket*{\phi_{|1}(0)}=\int_{-\infty}^{\infty} \dd \mu \left(e^{-i\mu H_S}\ket{\phi_S}\right) \otimes \left(\sqrt{N}\braket{0}{\mu}_1 e^{-i\mu(H_2+H_3)}\ket{f}_2\ket{g}_3\right)\\
    &\qquad\implies\nonumber\\
    \label{eq: QFT state C1}
    &\ket{\phi^{\mu}_S(t)}
    :=e^{-i(\mu+t) H_S}\ket{\phi_S}
    =\ket{\phi(\mu+t)}_S,\\
    \label{eq: frame state C1}
    &\ket*{\tilde{\phi}^{\mu}_{C_2C_3}(t)}
    :=\sqrt{N}\braket{0}{\mu}_1 e^{-i(\mu+t)(H_2+H_3)}\ket{f}_2\ket{g}_3.
 \end{align}
It will be useful to work out the central time integral that appears in the density operator expression \eqref{eq: rho without R_ibar} using the state \eqref{eq: frame state C1}.  With the abbreviation $\bar{H}'_1:=H_S-H_2-H_3$, this gives
\begin{equation}\begin{split}\label{eq: time integral rho1}
    \int_{-\infty}^{\infty}\dd t
    e^{i\bar{H}'_1t}&\bra*{\tilde{\phi}_{C_2C_3}^{\nu}(t)}\ket*{\tilde{\phi}_{C_2C_3}^{\mu}(0)}
    =N\int_{-\infty}^{\infty} \dd t e^{i\bar{H}'_1 t} \braket{\nu}{0}_1\braket{0}{\mu}_1
    \bra{f}_2\bra{g}_3 e^{i(\nu+t-\mu)(H_2+H_3)} \ket{f}_2 \ket{g}_3\\
    &=\frac{N}{2\pi}\int \dd \epsilon_1 \dd \epsilon'_1 \dd \epsilon_2 \Pi(\bar{H}'_1+\epsilon_2,-\sigma_3) |f(\epsilon_2)|^2|g(-\bar{H}'_1-\epsilon_2)|^2
    e^{i\nu(\epsilon_1-\bar{H}'_1)}
    e^{-i\mu(\epsilon'_1-\bar{H}'_1)}.\\
\end{split}\end{equation}
In the last line, $\Pi(\mathcal{O},\sigma)$ denotes the projector that restricts the operator $\mathcal{O}$ to the spectral range $\sigma$. Each energy integral in this expression ranges over its full spectra, implicitly specified by the subscripts on the integration parameters $\epsilon_i$.  We now insert this expression into the full density operator, leading to
\begin{equation}\begin{split}\label{eq: final density op SC1}
  \rho_{\phi|1}&=
  \frac{N}{2\pi}e^{-S_{0,1}-\beta (H_2+H_3)}\Pi_{|1}\int_{-\infty}^\infty\dd{\mu}\dd{\nu}~ 
  S_{\phi_S(\nu)|\psi_{S}}^\dagger \\
  &\quad\times\bigg(\int \dd \epsilon_1 \dd \epsilon'_1 \dd \epsilon_2 \Pi(\bar{H}'_1+\epsilon_2,-\sigma_3) |f(\epsilon_2)|^2|g(-\bar{H}'_1-\epsilon_2)|^2
    e^{i\nu(\epsilon_1-\bar{H}'_1)}
    e^{-i\mu(\epsilon'_1-\bar{H}'_1)}
  \bigg)S_{\phi_S(\mu)|\psi_{S}}\\
  &=\frac{N}{2\pi}e^{-S_{0,1}-\beta (H_2+H_3)}\Pi_{|1}
  \left(\int_{-\infty}^{\infty}\dd\nu \int_{\sigma_1}\dd \epsilon_1e^{-i\nu(\epsilon_1+H_S+H_2+H_3)}\right)S_{\phi_S(0)|\psi_{S}}^\dagger\\
  &\quad\times\int_{\sigma_2}\dd\epsilon_2\Pi(\bar{H}'_1+\epsilon_2,-\sigma_3) |f(\epsilon_2)|^2|g(-\bar{H}'_1-\epsilon_2)|^2
  S_{\phi_S(0)|\psi_{S}}
  \left(\int_{-\infty}^{\infty}\dd\mu \int_{\sigma_1}\dd \epsilon'_1e^{i\mu(\epsilon'_1+H_S+H_2+H_3)}\right)\\
  &=2\pi Ne^{-S_{0,1}-\beta (H_2+H_3)}
  \Pi_{|1} S_{\phi_S|\psi_{S}}^\dagger
  \int_{\sigma_2}\dd\epsilon_2\Pi(\bar{H}'_1+\epsilon_2,-\sigma_3) |f(\epsilon_2)|^2|g(-\bar{H}'_1-\epsilon_2)|^2
  S_{\phi_S|\psi_{S}}
  \Pi_{|1}.\\
\end{split}\end{equation}
Moving to the second expression we've brought the $\mu$ and $\nu$-dependent terms outside of the relative Tomitas, using the fact that $e^{-iH_S \nu}S^\dagger_{\phi_S(0)}e^{iH_S\nu} = S^\dagger_{\phi_S(\nu)}$.  The integrals in round brackets then reduce to projectors $\Pi_{|1}:=\Pi(H_S+H_2+H_3,-\sigma_1)$, simplifying the final expression.

We now repeat the above steps for the $r(\mathcal{A}_{SC_2}^H)$ algebra.   This means we now take $R=\{C_2\}$ and $R^c=\{C_1,C_3\}$. We must express the state from clock $C_2$ perspective, equation $\eqref{eq: unentangled state, change perspective}$, in terms of a decomposition across frame and QFT factors:
\begin{align}
    \label{eq: full state C2}
    &\ket*{\phi_{|2}(0)}
    =\int_{-\infty}^{\infty} \dd \mu \left(e^{-i \mu H_S}\ket{\phi_S}\right) \otimes 
    \left(\sqrt{N} \ket{\mu}_1 \braket{-\mu}{f}_2 e^{-i\mu H_3}\ket{g}_3\right)\\
    &\qquad\implies\nonumber\\
    \label{eq: QFT state C2}
    &\ket{\phi^{\mu}_S(t)}
    :=e^{-i (\mu+t) H_S}\ket{\phi_S}
    =\ket{\phi(\mu+t)}_S\\
    \label{eq: frame state C2}
    &\ket*{\tilde{\phi}_{C_1C_3}^{\mu}(t)}
    :=\sqrt{N}\ket{\mu+t}_1 \braket{-\mu}{f}_2 e^{-i(\mu+t) H_3}\ket{g}_3.
 \end{align}
 Employing \eqref{eq: frame state C2} in the central time integral, with the abbreviation $\bar{H}'_2:=H_S-H_1-H_3$, we have
\begin{equation}\begin{split}\label{eq: time integral rho C2}
    \int_{-\infty}^{\infty} \dd t e^{i\bar{H}'_2t}&\bra*{\tilde{\phi}_{C_1C_3}^{\nu}(t)}\ket*{\tilde{\phi}_{C_1C_3}^{\mu}(0)}
    =N \int_{-\infty}^{\infty} \dd t e^{i\bar{H}'_2 t} \braket{f}{-\nu}_2\braket{-\mu}{f}_2 \braket{\nu+t}{\mu}_1
    \bra{g}_3 e^{i(\nu+t-\mu)H_3}\ket{g}_3\\
    &=\frac{N}{2\pi}
    \int \dd \epsilon_1 \dd \epsilon_2 \dd \epsilon'_2
    \Pi(\bar{H}'_2+\epsilon_1,-\sigma_3)
    f^*(\epsilon_2)f(\epsilon'_2) |g(-\bar{H}'_2-\epsilon_1)|^2
    e^{i\nu (\epsilon_2-\bar{H}'_2)}e^{-i\mu (\epsilon'_2-\bar{H}'_2)}.\\
\end{split}\end{equation}
Plugging this into the density operator for the case of $r(\mathcal{A}_{SC_2}^H)$ gives
\begin{equation}\begin{split}\label{eq: final density op SC2}
  \rho_{\phi|2}
  &=\frac{N}{2\pi}e^{-S_{0,2}-\beta (H_1+H_3))}\Pi_{|2}
  \left(\int_{-\infty}^{\infty}\dd\nu \int_{\sigma_2}\dd \epsilon_2 f(\epsilon_2)e^{-i\nu(\epsilon_2+H_S+H_1+H_3)}\right)S_{\phi_S(0)|\psi_{S}}^\dagger\\
  &\qquad
  \int_{\sigma_1}\dd\epsilon_1\Pi(\bar{H}'_2+\epsilon_1,-\sigma_3)|g(-\bar{H}'_2-\epsilon_1)|^2
  S_{\phi_S(0)|\psi_{S}}
  \left(\int_{-\infty}^{\infty}\dd\mu \int_{\sigma_2}\dd \epsilon'_2f^*(\epsilon'_2)e^{i\mu(\epsilon'_2+H_S+H_1+H_3)}\right)\\
  &=2\pi Ne^{-S_{0,2}-\beta (H_1+H_3)}
  \Pi_{|2}f(\bar{H}_2) S_{\phi_S|\psi_{S}}^\dagger
  \int_{\sigma_2}\dd\epsilon_1\Pi(\bar{H}'_2+\epsilon_1,-\sigma_3) |g(-\bar{H}'_2-\epsilon_1)|^2 S_{\phi_S|\psi_{S}}f^*(\bar{H}_2) \Pi_{|2}.\\
\end{split}\end{equation}
In the last line, we've also defined $\bar{H}_2:=-H_S-H_1-H_3$ as the constraint-equivalent of $H_2$ on $\mathcal{H}_{|2}$.

Collecting results from \eqref{eq: final density op SC1} and \eqref{eq: final density op SC2} we have:

\begin{equation}\begin{split}\label{eq: rho1 and rho2}
    \rho_{\phi|1}
    &=2\pi Ne^{-S_{0,1}-\beta (H_2+H_3))}
    \Pi_{|1} S_{\phi_S|\psi_{S}}^\dagger
    \int_{\sigma_2}\dd\epsilon_2\Pi(\bar{H}'_1+\epsilon_2,-\sigma_3) |f(\epsilon_2)|^2|g(-\bar{H}'_1-\epsilon_2)|^2
  S_{\phi_S|\psi_{S}}
  \Pi_{|1},\\
  \rho_{\phi|2}
  &=2\pi Ne^{-S_{0,2}-\beta (H_1+H_3)}
  \Pi_{|2}f(\bar{H}_2) S_{\phi_S|\psi_{S}}^\dagger
  \int_{\sigma_1}\dd\epsilon_1\Pi(\bar{H}'_2+\epsilon_1,-\sigma_3) |g(-\bar{H}'_2-\epsilon_1)|^2 S_{\phi_S|\psi_{S}}f^*(\bar{H}_2) \Pi_{|2},\\
\end{split}\end{equation}
where $\bar{H}'_1:=H_S-H_2-H_3$, $\bar{H}'_2:=H_S-H_1-H_3$, and $\bar{H_2}:=-H_S-H_1-H_3$.  Recall that these are the density operators for algebras $r(\mathcal{A}^H_{SC_1})$ and $r(\mathcal{A}^H_{SC_2})$, respectively, for the same physical state, which is distinguished by the fact that in the perspective of clock $C_1$ it looks approximately unentangled (equation \eqref{eq: unentangled state}).

\section{Linear expansion of the symmetric BCH-formula} \label{sect:BCH}

The density operators in Subsect.~\ref{sect:Shapiro} in general contain non-commuting operators. Essentially, these are the relative modular Hamiltonian $H_{\phi_S \vert \psi_S}$ and the modular Hamiltonian $H_S$ related to the KMS-state $\psi_S$. To calculate the entropy, we need to take the logarithm of a product of exponentials of such operators, which can be done with the BCH formula. A direct application of this will involve nested commutators of $H_S$ and $H_{\phi_S \vert \psi_S}$, but without further conditions on these operators, one cannot be certain that the BCH formula converges.

However, a useful expansion can be performed when one of the operators can be expanded in a small parameter, allowing us to consider corrections to the leading order/commuting piece. This is true for the second density operator in the gravitational interferometer we consider in Subsect.~\ref{sect:Shapiro}

For this purpose, we make use of the following well-known form of the BCH-formula for $\log(e^X e^Y)$ which one writes as a nested commutator expansion acting on powers of $Y$
\begin{align}
    \log(e^X e^Y) = X + \frac{\text{ad}_X}{1 - e^{- \text{ad}_X}} Y + \order{Y^2}, \quad \text{ad}_X(Y) := \comm*{X}{Y},
    \label{eq:BCHlinear}
\end{align}
in which $\text{ad}_X$ is the adjoint action by $X$. 
The second term here may be functionally expanded as
\begin{equation}
    \frac{x}{1 - e^{- x}} = 1 + \frac{x}{2} + \frac{x^2}{12} - ...,
    \label{eq:AdExpand}
\end{equation}
The leading order piece here gives for commuting operators $\log(e^Xe^Y) = X + Y$ as expected. In the more general case, one can omit terms in~\eqref{eq:BCHlinear} of order $Y^2$, to obtain an expansion of $\log(e^Xe^Y)$ in terms of nested commutators of $X$ with $Y$.

We will also need the following general formula in order to switch orders
\begin{equation}
    \log(e^{X} e^{Y}) = - \log(e^{-Y} e^{-X}),
    \label{eq:BCHswitch}
\end{equation}
which is a special case of $\log(x)=-\log(x^{-1})$.

We now turn our attention towards the structure we are interested in, namely $e^Y e^X e^{Y^\dagger}$. We start by applying (\ref{eq:BCHlinear}) to $\log(e^X e^{Y^\dagger})$ and then switch the order through (\ref{eq:BCHswitch})
to apply it a second time. This leads to
\begin{align}
    \log(e^Y e^X e^{Y^\dagger}) &= - \log(e^{- X - \frac{\text{ad}_{X}}{1 - e^{- \text{ad}_{X}}} Y^\dagger + \mathcal{O}(Y^{\dagger 2}) } e^{-Y} ) \nonumber \\
    &= X + \frac{\text{ad}_{X}}{1 - e^{- \text{ad}_{X}}} Y^\dagger - \frac{\text{ad}_{X}}{1 - e^{ \text{ad}_{X}}} Y + \mathcal{O}(Y^2).
\end{align}
Where in the last line we made the approximation
\begin{equation}
    - X - \frac{\text{ad}_{X}}{1 - e^{- \text{ad}_{X}}} Y^\dagger = -X + \mathcal{O}(Y),
\end{equation}
since this will act again on $Y$ through the adjoint action.
This expression can be rewritten in a slightly more useful form
\begin{equation}
    \log(e^Y e^X e^{Y^\dagger}) \approx X + 
    \frac{\text{ad}_{X}}{2} \coth(\frac{\text{ad}_{X}}{2}) (Y + Y^\dagger) - \frac{1}{2} \text{ad}_{X} (Y - Y^\dagger),
    \label{eq:linearBCH}
\end{equation}
which is the one mentioned in the main body for the gravitational interferometer in Subsect.~\ref{sect:Shapiro}.

\printbibliography

\end{document}